\documentclass[acmsmall, nonacm]{acmart}

\usepackage{graphicx}

\usepackage{amsmath}
\usepackage{algorithm}
\usepackage[noend]{algpseudocode}
\usepackage{booktabs}
\usepackage{caption}
\usepackage{subcaption}
\usepackage{url}
\usepackage{xspace}
\usepackage{microtype}
\usepackage{tikz}
\usepackage{pgfplots}
\usepackage{pgfplotstable}
\usepackage[binary-units=true]{siunitx}
\usepackage{paralist}
\usepackage{listings}
\usepackage{lipsum}
\usepackage[noabbrev]{cleveref}
\usepackage{textcomp}

\usetikzlibrary{shapes,arrows}
\usepgflibrary{patterns}
\definecolor{YlGn-3-1}{RGB}{247,252,185}
\definecolor{YlGn-3-C}{RGB}{247,252,185}
\definecolor{YlGn-3-2}{RGB}{173,221,142}
\definecolor{YlGn-3-F}{RGB}{173,221,142}
\definecolor{YlGn-3-3}{RGB}{49,163,84}
\definecolor{YlGn-3-I}{RGB}{49,163,84}
\definecolor{YlGn-4-1}{RGB}{255,255,204}
\definecolor{YlGn-4-B}{RGB}{255,255,204}
\definecolor{YlGn-4-2}{RGB}{194,230,153}
\definecolor{YlGn-4-E}{RGB}{194,230,153}
\definecolor{YlGn-4-3}{RGB}{120,198,121}
\definecolor{YlGn-4-G}{RGB}{120,198,121}
\definecolor{YlGn-4-4}{RGB}{35,132,67}
\definecolor{YlGn-4-J}{RGB}{35,132,67}
\definecolor{YlGn-5-1}{RGB}{255,255,204}
\definecolor{YlGn-5-B}{RGB}{255,255,204}
\definecolor{YlGn-5-2}{RGB}{194,230,153}
\definecolor{YlGn-5-E}{RGB}{194,230,153}
\definecolor{YlGn-5-3}{RGB}{120,198,121}
\definecolor{YlGn-5-G}{RGB}{120,198,121}
\definecolor{YlGn-5-4}{RGB}{49,163,84}
\definecolor{YlGn-5-I}{RGB}{49,163,84}
\definecolor{YlGn-5-5}{RGB}{0,104,55}
\definecolor{YlGn-5-K}{RGB}{0,104,55}
\definecolor{YlGn-6-1}{RGB}{255,255,204}
\definecolor{YlGn-6-B}{RGB}{255,255,204}
\definecolor{YlGn-6-2}{RGB}{217,240,163}
\definecolor{YlGn-6-D}{RGB}{217,240,163}
\definecolor{YlGn-6-3}{RGB}{173,221,142}
\definecolor{YlGn-6-F}{RGB}{173,221,142}
\definecolor{YlGn-6-4}{RGB}{120,198,121}
\definecolor{YlGn-6-G}{RGB}{120,198,121}
\definecolor{YlGn-6-5}{RGB}{49,163,84}
\definecolor{YlGn-6-I}{RGB}{49,163,84}
\definecolor{YlGn-6-6}{RGB}{0,104,55}
\definecolor{YlGn-6-K}{RGB}{0,104,55}
\definecolor{YlGn-7-1}{RGB}{255,255,204}
\definecolor{YlGn-7-B}{RGB}{255,255,204}
\definecolor{YlGn-7-2}{RGB}{217,240,163}
\definecolor{YlGn-7-D}{RGB}{217,240,163}
\definecolor{YlGn-7-3}{RGB}{173,221,142}
\definecolor{YlGn-7-F}{RGB}{173,221,142}
\definecolor{YlGn-7-4}{RGB}{120,198,121}
\definecolor{YlGn-7-G}{RGB}{120,198,121}
\definecolor{YlGn-7-5}{RGB}{65,171,93}
\definecolor{YlGn-7-H}{RGB}{65,171,93}
\definecolor{YlGn-7-6}{RGB}{35,132,67}
\definecolor{YlGn-7-J}{RGB}{35,132,67}
\definecolor{YlGn-7-7}{RGB}{0,90,50}
\definecolor{YlGn-7-L}{RGB}{0,90,50}
\definecolor{YlGn-8-1}{RGB}{255,255,229}
\definecolor{YlGn-8-A}{RGB}{255,255,229}
\definecolor{YlGn-8-2}{RGB}{247,252,185}
\definecolor{YlGn-8-C}{RGB}{247,252,185}
\definecolor{YlGn-8-3}{RGB}{217,240,163}
\definecolor{YlGn-8-D}{RGB}{217,240,163}
\definecolor{YlGn-8-4}{RGB}{173,221,142}
\definecolor{YlGn-8-F}{RGB}{173,221,142}
\definecolor{YlGn-8-5}{RGB}{120,198,121}
\definecolor{YlGn-8-G}{RGB}{120,198,121}
\definecolor{YlGn-8-6}{RGB}{65,171,93}
\definecolor{YlGn-8-H}{RGB}{65,171,93}
\definecolor{YlGn-8-7}{RGB}{35,132,67}
\definecolor{YlGn-8-J}{RGB}{35,132,67}
\definecolor{YlGn-8-8}{RGB}{0,90,50}
\definecolor{YlGn-8-L}{RGB}{0,90,50}
\definecolor{YlGn-9-1}{RGB}{255,255,229}
\definecolor{YlGn-9-A}{RGB}{255,255,229}
\definecolor{YlGn-9-2}{RGB}{247,252,185}
\definecolor{YlGn-9-C}{RGB}{247,252,185}
\definecolor{YlGn-9-3}{RGB}{217,240,163}
\definecolor{YlGn-9-D}{RGB}{217,240,163}
\definecolor{YlGn-9-4}{RGB}{173,221,142}
\definecolor{YlGn-9-F}{RGB}{173,221,142}
\definecolor{YlGn-9-5}{RGB}{120,198,121}
\definecolor{YlGn-9-G}{RGB}{120,198,121}
\definecolor{YlGn-9-6}{RGB}{65,171,93}
\definecolor{YlGn-9-H}{RGB}{65,171,93}
\definecolor{YlGn-9-7}{RGB}{35,132,67}
\definecolor{YlGn-9-J}{RGB}{35,132,67}
\definecolor{YlGn-9-8}{RGB}{0,104,55}
\definecolor{YlGn-9-K}{RGB}{0,104,55}
\definecolor{YlGn-9-9}{RGB}{0,69,41}
\definecolor{YlGn-9-M}{RGB}{0,69,41}
\definecolor{YlGnBu-3-1}{RGB}{237,248,177}
\definecolor{YlGnBu-3-C}{RGB}{237,248,177}
\definecolor{YlGnBu-3-2}{RGB}{127,205,187}
\definecolor{YlGnBu-3-F}{RGB}{127,205,187}
\definecolor{YlGnBu-3-3}{RGB}{44,127,184}
\definecolor{YlGnBu-3-I}{RGB}{44,127,184}
\definecolor{YlGnBu-4-1}{RGB}{255,255,204}
\definecolor{YlGnBu-4-B}{RGB}{255,255,204}
\definecolor{YlGnBu-4-2}{RGB}{161,218,180}
\definecolor{YlGnBu-4-E}{RGB}{161,218,180}
\definecolor{YlGnBu-4-3}{RGB}{65,182,196}
\definecolor{YlGnBu-4-G}{RGB}{65,182,196}
\definecolor{YlGnBu-4-4}{RGB}{34,94,168}
\definecolor{YlGnBu-4-J}{RGB}{34,94,168}
\definecolor{YlGnBu-5-1}{RGB}{255,255,204}
\definecolor{YlGnBu-5-B}{RGB}{255,255,204}
\definecolor{YlGnBu-5-2}{RGB}{161,218,180}
\definecolor{YlGnBu-5-E}{RGB}{161,218,180}
\definecolor{YlGnBu-5-3}{RGB}{65,182,196}
\definecolor{YlGnBu-5-G}{RGB}{65,182,196}
\definecolor{YlGnBu-5-4}{RGB}{44,127,184}
\definecolor{YlGnBu-5-I}{RGB}{44,127,184}
\definecolor{YlGnBu-5-5}{RGB}{37,52,148}
\definecolor{YlGnBu-5-K}{RGB}{37,52,148}
\definecolor{YlGnBu-6-1}{RGB}{255,255,204}
\definecolor{YlGnBu-6-B}{RGB}{255,255,204}
\definecolor{YlGnBu-6-2}{RGB}{199,233,180}
\definecolor{YlGnBu-6-D}{RGB}{199,233,180}
\definecolor{YlGnBu-6-3}{RGB}{127,205,187}
\definecolor{YlGnBu-6-F}{RGB}{127,205,187}
\definecolor{YlGnBu-6-4}{RGB}{65,182,196}
\definecolor{YlGnBu-6-G}{RGB}{65,182,196}
\definecolor{YlGnBu-6-5}{RGB}{44,127,184}
\definecolor{YlGnBu-6-I}{RGB}{44,127,184}
\definecolor{YlGnBu-6-6}{RGB}{37,52,148}
\definecolor{YlGnBu-6-K}{RGB}{37,52,148}
\definecolor{YlGnBu-7-1}{RGB}{255,255,204}
\definecolor{YlGnBu-7-B}{RGB}{255,255,204}
\definecolor{YlGnBu-7-2}{RGB}{199,233,180}
\definecolor{YlGnBu-7-D}{RGB}{199,233,180}
\definecolor{YlGnBu-7-3}{RGB}{127,205,187}
\definecolor{YlGnBu-7-F}{RGB}{127,205,187}
\definecolor{YlGnBu-7-4}{RGB}{65,182,196}
\definecolor{YlGnBu-7-G}{RGB}{65,182,196}
\definecolor{YlGnBu-7-5}{RGB}{29,145,192}
\definecolor{YlGnBu-7-H}{RGB}{29,145,192}
\definecolor{YlGnBu-7-6}{RGB}{34,94,168}
\definecolor{YlGnBu-7-J}{RGB}{34,94,168}
\definecolor{YlGnBu-7-7}{RGB}{12,44,132}
\definecolor{YlGnBu-7-L}{RGB}{12,44,132}
\definecolor{YlGnBu-8-1}{RGB}{255,255,217}
\definecolor{YlGnBu-8-A}{RGB}{255,255,217}
\definecolor{YlGnBu-8-2}{RGB}{237,248,177}
\definecolor{YlGnBu-8-C}{RGB}{237,248,177}
\definecolor{YlGnBu-8-3}{RGB}{199,233,180}
\definecolor{YlGnBu-8-D}{RGB}{199,233,180}
\definecolor{YlGnBu-8-4}{RGB}{127,205,187}
\definecolor{YlGnBu-8-F}{RGB}{127,205,187}
\definecolor{YlGnBu-8-5}{RGB}{65,182,196}
\definecolor{YlGnBu-8-G}{RGB}{65,182,196}
\definecolor{YlGnBu-8-6}{RGB}{29,145,192}
\definecolor{YlGnBu-8-H}{RGB}{29,145,192}
\definecolor{YlGnBu-8-7}{RGB}{34,94,168}
\definecolor{YlGnBu-8-J}{RGB}{34,94,168}
\definecolor{YlGnBu-8-8}{RGB}{12,44,132}
\definecolor{YlGnBu-8-L}{RGB}{12,44,132}
\definecolor{YlGnBu-9-1}{RGB}{255,255,217}
\definecolor{YlGnBu-9-A}{RGB}{255,255,217}
\definecolor{YlGnBu-9-2}{RGB}{237,248,177}
\definecolor{YlGnBu-9-C}{RGB}{237,248,177}
\definecolor{YlGnBu-9-3}{RGB}{199,233,180}
\definecolor{YlGnBu-9-D}{RGB}{199,233,180}
\definecolor{YlGnBu-9-4}{RGB}{127,205,187}
\definecolor{YlGnBu-9-F}{RGB}{127,205,187}
\definecolor{YlGnBu-9-5}{RGB}{65,182,196}
\definecolor{YlGnBu-9-G}{RGB}{65,182,196}
\definecolor{YlGnBu-9-6}{RGB}{29,145,192}
\definecolor{YlGnBu-9-H}{RGB}{29,145,192}
\definecolor{YlGnBu-9-7}{RGB}{34,94,168}
\definecolor{YlGnBu-9-J}{RGB}{34,94,168}
\definecolor{YlGnBu-9-8}{RGB}{37,52,148}
\definecolor{YlGnBu-9-K}{RGB}{37,52,148}
\definecolor{YlGnBu-9-9}{RGB}{8,29,88}
\definecolor{YlGnBu-9-M}{RGB}{8,29,88}
\definecolor{GnBu-3-1}{RGB}{224,243,219}
\definecolor{GnBu-3-C}{RGB}{224,243,219}
\definecolor{GnBu-3-2}{RGB}{168,221,181}
\definecolor{GnBu-3-F}{RGB}{168,221,181}
\definecolor{GnBu-3-3}{RGB}{67,162,202}
\definecolor{GnBu-3-I}{RGB}{67,162,202}
\definecolor{GnBu-4-1}{RGB}{240,249,232}
\definecolor{GnBu-4-B}{RGB}{240,249,232}
\definecolor{GnBu-4-2}{RGB}{186,228,188}
\definecolor{GnBu-4-E}{RGB}{186,228,188}
\definecolor{GnBu-4-3}{RGB}{123,204,196}
\definecolor{GnBu-4-G}{RGB}{123,204,196}
\definecolor{GnBu-4-4}{RGB}{43,140,190}
\definecolor{GnBu-4-J}{RGB}{43,140,190}
\definecolor{GnBu-5-1}{RGB}{240,249,232}
\definecolor{GnBu-5-B}{RGB}{240,249,232}
\definecolor{GnBu-5-2}{RGB}{186,228,188}
\definecolor{GnBu-5-E}{RGB}{186,228,188}
\definecolor{GnBu-5-3}{RGB}{123,204,196}
\definecolor{GnBu-5-G}{RGB}{123,204,196}
\definecolor{GnBu-5-4}{RGB}{67,162,202}
\definecolor{GnBu-5-I}{RGB}{67,162,202}
\definecolor{GnBu-5-5}{RGB}{8,104,172}
\definecolor{GnBu-5-K}{RGB}{8,104,172}
\definecolor{GnBu-6-1}{RGB}{240,249,232}
\definecolor{GnBu-6-B}{RGB}{240,249,232}
\definecolor{GnBu-6-2}{RGB}{204,235,197}
\definecolor{GnBu-6-D}{RGB}{204,235,197}
\definecolor{GnBu-6-3}{RGB}{168,221,181}
\definecolor{GnBu-6-F}{RGB}{168,221,181}
\definecolor{GnBu-6-4}{RGB}{123,204,196}
\definecolor{GnBu-6-G}{RGB}{123,204,196}
\definecolor{GnBu-6-5}{RGB}{67,162,202}
\definecolor{GnBu-6-I}{RGB}{67,162,202}
\definecolor{GnBu-6-6}{RGB}{8,104,172}
\definecolor{GnBu-6-K}{RGB}{8,104,172}
\definecolor{GnBu-7-1}{RGB}{240,249,232}
\definecolor{GnBu-7-B}{RGB}{240,249,232}
\definecolor{GnBu-7-2}{RGB}{204,235,197}
\definecolor{GnBu-7-D}{RGB}{204,235,197}
\definecolor{GnBu-7-3}{RGB}{168,221,181}
\definecolor{GnBu-7-F}{RGB}{168,221,181}
\definecolor{GnBu-7-4}{RGB}{123,204,196}
\definecolor{GnBu-7-G}{RGB}{123,204,196}
\definecolor{GnBu-7-5}{RGB}{78,179,211}
\definecolor{GnBu-7-H}{RGB}{78,179,211}
\definecolor{GnBu-7-6}{RGB}{43,140,190}
\definecolor{GnBu-7-J}{RGB}{43,140,190}
\definecolor{GnBu-7-7}{RGB}{8,88,158}
\definecolor{GnBu-7-L}{RGB}{8,88,158}
\definecolor{GnBu-8-1}{RGB}{247,252,240}
\definecolor{GnBu-8-A}{RGB}{247,252,240}
\definecolor{GnBu-8-2}{RGB}{224,243,219}
\definecolor{GnBu-8-C}{RGB}{224,243,219}
\definecolor{GnBu-8-3}{RGB}{204,235,197}
\definecolor{GnBu-8-D}{RGB}{204,235,197}
\definecolor{GnBu-8-4}{RGB}{168,221,181}
\definecolor{GnBu-8-F}{RGB}{168,221,181}
\definecolor{GnBu-8-5}{RGB}{123,204,196}
\definecolor{GnBu-8-G}{RGB}{123,204,196}
\definecolor{GnBu-8-6}{RGB}{78,179,211}
\definecolor{GnBu-8-H}{RGB}{78,179,211}
\definecolor{GnBu-8-7}{RGB}{43,140,190}
\definecolor{GnBu-8-J}{RGB}{43,140,190}
\definecolor{GnBu-8-8}{RGB}{8,88,158}
\definecolor{GnBu-8-L}{RGB}{8,88,158}
\definecolor{GnBu-9-1}{RGB}{247,252,240}
\definecolor{GnBu-9-A}{RGB}{247,252,240}
\definecolor{GnBu-9-2}{RGB}{224,243,219}
\definecolor{GnBu-9-C}{RGB}{224,243,219}
\definecolor{GnBu-9-3}{RGB}{204,235,197}
\definecolor{GnBu-9-D}{RGB}{204,235,197}
\definecolor{GnBu-9-4}{RGB}{168,221,181}
\definecolor{GnBu-9-F}{RGB}{168,221,181}
\definecolor{GnBu-9-5}{RGB}{123,204,196}
\definecolor{GnBu-9-G}{RGB}{123,204,196}
\definecolor{GnBu-9-6}{RGB}{78,179,211}
\definecolor{GnBu-9-H}{RGB}{78,179,211}
\definecolor{GnBu-9-7}{RGB}{43,140,190}
\definecolor{GnBu-9-J}{RGB}{43,140,190}
\definecolor{GnBu-9-8}{RGB}{8,104,172}
\definecolor{GnBu-9-K}{RGB}{8,104,172}
\definecolor{GnBu-9-9}{RGB}{8,64,129}
\definecolor{GnBu-9-M}{RGB}{8,64,129}
\definecolor{BuGn-3-1}{RGB}{229,245,249}
\definecolor{BuGn-3-C}{RGB}{229,245,249}
\definecolor{BuGn-3-2}{RGB}{153,216,201}
\definecolor{BuGn-3-F}{RGB}{153,216,201}
\definecolor{BuGn-3-3}{RGB}{44,162,95}
\definecolor{BuGn-3-I}{RGB}{44,162,95}
\definecolor{BuGn-4-1}{RGB}{237,248,251}
\definecolor{BuGn-4-B}{RGB}{237,248,251}
\definecolor{BuGn-4-2}{RGB}{178,226,226}
\definecolor{BuGn-4-E}{RGB}{178,226,226}
\definecolor{BuGn-4-3}{RGB}{102,194,164}
\definecolor{BuGn-4-G}{RGB}{102,194,164}
\definecolor{BuGn-4-4}{RGB}{35,139,69}
\definecolor{BuGn-4-J}{RGB}{35,139,69}
\definecolor{BuGn-5-1}{RGB}{237,248,251}
\definecolor{BuGn-5-B}{RGB}{237,248,251}
\definecolor{BuGn-5-2}{RGB}{178,226,226}
\definecolor{BuGn-5-E}{RGB}{178,226,226}
\definecolor{BuGn-5-3}{RGB}{102,194,164}
\definecolor{BuGn-5-G}{RGB}{102,194,164}
\definecolor{BuGn-5-4}{RGB}{44,162,95}
\definecolor{BuGn-5-I}{RGB}{44,162,95}
\definecolor{BuGn-5-5}{RGB}{0,109,44}
\definecolor{BuGn-5-K}{RGB}{0,109,44}
\definecolor{BuGn-6-1}{RGB}{237,248,251}
\definecolor{BuGn-6-B}{RGB}{237,248,251}
\definecolor{BuGn-6-2}{RGB}{204,236,230}
\definecolor{BuGn-6-D}{RGB}{204,236,230}
\definecolor{BuGn-6-3}{RGB}{153,216,201}
\definecolor{BuGn-6-F}{RGB}{153,216,201}
\definecolor{BuGn-6-4}{RGB}{102,194,164}
\definecolor{BuGn-6-G}{RGB}{102,194,164}
\definecolor{BuGn-6-5}{RGB}{44,162,95}
\definecolor{BuGn-6-I}{RGB}{44,162,95}
\definecolor{BuGn-6-6}{RGB}{0,109,44}
\definecolor{BuGn-6-K}{RGB}{0,109,44}
\definecolor{BuGn-7-1}{RGB}{237,248,251}
\definecolor{BuGn-7-B}{RGB}{237,248,251}
\definecolor{BuGn-7-2}{RGB}{204,236,230}
\definecolor{BuGn-7-D}{RGB}{204,236,230}
\definecolor{BuGn-7-3}{RGB}{153,216,201}
\definecolor{BuGn-7-F}{RGB}{153,216,201}
\definecolor{BuGn-7-4}{RGB}{102,194,164}
\definecolor{BuGn-7-G}{RGB}{102,194,164}
\definecolor{BuGn-7-5}{RGB}{65,174,118}
\definecolor{BuGn-7-H}{RGB}{65,174,118}
\definecolor{BuGn-7-6}{RGB}{35,139,69}
\definecolor{BuGn-7-J}{RGB}{35,139,69}
\definecolor{BuGn-7-7}{RGB}{0,88,36}
\definecolor{BuGn-7-L}{RGB}{0,88,36}
\definecolor{BuGn-8-1}{RGB}{247,252,253}
\definecolor{BuGn-8-A}{RGB}{247,252,253}
\definecolor{BuGn-8-2}{RGB}{229,245,249}
\definecolor{BuGn-8-C}{RGB}{229,245,249}
\definecolor{BuGn-8-3}{RGB}{204,236,230}
\definecolor{BuGn-8-D}{RGB}{204,236,230}
\definecolor{BuGn-8-4}{RGB}{153,216,201}
\definecolor{BuGn-8-F}{RGB}{153,216,201}
\definecolor{BuGn-8-5}{RGB}{102,194,164}
\definecolor{BuGn-8-G}{RGB}{102,194,164}
\definecolor{BuGn-8-6}{RGB}{65,174,118}
\definecolor{BuGn-8-H}{RGB}{65,174,118}
\definecolor{BuGn-8-7}{RGB}{35,139,69}
\definecolor{BuGn-8-J}{RGB}{35,139,69}
\definecolor{BuGn-8-8}{RGB}{0,88,36}
\definecolor{BuGn-8-L}{RGB}{0,88,36}
\definecolor{BuGn-9-1}{RGB}{247,252,253}
\definecolor{BuGn-9-A}{RGB}{247,252,253}
\definecolor{BuGn-9-2}{RGB}{229,245,249}
\definecolor{BuGn-9-C}{RGB}{229,245,249}
\definecolor{BuGn-9-3}{RGB}{204,236,230}
\definecolor{BuGn-9-D}{RGB}{204,236,230}
\definecolor{BuGn-9-4}{RGB}{153,216,201}
\definecolor{BuGn-9-F}{RGB}{153,216,201}
\definecolor{BuGn-9-5}{RGB}{102,194,164}
\definecolor{BuGn-9-G}{RGB}{102,194,164}
\definecolor{BuGn-9-6}{RGB}{65,174,118}
\definecolor{BuGn-9-H}{RGB}{65,174,118}
\definecolor{BuGn-9-7}{RGB}{35,139,69}
\definecolor{BuGn-9-J}{RGB}{35,139,69}
\definecolor{BuGn-9-8}{RGB}{0,109,44}
\definecolor{BuGn-9-K}{RGB}{0,109,44}
\definecolor{BuGn-9-9}{RGB}{0,68,27}
\definecolor{BuGn-9-M}{RGB}{0,68,27}
\definecolor{PuBuGn-3-1}{RGB}{236,226,240}
\definecolor{PuBuGn-3-C}{RGB}{236,226,240}
\definecolor{PuBuGn-3-2}{RGB}{166,189,219}
\definecolor{PuBuGn-3-F}{RGB}{166,189,219}
\definecolor{PuBuGn-3-3}{RGB}{28,144,153}
\definecolor{PuBuGn-3-I}{RGB}{28,144,153}
\definecolor{PuBuGn-4-1}{RGB}{246,239,247}
\definecolor{PuBuGn-4-B}{RGB}{246,239,247}
\definecolor{PuBuGn-4-2}{RGB}{189,201,225}
\definecolor{PuBuGn-4-E}{RGB}{189,201,225}
\definecolor{PuBuGn-4-3}{RGB}{103,169,207}
\definecolor{PuBuGn-4-G}{RGB}{103,169,207}
\definecolor{PuBuGn-4-4}{RGB}{2,129,138}
\definecolor{PuBuGn-4-J}{RGB}{2,129,138}
\definecolor{PuBuGn-5-1}{RGB}{246,239,247}
\definecolor{PuBuGn-5-B}{RGB}{246,239,247}
\definecolor{PuBuGn-5-2}{RGB}{189,201,225}
\definecolor{PuBuGn-5-E}{RGB}{189,201,225}
\definecolor{PuBuGn-5-3}{RGB}{103,169,207}
\definecolor{PuBuGn-5-G}{RGB}{103,169,207}
\definecolor{PuBuGn-5-4}{RGB}{28,144,153}
\definecolor{PuBuGn-5-I}{RGB}{28,144,153}
\definecolor{PuBuGn-5-5}{RGB}{1,108,89}
\definecolor{PuBuGn-5-K}{RGB}{1,108,89}
\definecolor{PuBuGn-6-1}{RGB}{246,239,247}
\definecolor{PuBuGn-6-B}{RGB}{246,239,247}
\definecolor{PuBuGn-6-2}{RGB}{208,209,230}
\definecolor{PuBuGn-6-D}{RGB}{208,209,230}
\definecolor{PuBuGn-6-3}{RGB}{166,189,219}
\definecolor{PuBuGn-6-F}{RGB}{166,189,219}
\definecolor{PuBuGn-6-4}{RGB}{103,169,207}
\definecolor{PuBuGn-6-G}{RGB}{103,169,207}
\definecolor{PuBuGn-6-5}{RGB}{28,144,153}
\definecolor{PuBuGn-6-I}{RGB}{28,144,153}
\definecolor{PuBuGn-6-6}{RGB}{1,108,89}
\definecolor{PuBuGn-6-K}{RGB}{1,108,89}
\definecolor{PuBuGn-7-1}{RGB}{246,239,247}
\definecolor{PuBuGn-7-B}{RGB}{246,239,247}
\definecolor{PuBuGn-7-2}{RGB}{208,209,230}
\definecolor{PuBuGn-7-D}{RGB}{208,209,230}
\definecolor{PuBuGn-7-3}{RGB}{166,189,219}
\definecolor{PuBuGn-7-F}{RGB}{166,189,219}
\definecolor{PuBuGn-7-4}{RGB}{103,169,207}
\definecolor{PuBuGn-7-G}{RGB}{103,169,207}
\definecolor{PuBuGn-7-5}{RGB}{54,144,192}
\definecolor{PuBuGn-7-H}{RGB}{54,144,192}
\definecolor{PuBuGn-7-6}{RGB}{2,129,138}
\definecolor{PuBuGn-7-J}{RGB}{2,129,138}
\definecolor{PuBuGn-7-7}{RGB}{1,100,80}
\definecolor{PuBuGn-7-L}{RGB}{1,100,80}
\definecolor{PuBuGn-8-1}{RGB}{255,247,251}
\definecolor{PuBuGn-8-A}{RGB}{255,247,251}
\definecolor{PuBuGn-8-2}{RGB}{236,226,240}
\definecolor{PuBuGn-8-C}{RGB}{236,226,240}
\definecolor{PuBuGn-8-3}{RGB}{208,209,230}
\definecolor{PuBuGn-8-D}{RGB}{208,209,230}
\definecolor{PuBuGn-8-4}{RGB}{166,189,219}
\definecolor{PuBuGn-8-F}{RGB}{166,189,219}
\definecolor{PuBuGn-8-5}{RGB}{103,169,207}
\definecolor{PuBuGn-8-G}{RGB}{103,169,207}
\definecolor{PuBuGn-8-6}{RGB}{54,144,192}
\definecolor{PuBuGn-8-H}{RGB}{54,144,192}
\definecolor{PuBuGn-8-7}{RGB}{2,129,138}
\definecolor{PuBuGn-8-J}{RGB}{2,129,138}
\definecolor{PuBuGn-8-8}{RGB}{1,100,80}
\definecolor{PuBuGn-8-L}{RGB}{1,100,80}
\definecolor{PuBuGn-9-1}{RGB}{255,247,251}
\definecolor{PuBuGn-9-A}{RGB}{255,247,251}
\definecolor{PuBuGn-9-2}{RGB}{236,226,240}
\definecolor{PuBuGn-9-C}{RGB}{236,226,240}
\definecolor{PuBuGn-9-3}{RGB}{208,209,230}
\definecolor{PuBuGn-9-D}{RGB}{208,209,230}
\definecolor{PuBuGn-9-4}{RGB}{166,189,219}
\definecolor{PuBuGn-9-F}{RGB}{166,189,219}
\definecolor{PuBuGn-9-5}{RGB}{103,169,207}
\definecolor{PuBuGn-9-G}{RGB}{103,169,207}
\definecolor{PuBuGn-9-6}{RGB}{54,144,192}
\definecolor{PuBuGn-9-H}{RGB}{54,144,192}
\definecolor{PuBuGn-9-7}{RGB}{2,129,138}
\definecolor{PuBuGn-9-J}{RGB}{2,129,138}
\definecolor{PuBuGn-9-8}{RGB}{1,108,89}
\definecolor{PuBuGn-9-K}{RGB}{1,108,89}
\definecolor{PuBuGn-9-9}{RGB}{1,70,54}
\definecolor{PuBuGn-9-M}{RGB}{1,70,54}
\definecolor{PuBu-3-1}{RGB}{236,231,242}
\definecolor{PuBu-3-C}{RGB}{236,231,242}
\definecolor{PuBu-3-2}{RGB}{166,189,219}
\definecolor{PuBu-3-F}{RGB}{166,189,219}
\definecolor{PuBu-3-3}{RGB}{43,140,190}
\definecolor{PuBu-3-I}{RGB}{43,140,190}
\definecolor{PuBu-4-1}{RGB}{241,238,246}
\definecolor{PuBu-4-B}{RGB}{241,238,246}
\definecolor{PuBu-4-2}{RGB}{189,201,225}
\definecolor{PuBu-4-E}{RGB}{189,201,225}
\definecolor{PuBu-4-3}{RGB}{116,169,207}
\definecolor{PuBu-4-G}{RGB}{116,169,207}
\definecolor{PuBu-4-4}{RGB}{5,112,176}
\definecolor{PuBu-4-J}{RGB}{5,112,176}
\definecolor{PuBu-5-1}{RGB}{241,238,246}
\definecolor{PuBu-5-B}{RGB}{241,238,246}
\definecolor{PuBu-5-2}{RGB}{189,201,225}
\definecolor{PuBu-5-E}{RGB}{189,201,225}
\definecolor{PuBu-5-3}{RGB}{116,169,207}
\definecolor{PuBu-5-G}{RGB}{116,169,207}
\definecolor{PuBu-5-4}{RGB}{43,140,190}
\definecolor{PuBu-5-I}{RGB}{43,140,190}
\definecolor{PuBu-5-5}{RGB}{4,90,141}
\definecolor{PuBu-5-K}{RGB}{4,90,141}
\definecolor{PuBu-6-1}{RGB}{241,238,246}
\definecolor{PuBu-6-B}{RGB}{241,238,246}
\definecolor{PuBu-6-2}{RGB}{208,209,230}
\definecolor{PuBu-6-D}{RGB}{208,209,230}
\definecolor{PuBu-6-3}{RGB}{166,189,219}
\definecolor{PuBu-6-F}{RGB}{166,189,219}
\definecolor{PuBu-6-4}{RGB}{116,169,207}
\definecolor{PuBu-6-G}{RGB}{116,169,207}
\definecolor{PuBu-6-5}{RGB}{43,140,190}
\definecolor{PuBu-6-I}{RGB}{43,140,190}
\definecolor{PuBu-6-6}{RGB}{4,90,141}
\definecolor{PuBu-6-K}{RGB}{4,90,141}
\definecolor{PuBu-7-1}{RGB}{241,238,246}
\definecolor{PuBu-7-B}{RGB}{241,238,246}
\definecolor{PuBu-7-2}{RGB}{208,209,230}
\definecolor{PuBu-7-D}{RGB}{208,209,230}
\definecolor{PuBu-7-3}{RGB}{166,189,219}
\definecolor{PuBu-7-F}{RGB}{166,189,219}
\definecolor{PuBu-7-4}{RGB}{116,169,207}
\definecolor{PuBu-7-G}{RGB}{116,169,207}
\definecolor{PuBu-7-5}{RGB}{54,144,192}
\definecolor{PuBu-7-H}{RGB}{54,144,192}
\definecolor{PuBu-7-6}{RGB}{5,112,176}
\definecolor{PuBu-7-J}{RGB}{5,112,176}
\definecolor{PuBu-7-7}{RGB}{3,78,123}
\definecolor{PuBu-7-L}{RGB}{3,78,123}
\definecolor{PuBu-8-1}{RGB}{255,247,251}
\definecolor{PuBu-8-A}{RGB}{255,247,251}
\definecolor{PuBu-8-2}{RGB}{236,231,242}
\definecolor{PuBu-8-C}{RGB}{236,231,242}
\definecolor{PuBu-8-3}{RGB}{208,209,230}
\definecolor{PuBu-8-D}{RGB}{208,209,230}
\definecolor{PuBu-8-4}{RGB}{166,189,219}
\definecolor{PuBu-8-F}{RGB}{166,189,219}
\definecolor{PuBu-8-5}{RGB}{116,169,207}
\definecolor{PuBu-8-G}{RGB}{116,169,207}
\definecolor{PuBu-8-6}{RGB}{54,144,192}
\definecolor{PuBu-8-H}{RGB}{54,144,192}
\definecolor{PuBu-8-7}{RGB}{5,112,176}
\definecolor{PuBu-8-J}{RGB}{5,112,176}
\definecolor{PuBu-8-8}{RGB}{3,78,123}
\definecolor{PuBu-8-L}{RGB}{3,78,123}
\definecolor{PuBu-9-1}{RGB}{255,247,251}
\definecolor{PuBu-9-A}{RGB}{255,247,251}
\definecolor{PuBu-9-2}{RGB}{236,231,242}
\definecolor{PuBu-9-C}{RGB}{236,231,242}
\definecolor{PuBu-9-3}{RGB}{208,209,230}
\definecolor{PuBu-9-D}{RGB}{208,209,230}
\definecolor{PuBu-9-4}{RGB}{166,189,219}
\definecolor{PuBu-9-F}{RGB}{166,189,219}
\definecolor{PuBu-9-5}{RGB}{116,169,207}
\definecolor{PuBu-9-G}{RGB}{116,169,207}
\definecolor{PuBu-9-6}{RGB}{54,144,192}
\definecolor{PuBu-9-H}{RGB}{54,144,192}
\definecolor{PuBu-9-7}{RGB}{5,112,176}
\definecolor{PuBu-9-J}{RGB}{5,112,176}
\definecolor{PuBu-9-8}{RGB}{4,90,141}
\definecolor{PuBu-9-K}{RGB}{4,90,141}
\definecolor{PuBu-9-9}{RGB}{2,56,88}
\definecolor{PuBu-9-M}{RGB}{2,56,88}
\definecolor{BuPu-3-1}{RGB}{224,236,244}
\definecolor{BuPu-3-C}{RGB}{224,236,244}
\definecolor{BuPu-3-2}{RGB}{158,188,218}
\definecolor{BuPu-3-F}{RGB}{158,188,218}
\definecolor{BuPu-3-3}{RGB}{136,86,167}
\definecolor{BuPu-3-I}{RGB}{136,86,167}
\definecolor{BuPu-4-1}{RGB}{237,248,251}
\definecolor{BuPu-4-B}{RGB}{237,248,251}
\definecolor{BuPu-4-2}{RGB}{179,205,227}
\definecolor{BuPu-4-E}{RGB}{179,205,227}
\definecolor{BuPu-4-3}{RGB}{140,150,198}
\definecolor{BuPu-4-G}{RGB}{140,150,198}
\definecolor{BuPu-4-4}{RGB}{136,65,157}
\definecolor{BuPu-4-J}{RGB}{136,65,157}
\definecolor{BuPu-5-1}{RGB}{237,248,251}
\definecolor{BuPu-5-B}{RGB}{237,248,251}
\definecolor{BuPu-5-2}{RGB}{179,205,227}
\definecolor{BuPu-5-E}{RGB}{179,205,227}
\definecolor{BuPu-5-3}{RGB}{140,150,198}
\definecolor{BuPu-5-G}{RGB}{140,150,198}
\definecolor{BuPu-5-4}{RGB}{136,86,167}
\definecolor{BuPu-5-I}{RGB}{136,86,167}
\definecolor{BuPu-5-5}{RGB}{129,15,124}
\definecolor{BuPu-5-K}{RGB}{129,15,124}
\definecolor{BuPu-6-1}{RGB}{237,248,251}
\definecolor{BuPu-6-B}{RGB}{237,248,251}
\definecolor{BuPu-6-2}{RGB}{191,211,230}
\definecolor{BuPu-6-D}{RGB}{191,211,230}
\definecolor{BuPu-6-3}{RGB}{158,188,218}
\definecolor{BuPu-6-F}{RGB}{158,188,218}
\definecolor{BuPu-6-4}{RGB}{140,150,198}
\definecolor{BuPu-6-G}{RGB}{140,150,198}
\definecolor{BuPu-6-5}{RGB}{136,86,167}
\definecolor{BuPu-6-I}{RGB}{136,86,167}
\definecolor{BuPu-6-6}{RGB}{129,15,124}
\definecolor{BuPu-6-K}{RGB}{129,15,124}
\definecolor{BuPu-7-1}{RGB}{237,248,251}
\definecolor{BuPu-7-B}{RGB}{237,248,251}
\definecolor{BuPu-7-2}{RGB}{191,211,230}
\definecolor{BuPu-7-D}{RGB}{191,211,230}
\definecolor{BuPu-7-3}{RGB}{158,188,218}
\definecolor{BuPu-7-F}{RGB}{158,188,218}
\definecolor{BuPu-7-4}{RGB}{140,150,198}
\definecolor{BuPu-7-G}{RGB}{140,150,198}
\definecolor{BuPu-7-5}{RGB}{140,107,177}
\definecolor{BuPu-7-H}{RGB}{140,107,177}
\definecolor{BuPu-7-6}{RGB}{136,65,157}
\definecolor{BuPu-7-J}{RGB}{136,65,157}
\definecolor{BuPu-7-7}{RGB}{110,1,107}
\definecolor{BuPu-7-L}{RGB}{110,1,107}
\definecolor{BuPu-8-1}{RGB}{247,252,253}
\definecolor{BuPu-8-A}{RGB}{247,252,253}
\definecolor{BuPu-8-2}{RGB}{224,236,244}
\definecolor{BuPu-8-C}{RGB}{224,236,244}
\definecolor{BuPu-8-3}{RGB}{191,211,230}
\definecolor{BuPu-8-D}{RGB}{191,211,230}
\definecolor{BuPu-8-4}{RGB}{158,188,218}
\definecolor{BuPu-8-F}{RGB}{158,188,218}
\definecolor{BuPu-8-5}{RGB}{140,150,198}
\definecolor{BuPu-8-G}{RGB}{140,150,198}
\definecolor{BuPu-8-6}{RGB}{140,107,177}
\definecolor{BuPu-8-H}{RGB}{140,107,177}
\definecolor{BuPu-8-7}{RGB}{136,65,157}
\definecolor{BuPu-8-J}{RGB}{136,65,157}
\definecolor{BuPu-8-8}{RGB}{110,1,107}
\definecolor{BuPu-8-L}{RGB}{110,1,107}
\definecolor{BuPu-9-1}{RGB}{247,252,253}
\definecolor{BuPu-9-A}{RGB}{247,252,253}
\definecolor{BuPu-9-2}{RGB}{224,236,244}
\definecolor{BuPu-9-C}{RGB}{224,236,244}
\definecolor{BuPu-9-3}{RGB}{191,211,230}
\definecolor{BuPu-9-D}{RGB}{191,211,230}
\definecolor{BuPu-9-4}{RGB}{158,188,218}
\definecolor{BuPu-9-F}{RGB}{158,188,218}
\definecolor{BuPu-9-5}{RGB}{140,150,198}
\definecolor{BuPu-9-G}{RGB}{140,150,198}
\definecolor{BuPu-9-6}{RGB}{140,107,177}
\definecolor{BuPu-9-H}{RGB}{140,107,177}
\definecolor{BuPu-9-7}{RGB}{136,65,157}
\definecolor{BuPu-9-J}{RGB}{136,65,157}
\definecolor{BuPu-9-8}{RGB}{129,15,124}
\definecolor{BuPu-9-K}{RGB}{129,15,124}
\definecolor{BuPu-9-9}{RGB}{77,0,75}
\definecolor{BuPu-9-M}{RGB}{77,0,75}
\definecolor{RdPu-3-1}{RGB}{253,224,221}
\definecolor{RdPu-3-C}{RGB}{253,224,221}
\definecolor{RdPu-3-2}{RGB}{250,159,181}
\definecolor{RdPu-3-F}{RGB}{250,159,181}
\definecolor{RdPu-3-3}{RGB}{197,27,138}
\definecolor{RdPu-3-I}{RGB}{197,27,138}
\definecolor{RdPu-4-1}{RGB}{254,235,226}
\definecolor{RdPu-4-B}{RGB}{254,235,226}
\definecolor{RdPu-4-2}{RGB}{251,180,185}
\definecolor{RdPu-4-E}{RGB}{251,180,185}
\definecolor{RdPu-4-3}{RGB}{247,104,161}
\definecolor{RdPu-4-G}{RGB}{247,104,161}
\definecolor{RdPu-4-4}{RGB}{174,1,126}
\definecolor{RdPu-4-J}{RGB}{174,1,126}
\definecolor{RdPu-5-1}{RGB}{254,235,226}
\definecolor{RdPu-5-B}{RGB}{254,235,226}
\definecolor{RdPu-5-2}{RGB}{251,180,185}
\definecolor{RdPu-5-E}{RGB}{251,180,185}
\definecolor{RdPu-5-3}{RGB}{247,104,161}
\definecolor{RdPu-5-G}{RGB}{247,104,161}
\definecolor{RdPu-5-4}{RGB}{197,27,138}
\definecolor{RdPu-5-I}{RGB}{197,27,138}
\definecolor{RdPu-5-5}{RGB}{122,1,119}
\definecolor{RdPu-5-K}{RGB}{122,1,119}
\definecolor{RdPu-6-1}{RGB}{254,235,226}
\definecolor{RdPu-6-B}{RGB}{254,235,226}
\definecolor{RdPu-6-2}{RGB}{252,197,192}
\definecolor{RdPu-6-D}{RGB}{252,197,192}
\definecolor{RdPu-6-3}{RGB}{250,159,181}
\definecolor{RdPu-6-F}{RGB}{250,159,181}
\definecolor{RdPu-6-4}{RGB}{247,104,161}
\definecolor{RdPu-6-G}{RGB}{247,104,161}
\definecolor{RdPu-6-5}{RGB}{197,27,138}
\definecolor{RdPu-6-I}{RGB}{197,27,138}
\definecolor{RdPu-6-6}{RGB}{122,1,119}
\definecolor{RdPu-6-K}{RGB}{122,1,119}
\definecolor{RdPu-7-1}{RGB}{254,235,226}
\definecolor{RdPu-7-B}{RGB}{254,235,226}
\definecolor{RdPu-7-2}{RGB}{252,197,192}
\definecolor{RdPu-7-D}{RGB}{252,197,192}
\definecolor{RdPu-7-3}{RGB}{250,159,181}
\definecolor{RdPu-7-F}{RGB}{250,159,181}
\definecolor{RdPu-7-4}{RGB}{247,104,161}
\definecolor{RdPu-7-G}{RGB}{247,104,161}
\definecolor{RdPu-7-5}{RGB}{221,52,151}
\definecolor{RdPu-7-H}{RGB}{221,52,151}
\definecolor{RdPu-7-6}{RGB}{174,1,126}
\definecolor{RdPu-7-J}{RGB}{174,1,126}
\definecolor{RdPu-7-7}{RGB}{122,1,119}
\definecolor{RdPu-7-L}{RGB}{122,1,119}
\definecolor{RdPu-8-1}{RGB}{255,247,243}
\definecolor{RdPu-8-A}{RGB}{255,247,243}
\definecolor{RdPu-8-2}{RGB}{253,224,221}
\definecolor{RdPu-8-C}{RGB}{253,224,221}
\definecolor{RdPu-8-3}{RGB}{252,197,192}
\definecolor{RdPu-8-D}{RGB}{252,197,192}
\definecolor{RdPu-8-4}{RGB}{250,159,181}
\definecolor{RdPu-8-F}{RGB}{250,159,181}
\definecolor{RdPu-8-5}{RGB}{247,104,161}
\definecolor{RdPu-8-G}{RGB}{247,104,161}
\definecolor{RdPu-8-6}{RGB}{221,52,151}
\definecolor{RdPu-8-H}{RGB}{221,52,151}
\definecolor{RdPu-8-7}{RGB}{174,1,126}
\definecolor{RdPu-8-J}{RGB}{174,1,126}
\definecolor{RdPu-8-8}{RGB}{122,1,119}
\definecolor{RdPu-8-L}{RGB}{122,1,119}
\definecolor{RdPu-9-1}{RGB}{255,247,243}
\definecolor{RdPu-9-A}{RGB}{255,247,243}
\definecolor{RdPu-9-2}{RGB}{253,224,221}
\definecolor{RdPu-9-C}{RGB}{253,224,221}
\definecolor{RdPu-9-3}{RGB}{252,197,192}
\definecolor{RdPu-9-D}{RGB}{252,197,192}
\definecolor{RdPu-9-4}{RGB}{250,159,181}
\definecolor{RdPu-9-F}{RGB}{250,159,181}
\definecolor{RdPu-9-5}{RGB}{247,104,161}
\definecolor{RdPu-9-G}{RGB}{247,104,161}
\definecolor{RdPu-9-6}{RGB}{221,52,151}
\definecolor{RdPu-9-H}{RGB}{221,52,151}
\definecolor{RdPu-9-7}{RGB}{174,1,126}
\definecolor{RdPu-9-J}{RGB}{174,1,126}
\definecolor{RdPu-9-8}{RGB}{122,1,119}
\definecolor{RdPu-9-K}{RGB}{122,1,119}
\definecolor{RdPu-9-9}{RGB}{73,0,106}
\definecolor{RdPu-9-M}{RGB}{73,0,106}
\definecolor{PuRd-3-1}{RGB}{231,225,239}
\definecolor{PuRd-3-C}{RGB}{231,225,239}
\definecolor{PuRd-3-2}{RGB}{201,148,199}
\definecolor{PuRd-3-F}{RGB}{201,148,199}
\definecolor{PuRd-3-3}{RGB}{221,28,119}
\definecolor{PuRd-3-I}{RGB}{221,28,119}
\definecolor{PuRd-4-1}{RGB}{241,238,246}
\definecolor{PuRd-4-B}{RGB}{241,238,246}
\definecolor{PuRd-4-2}{RGB}{215,181,216}
\definecolor{PuRd-4-E}{RGB}{215,181,216}
\definecolor{PuRd-4-3}{RGB}{223,101,176}
\definecolor{PuRd-4-G}{RGB}{223,101,176}
\definecolor{PuRd-4-4}{RGB}{206,18,86}
\definecolor{PuRd-4-J}{RGB}{206,18,86}
\definecolor{PuRd-5-1}{RGB}{241,238,246}
\definecolor{PuRd-5-B}{RGB}{241,238,246}
\definecolor{PuRd-5-2}{RGB}{215,181,216}
\definecolor{PuRd-5-E}{RGB}{215,181,216}
\definecolor{PuRd-5-3}{RGB}{223,101,176}
\definecolor{PuRd-5-G}{RGB}{223,101,176}
\definecolor{PuRd-5-4}{RGB}{221,28,119}
\definecolor{PuRd-5-I}{RGB}{221,28,119}
\definecolor{PuRd-5-5}{RGB}{152,0,67}
\definecolor{PuRd-5-K}{RGB}{152,0,67}
\definecolor{PuRd-6-1}{RGB}{241,238,246}
\definecolor{PuRd-6-B}{RGB}{241,238,246}
\definecolor{PuRd-6-2}{RGB}{212,185,218}
\definecolor{PuRd-6-D}{RGB}{212,185,218}
\definecolor{PuRd-6-3}{RGB}{201,148,199}
\definecolor{PuRd-6-F}{RGB}{201,148,199}
\definecolor{PuRd-6-4}{RGB}{223,101,176}
\definecolor{PuRd-6-G}{RGB}{223,101,176}
\definecolor{PuRd-6-5}{RGB}{221,28,119}
\definecolor{PuRd-6-I}{RGB}{221,28,119}
\definecolor{PuRd-6-6}{RGB}{152,0,67}
\definecolor{PuRd-6-K}{RGB}{152,0,67}
\definecolor{PuRd-7-1}{RGB}{241,238,246}
\definecolor{PuRd-7-B}{RGB}{241,238,246}
\definecolor{PuRd-7-2}{RGB}{212,185,218}
\definecolor{PuRd-7-D}{RGB}{212,185,218}
\definecolor{PuRd-7-3}{RGB}{201,148,199}
\definecolor{PuRd-7-F}{RGB}{201,148,199}
\definecolor{PuRd-7-4}{RGB}{223,101,176}
\definecolor{PuRd-7-G}{RGB}{223,101,176}
\definecolor{PuRd-7-5}{RGB}{231,41,138}
\definecolor{PuRd-7-H}{RGB}{231,41,138}
\definecolor{PuRd-7-6}{RGB}{206,18,86}
\definecolor{PuRd-7-J}{RGB}{206,18,86}
\definecolor{PuRd-7-7}{RGB}{145,0,63}
\definecolor{PuRd-7-L}{RGB}{145,0,63}
\definecolor{PuRd-8-1}{RGB}{247,244,249}
\definecolor{PuRd-8-A}{RGB}{247,244,249}
\definecolor{PuRd-8-2}{RGB}{231,225,239}
\definecolor{PuRd-8-C}{RGB}{231,225,239}
\definecolor{PuRd-8-3}{RGB}{212,185,218}
\definecolor{PuRd-8-D}{RGB}{212,185,218}
\definecolor{PuRd-8-4}{RGB}{201,148,199}
\definecolor{PuRd-8-F}{RGB}{201,148,199}
\definecolor{PuRd-8-5}{RGB}{223,101,176}
\definecolor{PuRd-8-G}{RGB}{223,101,176}
\definecolor{PuRd-8-6}{RGB}{231,41,138}
\definecolor{PuRd-8-H}{RGB}{231,41,138}
\definecolor{PuRd-8-7}{RGB}{206,18,86}
\definecolor{PuRd-8-J}{RGB}{206,18,86}
\definecolor{PuRd-8-8}{RGB}{145,0,63}
\definecolor{PuRd-8-L}{RGB}{145,0,63}
\definecolor{PuRd-9-1}{RGB}{247,244,249}
\definecolor{PuRd-9-A}{RGB}{247,244,249}
\definecolor{PuRd-9-2}{RGB}{231,225,239}
\definecolor{PuRd-9-C}{RGB}{231,225,239}
\definecolor{PuRd-9-3}{RGB}{212,185,218}
\definecolor{PuRd-9-D}{RGB}{212,185,218}
\definecolor{PuRd-9-4}{RGB}{201,148,199}
\definecolor{PuRd-9-F}{RGB}{201,148,199}
\definecolor{PuRd-9-5}{RGB}{223,101,176}
\definecolor{PuRd-9-G}{RGB}{223,101,176}
\definecolor{PuRd-9-6}{RGB}{231,41,138}
\definecolor{PuRd-9-H}{RGB}{231,41,138}
\definecolor{PuRd-9-7}{RGB}{206,18,86}
\definecolor{PuRd-9-J}{RGB}{206,18,86}
\definecolor{PuRd-9-8}{RGB}{152,0,67}
\definecolor{PuRd-9-K}{RGB}{152,0,67}
\definecolor{PuRd-9-9}{RGB}{103,0,31}
\definecolor{PuRd-9-M}{RGB}{103,0,31}
\definecolor{OrRd-3-1}{RGB}{254,232,200}
\definecolor{OrRd-3-C}{RGB}{254,232,200}
\definecolor{OrRd-3-2}{RGB}{253,187,132}
\definecolor{OrRd-3-F}{RGB}{253,187,132}
\definecolor{OrRd-3-3}{RGB}{227,74,51}
\definecolor{OrRd-3-I}{RGB}{227,74,51}
\definecolor{OrRd-4-1}{RGB}{254,240,217}
\definecolor{OrRd-4-B}{RGB}{254,240,217}
\definecolor{OrRd-4-2}{RGB}{253,204,138}
\definecolor{OrRd-4-E}{RGB}{253,204,138}
\definecolor{OrRd-4-3}{RGB}{252,141,89}
\definecolor{OrRd-4-G}{RGB}{252,141,89}
\definecolor{OrRd-4-4}{RGB}{215,48,31}
\definecolor{OrRd-4-J}{RGB}{215,48,31}
\definecolor{OrRd-5-1}{RGB}{254,240,217}
\definecolor{OrRd-5-B}{RGB}{254,240,217}
\definecolor{OrRd-5-2}{RGB}{253,204,138}
\definecolor{OrRd-5-E}{RGB}{253,204,138}
\definecolor{OrRd-5-3}{RGB}{252,141,89}
\definecolor{OrRd-5-G}{RGB}{252,141,89}
\definecolor{OrRd-5-4}{RGB}{227,74,51}
\definecolor{OrRd-5-I}{RGB}{227,74,51}
\definecolor{OrRd-5-5}{RGB}{179,0,0}
\definecolor{OrRd-5-K}{RGB}{179,0,0}
\definecolor{OrRd-6-1}{RGB}{254,240,217}
\definecolor{OrRd-6-B}{RGB}{254,240,217}
\definecolor{OrRd-6-2}{RGB}{253,212,158}
\definecolor{OrRd-6-D}{RGB}{253,212,158}
\definecolor{OrRd-6-3}{RGB}{253,187,132}
\definecolor{OrRd-6-F}{RGB}{253,187,132}
\definecolor{OrRd-6-4}{RGB}{252,141,89}
\definecolor{OrRd-6-G}{RGB}{252,141,89}
\definecolor{OrRd-6-5}{RGB}{227,74,51}
\definecolor{OrRd-6-I}{RGB}{227,74,51}
\definecolor{OrRd-6-6}{RGB}{179,0,0}
\definecolor{OrRd-6-K}{RGB}{179,0,0}
\definecolor{OrRd-7-1}{RGB}{254,240,217}
\definecolor{OrRd-7-B}{RGB}{254,240,217}
\definecolor{OrRd-7-2}{RGB}{253,212,158}
\definecolor{OrRd-7-D}{RGB}{253,212,158}
\definecolor{OrRd-7-3}{RGB}{253,187,132}
\definecolor{OrRd-7-F}{RGB}{253,187,132}
\definecolor{OrRd-7-4}{RGB}{252,141,89}
\definecolor{OrRd-7-G}{RGB}{252,141,89}
\definecolor{OrRd-7-5}{RGB}{239,101,72}
\definecolor{OrRd-7-H}{RGB}{239,101,72}
\definecolor{OrRd-7-6}{RGB}{215,48,31}
\definecolor{OrRd-7-J}{RGB}{215,48,31}
\definecolor{OrRd-7-7}{RGB}{153,0,0}
\definecolor{OrRd-7-L}{RGB}{153,0,0}
\definecolor{OrRd-8-1}{RGB}{255,247,236}
\definecolor{OrRd-8-A}{RGB}{255,247,236}
\definecolor{OrRd-8-2}{RGB}{254,232,200}
\definecolor{OrRd-8-C}{RGB}{254,232,200}
\definecolor{OrRd-8-3}{RGB}{253,212,158}
\definecolor{OrRd-8-D}{RGB}{253,212,158}
\definecolor{OrRd-8-4}{RGB}{253,187,132}
\definecolor{OrRd-8-F}{RGB}{253,187,132}
\definecolor{OrRd-8-5}{RGB}{252,141,89}
\definecolor{OrRd-8-G}{RGB}{252,141,89}
\definecolor{OrRd-8-6}{RGB}{239,101,72}
\definecolor{OrRd-8-H}{RGB}{239,101,72}
\definecolor{OrRd-8-7}{RGB}{215,48,31}
\definecolor{OrRd-8-J}{RGB}{215,48,31}
\definecolor{OrRd-8-8}{RGB}{153,0,0}
\definecolor{OrRd-8-L}{RGB}{153,0,0}
\definecolor{OrRd-9-1}{RGB}{255,247,236}
\definecolor{OrRd-9-A}{RGB}{255,247,236}
\definecolor{OrRd-9-2}{RGB}{254,232,200}
\definecolor{OrRd-9-C}{RGB}{254,232,200}
\definecolor{OrRd-9-3}{RGB}{253,212,158}
\definecolor{OrRd-9-D}{RGB}{253,212,158}
\definecolor{OrRd-9-4}{RGB}{253,187,132}
\definecolor{OrRd-9-F}{RGB}{253,187,132}
\definecolor{OrRd-9-5}{RGB}{252,141,89}
\definecolor{OrRd-9-G}{RGB}{252,141,89}
\definecolor{OrRd-9-6}{RGB}{239,101,72}
\definecolor{OrRd-9-H}{RGB}{239,101,72}
\definecolor{OrRd-9-7}{RGB}{215,48,31}
\definecolor{OrRd-9-J}{RGB}{215,48,31}
\definecolor{OrRd-9-8}{RGB}{179,0,0}
\definecolor{OrRd-9-K}{RGB}{179,0,0}
\definecolor{OrRd-9-9}{RGB}{127,0,0}
\definecolor{OrRd-9-M}{RGB}{127,0,0}
\definecolor{YlOrRd-3-1}{RGB}{255,237,160}
\definecolor{YlOrRd-3-C}{RGB}{255,237,160}
\definecolor{YlOrRd-3-2}{RGB}{254,178,76}
\definecolor{YlOrRd-3-F}{RGB}{254,178,76}
\definecolor{YlOrRd-3-3}{RGB}{240,59,32}
\definecolor{YlOrRd-3-I}{RGB}{240,59,32}
\definecolor{YlOrRd-4-1}{RGB}{255,255,178}
\definecolor{YlOrRd-4-B}{RGB}{255,255,178}
\definecolor{YlOrRd-4-2}{RGB}{254,204,92}
\definecolor{YlOrRd-4-E}{RGB}{254,204,92}
\definecolor{YlOrRd-4-3}{RGB}{253,141,60}
\definecolor{YlOrRd-4-G}{RGB}{253,141,60}
\definecolor{YlOrRd-4-4}{RGB}{227,26,28}
\definecolor{YlOrRd-4-J}{RGB}{227,26,28}
\definecolor{YlOrRd-5-1}{RGB}{255,255,178}
\definecolor{YlOrRd-5-B}{RGB}{255,255,178}
\definecolor{YlOrRd-5-2}{RGB}{254,204,92}
\definecolor{YlOrRd-5-E}{RGB}{254,204,92}
\definecolor{YlOrRd-5-3}{RGB}{253,141,60}
\definecolor{YlOrRd-5-G}{RGB}{253,141,60}
\definecolor{YlOrRd-5-4}{RGB}{240,59,32}
\definecolor{YlOrRd-5-I}{RGB}{240,59,32}
\definecolor{YlOrRd-5-5}{RGB}{189,0,38}
\definecolor{YlOrRd-5-K}{RGB}{189,0,38}
\definecolor{YlOrRd-6-1}{RGB}{255,255,178}
\definecolor{YlOrRd-6-B}{RGB}{255,255,178}
\definecolor{YlOrRd-6-2}{RGB}{254,217,118}
\definecolor{YlOrRd-6-D}{RGB}{254,217,118}
\definecolor{YlOrRd-6-3}{RGB}{254,178,76}
\definecolor{YlOrRd-6-F}{RGB}{254,178,76}
\definecolor{YlOrRd-6-4}{RGB}{253,141,60}
\definecolor{YlOrRd-6-G}{RGB}{253,141,60}
\definecolor{YlOrRd-6-5}{RGB}{240,59,32}
\definecolor{YlOrRd-6-I}{RGB}{240,59,32}
\definecolor{YlOrRd-6-6}{RGB}{189,0,38}
\definecolor{YlOrRd-6-K}{RGB}{189,0,38}
\definecolor{YlOrRd-7-1}{RGB}{255,255,178}
\definecolor{YlOrRd-7-B}{RGB}{255,255,178}
\definecolor{YlOrRd-7-2}{RGB}{254,217,118}
\definecolor{YlOrRd-7-D}{RGB}{254,217,118}
\definecolor{YlOrRd-7-3}{RGB}{254,178,76}
\definecolor{YlOrRd-7-F}{RGB}{254,178,76}
\definecolor{YlOrRd-7-4}{RGB}{253,141,60}
\definecolor{YlOrRd-7-G}{RGB}{253,141,60}
\definecolor{YlOrRd-7-5}{RGB}{252,78,42}
\definecolor{YlOrRd-7-H}{RGB}{252,78,42}
\definecolor{YlOrRd-7-6}{RGB}{227,26,28}
\definecolor{YlOrRd-7-J}{RGB}{227,26,28}
\definecolor{YlOrRd-7-7}{RGB}{177,0,38}
\definecolor{YlOrRd-7-L}{RGB}{177,0,38}
\definecolor{YlOrRd-8-1}{RGB}{255,255,204}
\definecolor{YlOrRd-8-A}{RGB}{255,255,204}
\definecolor{YlOrRd-8-2}{RGB}{255,237,160}
\definecolor{YlOrRd-8-C}{RGB}{255,237,160}
\definecolor{YlOrRd-8-3}{RGB}{254,217,118}
\definecolor{YlOrRd-8-D}{RGB}{254,217,118}
\definecolor{YlOrRd-8-4}{RGB}{254,178,76}
\definecolor{YlOrRd-8-F}{RGB}{254,178,76}
\definecolor{YlOrRd-8-5}{RGB}{253,141,60}
\definecolor{YlOrRd-8-G}{RGB}{253,141,60}
\definecolor{YlOrRd-8-6}{RGB}{252,78,42}
\definecolor{YlOrRd-8-H}{RGB}{252,78,42}
\definecolor{YlOrRd-8-7}{RGB}{227,26,28}
\definecolor{YlOrRd-8-J}{RGB}{227,26,28}
\definecolor{YlOrRd-8-8}{RGB}{177,0,38}
\definecolor{YlOrRd-8-L}{RGB}{177,0,38}
\definecolor{YlOrRd-9-1}{RGB}{255,255,204}
\definecolor{YlOrRd-9-A}{RGB}{255,255,204}
\definecolor{YlOrRd-9-2}{RGB}{255,237,160}
\definecolor{YlOrRd-9-C}{RGB}{255,237,160}
\definecolor{YlOrRd-9-3}{RGB}{254,217,118}
\definecolor{YlOrRd-9-D}{RGB}{254,217,118}
\definecolor{YlOrRd-9-4}{RGB}{254,178,76}
\definecolor{YlOrRd-9-F}{RGB}{254,178,76}
\definecolor{YlOrRd-9-5}{RGB}{253,141,60}
\definecolor{YlOrRd-9-G}{RGB}{253,141,60}
\definecolor{YlOrRd-9-6}{RGB}{252,78,42}
\definecolor{YlOrRd-9-H}{RGB}{252,78,42}
\definecolor{YlOrRd-9-7}{RGB}{227,26,28}
\definecolor{YlOrRd-9-J}{RGB}{227,26,28}
\definecolor{YlOrRd-9-8}{RGB}{189,0,38}
\definecolor{YlOrRd-9-K}{RGB}{189,0,38}
\definecolor{YlOrRd-9-9}{RGB}{128,0,38}
\definecolor{YlOrRd-9-M}{RGB}{128,0,38}
\definecolor{YlOrBr-3-1}{RGB}{255,247,188}
\definecolor{YlOrBr-3-C}{RGB}{255,247,188}
\definecolor{YlOrBr-3-2}{RGB}{254,196,79}
\definecolor{YlOrBr-3-F}{RGB}{254,196,79}
\definecolor{YlOrBr-3-3}{RGB}{217,95,14}
\definecolor{YlOrBr-3-I}{RGB}{217,95,14}
\definecolor{YlOrBr-4-1}{RGB}{255,255,212}
\definecolor{YlOrBr-4-B}{RGB}{255,255,212}
\definecolor{YlOrBr-4-2}{RGB}{254,217,142}
\definecolor{YlOrBr-4-E}{RGB}{254,217,142}
\definecolor{YlOrBr-4-3}{RGB}{254,153,41}
\definecolor{YlOrBr-4-G}{RGB}{254,153,41}
\definecolor{YlOrBr-4-4}{RGB}{204,76,2}
\definecolor{YlOrBr-4-J}{RGB}{204,76,2}
\definecolor{YlOrBr-5-1}{RGB}{255,255,212}
\definecolor{YlOrBr-5-B}{RGB}{255,255,212}
\definecolor{YlOrBr-5-2}{RGB}{254,217,142}
\definecolor{YlOrBr-5-E}{RGB}{254,217,142}
\definecolor{YlOrBr-5-3}{RGB}{254,153,41}
\definecolor{YlOrBr-5-G}{RGB}{254,153,41}
\definecolor{YlOrBr-5-4}{RGB}{217,95,14}
\definecolor{YlOrBr-5-I}{RGB}{217,95,14}
\definecolor{YlOrBr-5-5}{RGB}{153,52,4}
\definecolor{YlOrBr-5-K}{RGB}{153,52,4}
\definecolor{YlOrBr-6-1}{RGB}{255,255,212}
\definecolor{YlOrBr-6-B}{RGB}{255,255,212}
\definecolor{YlOrBr-6-2}{RGB}{254,227,145}
\definecolor{YlOrBr-6-D}{RGB}{254,227,145}
\definecolor{YlOrBr-6-3}{RGB}{254,196,79}
\definecolor{YlOrBr-6-F}{RGB}{254,196,79}
\definecolor{YlOrBr-6-4}{RGB}{254,153,41}
\definecolor{YlOrBr-6-G}{RGB}{254,153,41}
\definecolor{YlOrBr-6-5}{RGB}{217,95,14}
\definecolor{YlOrBr-6-I}{RGB}{217,95,14}
\definecolor{YlOrBr-6-6}{RGB}{153,52,4}
\definecolor{YlOrBr-6-K}{RGB}{153,52,4}
\definecolor{YlOrBr-7-1}{RGB}{255,255,212}
\definecolor{YlOrBr-7-B}{RGB}{255,255,212}
\definecolor{YlOrBr-7-2}{RGB}{254,227,145}
\definecolor{YlOrBr-7-D}{RGB}{254,227,145}
\definecolor{YlOrBr-7-3}{RGB}{254,196,79}
\definecolor{YlOrBr-7-F}{RGB}{254,196,79}
\definecolor{YlOrBr-7-4}{RGB}{254,153,41}
\definecolor{YlOrBr-7-G}{RGB}{254,153,41}
\definecolor{YlOrBr-7-5}{RGB}{236,112,20}
\definecolor{YlOrBr-7-H}{RGB}{236,112,20}
\definecolor{YlOrBr-7-6}{RGB}{204,76,2}
\definecolor{YlOrBr-7-J}{RGB}{204,76,2}
\definecolor{YlOrBr-7-7}{RGB}{140,45,4}
\definecolor{YlOrBr-7-L}{RGB}{140,45,4}
\definecolor{YlOrBr-8-1}{RGB}{255,255,229}
\definecolor{YlOrBr-8-A}{RGB}{255,255,229}
\definecolor{YlOrBr-8-2}{RGB}{255,247,188}
\definecolor{YlOrBr-8-C}{RGB}{255,247,188}
\definecolor{YlOrBr-8-3}{RGB}{254,227,145}
\definecolor{YlOrBr-8-D}{RGB}{254,227,145}
\definecolor{YlOrBr-8-4}{RGB}{254,196,79}
\definecolor{YlOrBr-8-F}{RGB}{254,196,79}
\definecolor{YlOrBr-8-5}{RGB}{254,153,41}
\definecolor{YlOrBr-8-G}{RGB}{254,153,41}
\definecolor{YlOrBr-8-6}{RGB}{236,112,20}
\definecolor{YlOrBr-8-H}{RGB}{236,112,20}
\definecolor{YlOrBr-8-7}{RGB}{204,76,2}
\definecolor{YlOrBr-8-J}{RGB}{204,76,2}
\definecolor{YlOrBr-8-8}{RGB}{140,45,4}
\definecolor{YlOrBr-8-L}{RGB}{140,45,4}
\definecolor{YlOrBr-9-1}{RGB}{255,255,229}
\definecolor{YlOrBr-9-A}{RGB}{255,255,229}
\definecolor{YlOrBr-9-2}{RGB}{255,247,188}
\definecolor{YlOrBr-9-C}{RGB}{255,247,188}
\definecolor{YlOrBr-9-3}{RGB}{254,227,145}
\definecolor{YlOrBr-9-D}{RGB}{254,227,145}
\definecolor{YlOrBr-9-4}{RGB}{254,196,79}
\definecolor{YlOrBr-9-F}{RGB}{254,196,79}
\definecolor{YlOrBr-9-5}{RGB}{254,153,41}
\definecolor{YlOrBr-9-G}{RGB}{254,153,41}
\definecolor{YlOrBr-9-6}{RGB}{236,112,20}
\definecolor{YlOrBr-9-H}{RGB}{236,112,20}
\definecolor{YlOrBr-9-7}{RGB}{204,76,2}
\definecolor{YlOrBr-9-J}{RGB}{204,76,2}
\definecolor{YlOrBr-9-8}{RGB}{153,52,4}
\definecolor{YlOrBr-9-K}{RGB}{153,52,4}
\definecolor{YlOrBr-9-9}{RGB}{102,37,6}
\definecolor{YlOrBr-9-M}{RGB}{102,37,6}
\definecolor{Purples-3-1}{RGB}{239,237,245}
\definecolor{Purples-3-C}{RGB}{239,237,245}
\definecolor{Purples-3-2}{RGB}{188,189,220}
\definecolor{Purples-3-F}{RGB}{188,189,220}
\definecolor{Purples-3-3}{RGB}{117,107,177}
\definecolor{Purples-3-I}{RGB}{117,107,177}
\definecolor{Purples-4-1}{RGB}{242,240,247}
\definecolor{Purples-4-B}{RGB}{242,240,247}
\definecolor{Purples-4-2}{RGB}{203,201,226}
\definecolor{Purples-4-E}{RGB}{203,201,226}
\definecolor{Purples-4-3}{RGB}{158,154,200}
\definecolor{Purples-4-G}{RGB}{158,154,200}
\definecolor{Purples-4-4}{RGB}{106,81,163}
\definecolor{Purples-4-J}{RGB}{106,81,163}
\definecolor{Purples-5-1}{RGB}{242,240,247}
\definecolor{Purples-5-B}{RGB}{242,240,247}
\definecolor{Purples-5-2}{RGB}{203,201,226}
\definecolor{Purples-5-E}{RGB}{203,201,226}
\definecolor{Purples-5-3}{RGB}{158,154,200}
\definecolor{Purples-5-G}{RGB}{158,154,200}
\definecolor{Purples-5-4}{RGB}{117,107,177}
\definecolor{Purples-5-I}{RGB}{117,107,177}
\definecolor{Purples-5-5}{RGB}{84,39,143}
\definecolor{Purples-5-K}{RGB}{84,39,143}
\definecolor{Purples-6-1}{RGB}{242,240,247}
\definecolor{Purples-6-B}{RGB}{242,240,247}
\definecolor{Purples-6-2}{RGB}{218,218,235}
\definecolor{Purples-6-D}{RGB}{218,218,235}
\definecolor{Purples-6-3}{RGB}{188,189,220}
\definecolor{Purples-6-F}{RGB}{188,189,220}
\definecolor{Purples-6-4}{RGB}{158,154,200}
\definecolor{Purples-6-G}{RGB}{158,154,200}
\definecolor{Purples-6-5}{RGB}{117,107,177}
\definecolor{Purples-6-I}{RGB}{117,107,177}
\definecolor{Purples-6-6}{RGB}{84,39,143}
\definecolor{Purples-6-K}{RGB}{84,39,143}
\definecolor{Purples-7-1}{RGB}{242,240,247}
\definecolor{Purples-7-B}{RGB}{242,240,247}
\definecolor{Purples-7-2}{RGB}{218,218,235}
\definecolor{Purples-7-D}{RGB}{218,218,235}
\definecolor{Purples-7-3}{RGB}{188,189,220}
\definecolor{Purples-7-F}{RGB}{188,189,220}
\definecolor{Purples-7-4}{RGB}{158,154,200}
\definecolor{Purples-7-G}{RGB}{158,154,200}
\definecolor{Purples-7-5}{RGB}{128,125,186}
\definecolor{Purples-7-H}{RGB}{128,125,186}
\definecolor{Purples-7-6}{RGB}{106,81,163}
\definecolor{Purples-7-J}{RGB}{106,81,163}
\definecolor{Purples-7-7}{RGB}{74,20,134}
\definecolor{Purples-7-L}{RGB}{74,20,134}
\definecolor{Purples-8-1}{RGB}{252,251,253}
\definecolor{Purples-8-A}{RGB}{252,251,253}
\definecolor{Purples-8-2}{RGB}{239,237,245}
\definecolor{Purples-8-C}{RGB}{239,237,245}
\definecolor{Purples-8-3}{RGB}{218,218,235}
\definecolor{Purples-8-D}{RGB}{218,218,235}
\definecolor{Purples-8-4}{RGB}{188,189,220}
\definecolor{Purples-8-F}{RGB}{188,189,220}
\definecolor{Purples-8-5}{RGB}{158,154,200}
\definecolor{Purples-8-G}{RGB}{158,154,200}
\definecolor{Purples-8-6}{RGB}{128,125,186}
\definecolor{Purples-8-H}{RGB}{128,125,186}
\definecolor{Purples-8-7}{RGB}{106,81,163}
\definecolor{Purples-8-J}{RGB}{106,81,163}
\definecolor{Purples-8-8}{RGB}{74,20,134}
\definecolor{Purples-8-L}{RGB}{74,20,134}
\definecolor{Purples-9-1}{RGB}{252,251,253}
\definecolor{Purples-9-A}{RGB}{252,251,253}
\definecolor{Purples-9-2}{RGB}{239,237,245}
\definecolor{Purples-9-C}{RGB}{239,237,245}
\definecolor{Purples-9-3}{RGB}{218,218,235}
\definecolor{Purples-9-D}{RGB}{218,218,235}
\definecolor{Purples-9-4}{RGB}{188,189,220}
\definecolor{Purples-9-F}{RGB}{188,189,220}
\definecolor{Purples-9-5}{RGB}{158,154,200}
\definecolor{Purples-9-G}{RGB}{158,154,200}
\definecolor{Purples-9-6}{RGB}{128,125,186}
\definecolor{Purples-9-H}{RGB}{128,125,186}
\definecolor{Purples-9-7}{RGB}{106,81,163}
\definecolor{Purples-9-J}{RGB}{106,81,163}
\definecolor{Purples-9-8}{RGB}{84,39,143}
\definecolor{Purples-9-K}{RGB}{84,39,143}
\definecolor{Purples-9-9}{RGB}{63,0,125}
\definecolor{Purples-9-M}{RGB}{63,0,125}
\definecolor{Blues-3-1}{RGB}{222,235,247}
\definecolor{Blues-3-C}{RGB}{222,235,247}
\definecolor{Blues-3-2}{RGB}{158,202,225}
\definecolor{Blues-3-F}{RGB}{158,202,225}
\definecolor{Blues-3-3}{RGB}{49,130,189}
\definecolor{Blues-3-I}{RGB}{49,130,189}
\definecolor{Blues-4-1}{RGB}{239,243,255}
\definecolor{Blues-4-B}{RGB}{239,243,255}
\definecolor{Blues-4-2}{RGB}{189,215,231}
\definecolor{Blues-4-E}{RGB}{189,215,231}
\definecolor{Blues-4-3}{RGB}{107,174,214}
\definecolor{Blues-4-G}{RGB}{107,174,214}
\definecolor{Blues-4-4}{RGB}{33,113,181}
\definecolor{Blues-4-J}{RGB}{33,113,181}
\definecolor{Blues-5-1}{RGB}{239,243,255}
\definecolor{Blues-5-B}{RGB}{239,243,255}
\definecolor{Blues-5-2}{RGB}{189,215,231}
\definecolor{Blues-5-E}{RGB}{189,215,231}
\definecolor{Blues-5-3}{RGB}{107,174,214}
\definecolor{Blues-5-G}{RGB}{107,174,214}
\definecolor{Blues-5-4}{RGB}{49,130,189}
\definecolor{Blues-5-I}{RGB}{49,130,189}
\definecolor{Blues-5-5}{RGB}{8,81,156}
\definecolor{Blues-5-K}{RGB}{8,81,156}
\definecolor{Blues-6-1}{RGB}{239,243,255}
\definecolor{Blues-6-B}{RGB}{239,243,255}
\definecolor{Blues-6-2}{RGB}{198,219,239}
\definecolor{Blues-6-D}{RGB}{198,219,239}
\definecolor{Blues-6-3}{RGB}{158,202,225}
\definecolor{Blues-6-F}{RGB}{158,202,225}
\definecolor{Blues-6-4}{RGB}{107,174,214}
\definecolor{Blues-6-G}{RGB}{107,174,214}
\definecolor{Blues-6-5}{RGB}{49,130,189}
\definecolor{Blues-6-I}{RGB}{49,130,189}
\definecolor{Blues-6-6}{RGB}{8,81,156}
\definecolor{Blues-6-K}{RGB}{8,81,156}
\definecolor{Blues-7-1}{RGB}{239,243,255}
\definecolor{Blues-7-B}{RGB}{239,243,255}
\definecolor{Blues-7-2}{RGB}{198,219,239}
\definecolor{Blues-7-D}{RGB}{198,219,239}
\definecolor{Blues-7-3}{RGB}{158,202,225}
\definecolor{Blues-7-F}{RGB}{158,202,225}
\definecolor{Blues-7-4}{RGB}{107,174,214}
\definecolor{Blues-7-G}{RGB}{107,174,214}
\definecolor{Blues-7-5}{RGB}{66,146,198}
\definecolor{Blues-7-H}{RGB}{66,146,198}
\definecolor{Blues-7-6}{RGB}{33,113,181}
\definecolor{Blues-7-J}{RGB}{33,113,181}
\definecolor{Blues-7-7}{RGB}{8,69,148}
\definecolor{Blues-7-L}{RGB}{8,69,148}
\definecolor{Blues-8-1}{RGB}{247,251,255}
\definecolor{Blues-8-A}{RGB}{247,251,255}
\definecolor{Blues-8-2}{RGB}{222,235,247}
\definecolor{Blues-8-C}{RGB}{222,235,247}
\definecolor{Blues-8-3}{RGB}{198,219,239}
\definecolor{Blues-8-D}{RGB}{198,219,239}
\definecolor{Blues-8-4}{RGB}{158,202,225}
\definecolor{Blues-8-F}{RGB}{158,202,225}
\definecolor{Blues-8-5}{RGB}{107,174,214}
\definecolor{Blues-8-G}{RGB}{107,174,214}
\definecolor{Blues-8-6}{RGB}{66,146,198}
\definecolor{Blues-8-H}{RGB}{66,146,198}
\definecolor{Blues-8-7}{RGB}{33,113,181}
\definecolor{Blues-8-J}{RGB}{33,113,181}
\definecolor{Blues-8-8}{RGB}{8,69,148}
\definecolor{Blues-8-L}{RGB}{8,69,148}
\definecolor{Blues-9-1}{RGB}{247,251,255}
\definecolor{Blues-9-A}{RGB}{247,251,255}
\definecolor{Blues-9-2}{RGB}{222,235,247}
\definecolor{Blues-9-C}{RGB}{222,235,247}
\definecolor{Blues-9-3}{RGB}{198,219,239}
\definecolor{Blues-9-D}{RGB}{198,219,239}
\definecolor{Blues-9-4}{RGB}{158,202,225}
\definecolor{Blues-9-F}{RGB}{158,202,225}
\definecolor{Blues-9-5}{RGB}{107,174,214}
\definecolor{Blues-9-G}{RGB}{107,174,214}
\definecolor{Blues-9-6}{RGB}{66,146,198}
\definecolor{Blues-9-H}{RGB}{66,146,198}
\definecolor{Blues-9-7}{RGB}{33,113,181}
\definecolor{Blues-9-J}{RGB}{33,113,181}
\definecolor{Blues-9-8}{RGB}{8,81,156}
\definecolor{Blues-9-K}{RGB}{8,81,156}
\definecolor{Blues-9-9}{RGB}{8,48,107}
\definecolor{Blues-9-M}{RGB}{8,48,107}
\definecolor{Greens-3-1}{RGB}{229,245,224}
\definecolor{Greens-3-C}{RGB}{229,245,224}
\definecolor{Greens-3-2}{RGB}{161,217,155}
\definecolor{Greens-3-F}{RGB}{161,217,155}
\definecolor{Greens-3-3}{RGB}{49,163,84}
\definecolor{Greens-3-I}{RGB}{49,163,84}
\definecolor{Greens-4-1}{RGB}{237,248,233}
\definecolor{Greens-4-B}{RGB}{237,248,233}
\definecolor{Greens-4-2}{RGB}{186,228,179}
\definecolor{Greens-4-E}{RGB}{186,228,179}
\definecolor{Greens-4-3}{RGB}{116,196,118}
\definecolor{Greens-4-G}{RGB}{116,196,118}
\definecolor{Greens-4-4}{RGB}{35,139,69}
\definecolor{Greens-4-J}{RGB}{35,139,69}
\definecolor{Greens-5-1}{RGB}{237,248,233}
\definecolor{Greens-5-B}{RGB}{237,248,233}
\definecolor{Greens-5-2}{RGB}{186,228,179}
\definecolor{Greens-5-E}{RGB}{186,228,179}
\definecolor{Greens-5-3}{RGB}{116,196,118}
\definecolor{Greens-5-G}{RGB}{116,196,118}
\definecolor{Greens-5-4}{RGB}{49,163,84}
\definecolor{Greens-5-I}{RGB}{49,163,84}
\definecolor{Greens-5-5}{RGB}{0,109,44}
\definecolor{Greens-5-K}{RGB}{0,109,44}
\definecolor{Greens-6-1}{RGB}{237,248,233}
\definecolor{Greens-6-B}{RGB}{237,248,233}
\definecolor{Greens-6-2}{RGB}{199,233,192}
\definecolor{Greens-6-D}{RGB}{199,233,192}
\definecolor{Greens-6-3}{RGB}{161,217,155}
\definecolor{Greens-6-F}{RGB}{161,217,155}
\definecolor{Greens-6-4}{RGB}{116,196,118}
\definecolor{Greens-6-G}{RGB}{116,196,118}
\definecolor{Greens-6-5}{RGB}{49,163,84}
\definecolor{Greens-6-I}{RGB}{49,163,84}
\definecolor{Greens-6-6}{RGB}{0,109,44}
\definecolor{Greens-6-K}{RGB}{0,109,44}
\definecolor{Greens-7-1}{RGB}{237,248,233}
\definecolor{Greens-7-B}{RGB}{237,248,233}
\definecolor{Greens-7-2}{RGB}{199,233,192}
\definecolor{Greens-7-D}{RGB}{199,233,192}
\definecolor{Greens-7-3}{RGB}{161,217,155}
\definecolor{Greens-7-F}{RGB}{161,217,155}
\definecolor{Greens-7-4}{RGB}{116,196,118}
\definecolor{Greens-7-G}{RGB}{116,196,118}
\definecolor{Greens-7-5}{RGB}{65,171,93}
\definecolor{Greens-7-H}{RGB}{65,171,93}
\definecolor{Greens-7-6}{RGB}{35,139,69}
\definecolor{Greens-7-J}{RGB}{35,139,69}
\definecolor{Greens-7-7}{RGB}{0,90,50}
\definecolor{Greens-7-L}{RGB}{0,90,50}
\definecolor{Greens-8-1}{RGB}{247,252,245}
\definecolor{Greens-8-A}{RGB}{247,252,245}
\definecolor{Greens-8-2}{RGB}{229,245,224}
\definecolor{Greens-8-C}{RGB}{229,245,224}
\definecolor{Greens-8-3}{RGB}{199,233,192}
\definecolor{Greens-8-D}{RGB}{199,233,192}
\definecolor{Greens-8-4}{RGB}{161,217,155}
\definecolor{Greens-8-F}{RGB}{161,217,155}
\definecolor{Greens-8-5}{RGB}{116,196,118}
\definecolor{Greens-8-G}{RGB}{116,196,118}
\definecolor{Greens-8-6}{RGB}{65,171,93}
\definecolor{Greens-8-H}{RGB}{65,171,93}
\definecolor{Greens-8-7}{RGB}{35,139,69}
\definecolor{Greens-8-J}{RGB}{35,139,69}
\definecolor{Greens-8-8}{RGB}{0,90,50}
\definecolor{Greens-8-L}{RGB}{0,90,50}
\definecolor{Greens-9-1}{RGB}{247,252,245}
\definecolor{Greens-9-A}{RGB}{247,252,245}
\definecolor{Greens-9-2}{RGB}{229,245,224}
\definecolor{Greens-9-C}{RGB}{229,245,224}
\definecolor{Greens-9-3}{RGB}{199,233,192}
\definecolor{Greens-9-D}{RGB}{199,233,192}
\definecolor{Greens-9-4}{RGB}{161,217,155}
\definecolor{Greens-9-F}{RGB}{161,217,155}
\definecolor{Greens-9-5}{RGB}{116,196,118}
\definecolor{Greens-9-G}{RGB}{116,196,118}
\definecolor{Greens-9-6}{RGB}{65,171,93}
\definecolor{Greens-9-H}{RGB}{65,171,93}
\definecolor{Greens-9-7}{RGB}{35,139,69}
\definecolor{Greens-9-J}{RGB}{35,139,69}
\definecolor{Greens-9-8}{RGB}{0,109,44}
\definecolor{Greens-9-K}{RGB}{0,109,44}
\definecolor{Greens-9-9}{RGB}{0,68,27}
\definecolor{Greens-9-M}{RGB}{0,68,27}
\definecolor{Oranges-3-1}{RGB}{254,230,206}
\definecolor{Oranges-3-C}{RGB}{254,230,206}
\definecolor{Oranges-3-2}{RGB}{253,174,107}
\definecolor{Oranges-3-F}{RGB}{253,174,107}
\definecolor{Oranges-3-3}{RGB}{230,85,13}
\definecolor{Oranges-3-I}{RGB}{230,85,13}
\definecolor{Oranges-4-1}{RGB}{254,237,222}
\definecolor{Oranges-4-B}{RGB}{254,237,222}
\definecolor{Oranges-4-2}{RGB}{253,190,133}
\definecolor{Oranges-4-E}{RGB}{253,190,133}
\definecolor{Oranges-4-3}{RGB}{253,141,60}
\definecolor{Oranges-4-G}{RGB}{253,141,60}
\definecolor{Oranges-4-4}{RGB}{217,71,1}
\definecolor{Oranges-4-J}{RGB}{217,71,1}
\definecolor{Oranges-5-1}{RGB}{254,237,222}
\definecolor{Oranges-5-B}{RGB}{254,237,222}
\definecolor{Oranges-5-2}{RGB}{253,190,133}
\definecolor{Oranges-5-E}{RGB}{253,190,133}
\definecolor{Oranges-5-3}{RGB}{253,141,60}
\definecolor{Oranges-5-G}{RGB}{253,141,60}
\definecolor{Oranges-5-4}{RGB}{230,85,13}
\definecolor{Oranges-5-I}{RGB}{230,85,13}
\definecolor{Oranges-5-5}{RGB}{166,54,3}
\definecolor{Oranges-5-K}{RGB}{166,54,3}
\definecolor{Oranges-6-1}{RGB}{254,237,222}
\definecolor{Oranges-6-B}{RGB}{254,237,222}
\definecolor{Oranges-6-2}{RGB}{253,208,162}
\definecolor{Oranges-6-D}{RGB}{253,208,162}
\definecolor{Oranges-6-3}{RGB}{253,174,107}
\definecolor{Oranges-6-F}{RGB}{253,174,107}
\definecolor{Oranges-6-4}{RGB}{253,141,60}
\definecolor{Oranges-6-G}{RGB}{253,141,60}
\definecolor{Oranges-6-5}{RGB}{230,85,13}
\definecolor{Oranges-6-I}{RGB}{230,85,13}
\definecolor{Oranges-6-6}{RGB}{166,54,3}
\definecolor{Oranges-6-K}{RGB}{166,54,3}
\definecolor{Oranges-7-1}{RGB}{254,237,222}
\definecolor{Oranges-7-B}{RGB}{254,237,222}
\definecolor{Oranges-7-2}{RGB}{253,208,162}
\definecolor{Oranges-7-D}{RGB}{253,208,162}
\definecolor{Oranges-7-3}{RGB}{253,174,107}
\definecolor{Oranges-7-F}{RGB}{253,174,107}
\definecolor{Oranges-7-4}{RGB}{253,141,60}
\definecolor{Oranges-7-G}{RGB}{253,141,60}
\definecolor{Oranges-7-5}{RGB}{241,105,19}
\definecolor{Oranges-7-H}{RGB}{241,105,19}
\definecolor{Oranges-7-6}{RGB}{217,72,1}
\definecolor{Oranges-7-J}{RGB}{217,72,1}
\definecolor{Oranges-7-7}{RGB}{140,45,4}
\definecolor{Oranges-7-L}{RGB}{140,45,4}
\definecolor{Oranges-8-1}{RGB}{255,245,235}
\definecolor{Oranges-8-A}{RGB}{255,245,235}
\definecolor{Oranges-8-2}{RGB}{254,230,206}
\definecolor{Oranges-8-C}{RGB}{254,230,206}
\definecolor{Oranges-8-3}{RGB}{253,208,162}
\definecolor{Oranges-8-D}{RGB}{253,208,162}
\definecolor{Oranges-8-4}{RGB}{253,174,107}
\definecolor{Oranges-8-F}{RGB}{253,174,107}
\definecolor{Oranges-8-5}{RGB}{253,141,60}
\definecolor{Oranges-8-G}{RGB}{253,141,60}
\definecolor{Oranges-8-6}{RGB}{241,105,19}
\definecolor{Oranges-8-H}{RGB}{241,105,19}
\definecolor{Oranges-8-7}{RGB}{217,72,1}
\definecolor{Oranges-8-J}{RGB}{217,72,1}
\definecolor{Oranges-8-8}{RGB}{140,45,4}
\definecolor{Oranges-8-L}{RGB}{140,45,4}
\definecolor{Oranges-9-1}{RGB}{255,245,235}
\definecolor{Oranges-9-A}{RGB}{255,245,235}
\definecolor{Oranges-9-2}{RGB}{254,230,206}
\definecolor{Oranges-9-C}{RGB}{254,230,206}
\definecolor{Oranges-9-3}{RGB}{253,208,162}
\definecolor{Oranges-9-D}{RGB}{253,208,162}
\definecolor{Oranges-9-4}{RGB}{253,174,107}
\definecolor{Oranges-9-F}{RGB}{253,174,107}
\definecolor{Oranges-9-5}{RGB}{253,141,60}
\definecolor{Oranges-9-G}{RGB}{253,141,60}
\definecolor{Oranges-9-6}{RGB}{241,105,19}
\definecolor{Oranges-9-H}{RGB}{241,105,19}
\definecolor{Oranges-9-7}{RGB}{217,72,1}
\definecolor{Oranges-9-J}{RGB}{217,72,1}
\definecolor{Oranges-9-8}{RGB}{166,54,3}
\definecolor{Oranges-9-K}{RGB}{166,54,3}
\definecolor{Oranges-9-9}{RGB}{127,39,4}
\definecolor{Oranges-9-M}{RGB}{127,39,4}
\definecolor{Reds-3-1}{RGB}{254,224,210}
\definecolor{Reds-3-C}{RGB}{254,224,210}
\definecolor{Reds-3-2}{RGB}{252,146,114}
\definecolor{Reds-3-F}{RGB}{252,146,114}
\definecolor{Reds-3-3}{RGB}{222,45,38}
\definecolor{Reds-3-I}{RGB}{222,45,38}
\definecolor{Reds-4-1}{RGB}{254,229,217}
\definecolor{Reds-4-B}{RGB}{254,229,217}
\definecolor{Reds-4-2}{RGB}{252,174,145}
\definecolor{Reds-4-E}{RGB}{252,174,145}
\definecolor{Reds-4-3}{RGB}{251,106,74}
\definecolor{Reds-4-G}{RGB}{251,106,74}
\definecolor{Reds-4-4}{RGB}{203,24,29}
\definecolor{Reds-4-J}{RGB}{203,24,29}
\definecolor{Reds-5-1}{RGB}{254,229,217}
\definecolor{Reds-5-B}{RGB}{254,229,217}
\definecolor{Reds-5-2}{RGB}{252,174,145}
\definecolor{Reds-5-E}{RGB}{252,174,145}
\definecolor{Reds-5-3}{RGB}{251,106,74}
\definecolor{Reds-5-G}{RGB}{251,106,74}
\definecolor{Reds-5-4}{RGB}{222,45,38}
\definecolor{Reds-5-I}{RGB}{222,45,38}
\definecolor{Reds-5-5}{RGB}{165,15,21}
\definecolor{Reds-5-K}{RGB}{165,15,21}
\definecolor{Reds-6-1}{RGB}{254,229,217}
\definecolor{Reds-6-B}{RGB}{254,229,217}
\definecolor{Reds-6-2}{RGB}{252,187,161}
\definecolor{Reds-6-D}{RGB}{252,187,161}
\definecolor{Reds-6-3}{RGB}{252,146,114}
\definecolor{Reds-6-F}{RGB}{252,146,114}
\definecolor{Reds-6-4}{RGB}{251,106,74}
\definecolor{Reds-6-G}{RGB}{251,106,74}
\definecolor{Reds-6-5}{RGB}{222,45,38}
\definecolor{Reds-6-I}{RGB}{222,45,38}
\definecolor{Reds-6-6}{RGB}{165,15,21}
\definecolor{Reds-6-K}{RGB}{165,15,21}
\definecolor{Reds-7-1}{RGB}{254,229,217}
\definecolor{Reds-7-B}{RGB}{254,229,217}
\definecolor{Reds-7-2}{RGB}{252,187,161}
\definecolor{Reds-7-D}{RGB}{252,187,161}
\definecolor{Reds-7-3}{RGB}{252,146,114}
\definecolor{Reds-7-F}{RGB}{252,146,114}
\definecolor{Reds-7-4}{RGB}{251,106,74}
\definecolor{Reds-7-G}{RGB}{251,106,74}
\definecolor{Reds-7-5}{RGB}{239,59,44}
\definecolor{Reds-7-H}{RGB}{239,59,44}
\definecolor{Reds-7-6}{RGB}{203,24,29}
\definecolor{Reds-7-J}{RGB}{203,24,29}
\definecolor{Reds-7-7}{RGB}{153,0,13}
\definecolor{Reds-7-L}{RGB}{153,0,13}
\definecolor{Reds-8-1}{RGB}{255,245,240}
\definecolor{Reds-8-A}{RGB}{255,245,240}
\definecolor{Reds-8-2}{RGB}{254,224,210}
\definecolor{Reds-8-C}{RGB}{254,224,210}
\definecolor{Reds-8-3}{RGB}{252,187,161}
\definecolor{Reds-8-D}{RGB}{252,187,161}
\definecolor{Reds-8-4}{RGB}{252,146,114}
\definecolor{Reds-8-F}{RGB}{252,146,114}
\definecolor{Reds-8-5}{RGB}{251,106,74}
\definecolor{Reds-8-G}{RGB}{251,106,74}
\definecolor{Reds-8-6}{RGB}{239,59,44}
\definecolor{Reds-8-H}{RGB}{239,59,44}
\definecolor{Reds-8-7}{RGB}{203,24,29}
\definecolor{Reds-8-J}{RGB}{203,24,29}
\definecolor{Reds-8-8}{RGB}{153,0,13}
\definecolor{Reds-8-L}{RGB}{153,0,13}
\definecolor{Reds-9-1}{RGB}{255,245,240}
\definecolor{Reds-9-A}{RGB}{255,245,240}
\definecolor{Reds-9-2}{RGB}{254,224,210}
\definecolor{Reds-9-C}{RGB}{254,224,210}
\definecolor{Reds-9-3}{RGB}{252,187,161}
\definecolor{Reds-9-D}{RGB}{252,187,161}
\definecolor{Reds-9-4}{RGB}{252,146,114}
\definecolor{Reds-9-F}{RGB}{252,146,114}
\definecolor{Reds-9-5}{RGB}{251,106,74}
\definecolor{Reds-9-G}{RGB}{251,106,74}
\definecolor{Reds-9-6}{RGB}{239,59,44}
\definecolor{Reds-9-H}{RGB}{239,59,44}
\definecolor{Reds-9-7}{RGB}{203,24,29}
\definecolor{Reds-9-J}{RGB}{203,24,29}
\definecolor{Reds-9-8}{RGB}{165,15,21}
\definecolor{Reds-9-K}{RGB}{165,15,21}
\definecolor{Reds-9-9}{RGB}{103,0,13}
\definecolor{Reds-9-M}{RGB}{103,0,13}
\definecolor{Greys-3-1}{RGB}{240,240,240}
\definecolor{Greys-3-C}{RGB}{240,240,240}
\definecolor{Greys-3-2}{RGB}{189,189,189}
\definecolor{Greys-3-F}{RGB}{189,189,189}
\definecolor{Greys-3-3}{RGB}{99,99,99}
\definecolor{Greys-3-I}{RGB}{99,99,99}
\definecolor{Greys-4-1}{RGB}{247,247,247}
\definecolor{Greys-4-B}{RGB}{247,247,247}
\definecolor{Greys-4-2}{RGB}{204,204,204}
\definecolor{Greys-4-E}{RGB}{204,204,204}
\definecolor{Greys-4-3}{RGB}{150,150,150}
\definecolor{Greys-4-G}{RGB}{150,150,150}
\definecolor{Greys-4-4}{RGB}{82,82,82}
\definecolor{Greys-4-J}{RGB}{82,82,82}
\definecolor{Greys-5-1}{RGB}{247,247,247}
\definecolor{Greys-5-B}{RGB}{247,247,247}
\definecolor{Greys-5-2}{RGB}{204,204,204}
\definecolor{Greys-5-E}{RGB}{204,204,204}
\definecolor{Greys-5-3}{RGB}{150,150,150}
\definecolor{Greys-5-G}{RGB}{150,150,150}
\definecolor{Greys-5-4}{RGB}{99,99,99}
\definecolor{Greys-5-I}{RGB}{99,99,99}
\definecolor{Greys-5-5}{RGB}{37,37,37}
\definecolor{Greys-5-K}{RGB}{37,37,37}
\definecolor{Greys-6-1}{RGB}{247,247,247}
\definecolor{Greys-6-B}{RGB}{247,247,247}
\definecolor{Greys-6-2}{RGB}{217,217,217}
\definecolor{Greys-6-D}{RGB}{217,217,217}
\definecolor{Greys-6-3}{RGB}{189,189,189}
\definecolor{Greys-6-F}{RGB}{189,189,189}
\definecolor{Greys-6-4}{RGB}{150,150,150}
\definecolor{Greys-6-G}{RGB}{150,150,150}
\definecolor{Greys-6-5}{RGB}{99,99,99}
\definecolor{Greys-6-I}{RGB}{99,99,99}
\definecolor{Greys-6-6}{RGB}{37,37,37}
\definecolor{Greys-6-K}{RGB}{37,37,37}
\definecolor{Greys-7-1}{RGB}{247,247,247}
\definecolor{Greys-7-B}{RGB}{247,247,247}
\definecolor{Greys-7-2}{RGB}{217,217,217}
\definecolor{Greys-7-D}{RGB}{217,217,217}
\definecolor{Greys-7-3}{RGB}{189,189,189}
\definecolor{Greys-7-F}{RGB}{189,189,189}
\definecolor{Greys-7-4}{RGB}{150,150,150}
\definecolor{Greys-7-G}{RGB}{150,150,150}
\definecolor{Greys-7-5}{RGB}{115,115,115}
\definecolor{Greys-7-H}{RGB}{115,115,115}
\definecolor{Greys-7-6}{RGB}{82,82,82}
\definecolor{Greys-7-J}{RGB}{82,82,82}
\definecolor{Greys-7-7}{RGB}{37,37,37}
\definecolor{Greys-7-L}{RGB}{37,37,37}
\definecolor{Greys-8-1}{RGB}{255,255,255}
\definecolor{Greys-8-A}{RGB}{255,255,255}
\definecolor{Greys-8-2}{RGB}{240,240,240}
\definecolor{Greys-8-C}{RGB}{240,240,240}
\definecolor{Greys-8-3}{RGB}{217,217,217}
\definecolor{Greys-8-D}{RGB}{217,217,217}
\definecolor{Greys-8-4}{RGB}{189,189,189}
\definecolor{Greys-8-F}{RGB}{189,189,189}
\definecolor{Greys-8-5}{RGB}{150,150,150}
\definecolor{Greys-8-G}{RGB}{150,150,150}
\definecolor{Greys-8-6}{RGB}{115,115,115}
\definecolor{Greys-8-H}{RGB}{115,115,115}
\definecolor{Greys-8-7}{RGB}{82,82,82}
\definecolor{Greys-8-J}{RGB}{82,82,82}
\definecolor{Greys-8-8}{RGB}{37,37,37}
\definecolor{Greys-8-L}{RGB}{37,37,37}
\definecolor{Greys-9-1}{RGB}{255,255,255}
\definecolor{Greys-9-A}{RGB}{255,255,255}
\definecolor{Greys-9-2}{RGB}{240,240,240}
\definecolor{Greys-9-C}{RGB}{240,240,240}
\definecolor{Greys-9-3}{RGB}{217,217,217}
\definecolor{Greys-9-D}{RGB}{217,217,217}
\definecolor{Greys-9-4}{RGB}{189,189,189}
\definecolor{Greys-9-F}{RGB}{189,189,189}
\definecolor{Greys-9-5}{RGB}{150,150,150}
\definecolor{Greys-9-G}{RGB}{150,150,150}
\definecolor{Greys-9-6}{RGB}{115,115,115}
\definecolor{Greys-9-H}{RGB}{115,115,115}
\definecolor{Greys-9-7}{RGB}{82,82,82}
\definecolor{Greys-9-J}{RGB}{82,82,82}
\definecolor{Greys-9-8}{RGB}{37,37,37}
\definecolor{Greys-9-K}{RGB}{37,37,37}
\definecolor{Greys-9-9}{RGB}{0,0,0}
\definecolor{Greys-9-M}{RGB}{0,0,0}
\definecolor{PuOr-3-1}{RGB}{241,163,64}
\definecolor{PuOr-3-E}{RGB}{241,163,64}
\definecolor{PuOr-3-2}{RGB}{247,247,247}
\definecolor{PuOr-3-H}{RGB}{247,247,247}
\definecolor{PuOr-3-3}{RGB}{153,142,195}
\definecolor{PuOr-3-K}{RGB}{153,142,195}
\definecolor{PuOr-4-1}{RGB}{230,97,1}
\definecolor{PuOr-4-C}{RGB}{230,97,1}
\definecolor{PuOr-4-2}{RGB}{253,184,99}
\definecolor{PuOr-4-F}{RGB}{253,184,99}
\definecolor{PuOr-4-3}{RGB}{178,171,210}
\definecolor{PuOr-4-J}{RGB}{178,171,210}
\definecolor{PuOr-4-4}{RGB}{94,60,153}
\definecolor{PuOr-4-M}{RGB}{94,60,153}
\definecolor{PuOr-5-1}{RGB}{230,97,1}
\definecolor{PuOr-5-C}{RGB}{230,97,1}
\definecolor{PuOr-5-2}{RGB}{253,184,99}
\definecolor{PuOr-5-F}{RGB}{253,184,99}
\definecolor{PuOr-5-3}{RGB}{247,247,247}
\definecolor{PuOr-5-H}{RGB}{247,247,247}
\definecolor{PuOr-5-4}{RGB}{178,171,210}
\definecolor{PuOr-5-J}{RGB}{178,171,210}
\definecolor{PuOr-5-5}{RGB}{94,60,153}
\definecolor{PuOr-5-M}{RGB}{94,60,153}
\definecolor{PuOr-6-1}{RGB}{179,88,6}
\definecolor{PuOr-6-B}{RGB}{179,88,6}
\definecolor{PuOr-6-2}{RGB}{241,163,64}
\definecolor{PuOr-6-E}{RGB}{241,163,64}
\definecolor{PuOr-6-3}{RGB}{254,224,182}
\definecolor{PuOr-6-G}{RGB}{254,224,182}
\definecolor{PuOr-6-4}{RGB}{216,218,235}
\definecolor{PuOr-6-I}{RGB}{216,218,235}
\definecolor{PuOr-6-5}{RGB}{153,142,195}
\definecolor{PuOr-6-K}{RGB}{153,142,195}
\definecolor{PuOr-6-6}{RGB}{84,39,136}
\definecolor{PuOr-6-N}{RGB}{84,39,136}
\definecolor{PuOr-7-1}{RGB}{179,88,6}
\definecolor{PuOr-7-B}{RGB}{179,88,6}
\definecolor{PuOr-7-2}{RGB}{241,163,64}
\definecolor{PuOr-7-E}{RGB}{241,163,64}
\definecolor{PuOr-7-3}{RGB}{254,224,182}
\definecolor{PuOr-7-G}{RGB}{254,224,182}
\definecolor{PuOr-7-4}{RGB}{247,247,247}
\definecolor{PuOr-7-H}{RGB}{247,247,247}
\definecolor{PuOr-7-5}{RGB}{216,218,235}
\definecolor{PuOr-7-I}{RGB}{216,218,235}
\definecolor{PuOr-7-6}{RGB}{153,142,195}
\definecolor{PuOr-7-K}{RGB}{153,142,195}
\definecolor{PuOr-7-7}{RGB}{84,39,136}
\definecolor{PuOr-7-N}{RGB}{84,39,136}
\definecolor{PuOr-8-1}{RGB}{179,88,6}
\definecolor{PuOr-8-B}{RGB}{179,88,6}
\definecolor{PuOr-8-2}{RGB}{224,130,20}
\definecolor{PuOr-8-D}{RGB}{224,130,20}
\definecolor{PuOr-8-3}{RGB}{253,184,99}
\definecolor{PuOr-8-F}{RGB}{253,184,99}
\definecolor{PuOr-8-4}{RGB}{254,224,182}
\definecolor{PuOr-8-G}{RGB}{254,224,182}
\definecolor{PuOr-8-5}{RGB}{216,218,235}
\definecolor{PuOr-8-I}{RGB}{216,218,235}
\definecolor{PuOr-8-6}{RGB}{178,171,210}
\definecolor{PuOr-8-J}{RGB}{178,171,210}
\definecolor{PuOr-8-7}{RGB}{128,115,172}
\definecolor{PuOr-8-L}{RGB}{128,115,172}
\definecolor{PuOr-8-8}{RGB}{84,39,136}
\definecolor{PuOr-8-N}{RGB}{84,39,136}
\definecolor{PuOr-9-1}{RGB}{179,88,6}
\definecolor{PuOr-9-B}{RGB}{179,88,6}
\definecolor{PuOr-9-2}{RGB}{224,130,20}
\definecolor{PuOr-9-D}{RGB}{224,130,20}
\definecolor{PuOr-9-3}{RGB}{253,184,99}
\definecolor{PuOr-9-F}{RGB}{253,184,99}
\definecolor{PuOr-9-4}{RGB}{254,224,182}
\definecolor{PuOr-9-G}{RGB}{254,224,182}
\definecolor{PuOr-9-5}{RGB}{247,247,247}
\definecolor{PuOr-9-H}{RGB}{247,247,247}
\definecolor{PuOr-9-6}{RGB}{216,218,235}
\definecolor{PuOr-9-I}{RGB}{216,218,235}
\definecolor{PuOr-9-7}{RGB}{178,171,210}
\definecolor{PuOr-9-J}{RGB}{178,171,210}
\definecolor{PuOr-9-8}{RGB}{128,115,172}
\definecolor{PuOr-9-L}{RGB}{128,115,172}
\definecolor{PuOr-9-9}{RGB}{84,39,136}
\definecolor{PuOr-9-N}{RGB}{84,39,136}
\definecolor{PuOr-10-1}{RGB}{127,59,8}
\definecolor{PuOr-10-A}{RGB}{127,59,8}
\definecolor{PuOr-10-2}{RGB}{179,88,6}
\definecolor{PuOr-10-B}{RGB}{179,88,6}
\definecolor{PuOr-10-3}{RGB}{224,130,20}
\definecolor{PuOr-10-D}{RGB}{224,130,20}
\definecolor{PuOr-10-4}{RGB}{253,184,99}
\definecolor{PuOr-10-F}{RGB}{253,184,99}
\definecolor{PuOr-10-5}{RGB}{254,224,182}
\definecolor{PuOr-10-G}{RGB}{254,224,182}
\definecolor{PuOr-10-6}{RGB}{216,218,235}
\definecolor{PuOr-10-I}{RGB}{216,218,235}
\definecolor{PuOr-10-7}{RGB}{178,171,210}
\definecolor{PuOr-10-J}{RGB}{178,171,210}
\definecolor{PuOr-10-8}{RGB}{128,115,172}
\definecolor{PuOr-10-L}{RGB}{128,115,172}
\definecolor{PuOr-10-9}{RGB}{84,39,136}
\definecolor{PuOr-10-N}{RGB}{84,39,136}
\definecolor{PuOr-10-10}{RGB}{45,0,75}
\definecolor{PuOr-10-O}{RGB}{45,0,75}
\definecolor{PuOr-11-1}{RGB}{127,59,8}
\definecolor{PuOr-11-A}{RGB}{127,59,8}
\definecolor{PuOr-11-2}{RGB}{179,88,6}
\definecolor{PuOr-11-B}{RGB}{179,88,6}
\definecolor{PuOr-11-3}{RGB}{224,130,20}
\definecolor{PuOr-11-D}{RGB}{224,130,20}
\definecolor{PuOr-11-4}{RGB}{253,184,99}
\definecolor{PuOr-11-F}{RGB}{253,184,99}
\definecolor{PuOr-11-5}{RGB}{254,224,182}
\definecolor{PuOr-11-G}{RGB}{254,224,182}
\definecolor{PuOr-11-6}{RGB}{247,247,247}
\definecolor{PuOr-11-H}{RGB}{247,247,247}
\definecolor{PuOr-11-7}{RGB}{216,218,235}
\definecolor{PuOr-11-I}{RGB}{216,218,235}
\definecolor{PuOr-11-8}{RGB}{178,171,210}
\definecolor{PuOr-11-J}{RGB}{178,171,210}
\definecolor{PuOr-11-9}{RGB}{128,115,172}
\definecolor{PuOr-11-L}{RGB}{128,115,172}
\definecolor{PuOr-11-10}{RGB}{84,39,136}
\definecolor{PuOr-11-N}{RGB}{84,39,136}
\definecolor{PuOr-11-11}{RGB}{45,0,75}
\definecolor{PuOr-11-O}{RGB}{45,0,75}
\definecolor{BrBG-3-1}{RGB}{216,179,101}
\definecolor{BrBG-3-E}{RGB}{216,179,101}
\definecolor{BrBG-3-2}{RGB}{245,245,245}
\definecolor{BrBG-3-H}{RGB}{245,245,245}
\definecolor{BrBG-3-3}{RGB}{90,180,172}
\definecolor{BrBG-3-K}{RGB}{90,180,172}
\definecolor{BrBG-4-1}{RGB}{166,97,26}
\definecolor{BrBG-4-C}{RGB}{166,97,26}
\definecolor{BrBG-4-2}{RGB}{223,194,125}
\definecolor{BrBG-4-F}{RGB}{223,194,125}
\definecolor{BrBG-4-3}{RGB}{128,205,193}
\definecolor{BrBG-4-J}{RGB}{128,205,193}
\definecolor{BrBG-4-4}{RGB}{1,133,113}
\definecolor{BrBG-4-M}{RGB}{1,133,113}
\definecolor{BrBG-5-1}{RGB}{166,97,26}
\definecolor{BrBG-5-C}{RGB}{166,97,26}
\definecolor{BrBG-5-2}{RGB}{223,194,125}
\definecolor{BrBG-5-F}{RGB}{223,194,125}
\definecolor{BrBG-5-3}{RGB}{245,245,245}
\definecolor{BrBG-5-H}{RGB}{245,245,245}
\definecolor{BrBG-5-4}{RGB}{128,205,193}
\definecolor{BrBG-5-J}{RGB}{128,205,193}
\definecolor{BrBG-5-5}{RGB}{1,133,113}
\definecolor{BrBG-5-M}{RGB}{1,133,113}
\definecolor{BrBG-6-1}{RGB}{140,81,10}
\definecolor{BrBG-6-B}{RGB}{140,81,10}
\definecolor{BrBG-6-2}{RGB}{216,179,101}
\definecolor{BrBG-6-E}{RGB}{216,179,101}
\definecolor{BrBG-6-3}{RGB}{246,232,195}
\definecolor{BrBG-6-G}{RGB}{246,232,195}
\definecolor{BrBG-6-4}{RGB}{199,234,229}
\definecolor{BrBG-6-I}{RGB}{199,234,229}
\definecolor{BrBG-6-5}{RGB}{90,180,172}
\definecolor{BrBG-6-K}{RGB}{90,180,172}
\definecolor{BrBG-6-6}{RGB}{1,102,94}
\definecolor{BrBG-6-N}{RGB}{1,102,94}
\definecolor{BrBG-7-1}{RGB}{140,81,10}
\definecolor{BrBG-7-B}{RGB}{140,81,10}
\definecolor{BrBG-7-2}{RGB}{216,179,101}
\definecolor{BrBG-7-E}{RGB}{216,179,101}
\definecolor{BrBG-7-3}{RGB}{246,232,195}
\definecolor{BrBG-7-G}{RGB}{246,232,195}
\definecolor{BrBG-7-4}{RGB}{245,245,245}
\definecolor{BrBG-7-H}{RGB}{245,245,245}
\definecolor{BrBG-7-5}{RGB}{199,234,229}
\definecolor{BrBG-7-I}{RGB}{199,234,229}
\definecolor{BrBG-7-6}{RGB}{90,180,172}
\definecolor{BrBG-7-K}{RGB}{90,180,172}
\definecolor{BrBG-7-7}{RGB}{1,102,94}
\definecolor{BrBG-7-N}{RGB}{1,102,94}
\definecolor{BrBG-8-1}{RGB}{140,81,10}
\definecolor{BrBG-8-B}{RGB}{140,81,10}
\definecolor{BrBG-8-2}{RGB}{191,129,45}
\definecolor{BrBG-8-D}{RGB}{191,129,45}
\definecolor{BrBG-8-3}{RGB}{223,194,125}
\definecolor{BrBG-8-F}{RGB}{223,194,125}
\definecolor{BrBG-8-4}{RGB}{246,232,195}
\definecolor{BrBG-8-G}{RGB}{246,232,195}
\definecolor{BrBG-8-5}{RGB}{199,234,229}
\definecolor{BrBG-8-I}{RGB}{199,234,229}
\definecolor{BrBG-8-6}{RGB}{128,205,193}
\definecolor{BrBG-8-J}{RGB}{128,205,193}
\definecolor{BrBG-8-7}{RGB}{53,151,143}
\definecolor{BrBG-8-L}{RGB}{53,151,143}
\definecolor{BrBG-8-8}{RGB}{1,102,94}
\definecolor{BrBG-8-N}{RGB}{1,102,94}
\definecolor{BrBG-9-1}{RGB}{140,81,10}
\definecolor{BrBG-9-B}{RGB}{140,81,10}
\definecolor{BrBG-9-2}{RGB}{191,129,45}
\definecolor{BrBG-9-D}{RGB}{191,129,45}
\definecolor{BrBG-9-3}{RGB}{223,194,125}
\definecolor{BrBG-9-F}{RGB}{223,194,125}
\definecolor{BrBG-9-4}{RGB}{246,232,195}
\definecolor{BrBG-9-G}{RGB}{246,232,195}
\definecolor{BrBG-9-5}{RGB}{245,245,245}
\definecolor{BrBG-9-H}{RGB}{245,245,245}
\definecolor{BrBG-9-6}{RGB}{199,234,229}
\definecolor{BrBG-9-I}{RGB}{199,234,229}
\definecolor{BrBG-9-7}{RGB}{128,205,193}
\definecolor{BrBG-9-J}{RGB}{128,205,193}
\definecolor{BrBG-9-8}{RGB}{53,151,143}
\definecolor{BrBG-9-L}{RGB}{53,151,143}
\definecolor{BrBG-9-9}{RGB}{1,102,94}
\definecolor{BrBG-9-N}{RGB}{1,102,94}
\definecolor{BrBG-10-1}{RGB}{84,48,5}
\definecolor{BrBG-10-A}{RGB}{84,48,5}
\definecolor{BrBG-10-2}{RGB}{140,81,10}
\definecolor{BrBG-10-B}{RGB}{140,81,10}
\definecolor{BrBG-10-3}{RGB}{191,129,45}
\definecolor{BrBG-10-D}{RGB}{191,129,45}
\definecolor{BrBG-10-4}{RGB}{223,194,125}
\definecolor{BrBG-10-F}{RGB}{223,194,125}
\definecolor{BrBG-10-5}{RGB}{246,232,195}
\definecolor{BrBG-10-G}{RGB}{246,232,195}
\definecolor{BrBG-10-6}{RGB}{199,234,229}
\definecolor{BrBG-10-I}{RGB}{199,234,229}
\definecolor{BrBG-10-7}{RGB}{128,205,193}
\definecolor{BrBG-10-J}{RGB}{128,205,193}
\definecolor{BrBG-10-8}{RGB}{53,151,143}
\definecolor{BrBG-10-L}{RGB}{53,151,143}
\definecolor{BrBG-10-9}{RGB}{1,102,94}
\definecolor{BrBG-10-N}{RGB}{1,102,94}
\definecolor{BrBG-10-10}{RGB}{0,60,48}
\definecolor{BrBG-10-O}{RGB}{0,60,48}
\definecolor{BrBG-11-1}{RGB}{84,48,5}
\definecolor{BrBG-11-A}{RGB}{84,48,5}
\definecolor{BrBG-11-2}{RGB}{140,81,10}
\definecolor{BrBG-11-B}{RGB}{140,81,10}
\definecolor{BrBG-11-3}{RGB}{191,129,45}
\definecolor{BrBG-11-D}{RGB}{191,129,45}
\definecolor{BrBG-11-4}{RGB}{223,194,125}
\definecolor{BrBG-11-F}{RGB}{223,194,125}
\definecolor{BrBG-11-5}{RGB}{246,232,195}
\definecolor{BrBG-11-G}{RGB}{246,232,195}
\definecolor{BrBG-11-6}{RGB}{245,245,245}
\definecolor{BrBG-11-H}{RGB}{245,245,245}
\definecolor{BrBG-11-7}{RGB}{199,234,229}
\definecolor{BrBG-11-I}{RGB}{199,234,229}
\definecolor{BrBG-11-8}{RGB}{128,205,193}
\definecolor{BrBG-11-J}{RGB}{128,205,193}
\definecolor{BrBG-11-9}{RGB}{53,151,143}
\definecolor{BrBG-11-L}{RGB}{53,151,143}
\definecolor{BrBG-11-10}{RGB}{1,102,94}
\definecolor{BrBG-11-N}{RGB}{1,102,94}
\definecolor{BrBG-11-11}{RGB}{0,60,48}
\definecolor{BrBG-11-O}{RGB}{0,60,48}
\definecolor{PRGn-3-1}{RGB}{175,141,195}
\definecolor{PRGn-3-E}{RGB}{175,141,195}
\definecolor{PRGn-3-2}{RGB}{247,247,247}
\definecolor{PRGn-3-H}{RGB}{247,247,247}
\definecolor{PRGn-3-3}{RGB}{127,191,123}
\definecolor{PRGn-3-K}{RGB}{127,191,123}
\definecolor{PRGn-4-1}{RGB}{123,50,148}
\definecolor{PRGn-4-C}{RGB}{123,50,148}
\definecolor{PRGn-4-2}{RGB}{194,165,207}
\definecolor{PRGn-4-F}{RGB}{194,165,207}
\definecolor{PRGn-4-3}{RGB}{166,219,160}
\definecolor{PRGn-4-J}{RGB}{166,219,160}
\definecolor{PRGn-4-4}{RGB}{0,136,55}
\definecolor{PRGn-4-M}{RGB}{0,136,55}
\definecolor{PRGn-5-1}{RGB}{123,50,148}
\definecolor{PRGn-5-C}{RGB}{123,50,148}
\definecolor{PRGn-5-2}{RGB}{194,165,207}
\definecolor{PRGn-5-F}{RGB}{194,165,207}
\definecolor{PRGn-5-3}{RGB}{247,247,247}
\definecolor{PRGn-5-H}{RGB}{247,247,247}
\definecolor{PRGn-5-4}{RGB}{166,219,160}
\definecolor{PRGn-5-J}{RGB}{166,219,160}
\definecolor{PRGn-5-5}{RGB}{0,136,55}
\definecolor{PRGn-5-M}{RGB}{0,136,55}
\definecolor{PRGn-6-1}{RGB}{118,42,131}
\definecolor{PRGn-6-B}{RGB}{118,42,131}
\definecolor{PRGn-6-2}{RGB}{175,141,195}
\definecolor{PRGn-6-E}{RGB}{175,141,195}
\definecolor{PRGn-6-3}{RGB}{231,212,232}
\definecolor{PRGn-6-G}{RGB}{231,212,232}
\definecolor{PRGn-6-4}{RGB}{217,240,211}
\definecolor{PRGn-6-I}{RGB}{217,240,211}
\definecolor{PRGn-6-5}{RGB}{127,191,123}
\definecolor{PRGn-6-K}{RGB}{127,191,123}
\definecolor{PRGn-6-6}{RGB}{27,120,55}
\definecolor{PRGn-6-N}{RGB}{27,120,55}
\definecolor{PRGn-7-1}{RGB}{118,42,131}
\definecolor{PRGn-7-B}{RGB}{118,42,131}
\definecolor{PRGn-7-2}{RGB}{175,141,195}
\definecolor{PRGn-7-E}{RGB}{175,141,195}
\definecolor{PRGn-7-3}{RGB}{231,212,232}
\definecolor{PRGn-7-G}{RGB}{231,212,232}
\definecolor{PRGn-7-4}{RGB}{247,247,247}
\definecolor{PRGn-7-H}{RGB}{247,247,247}
\definecolor{PRGn-7-5}{RGB}{217,240,211}
\definecolor{PRGn-7-I}{RGB}{217,240,211}
\definecolor{PRGn-7-6}{RGB}{127,191,123}
\definecolor{PRGn-7-K}{RGB}{127,191,123}
\definecolor{PRGn-7-7}{RGB}{27,120,55}
\definecolor{PRGn-7-N}{RGB}{27,120,55}
\definecolor{PRGn-8-1}{RGB}{118,42,131}
\definecolor{PRGn-8-B}{RGB}{118,42,131}
\definecolor{PRGn-8-2}{RGB}{153,112,171}
\definecolor{PRGn-8-D}{RGB}{153,112,171}
\definecolor{PRGn-8-3}{RGB}{194,165,207}
\definecolor{PRGn-8-F}{RGB}{194,165,207}
\definecolor{PRGn-8-4}{RGB}{231,212,232}
\definecolor{PRGn-8-G}{RGB}{231,212,232}
\definecolor{PRGn-8-5}{RGB}{217,240,211}
\definecolor{PRGn-8-I}{RGB}{217,240,211}
\definecolor{PRGn-8-6}{RGB}{166,219,160}
\definecolor{PRGn-8-J}{RGB}{166,219,160}
\definecolor{PRGn-8-7}{RGB}{90,174,97}
\definecolor{PRGn-8-L}{RGB}{90,174,97}
\definecolor{PRGn-8-8}{RGB}{27,120,55}
\definecolor{PRGn-8-N}{RGB}{27,120,55}
\definecolor{PRGn-9-1}{RGB}{118,42,131}
\definecolor{PRGn-9-B}{RGB}{118,42,131}
\definecolor{PRGn-9-2}{RGB}{153,112,171}
\definecolor{PRGn-9-D}{RGB}{153,112,171}
\definecolor{PRGn-9-3}{RGB}{194,165,207}
\definecolor{PRGn-9-F}{RGB}{194,165,207}
\definecolor{PRGn-9-4}{RGB}{231,212,232}
\definecolor{PRGn-9-G}{RGB}{231,212,232}
\definecolor{PRGn-9-5}{RGB}{247,247,247}
\definecolor{PRGn-9-H}{RGB}{247,247,247}
\definecolor{PRGn-9-6}{RGB}{217,240,211}
\definecolor{PRGn-9-I}{RGB}{217,240,211}
\definecolor{PRGn-9-7}{RGB}{166,219,160}
\definecolor{PRGn-9-J}{RGB}{166,219,160}
\definecolor{PRGn-9-8}{RGB}{90,174,97}
\definecolor{PRGn-9-L}{RGB}{90,174,97}
\definecolor{PRGn-9-9}{RGB}{27,120,55}
\definecolor{PRGn-9-N}{RGB}{27,120,55}
\definecolor{PRGn-10-1}{RGB}{64,0,75}
\definecolor{PRGn-10-A}{RGB}{64,0,75}
\definecolor{PRGn-10-2}{RGB}{118,42,131}
\definecolor{PRGn-10-B}{RGB}{118,42,131}
\definecolor{PRGn-10-3}{RGB}{153,112,171}
\definecolor{PRGn-10-D}{RGB}{153,112,171}
\definecolor{PRGn-10-4}{RGB}{194,165,207}
\definecolor{PRGn-10-F}{RGB}{194,165,207}
\definecolor{PRGn-10-5}{RGB}{231,212,232}
\definecolor{PRGn-10-G}{RGB}{231,212,232}
\definecolor{PRGn-10-6}{RGB}{217,240,211}
\definecolor{PRGn-10-I}{RGB}{217,240,211}
\definecolor{PRGn-10-7}{RGB}{166,219,160}
\definecolor{PRGn-10-J}{RGB}{166,219,160}
\definecolor{PRGn-10-8}{RGB}{90,174,97}
\definecolor{PRGn-10-L}{RGB}{90,174,97}
\definecolor{PRGn-10-9}{RGB}{27,120,55}
\definecolor{PRGn-10-N}{RGB}{27,120,55}
\definecolor{PRGn-10-10}{RGB}{0,68,27}
\definecolor{PRGn-10-O}{RGB}{0,68,27}
\definecolor{PRGn-11-1}{RGB}{64,0,75}
\definecolor{PRGn-11-A}{RGB}{64,0,75}
\definecolor{PRGn-11-2}{RGB}{118,42,131}
\definecolor{PRGn-11-B}{RGB}{118,42,131}
\definecolor{PRGn-11-3}{RGB}{153,112,171}
\definecolor{PRGn-11-D}{RGB}{153,112,171}
\definecolor{PRGn-11-4}{RGB}{194,165,207}
\definecolor{PRGn-11-F}{RGB}{194,165,207}
\definecolor{PRGn-11-5}{RGB}{231,212,232}
\definecolor{PRGn-11-G}{RGB}{231,212,232}
\definecolor{PRGn-11-6}{RGB}{247,247,247}
\definecolor{PRGn-11-H}{RGB}{247,247,247}
\definecolor{PRGn-11-7}{RGB}{217,240,211}
\definecolor{PRGn-11-I}{RGB}{217,240,211}
\definecolor{PRGn-11-8}{RGB}{166,219,160}
\definecolor{PRGn-11-J}{RGB}{166,219,160}
\definecolor{PRGn-11-9}{RGB}{90,174,97}
\definecolor{PRGn-11-L}{RGB}{90,174,97}
\definecolor{PRGn-11-10}{RGB}{27,120,55}
\definecolor{PRGn-11-N}{RGB}{27,120,55}
\definecolor{PRGn-11-11}{RGB}{0,68,27}
\definecolor{PRGn-11-O}{RGB}{0,68,27}
\definecolor{PiYG-3-1}{RGB}{233,163,201}
\definecolor{PiYG-3-E}{RGB}{233,163,201}
\definecolor{PiYG-3-2}{RGB}{247,247,247}
\definecolor{PiYG-3-H}{RGB}{247,247,247}
\definecolor{PiYG-3-3}{RGB}{161,215,106}
\definecolor{PiYG-3-K}{RGB}{161,215,106}
\definecolor{PiYG-4-1}{RGB}{208,28,139}
\definecolor{PiYG-4-C}{RGB}{208,28,139}
\definecolor{PiYG-4-2}{RGB}{241,182,218}
\definecolor{PiYG-4-F}{RGB}{241,182,218}
\definecolor{PiYG-4-3}{RGB}{184,225,134}
\definecolor{PiYG-4-J}{RGB}{184,225,134}
\definecolor{PiYG-4-4}{RGB}{77,172,38}
\definecolor{PiYG-4-M}{RGB}{77,172,38}
\definecolor{PiYG-5-1}{RGB}{208,28,139}
\definecolor{PiYG-5-C}{RGB}{208,28,139}
\definecolor{PiYG-5-2}{RGB}{241,182,218}
\definecolor{PiYG-5-F}{RGB}{241,182,218}
\definecolor{PiYG-5-3}{RGB}{247,247,247}
\definecolor{PiYG-5-H}{RGB}{247,247,247}
\definecolor{PiYG-5-4}{RGB}{184,225,134}
\definecolor{PiYG-5-J}{RGB}{184,225,134}
\definecolor{PiYG-5-5}{RGB}{77,172,38}
\definecolor{PiYG-5-M}{RGB}{77,172,38}
\definecolor{PiYG-6-1}{RGB}{197,27,125}
\definecolor{PiYG-6-B}{RGB}{197,27,125}
\definecolor{PiYG-6-2}{RGB}{233,163,201}
\definecolor{PiYG-6-E}{RGB}{233,163,201}
\definecolor{PiYG-6-3}{RGB}{253,224,239}
\definecolor{PiYG-6-G}{RGB}{253,224,239}
\definecolor{PiYG-6-4}{RGB}{230,245,208}
\definecolor{PiYG-6-I}{RGB}{230,245,208}
\definecolor{PiYG-6-5}{RGB}{161,215,106}
\definecolor{PiYG-6-K}{RGB}{161,215,106}
\definecolor{PiYG-6-6}{RGB}{77,146,33}
\definecolor{PiYG-6-N}{RGB}{77,146,33}
\definecolor{PiYG-7-1}{RGB}{197,27,125}
\definecolor{PiYG-7-B}{RGB}{197,27,125}
\definecolor{PiYG-7-2}{RGB}{233,163,201}
\definecolor{PiYG-7-E}{RGB}{233,163,201}
\definecolor{PiYG-7-3}{RGB}{253,224,239}
\definecolor{PiYG-7-G}{RGB}{253,224,239}
\definecolor{PiYG-7-4}{RGB}{247,247,247}
\definecolor{PiYG-7-H}{RGB}{247,247,247}
\definecolor{PiYG-7-5}{RGB}{230,245,208}
\definecolor{PiYG-7-I}{RGB}{230,245,208}
\definecolor{PiYG-7-6}{RGB}{161,215,106}
\definecolor{PiYG-7-K}{RGB}{161,215,106}
\definecolor{PiYG-7-7}{RGB}{77,146,33}
\definecolor{PiYG-7-N}{RGB}{77,146,33}
\definecolor{PiYG-8-1}{RGB}{197,27,125}
\definecolor{PiYG-8-B}{RGB}{197,27,125}
\definecolor{PiYG-8-2}{RGB}{222,119,174}
\definecolor{PiYG-8-D}{RGB}{222,119,174}
\definecolor{PiYG-8-3}{RGB}{241,182,218}
\definecolor{PiYG-8-F}{RGB}{241,182,218}
\definecolor{PiYG-8-4}{RGB}{253,224,239}
\definecolor{PiYG-8-G}{RGB}{253,224,239}
\definecolor{PiYG-8-5}{RGB}{230,245,208}
\definecolor{PiYG-8-I}{RGB}{230,245,208}
\definecolor{PiYG-8-6}{RGB}{184,225,134}
\definecolor{PiYG-8-J}{RGB}{184,225,134}
\definecolor{PiYG-8-7}{RGB}{127,188,65}
\definecolor{PiYG-8-L}{RGB}{127,188,65}
\definecolor{PiYG-8-8}{RGB}{77,146,33}
\definecolor{PiYG-8-N}{RGB}{77,146,33}
\definecolor{PiYG-9-1}{RGB}{197,27,125}
\definecolor{PiYG-9-B}{RGB}{197,27,125}
\definecolor{PiYG-9-2}{RGB}{222,119,174}
\definecolor{PiYG-9-D}{RGB}{222,119,174}
\definecolor{PiYG-9-3}{RGB}{241,182,218}
\definecolor{PiYG-9-F}{RGB}{241,182,218}
\definecolor{PiYG-9-4}{RGB}{253,224,239}
\definecolor{PiYG-9-G}{RGB}{253,224,239}
\definecolor{PiYG-9-5}{RGB}{247,247,247}
\definecolor{PiYG-9-H}{RGB}{247,247,247}
\definecolor{PiYG-9-6}{RGB}{230,245,208}
\definecolor{PiYG-9-I}{RGB}{230,245,208}
\definecolor{PiYG-9-7}{RGB}{184,225,134}
\definecolor{PiYG-9-J}{RGB}{184,225,134}
\definecolor{PiYG-9-8}{RGB}{127,188,65}
\definecolor{PiYG-9-L}{RGB}{127,188,65}
\definecolor{PiYG-9-9}{RGB}{77,146,33}
\definecolor{PiYG-9-N}{RGB}{77,146,33}
\definecolor{PiYG-10-1}{RGB}{142,1,82}
\definecolor{PiYG-10-A}{RGB}{142,1,82}
\definecolor{PiYG-10-2}{RGB}{197,27,125}
\definecolor{PiYG-10-B}{RGB}{197,27,125}
\definecolor{PiYG-10-3}{RGB}{222,119,174}
\definecolor{PiYG-10-D}{RGB}{222,119,174}
\definecolor{PiYG-10-4}{RGB}{241,182,218}
\definecolor{PiYG-10-F}{RGB}{241,182,218}
\definecolor{PiYG-10-5}{RGB}{253,224,239}
\definecolor{PiYG-10-G}{RGB}{253,224,239}
\definecolor{PiYG-10-6}{RGB}{230,245,208}
\definecolor{PiYG-10-I}{RGB}{230,245,208}
\definecolor{PiYG-10-7}{RGB}{184,225,134}
\definecolor{PiYG-10-J}{RGB}{184,225,134}
\definecolor{PiYG-10-8}{RGB}{127,188,65}
\definecolor{PiYG-10-L}{RGB}{127,188,65}
\definecolor{PiYG-10-9}{RGB}{77,146,33}
\definecolor{PiYG-10-N}{RGB}{77,146,33}
\definecolor{PiYG-10-10}{RGB}{39,100,25}
\definecolor{PiYG-10-O}{RGB}{39,100,25}
\definecolor{PiYG-11-1}{RGB}{142,1,82}
\definecolor{PiYG-11-A}{RGB}{142,1,82}
\definecolor{PiYG-11-2}{RGB}{197,27,125}
\definecolor{PiYG-11-B}{RGB}{197,27,125}
\definecolor{PiYG-11-3}{RGB}{222,119,174}
\definecolor{PiYG-11-D}{RGB}{222,119,174}
\definecolor{PiYG-11-4}{RGB}{241,182,218}
\definecolor{PiYG-11-F}{RGB}{241,182,218}
\definecolor{PiYG-11-5}{RGB}{253,224,239}
\definecolor{PiYG-11-G}{RGB}{253,224,239}
\definecolor{PiYG-11-6}{RGB}{247,247,247}
\definecolor{PiYG-11-H}{RGB}{247,247,247}
\definecolor{PiYG-11-7}{RGB}{230,245,208}
\definecolor{PiYG-11-I}{RGB}{230,245,208}
\definecolor{PiYG-11-8}{RGB}{184,225,134}
\definecolor{PiYG-11-J}{RGB}{184,225,134}
\definecolor{PiYG-11-9}{RGB}{127,188,65}
\definecolor{PiYG-11-L}{RGB}{127,188,65}
\definecolor{PiYG-11-10}{RGB}{77,146,33}
\definecolor{PiYG-11-N}{RGB}{77,146,33}
\definecolor{PiYG-11-11}{RGB}{39,100,25}
\definecolor{PiYG-11-O}{RGB}{39,100,25}
\definecolor{RdBu-3-1}{RGB}{239,138,98}
\definecolor{RdBu-3-E}{RGB}{239,138,98}
\definecolor{RdBu-3-2}{RGB}{247,247,247}
\definecolor{RdBu-3-H}{RGB}{247,247,247}
\definecolor{RdBu-3-3}{RGB}{103,169,207}
\definecolor{RdBu-3-K}{RGB}{103,169,207}
\definecolor{RdBu-4-1}{RGB}{202,0,32}
\definecolor{RdBu-4-C}{RGB}{202,0,32}
\definecolor{RdBu-4-2}{RGB}{244,165,130}
\definecolor{RdBu-4-F}{RGB}{244,165,130}
\definecolor{RdBu-4-3}{RGB}{146,197,222}
\definecolor{RdBu-4-J}{RGB}{146,197,222}
\definecolor{RdBu-4-4}{RGB}{5,113,176}
\definecolor{RdBu-4-M}{RGB}{5,113,176}
\definecolor{RdBu-5-1}{RGB}{202,0,32}
\definecolor{RdBu-5-C}{RGB}{202,0,32}
\definecolor{RdBu-5-2}{RGB}{244,165,130}
\definecolor{RdBu-5-F}{RGB}{244,165,130}
\definecolor{RdBu-5-3}{RGB}{247,247,247}
\definecolor{RdBu-5-H}{RGB}{247,247,247}
\definecolor{RdBu-5-4}{RGB}{146,197,222}
\definecolor{RdBu-5-J}{RGB}{146,197,222}
\definecolor{RdBu-5-5}{RGB}{5,113,176}
\definecolor{RdBu-5-M}{RGB}{5,113,176}
\definecolor{RdBu-6-1}{RGB}{178,24,43}
\definecolor{RdBu-6-B}{RGB}{178,24,43}
\definecolor{RdBu-6-2}{RGB}{239,138,98}
\definecolor{RdBu-6-E}{RGB}{239,138,98}
\definecolor{RdBu-6-3}{RGB}{253,219,199}
\definecolor{RdBu-6-G}{RGB}{253,219,199}
\definecolor{RdBu-6-4}{RGB}{209,229,240}
\definecolor{RdBu-6-I}{RGB}{209,229,240}
\definecolor{RdBu-6-5}{RGB}{103,169,207}
\definecolor{RdBu-6-K}{RGB}{103,169,207}
\definecolor{RdBu-6-6}{RGB}{33,102,172}
\definecolor{RdBu-6-N}{RGB}{33,102,172}
\definecolor{RdBu-7-1}{RGB}{178,24,43}
\definecolor{RdBu-7-B}{RGB}{178,24,43}
\definecolor{RdBu-7-2}{RGB}{239,138,98}
\definecolor{RdBu-7-E}{RGB}{239,138,98}
\definecolor{RdBu-7-3}{RGB}{253,219,199}
\definecolor{RdBu-7-G}{RGB}{253,219,199}
\definecolor{RdBu-7-4}{RGB}{247,247,247}
\definecolor{RdBu-7-H}{RGB}{247,247,247}
\definecolor{RdBu-7-5}{RGB}{209,229,240}
\definecolor{RdBu-7-I}{RGB}{209,229,240}
\definecolor{RdBu-7-6}{RGB}{103,169,207}
\definecolor{RdBu-7-K}{RGB}{103,169,207}
\definecolor{RdBu-7-7}{RGB}{33,102,172}
\definecolor{RdBu-7-N}{RGB}{33,102,172}
\definecolor{RdBu-8-1}{RGB}{178,24,43}
\definecolor{RdBu-8-B}{RGB}{178,24,43}
\definecolor{RdBu-8-2}{RGB}{214,96,77}
\definecolor{RdBu-8-D}{RGB}{214,96,77}
\definecolor{RdBu-8-3}{RGB}{244,165,130}
\definecolor{RdBu-8-F}{RGB}{244,165,130}
\definecolor{RdBu-8-4}{RGB}{253,219,199}
\definecolor{RdBu-8-G}{RGB}{253,219,199}
\definecolor{RdBu-8-5}{RGB}{209,229,240}
\definecolor{RdBu-8-I}{RGB}{209,229,240}
\definecolor{RdBu-8-6}{RGB}{146,197,222}
\definecolor{RdBu-8-J}{RGB}{146,197,222}
\definecolor{RdBu-8-7}{RGB}{67,147,195}
\definecolor{RdBu-8-L}{RGB}{67,147,195}
\definecolor{RdBu-8-8}{RGB}{33,102,172}
\definecolor{RdBu-8-N}{RGB}{33,102,172}
\definecolor{RdBu-9-1}{RGB}{178,24,43}
\definecolor{RdBu-9-B}{RGB}{178,24,43}
\definecolor{RdBu-9-2}{RGB}{214,96,77}
\definecolor{RdBu-9-D}{RGB}{214,96,77}
\definecolor{RdBu-9-3}{RGB}{244,165,130}
\definecolor{RdBu-9-F}{RGB}{244,165,130}
\definecolor{RdBu-9-4}{RGB}{253,219,199}
\definecolor{RdBu-9-G}{RGB}{253,219,199}
\definecolor{RdBu-9-5}{RGB}{247,247,247}
\definecolor{RdBu-9-H}{RGB}{247,247,247}
\definecolor{RdBu-9-6}{RGB}{209,229,240}
\definecolor{RdBu-9-I}{RGB}{209,229,240}
\definecolor{RdBu-9-7}{RGB}{146,197,222}
\definecolor{RdBu-9-J}{RGB}{146,197,222}
\definecolor{RdBu-9-8}{RGB}{67,147,195}
\definecolor{RdBu-9-L}{RGB}{67,147,195}
\definecolor{RdBu-9-9}{RGB}{33,102,172}
\definecolor{RdBu-9-N}{RGB}{33,102,172}
\definecolor{RdBu-10-1}{RGB}{103,0,31}
\definecolor{RdBu-10-A}{RGB}{103,0,31}
\definecolor{RdBu-10-2}{RGB}{178,24,43}
\definecolor{RdBu-10-B}{RGB}{178,24,43}
\definecolor{RdBu-10-3}{RGB}{214,96,77}
\definecolor{RdBu-10-D}{RGB}{214,96,77}
\definecolor{RdBu-10-4}{RGB}{244,165,130}
\definecolor{RdBu-10-F}{RGB}{244,165,130}
\definecolor{RdBu-10-5}{RGB}{253,219,199}
\definecolor{RdBu-10-G}{RGB}{253,219,199}
\definecolor{RdBu-10-6}{RGB}{209,229,240}
\definecolor{RdBu-10-I}{RGB}{209,229,240}
\definecolor{RdBu-10-7}{RGB}{146,197,222}
\definecolor{RdBu-10-J}{RGB}{146,197,222}
\definecolor{RdBu-10-8}{RGB}{67,147,195}
\definecolor{RdBu-10-L}{RGB}{67,147,195}
\definecolor{RdBu-10-9}{RGB}{33,102,172}
\definecolor{RdBu-10-N}{RGB}{33,102,172}
\definecolor{RdBu-10-10}{RGB}{5,48,97}
\definecolor{RdBu-10-O}{RGB}{5,48,97}
\definecolor{RdBu-11-1}{RGB}{103,0,31}
\definecolor{RdBu-11-A}{RGB}{103,0,31}
\definecolor{RdBu-11-2}{RGB}{178,24,43}
\definecolor{RdBu-11-B}{RGB}{178,24,43}
\definecolor{RdBu-11-3}{RGB}{214,96,77}
\definecolor{RdBu-11-D}{RGB}{214,96,77}
\definecolor{RdBu-11-4}{RGB}{244,165,130}
\definecolor{RdBu-11-F}{RGB}{244,165,130}
\definecolor{RdBu-11-5}{RGB}{253,219,199}
\definecolor{RdBu-11-G}{RGB}{253,219,199}
\definecolor{RdBu-11-6}{RGB}{247,247,247}
\definecolor{RdBu-11-H}{RGB}{247,247,247}
\definecolor{RdBu-11-7}{RGB}{209,229,240}
\definecolor{RdBu-11-I}{RGB}{209,229,240}
\definecolor{RdBu-11-8}{RGB}{146,197,222}
\definecolor{RdBu-11-J}{RGB}{146,197,222}
\definecolor{RdBu-11-9}{RGB}{67,147,195}
\definecolor{RdBu-11-L}{RGB}{67,147,195}
\definecolor{RdBu-11-10}{RGB}{33,102,172}
\definecolor{RdBu-11-N}{RGB}{33,102,172}
\definecolor{RdBu-11-11}{RGB}{5,48,97}
\definecolor{RdBu-11-O}{RGB}{5,48,97}
\definecolor{RdGy-3-1}{RGB}{239,138,98}
\definecolor{RdGy-3-E}{RGB}{239,138,98}
\definecolor{RdGy-3-2}{RGB}{255,255,255}
\definecolor{RdGy-3-H}{RGB}{255,255,255}
\definecolor{RdGy-3-3}{RGB}{153,153,153}
\definecolor{RdGy-3-K}{RGB}{153,153,153}
\definecolor{RdGy-4-1}{RGB}{202,0,32}
\definecolor{RdGy-4-C}{RGB}{202,0,32}
\definecolor{RdGy-4-2}{RGB}{244,165,130}
\definecolor{RdGy-4-F}{RGB}{244,165,130}
\definecolor{RdGy-4-3}{RGB}{186,186,186}
\definecolor{RdGy-4-J}{RGB}{186,186,186}
\definecolor{RdGy-4-4}{RGB}{64,64,64}
\definecolor{RdGy-4-M}{RGB}{64,64,64}
\definecolor{RdGy-5-1}{RGB}{202,0,32}
\definecolor{RdGy-5-C}{RGB}{202,0,32}
\definecolor{RdGy-5-2}{RGB}{244,165,130}
\definecolor{RdGy-5-F}{RGB}{244,165,130}
\definecolor{RdGy-5-3}{RGB}{255,255,255}
\definecolor{RdGy-5-H}{RGB}{255,255,255}
\definecolor{RdGy-5-4}{RGB}{186,186,186}
\definecolor{RdGy-5-J}{RGB}{186,186,186}
\definecolor{RdGy-5-5}{RGB}{64,64,64}
\definecolor{RdGy-5-M}{RGB}{64,64,64}
\definecolor{RdGy-6-1}{RGB}{178,24,43}
\definecolor{RdGy-6-B}{RGB}{178,24,43}
\definecolor{RdGy-6-2}{RGB}{239,138,98}
\definecolor{RdGy-6-E}{RGB}{239,138,98}
\definecolor{RdGy-6-3}{RGB}{253,219,199}
\definecolor{RdGy-6-G}{RGB}{253,219,199}
\definecolor{RdGy-6-4}{RGB}{224,224,224}
\definecolor{RdGy-6-I}{RGB}{224,224,224}
\definecolor{RdGy-6-5}{RGB}{153,153,153}
\definecolor{RdGy-6-K}{RGB}{153,153,153}
\definecolor{RdGy-6-6}{RGB}{77,77,77}
\definecolor{RdGy-6-N}{RGB}{77,77,77}
\definecolor{RdGy-7-1}{RGB}{178,24,43}
\definecolor{RdGy-7-B}{RGB}{178,24,43}
\definecolor{RdGy-7-2}{RGB}{239,138,98}
\definecolor{RdGy-7-E}{RGB}{239,138,98}
\definecolor{RdGy-7-3}{RGB}{253,219,199}
\definecolor{RdGy-7-G}{RGB}{253,219,199}
\definecolor{RdGy-7-4}{RGB}{255,255,255}
\definecolor{RdGy-7-H}{RGB}{255,255,255}
\definecolor{RdGy-7-5}{RGB}{224,224,224}
\definecolor{RdGy-7-I}{RGB}{224,224,224}
\definecolor{RdGy-7-6}{RGB}{153,153,153}
\definecolor{RdGy-7-K}{RGB}{153,153,153}
\definecolor{RdGy-7-7}{RGB}{77,77,77}
\definecolor{RdGy-7-N}{RGB}{77,77,77}
\definecolor{RdGy-8-1}{RGB}{178,24,43}
\definecolor{RdGy-8-B}{RGB}{178,24,43}
\definecolor{RdGy-8-2}{RGB}{214,96,77}
\definecolor{RdGy-8-D}{RGB}{214,96,77}
\definecolor{RdGy-8-3}{RGB}{244,165,130}
\definecolor{RdGy-8-F}{RGB}{244,165,130}
\definecolor{RdGy-8-4}{RGB}{253,219,199}
\definecolor{RdGy-8-G}{RGB}{253,219,199}
\definecolor{RdGy-8-5}{RGB}{224,224,224}
\definecolor{RdGy-8-I}{RGB}{224,224,224}
\definecolor{RdGy-8-6}{RGB}{186,186,186}
\definecolor{RdGy-8-J}{RGB}{186,186,186}
\definecolor{RdGy-8-7}{RGB}{135,135,135}
\definecolor{RdGy-8-L}{RGB}{135,135,135}
\definecolor{RdGy-8-8}{RGB}{77,77,77}
\definecolor{RdGy-8-N}{RGB}{77,77,77}
\definecolor{RdGy-9-1}{RGB}{178,24,43}
\definecolor{RdGy-9-B}{RGB}{178,24,43}
\definecolor{RdGy-9-2}{RGB}{214,96,77}
\definecolor{RdGy-9-D}{RGB}{214,96,77}
\definecolor{RdGy-9-3}{RGB}{244,165,130}
\definecolor{RdGy-9-F}{RGB}{244,165,130}
\definecolor{RdGy-9-4}{RGB}{253,219,199}
\definecolor{RdGy-9-G}{RGB}{253,219,199}
\definecolor{RdGy-9-5}{RGB}{255,255,255}
\definecolor{RdGy-9-H}{RGB}{255,255,255}
\definecolor{RdGy-9-6}{RGB}{224,224,224}
\definecolor{RdGy-9-I}{RGB}{224,224,224}
\definecolor{RdGy-9-7}{RGB}{186,186,186}
\definecolor{RdGy-9-J}{RGB}{186,186,186}
\definecolor{RdGy-9-8}{RGB}{135,135,135}
\definecolor{RdGy-9-L}{RGB}{135,135,135}
\definecolor{RdGy-9-9}{RGB}{77,77,77}
\definecolor{RdGy-9-N}{RGB}{77,77,77}
\definecolor{RdGy-10-1}{RGB}{103,0,31}
\definecolor{RdGy-10-A}{RGB}{103,0,31}
\definecolor{RdGy-10-2}{RGB}{178,24,43}
\definecolor{RdGy-10-B}{RGB}{178,24,43}
\definecolor{RdGy-10-3}{RGB}{214,96,77}
\definecolor{RdGy-10-D}{RGB}{214,96,77}
\definecolor{RdGy-10-4}{RGB}{244,165,130}
\definecolor{RdGy-10-F}{RGB}{244,165,130}
\definecolor{RdGy-10-5}{RGB}{253,219,199}
\definecolor{RdGy-10-G}{RGB}{253,219,199}
\definecolor{RdGy-10-6}{RGB}{224,224,224}
\definecolor{RdGy-10-I}{RGB}{224,224,224}
\definecolor{RdGy-10-7}{RGB}{186,186,186}
\definecolor{RdGy-10-J}{RGB}{186,186,186}
\definecolor{RdGy-10-8}{RGB}{135,135,135}
\definecolor{RdGy-10-L}{RGB}{135,135,135}
\definecolor{RdGy-10-9}{RGB}{77,77,77}
\definecolor{RdGy-10-N}{RGB}{77,77,77}
\definecolor{RdGy-10-10}{RGB}{26,26,26}
\definecolor{RdGy-10-O}{RGB}{26,26,26}
\definecolor{RdGy-11-1}{RGB}{103,0,31}
\definecolor{RdGy-11-A}{RGB}{103,0,31}
\definecolor{RdGy-11-2}{RGB}{178,24,43}
\definecolor{RdGy-11-B}{RGB}{178,24,43}
\definecolor{RdGy-11-3}{RGB}{214,96,77}
\definecolor{RdGy-11-D}{RGB}{214,96,77}
\definecolor{RdGy-11-4}{RGB}{244,165,130}
\definecolor{RdGy-11-F}{RGB}{244,165,130}
\definecolor{RdGy-11-5}{RGB}{253,219,199}
\definecolor{RdGy-11-G}{RGB}{253,219,199}
\definecolor{RdGy-11-6}{RGB}{255,255,255}
\definecolor{RdGy-11-H}{RGB}{255,255,255}
\definecolor{RdGy-11-7}{RGB}{224,224,224}
\definecolor{RdGy-11-I}{RGB}{224,224,224}
\definecolor{RdGy-11-8}{RGB}{186,186,186}
\definecolor{RdGy-11-J}{RGB}{186,186,186}
\definecolor{RdGy-11-9}{RGB}{135,135,135}
\definecolor{RdGy-11-L}{RGB}{135,135,135}
\definecolor{RdGy-11-10}{RGB}{77,77,77}
\definecolor{RdGy-11-N}{RGB}{77,77,77}
\definecolor{RdGy-11-11}{RGB}{26,26,26}
\definecolor{RdGy-11-O}{RGB}{26,26,26}
\definecolor{RdYlBu-3-1}{RGB}{252,141,89}
\definecolor{RdYlBu-3-E}{RGB}{252,141,89}
\definecolor{RdYlBu-3-2}{RGB}{255,255,191}
\definecolor{RdYlBu-3-H}{RGB}{255,255,191}
\definecolor{RdYlBu-3-3}{RGB}{145,191,219}
\definecolor{RdYlBu-3-K}{RGB}{145,191,219}
\definecolor{RdYlBu-4-1}{RGB}{215,25,28}
\definecolor{RdYlBu-4-C}{RGB}{215,25,28}
\definecolor{RdYlBu-4-2}{RGB}{253,174,97}
\definecolor{RdYlBu-4-F}{RGB}{253,174,97}
\definecolor{RdYlBu-4-3}{RGB}{171,217,233}
\definecolor{RdYlBu-4-J}{RGB}{171,217,233}
\definecolor{RdYlBu-4-4}{RGB}{44,123,182}
\definecolor{RdYlBu-4-M}{RGB}{44,123,182}
\definecolor{RdYlBu-5-1}{RGB}{215,25,28}
\definecolor{RdYlBu-5-C}{RGB}{215,25,28}
\definecolor{RdYlBu-5-2}{RGB}{253,174,97}
\definecolor{RdYlBu-5-F}{RGB}{253,174,97}
\definecolor{RdYlBu-5-3}{RGB}{255,255,191}
\definecolor{RdYlBu-5-H}{RGB}{255,255,191}
\definecolor{RdYlBu-5-4}{RGB}{171,217,233}
\definecolor{RdYlBu-5-J}{RGB}{171,217,233}
\definecolor{RdYlBu-5-5}{RGB}{44,123,182}
\definecolor{RdYlBu-5-M}{RGB}{44,123,182}
\definecolor{RdYlBu-6-1}{RGB}{215,48,39}
\definecolor{RdYlBu-6-B}{RGB}{215,48,39}
\definecolor{RdYlBu-6-2}{RGB}{252,141,89}
\definecolor{RdYlBu-6-E}{RGB}{252,141,89}
\definecolor{RdYlBu-6-3}{RGB}{254,224,144}
\definecolor{RdYlBu-6-G}{RGB}{254,224,144}
\definecolor{RdYlBu-6-4}{RGB}{224,243,248}
\definecolor{RdYlBu-6-I}{RGB}{224,243,248}
\definecolor{RdYlBu-6-5}{RGB}{145,191,219}
\definecolor{RdYlBu-6-K}{RGB}{145,191,219}
\definecolor{RdYlBu-6-6}{RGB}{69,117,180}
\definecolor{RdYlBu-6-N}{RGB}{69,117,180}
\definecolor{RdYlBu-7-1}{RGB}{215,48,39}
\definecolor{RdYlBu-7-B}{RGB}{215,48,39}
\definecolor{RdYlBu-7-2}{RGB}{252,141,89}
\definecolor{RdYlBu-7-E}{RGB}{252,141,89}
\definecolor{RdYlBu-7-3}{RGB}{254,224,144}
\definecolor{RdYlBu-7-G}{RGB}{254,224,144}
\definecolor{RdYlBu-7-4}{RGB}{255,255,191}
\definecolor{RdYlBu-7-H}{RGB}{255,255,191}
\definecolor{RdYlBu-7-5}{RGB}{224,243,248}
\definecolor{RdYlBu-7-I}{RGB}{224,243,248}
\definecolor{RdYlBu-7-6}{RGB}{145,191,219}
\definecolor{RdYlBu-7-K}{RGB}{145,191,219}
\definecolor{RdYlBu-7-7}{RGB}{69,117,180}
\definecolor{RdYlBu-7-N}{RGB}{69,117,180}
\definecolor{RdYlBu-8-1}{RGB}{215,48,39}
\definecolor{RdYlBu-8-B}{RGB}{215,48,39}
\definecolor{RdYlBu-8-2}{RGB}{244,109,67}
\definecolor{RdYlBu-8-D}{RGB}{244,109,67}
\definecolor{RdYlBu-8-3}{RGB}{253,174,97}
\definecolor{RdYlBu-8-F}{RGB}{253,174,97}
\definecolor{RdYlBu-8-4}{RGB}{254,224,144}
\definecolor{RdYlBu-8-G}{RGB}{254,224,144}
\definecolor{RdYlBu-8-5}{RGB}{224,243,248}
\definecolor{RdYlBu-8-I}{RGB}{224,243,248}
\definecolor{RdYlBu-8-6}{RGB}{171,217,233}
\definecolor{RdYlBu-8-J}{RGB}{171,217,233}
\definecolor{RdYlBu-8-7}{RGB}{116,173,209}
\definecolor{RdYlBu-8-L}{RGB}{116,173,209}
\definecolor{RdYlBu-8-8}{RGB}{69,117,180}
\definecolor{RdYlBu-8-N}{RGB}{69,117,180}
\definecolor{RdYlBu-9-1}{RGB}{215,48,39}
\definecolor{RdYlBu-9-B}{RGB}{215,48,39}
\definecolor{RdYlBu-9-2}{RGB}{244,109,67}
\definecolor{RdYlBu-9-D}{RGB}{244,109,67}
\definecolor{RdYlBu-9-3}{RGB}{253,174,97}
\definecolor{RdYlBu-9-F}{RGB}{253,174,97}
\definecolor{RdYlBu-9-4}{RGB}{254,224,144}
\definecolor{RdYlBu-9-G}{RGB}{254,224,144}
\definecolor{RdYlBu-9-5}{RGB}{255,255,191}
\definecolor{RdYlBu-9-H}{RGB}{255,255,191}
\definecolor{RdYlBu-9-6}{RGB}{224,243,248}
\definecolor{RdYlBu-9-I}{RGB}{224,243,248}
\definecolor{RdYlBu-9-7}{RGB}{171,217,233}
\definecolor{RdYlBu-9-J}{RGB}{171,217,233}
\definecolor{RdYlBu-9-8}{RGB}{116,173,209}
\definecolor{RdYlBu-9-L}{RGB}{116,173,209}
\definecolor{RdYlBu-9-9}{RGB}{69,117,180}
\definecolor{RdYlBu-9-N}{RGB}{69,117,180}
\definecolor{RdYlBu-10-1}{RGB}{165,0,38}
\definecolor{RdYlBu-10-A}{RGB}{165,0,38}
\definecolor{RdYlBu-10-2}{RGB}{215,48,39}
\definecolor{RdYlBu-10-B}{RGB}{215,48,39}
\definecolor{RdYlBu-10-3}{RGB}{244,109,67}
\definecolor{RdYlBu-10-D}{RGB}{244,109,67}
\definecolor{RdYlBu-10-4}{RGB}{253,174,97}
\definecolor{RdYlBu-10-F}{RGB}{253,174,97}
\definecolor{RdYlBu-10-5}{RGB}{254,224,144}
\definecolor{RdYlBu-10-G}{RGB}{254,224,144}
\definecolor{RdYlBu-10-6}{RGB}{224,243,248}
\definecolor{RdYlBu-10-I}{RGB}{224,243,248}
\definecolor{RdYlBu-10-7}{RGB}{171,217,233}
\definecolor{RdYlBu-10-J}{RGB}{171,217,233}
\definecolor{RdYlBu-10-8}{RGB}{116,173,209}
\definecolor{RdYlBu-10-L}{RGB}{116,173,209}
\definecolor{RdYlBu-10-9}{RGB}{69,117,180}
\definecolor{RdYlBu-10-N}{RGB}{69,117,180}
\definecolor{RdYlBu-10-10}{RGB}{49,54,149}
\definecolor{RdYlBu-10-O}{RGB}{49,54,149}
\definecolor{RdYlBu-11-1}{RGB}{165,0,38}
\definecolor{RdYlBu-11-A}{RGB}{165,0,38}
\definecolor{RdYlBu-11-2}{RGB}{215,48,39}
\definecolor{RdYlBu-11-B}{RGB}{215,48,39}
\definecolor{RdYlBu-11-3}{RGB}{244,109,67}
\definecolor{RdYlBu-11-D}{RGB}{244,109,67}
\definecolor{RdYlBu-11-4}{RGB}{253,174,97}
\definecolor{RdYlBu-11-F}{RGB}{253,174,97}
\definecolor{RdYlBu-11-5}{RGB}{254,224,144}
\definecolor{RdYlBu-11-G}{RGB}{254,224,144}
\definecolor{RdYlBu-11-6}{RGB}{255,255,191}
\definecolor{RdYlBu-11-H}{RGB}{255,255,191}
\definecolor{RdYlBu-11-7}{RGB}{224,243,248}
\definecolor{RdYlBu-11-I}{RGB}{224,243,248}
\definecolor{RdYlBu-11-8}{RGB}{171,217,233}
\definecolor{RdYlBu-11-J}{RGB}{171,217,233}
\definecolor{RdYlBu-11-9}{RGB}{116,173,209}
\definecolor{RdYlBu-11-L}{RGB}{116,173,209}
\definecolor{RdYlBu-11-10}{RGB}{69,117,180}
\definecolor{RdYlBu-11-N}{RGB}{69,117,180}
\definecolor{RdYlBu-11-11}{RGB}{49,54,149}
\definecolor{RdYlBu-11-O}{RGB}{49,54,149}
\definecolor{Spectral-3-1}{RGB}{252,141,89}
\definecolor{Spectral-3-E}{RGB}{252,141,89}
\definecolor{Spectral-3-2}{RGB}{255,255,191}
\definecolor{Spectral-3-H}{RGB}{255,255,191}
\definecolor{Spectral-3-3}{RGB}{153,213,148}
\definecolor{Spectral-3-K}{RGB}{153,213,148}
\definecolor{Spectral-4-1}{RGB}{215,25,28}
\definecolor{Spectral-4-C}{RGB}{215,25,28}
\definecolor{Spectral-4-2}{RGB}{253,174,97}
\definecolor{Spectral-4-F}{RGB}{253,174,97}
\definecolor{Spectral-4-3}{RGB}{171,221,164}
\definecolor{Spectral-4-J}{RGB}{171,221,164}
\definecolor{Spectral-4-4}{RGB}{43,131,186}
\definecolor{Spectral-4-M}{RGB}{43,131,186}
\definecolor{Spectral-5-1}{RGB}{215,25,28}
\definecolor{Spectral-5-C}{RGB}{215,25,28}
\definecolor{Spectral-5-2}{RGB}{253,174,97}
\definecolor{Spectral-5-F}{RGB}{253,174,97}
\definecolor{Spectral-5-3}{RGB}{255,255,191}
\definecolor{Spectral-5-H}{RGB}{255,255,191}
\definecolor{Spectral-5-4}{RGB}{171,221,164}
\definecolor{Spectral-5-J}{RGB}{171,221,164}
\definecolor{Spectral-5-5}{RGB}{43,131,186}
\definecolor{Spectral-5-M}{RGB}{43,131,186}
\definecolor{Spectral-6-1}{RGB}{213,62,79}
\definecolor{Spectral-6-B}{RGB}{213,62,79}
\definecolor{Spectral-6-2}{RGB}{252,141,89}
\definecolor{Spectral-6-E}{RGB}{252,141,89}
\definecolor{Spectral-6-3}{RGB}{254,224,139}
\definecolor{Spectral-6-G}{RGB}{254,224,139}
\definecolor{Spectral-6-4}{RGB}{230,245,152}
\definecolor{Spectral-6-I}{RGB}{230,245,152}
\definecolor{Spectral-6-5}{RGB}{153,213,148}
\definecolor{Spectral-6-K}{RGB}{153,213,148}
\definecolor{Spectral-6-6}{RGB}{50,136,189}
\definecolor{Spectral-6-N}{RGB}{50,136,189}
\definecolor{Spectral-7-1}{RGB}{213,62,79}
\definecolor{Spectral-7-B}{RGB}{213,62,79}
\definecolor{Spectral-7-2}{RGB}{252,141,89}
\definecolor{Spectral-7-E}{RGB}{252,141,89}
\definecolor{Spectral-7-3}{RGB}{254,224,139}
\definecolor{Spectral-7-G}{RGB}{254,224,139}
\definecolor{Spectral-7-4}{RGB}{255,255,191}
\definecolor{Spectral-7-H}{RGB}{255,255,191}
\definecolor{Spectral-7-5}{RGB}{230,245,152}
\definecolor{Spectral-7-I}{RGB}{230,245,152}
\definecolor{Spectral-7-6}{RGB}{153,213,148}
\definecolor{Spectral-7-K}{RGB}{153,213,148}
\definecolor{Spectral-7-7}{RGB}{50,136,189}
\definecolor{Spectral-7-N}{RGB}{50,136,189}
\definecolor{Spectral-8-1}{RGB}{213,62,79}
\definecolor{Spectral-8-B}{RGB}{213,62,79}
\definecolor{Spectral-8-2}{RGB}{244,109,67}
\definecolor{Spectral-8-D}{RGB}{244,109,67}
\definecolor{Spectral-8-3}{RGB}{253,174,97}
\definecolor{Spectral-8-F}{RGB}{253,174,97}
\definecolor{Spectral-8-4}{RGB}{254,224,139}
\definecolor{Spectral-8-G}{RGB}{254,224,139}
\definecolor{Spectral-8-5}{RGB}{230,245,152}
\definecolor{Spectral-8-I}{RGB}{230,245,152}
\definecolor{Spectral-8-6}{RGB}{171,221,164}
\definecolor{Spectral-8-J}{RGB}{171,221,164}
\definecolor{Spectral-8-7}{RGB}{102,194,165}
\definecolor{Spectral-8-L}{RGB}{102,194,165}
\definecolor{Spectral-8-8}{RGB}{50,136,189}
\definecolor{Spectral-8-N}{RGB}{50,136,189}
\definecolor{Spectral-9-1}{RGB}{213,62,79}
\definecolor{Spectral-9-B}{RGB}{213,62,79}
\definecolor{Spectral-9-2}{RGB}{244,109,67}
\definecolor{Spectral-9-D}{RGB}{244,109,67}
\definecolor{Spectral-9-3}{RGB}{253,174,97}
\definecolor{Spectral-9-F}{RGB}{253,174,97}
\definecolor{Spectral-9-4}{RGB}{254,224,139}
\definecolor{Spectral-9-G}{RGB}{254,224,139}
\definecolor{Spectral-9-5}{RGB}{255,255,191}
\definecolor{Spectral-9-H}{RGB}{255,255,191}
\definecolor{Spectral-9-6}{RGB}{230,245,152}
\definecolor{Spectral-9-I}{RGB}{230,245,152}
\definecolor{Spectral-9-7}{RGB}{171,221,164}
\definecolor{Spectral-9-J}{RGB}{171,221,164}
\definecolor{Spectral-9-8}{RGB}{102,194,165}
\definecolor{Spectral-9-L}{RGB}{102,194,165}
\definecolor{Spectral-9-9}{RGB}{50,136,189}
\definecolor{Spectral-9-N}{RGB}{50,136,189}
\definecolor{Spectral-10-1}{RGB}{158,1,66}
\definecolor{Spectral-10-A}{RGB}{158,1,66}
\definecolor{Spectral-10-2}{RGB}{213,62,79}
\definecolor{Spectral-10-B}{RGB}{213,62,79}
\definecolor{Spectral-10-3}{RGB}{244,109,67}
\definecolor{Spectral-10-D}{RGB}{244,109,67}
\definecolor{Spectral-10-4}{RGB}{253,174,97}
\definecolor{Spectral-10-F}{RGB}{253,174,97}
\definecolor{Spectral-10-5}{RGB}{254,224,139}
\definecolor{Spectral-10-G}{RGB}{254,224,139}
\definecolor{Spectral-10-6}{RGB}{230,245,152}
\definecolor{Spectral-10-I}{RGB}{230,245,152}
\definecolor{Spectral-10-7}{RGB}{171,221,164}
\definecolor{Spectral-10-J}{RGB}{171,221,164}
\definecolor{Spectral-10-8}{RGB}{102,194,165}
\definecolor{Spectral-10-L}{RGB}{102,194,165}
\definecolor{Spectral-10-9}{RGB}{50,136,189}
\definecolor{Spectral-10-N}{RGB}{50,136,189}
\definecolor{Spectral-10-10}{RGB}{94,79,162}
\definecolor{Spectral-10-O}{RGB}{94,79,162}
\definecolor{Spectral-11-1}{RGB}{158,1,66}
\definecolor{Spectral-11-A}{RGB}{158,1,66}
\definecolor{Spectral-11-2}{RGB}{213,62,79}
\definecolor{Spectral-11-B}{RGB}{213,62,79}
\definecolor{Spectral-11-3}{RGB}{244,109,67}
\definecolor{Spectral-11-D}{RGB}{244,109,67}
\definecolor{Spectral-11-4}{RGB}{253,174,97}
\definecolor{Spectral-11-F}{RGB}{253,174,97}
\definecolor{Spectral-11-5}{RGB}{254,224,139}
\definecolor{Spectral-11-G}{RGB}{254,224,139}
\definecolor{Spectral-11-6}{RGB}{255,255,191}
\definecolor{Spectral-11-H}{RGB}{255,255,191}
\definecolor{Spectral-11-7}{RGB}{230,245,152}
\definecolor{Spectral-11-I}{RGB}{230,245,152}
\definecolor{Spectral-11-8}{RGB}{171,221,164}
\definecolor{Spectral-11-J}{RGB}{171,221,164}
\definecolor{Spectral-11-9}{RGB}{102,194,165}
\definecolor{Spectral-11-L}{RGB}{102,194,165}
\definecolor{Spectral-11-10}{RGB}{50,136,189}
\definecolor{Spectral-11-N}{RGB}{50,136,189}
\definecolor{Spectral-11-11}{RGB}{94,79,162}
\definecolor{Spectral-11-O}{RGB}{94,79,162}
\definecolor{RdYlGn-3-1}{RGB}{252,141,89}
\definecolor{RdYlGn-3-E}{RGB}{252,141,89}
\definecolor{RdYlGn-3-2}{RGB}{255,255,191}
\definecolor{RdYlGn-3-H}{RGB}{255,255,191}
\definecolor{RdYlGn-3-3}{RGB}{145,207,96}
\definecolor{RdYlGn-3-K}{RGB}{145,207,96}
\definecolor{RdYlGn-4-1}{RGB}{215,25,28}
\definecolor{RdYlGn-4-C}{RGB}{215,25,28}
\definecolor{RdYlGn-4-2}{RGB}{253,174,97}
\definecolor{RdYlGn-4-F}{RGB}{253,174,97}
\definecolor{RdYlGn-4-3}{RGB}{166,217,106}
\definecolor{RdYlGn-4-J}{RGB}{166,217,106}
\definecolor{RdYlGn-4-4}{RGB}{26,150,65}
\definecolor{RdYlGn-4-M}{RGB}{26,150,65}
\definecolor{RdYlGn-5-1}{RGB}{215,25,28}
\definecolor{RdYlGn-5-C}{RGB}{215,25,28}
\definecolor{RdYlGn-5-2}{RGB}{253,174,97}
\definecolor{RdYlGn-5-F}{RGB}{253,174,97}
\definecolor{RdYlGn-5-3}{RGB}{255,255,191}
\definecolor{RdYlGn-5-H}{RGB}{255,255,191}
\definecolor{RdYlGn-5-4}{RGB}{166,217,106}
\definecolor{RdYlGn-5-J}{RGB}{166,217,106}
\definecolor{RdYlGn-5-5}{RGB}{26,150,65}
\definecolor{RdYlGn-5-M}{RGB}{26,150,65}
\definecolor{RdYlGn-6-1}{RGB}{215,48,39}
\definecolor{RdYlGn-6-B}{RGB}{215,48,39}
\definecolor{RdYlGn-6-2}{RGB}{252,141,89}
\definecolor{RdYlGn-6-E}{RGB}{252,141,89}
\definecolor{RdYlGn-6-3}{RGB}{254,224,139}
\definecolor{RdYlGn-6-G}{RGB}{254,224,139}
\definecolor{RdYlGn-6-4}{RGB}{217,239,139}
\definecolor{RdYlGn-6-I}{RGB}{217,239,139}
\definecolor{RdYlGn-6-5}{RGB}{145,207,96}
\definecolor{RdYlGn-6-K}{RGB}{145,207,96}
\definecolor{RdYlGn-6-6}{RGB}{26,152,80}
\definecolor{RdYlGn-6-N}{RGB}{26,152,80}
\definecolor{RdYlGn-7-1}{RGB}{215,48,39}
\definecolor{RdYlGn-7-B}{RGB}{215,48,39}
\definecolor{RdYlGn-7-2}{RGB}{252,141,89}
\definecolor{RdYlGn-7-E}{RGB}{252,141,89}
\definecolor{RdYlGn-7-3}{RGB}{254,224,139}
\definecolor{RdYlGn-7-G}{RGB}{254,224,139}
\definecolor{RdYlGn-7-4}{RGB}{255,255,191}
\definecolor{RdYlGn-7-H}{RGB}{255,255,191}
\definecolor{RdYlGn-7-5}{RGB}{217,239,139}
\definecolor{RdYlGn-7-I}{RGB}{217,239,139}
\definecolor{RdYlGn-7-6}{RGB}{145,207,96}
\definecolor{RdYlGn-7-K}{RGB}{145,207,96}
\definecolor{RdYlGn-7-7}{RGB}{26,152,80}
\definecolor{RdYlGn-7-N}{RGB}{26,152,80}
\definecolor{RdYlGn-8-1}{RGB}{215,48,39}
\definecolor{RdYlGn-8-B}{RGB}{215,48,39}
\definecolor{RdYlGn-8-2}{RGB}{244,109,67}
\definecolor{RdYlGn-8-D}{RGB}{244,109,67}
\definecolor{RdYlGn-8-3}{RGB}{253,174,97}
\definecolor{RdYlGn-8-F}{RGB}{253,174,97}
\definecolor{RdYlGn-8-4}{RGB}{254,224,139}
\definecolor{RdYlGn-8-G}{RGB}{254,224,139}
\definecolor{RdYlGn-8-5}{RGB}{217,239,139}
\definecolor{RdYlGn-8-I}{RGB}{217,239,139}
\definecolor{RdYlGn-8-6}{RGB}{166,217,106}
\definecolor{RdYlGn-8-J}{RGB}{166,217,106}
\definecolor{RdYlGn-8-7}{RGB}{102,189,99}
\definecolor{RdYlGn-8-L}{RGB}{102,189,99}
\definecolor{RdYlGn-8-8}{RGB}{26,152,80}
\definecolor{RdYlGn-8-N}{RGB}{26,152,80}
\definecolor{RdYlGn-9-1}{RGB}{215,48,39}
\definecolor{RdYlGn-9-B}{RGB}{215,48,39}
\definecolor{RdYlGn-9-2}{RGB}{244,109,67}
\definecolor{RdYlGn-9-D}{RGB}{244,109,67}
\definecolor{RdYlGn-9-3}{RGB}{253,174,97}
\definecolor{RdYlGn-9-F}{RGB}{253,174,97}
\definecolor{RdYlGn-9-4}{RGB}{254,224,139}
\definecolor{RdYlGn-9-G}{RGB}{254,224,139}
\definecolor{RdYlGn-9-5}{RGB}{255,255,191}
\definecolor{RdYlGn-9-H}{RGB}{255,255,191}
\definecolor{RdYlGn-9-6}{RGB}{217,239,139}
\definecolor{RdYlGn-9-I}{RGB}{217,239,139}
\definecolor{RdYlGn-9-7}{RGB}{166,217,106}
\definecolor{RdYlGn-9-J}{RGB}{166,217,106}
\definecolor{RdYlGn-9-8}{RGB}{102,189,99}
\definecolor{RdYlGn-9-L}{RGB}{102,189,99}
\definecolor{RdYlGn-9-9}{RGB}{26,152,80}
\definecolor{RdYlGn-9-N}{RGB}{26,152,80}
\definecolor{RdYlGn-10-1}{RGB}{165,0,38}
\definecolor{RdYlGn-10-A}{RGB}{165,0,38}
\definecolor{RdYlGn-10-2}{RGB}{215,48,39}
\definecolor{RdYlGn-10-B}{RGB}{215,48,39}
\definecolor{RdYlGn-10-3}{RGB}{244,109,67}
\definecolor{RdYlGn-10-D}{RGB}{244,109,67}
\definecolor{RdYlGn-10-4}{RGB}{253,174,97}
\definecolor{RdYlGn-10-F}{RGB}{253,174,97}
\definecolor{RdYlGn-10-5}{RGB}{254,224,139}
\definecolor{RdYlGn-10-G}{RGB}{254,224,139}
\definecolor{RdYlGn-10-6}{RGB}{217,239,139}
\definecolor{RdYlGn-10-I}{RGB}{217,239,139}
\definecolor{RdYlGn-10-7}{RGB}{166,217,106}
\definecolor{RdYlGn-10-J}{RGB}{166,217,106}
\definecolor{RdYlGn-10-8}{RGB}{102,189,99}
\definecolor{RdYlGn-10-L}{RGB}{102,189,99}
\definecolor{RdYlGn-10-9}{RGB}{26,152,80}
\definecolor{RdYlGn-10-N}{RGB}{26,152,80}
\definecolor{RdYlGn-10-10}{RGB}{0,104,55}
\definecolor{RdYlGn-10-O}{RGB}{0,104,55}
\definecolor{RdYlGn-11-1}{RGB}{165,0,38}
\definecolor{RdYlGn-11-A}{RGB}{165,0,38}
\definecolor{RdYlGn-11-2}{RGB}{215,48,39}
\definecolor{RdYlGn-11-B}{RGB}{215,48,39}
\definecolor{RdYlGn-11-3}{RGB}{244,109,67}
\definecolor{RdYlGn-11-D}{RGB}{244,109,67}
\definecolor{RdYlGn-11-4}{RGB}{253,174,97}
\definecolor{RdYlGn-11-F}{RGB}{253,174,97}
\definecolor{RdYlGn-11-5}{RGB}{254,224,139}
\definecolor{RdYlGn-11-G}{RGB}{254,224,139}
\definecolor{RdYlGn-11-6}{RGB}{255,255,191}
\definecolor{RdYlGn-11-H}{RGB}{255,255,191}
\definecolor{RdYlGn-11-7}{RGB}{217,239,139}
\definecolor{RdYlGn-11-I}{RGB}{217,239,139}
\definecolor{RdYlGn-11-8}{RGB}{166,217,106}
\definecolor{RdYlGn-11-J}{RGB}{166,217,106}
\definecolor{RdYlGn-11-9}{RGB}{102,189,99}
\definecolor{RdYlGn-11-L}{RGB}{102,189,99}
\definecolor{RdYlGn-11-10}{RGB}{26,152,80}
\definecolor{RdYlGn-11-N}{RGB}{26,152,80}
\definecolor{RdYlGn-11-11}{RGB}{0,104,55}
\definecolor{RdYlGn-11-O}{RGB}{0,104,55}
\definecolor{Set3-3-1}{RGB}{141,211,199}
\definecolor{Set3-3-A}{RGB}{141,211,199}
\definecolor{Set3-3-2}{RGB}{255,255,179}
\definecolor{Set3-3-B}{RGB}{255,255,179}
\definecolor{Set3-3-3}{RGB}{190,186,218}
\definecolor{Set3-3-C}{RGB}{190,186,218}
\definecolor{Set3-4-1}{RGB}{141,211,199}
\definecolor{Set3-4-A}{RGB}{141,211,199}
\definecolor{Set3-4-2}{RGB}{255,255,179}
\definecolor{Set3-4-B}{RGB}{255,255,179}
\definecolor{Set3-4-3}{RGB}{190,186,218}
\definecolor{Set3-4-C}{RGB}{190,186,218}
\definecolor{Set3-4-4}{RGB}{251,128,114}
\definecolor{Set3-4-D}{RGB}{251,128,114}
\definecolor{Set3-5-1}{RGB}{141,211,199}
\definecolor{Set3-5-A}{RGB}{141,211,199}
\definecolor{Set3-5-2}{RGB}{255,255,179}
\definecolor{Set3-5-B}{RGB}{255,255,179}
\definecolor{Set3-5-3}{RGB}{190,186,218}
\definecolor{Set3-5-C}{RGB}{190,186,218}
\definecolor{Set3-5-4}{RGB}{251,128,114}
\definecolor{Set3-5-D}{RGB}{251,128,114}
\definecolor{Set3-5-5}{RGB}{128,177,211}
\definecolor{Set3-5-E}{RGB}{128,177,211}
\definecolor{Set3-6-1}{RGB}{141,211,199}
\definecolor{Set3-6-A}{RGB}{141,211,199}
\definecolor{Set3-6-2}{RGB}{255,255,179}
\definecolor{Set3-6-B}{RGB}{255,255,179}
\definecolor{Set3-6-3}{RGB}{190,186,218}
\definecolor{Set3-6-C}{RGB}{190,186,218}
\definecolor{Set3-6-4}{RGB}{251,128,114}
\definecolor{Set3-6-D}{RGB}{251,128,114}
\definecolor{Set3-6-5}{RGB}{128,177,211}
\definecolor{Set3-6-E}{RGB}{128,177,211}
\definecolor{Set3-6-6}{RGB}{253,180,98}
\definecolor{Set3-6-F}{RGB}{253,180,98}
\definecolor{Set3-7-1}{RGB}{141,211,199}
\definecolor{Set3-7-A}{RGB}{141,211,199}
\definecolor{Set3-7-2}{RGB}{255,255,179}
\definecolor{Set3-7-B}{RGB}{255,255,179}
\definecolor{Set3-7-3}{RGB}{190,186,218}
\definecolor{Set3-7-C}{RGB}{190,186,218}
\definecolor{Set3-7-4}{RGB}{251,128,114}
\definecolor{Set3-7-D}{RGB}{251,128,114}
\definecolor{Set3-7-5}{RGB}{128,177,211}
\definecolor{Set3-7-E}{RGB}{128,177,211}
\definecolor{Set3-7-6}{RGB}{253,180,98}
\definecolor{Set3-7-F}{RGB}{253,180,98}
\definecolor{Set3-7-7}{RGB}{179,222,105}
\definecolor{Set3-7-G}{RGB}{179,222,105}
\definecolor{Set3-8-1}{RGB}{141,211,199}
\definecolor{Set3-8-A}{RGB}{141,211,199}
\definecolor{Set3-8-2}{RGB}{255,255,179}
\definecolor{Set3-8-B}{RGB}{255,255,179}
\definecolor{Set3-8-3}{RGB}{190,186,218}
\definecolor{Set3-8-C}{RGB}{190,186,218}
\definecolor{Set3-8-4}{RGB}{251,128,114}
\definecolor{Set3-8-D}{RGB}{251,128,114}
\definecolor{Set3-8-5}{RGB}{128,177,211}
\definecolor{Set3-8-E}{RGB}{128,177,211}
\definecolor{Set3-8-6}{RGB}{253,180,98}
\definecolor{Set3-8-F}{RGB}{253,180,98}
\definecolor{Set3-8-7}{RGB}{179,222,105}
\definecolor{Set3-8-G}{RGB}{179,222,105}
\definecolor{Set3-8-8}{RGB}{252,205,229}
\definecolor{Set3-8-H}{RGB}{252,205,229}
\definecolor{Set3-9-1}{RGB}{141,211,199}
\definecolor{Set3-9-A}{RGB}{141,211,199}
\definecolor{Set3-9-2}{RGB}{255,255,179}
\definecolor{Set3-9-B}{RGB}{255,255,179}
\definecolor{Set3-9-3}{RGB}{190,186,218}
\definecolor{Set3-9-C}{RGB}{190,186,218}
\definecolor{Set3-9-4}{RGB}{251,128,114}
\definecolor{Set3-9-D}{RGB}{251,128,114}
\definecolor{Set3-9-5}{RGB}{128,177,211}
\definecolor{Set3-9-E}{RGB}{128,177,211}
\definecolor{Set3-9-6}{RGB}{253,180,98}
\definecolor{Set3-9-F}{RGB}{253,180,98}
\definecolor{Set3-9-7}{RGB}{179,222,105}
\definecolor{Set3-9-G}{RGB}{179,222,105}
\definecolor{Set3-9-8}{RGB}{252,205,229}
\definecolor{Set3-9-H}{RGB}{252,205,229}
\definecolor{Set3-9-9}{RGB}{217,217,217}
\definecolor{Set3-9-I}{RGB}{217,217,217}
\definecolor{Set3-10-1}{RGB}{141,211,199}
\definecolor{Set3-10-A}{RGB}{141,211,199}
\definecolor{Set3-10-2}{RGB}{255,255,179}
\definecolor{Set3-10-B}{RGB}{255,255,179}
\definecolor{Set3-10-3}{RGB}{190,186,218}
\definecolor{Set3-10-C}{RGB}{190,186,218}
\definecolor{Set3-10-4}{RGB}{251,128,114}
\definecolor{Set3-10-D}{RGB}{251,128,114}
\definecolor{Set3-10-5}{RGB}{128,177,211}
\definecolor{Set3-10-E}{RGB}{128,177,211}
\definecolor{Set3-10-6}{RGB}{253,180,98}
\definecolor{Set3-10-F}{RGB}{253,180,98}
\definecolor{Set3-10-7}{RGB}{179,222,105}
\definecolor{Set3-10-G}{RGB}{179,222,105}
\definecolor{Set3-10-8}{RGB}{252,205,229}
\definecolor{Set3-10-H}{RGB}{252,205,229}
\definecolor{Set3-10-9}{RGB}{217,217,217}
\definecolor{Set3-10-I}{RGB}{217,217,217}
\definecolor{Set3-10-10}{RGB}{188,128,189}
\definecolor{Set3-10-J}{RGB}{188,128,189}
\definecolor{Set3-11-1}{RGB}{141,211,199}
\definecolor{Set3-11-A}{RGB}{141,211,199}
\definecolor{Set3-11-2}{RGB}{255,255,179}
\definecolor{Set3-11-B}{RGB}{255,255,179}
\definecolor{Set3-11-3}{RGB}{190,186,218}
\definecolor{Set3-11-C}{RGB}{190,186,218}
\definecolor{Set3-11-4}{RGB}{251,128,114}
\definecolor{Set3-11-D}{RGB}{251,128,114}
\definecolor{Set3-11-5}{RGB}{128,177,211}
\definecolor{Set3-11-E}{RGB}{128,177,211}
\definecolor{Set3-11-6}{RGB}{253,180,98}
\definecolor{Set3-11-F}{RGB}{253,180,98}
\definecolor{Set3-11-7}{RGB}{179,222,105}
\definecolor{Set3-11-G}{RGB}{179,222,105}
\definecolor{Set3-11-8}{RGB}{252,205,229}
\definecolor{Set3-11-H}{RGB}{252,205,229}
\definecolor{Set3-11-9}{RGB}{217,217,217}
\definecolor{Set3-11-I}{RGB}{217,217,217}
\definecolor{Set3-11-10}{RGB}{188,128,189}
\definecolor{Set3-11-J}{RGB}{188,128,189}
\definecolor{Set3-11-11}{RGB}{204,235,197}
\definecolor{Set3-11-K}{RGB}{204,235,197}
\definecolor{Set3-12-1}{RGB}{141,211,199}
\definecolor{Set3-12-A}{RGB}{141,211,199}
\definecolor{Set3-12-2}{RGB}{255,255,179}
\definecolor{Set3-12-B}{RGB}{255,255,179}
\definecolor{Set3-12-3}{RGB}{190,186,218}
\definecolor{Set3-12-C}{RGB}{190,186,218}
\definecolor{Set3-12-4}{RGB}{251,128,114}
\definecolor{Set3-12-D}{RGB}{251,128,114}
\definecolor{Set3-12-5}{RGB}{128,177,211}
\definecolor{Set3-12-E}{RGB}{128,177,211}
\definecolor{Set3-12-6}{RGB}{253,180,98}
\definecolor{Set3-12-F}{RGB}{253,180,98}
\definecolor{Set3-12-7}{RGB}{179,222,105}
\definecolor{Set3-12-G}{RGB}{179,222,105}
\definecolor{Set3-12-8}{RGB}{252,205,229}
\definecolor{Set3-12-H}{RGB}{252,205,229}
\definecolor{Set3-12-9}{RGB}{217,217,217}
\definecolor{Set3-12-I}{RGB}{217,217,217}
\definecolor{Set3-12-10}{RGB}{188,128,189}
\definecolor{Set3-12-J}{RGB}{188,128,189}
\definecolor{Set3-12-11}{RGB}{204,235,197}
\definecolor{Set3-12-K}{RGB}{204,235,197}
\definecolor{Set3-12-12}{RGB}{255,237,111}
\definecolor{Set3-12-L}{RGB}{255,237,111}
\definecolor{Pastel1-3-1}{RGB}{251,180,174}
\definecolor{Pastel1-3-A}{RGB}{251,180,174}
\definecolor{Pastel1-3-2}{RGB}{179,205,227}
\definecolor{Pastel1-3-B}{RGB}{179,205,227}
\definecolor{Pastel1-3-3}{RGB}{204,235,197}
\definecolor{Pastel1-3-C}{RGB}{204,235,197}
\definecolor{Pastel1-4-1}{RGB}{251,180,174}
\definecolor{Pastel1-4-A}{RGB}{251,180,174}
\definecolor{Pastel1-4-2}{RGB}{179,205,227}
\definecolor{Pastel1-4-B}{RGB}{179,205,227}
\definecolor{Pastel1-4-3}{RGB}{204,235,197}
\definecolor{Pastel1-4-C}{RGB}{204,235,197}
\definecolor{Pastel1-4-4}{RGB}{222,203,228}
\definecolor{Pastel1-4-D}{RGB}{222,203,228}
\definecolor{Pastel1-5-1}{RGB}{251,180,174}
\definecolor{Pastel1-5-A}{RGB}{251,180,174}
\definecolor{Pastel1-5-2}{RGB}{179,205,227}
\definecolor{Pastel1-5-B}{RGB}{179,205,227}
\definecolor{Pastel1-5-3}{RGB}{204,235,197}
\definecolor{Pastel1-5-C}{RGB}{204,235,197}
\definecolor{Pastel1-5-4}{RGB}{222,203,228}
\definecolor{Pastel1-5-D}{RGB}{222,203,228}
\definecolor{Pastel1-5-5}{RGB}{254,217,166}
\definecolor{Pastel1-5-E}{RGB}{254,217,166}
\definecolor{Pastel1-6-1}{RGB}{251,180,174}
\definecolor{Pastel1-6-A}{RGB}{251,180,174}
\definecolor{Pastel1-6-2}{RGB}{179,205,227}
\definecolor{Pastel1-6-B}{RGB}{179,205,227}
\definecolor{Pastel1-6-3}{RGB}{204,235,197}
\definecolor{Pastel1-6-C}{RGB}{204,235,197}
\definecolor{Pastel1-6-4}{RGB}{222,203,228}
\definecolor{Pastel1-6-D}{RGB}{222,203,228}
\definecolor{Pastel1-6-5}{RGB}{254,217,166}
\definecolor{Pastel1-6-E}{RGB}{254,217,166}
\definecolor{Pastel1-6-6}{RGB}{255,255,204}
\definecolor{Pastel1-6-F}{RGB}{255,255,204}
\definecolor{Pastel1-7-1}{RGB}{251,180,174}
\definecolor{Pastel1-7-A}{RGB}{251,180,174}
\definecolor{Pastel1-7-2}{RGB}{179,205,227}
\definecolor{Pastel1-7-B}{RGB}{179,205,227}
\definecolor{Pastel1-7-3}{RGB}{204,235,197}
\definecolor{Pastel1-7-C}{RGB}{204,235,197}
\definecolor{Pastel1-7-4}{RGB}{222,203,228}
\definecolor{Pastel1-7-D}{RGB}{222,203,228}
\definecolor{Pastel1-7-5}{RGB}{254,217,166}
\definecolor{Pastel1-7-E}{RGB}{254,217,166}
\definecolor{Pastel1-7-6}{RGB}{255,255,204}
\definecolor{Pastel1-7-F}{RGB}{255,255,204}
\definecolor{Pastel1-7-7}{RGB}{229,216,189}
\definecolor{Pastel1-7-G}{RGB}{229,216,189}
\definecolor{Pastel1-8-1}{RGB}{251,180,174}
\definecolor{Pastel1-8-A}{RGB}{251,180,174}
\definecolor{Pastel1-8-2}{RGB}{179,205,227}
\definecolor{Pastel1-8-B}{RGB}{179,205,227}
\definecolor{Pastel1-8-3}{RGB}{204,235,197}
\definecolor{Pastel1-8-C}{RGB}{204,235,197}
\definecolor{Pastel1-8-4}{RGB}{222,203,228}
\definecolor{Pastel1-8-D}{RGB}{222,203,228}
\definecolor{Pastel1-8-5}{RGB}{254,217,166}
\definecolor{Pastel1-8-E}{RGB}{254,217,166}
\definecolor{Pastel1-8-6}{RGB}{255,255,204}
\definecolor{Pastel1-8-F}{RGB}{255,255,204}
\definecolor{Pastel1-8-7}{RGB}{229,216,189}
\definecolor{Pastel1-8-G}{RGB}{229,216,189}
\definecolor{Pastel1-8-8}{RGB}{253,218,236}
\definecolor{Pastel1-8-H}{RGB}{253,218,236}
\definecolor{Pastel1-9-1}{RGB}{251,180,174}
\definecolor{Pastel1-9-A}{RGB}{251,180,174}
\definecolor{Pastel1-9-2}{RGB}{179,205,227}
\definecolor{Pastel1-9-B}{RGB}{179,205,227}
\definecolor{Pastel1-9-3}{RGB}{204,235,197}
\definecolor{Pastel1-9-C}{RGB}{204,235,197}
\definecolor{Pastel1-9-4}{RGB}{222,203,228}
\definecolor{Pastel1-9-D}{RGB}{222,203,228}
\definecolor{Pastel1-9-5}{RGB}{254,217,166}
\definecolor{Pastel1-9-E}{RGB}{254,217,166}
\definecolor{Pastel1-9-6}{RGB}{255,255,204}
\definecolor{Pastel1-9-F}{RGB}{255,255,204}
\definecolor{Pastel1-9-7}{RGB}{229,216,189}
\definecolor{Pastel1-9-G}{RGB}{229,216,189}
\definecolor{Pastel1-9-8}{RGB}{253,218,236}
\definecolor{Pastel1-9-H}{RGB}{253,218,236}
\definecolor{Pastel1-9-9}{RGB}{242,242,242}
\definecolor{Pastel1-9-I}{RGB}{242,242,242}
\definecolor{Set1-3-1}{RGB}{228,26,28}
\definecolor{Set1-3-A}{RGB}{228,26,28}
\definecolor{Set1-3-2}{RGB}{55,126,184}
\definecolor{Set1-3-B}{RGB}{55,126,184}
\definecolor{Set1-3-3}{RGB}{77,175,74}
\definecolor{Set1-3-C}{RGB}{77,175,74}
\definecolor{Set1-4-1}{RGB}{228,26,28}
\definecolor{Set1-4-A}{RGB}{228,26,28}
\definecolor{Set1-4-2}{RGB}{55,126,184}
\definecolor{Set1-4-B}{RGB}{55,126,184}
\definecolor{Set1-4-3}{RGB}{77,175,74}
\definecolor{Set1-4-C}{RGB}{77,175,74}
\definecolor{Set1-4-4}{RGB}{152,78,163}
\definecolor{Set1-4-D}{RGB}{152,78,163}
\definecolor{Set1-5-1}{RGB}{228,26,28}
\definecolor{Set1-5-A}{RGB}{228,26,28}
\definecolor{Set1-5-2}{RGB}{55,126,184}
\definecolor{Set1-5-B}{RGB}{55,126,184}
\definecolor{Set1-5-3}{RGB}{77,175,74}
\definecolor{Set1-5-C}{RGB}{77,175,74}
\definecolor{Set1-5-4}{RGB}{152,78,163}
\definecolor{Set1-5-D}{RGB}{152,78,163}
\definecolor{Set1-5-5}{RGB}{255,127,0}
\definecolor{Set1-5-E}{RGB}{255,127,0}
\definecolor{Set1-6-1}{RGB}{228,26,28}
\definecolor{Set1-6-A}{RGB}{228,26,28}
\definecolor{Set1-6-2}{RGB}{55,126,184}
\definecolor{Set1-6-B}{RGB}{55,126,184}
\definecolor{Set1-6-3}{RGB}{77,175,74}
\definecolor{Set1-6-C}{RGB}{77,175,74}
\definecolor{Set1-6-4}{RGB}{152,78,163}
\definecolor{Set1-6-D}{RGB}{152,78,163}
\definecolor{Set1-6-5}{RGB}{255,127,0}
\definecolor{Set1-6-E}{RGB}{255,127,0}
\definecolor{Set1-6-6}{RGB}{255,255,51}
\definecolor{Set1-6-F}{RGB}{255,255,51}
\definecolor{Set1-7-1}{RGB}{228,26,28}
\definecolor{Set1-7-A}{RGB}{228,26,28}
\definecolor{Set1-7-2}{RGB}{55,126,184}
\definecolor{Set1-7-B}{RGB}{55,126,184}
\definecolor{Set1-7-3}{RGB}{77,175,74}
\definecolor{Set1-7-C}{RGB}{77,175,74}
\definecolor{Set1-7-4}{RGB}{152,78,163}
\definecolor{Set1-7-D}{RGB}{152,78,163}
\definecolor{Set1-7-5}{RGB}{255,127,0}
\definecolor{Set1-7-E}{RGB}{255,127,0}
\definecolor{Set1-7-6}{RGB}{255,255,51}
\definecolor{Set1-7-F}{RGB}{255,255,51}
\definecolor{Set1-7-7}{RGB}{166,86,40}
\definecolor{Set1-7-G}{RGB}{166,86,40}
\definecolor{Set1-8-1}{RGB}{228,26,28}
\definecolor{Set1-8-A}{RGB}{228,26,28}
\definecolor{Set1-8-2}{RGB}{55,126,184}
\definecolor{Set1-8-B}{RGB}{55,126,184}
\definecolor{Set1-8-3}{RGB}{77,175,74}
\definecolor{Set1-8-C}{RGB}{77,175,74}
\definecolor{Set1-8-4}{RGB}{152,78,163}
\definecolor{Set1-8-D}{RGB}{152,78,163}
\definecolor{Set1-8-5}{RGB}{255,127,0}
\definecolor{Set1-8-E}{RGB}{255,127,0}
\definecolor{Set1-8-6}{RGB}{255,255,51}
\definecolor{Set1-8-F}{RGB}{255,255,51}
\definecolor{Set1-8-7}{RGB}{166,86,40}
\definecolor{Set1-8-G}{RGB}{166,86,40}
\definecolor{Set1-8-8}{RGB}{247,129,191}
\definecolor{Set1-8-H}{RGB}{247,129,191}
\definecolor{Set1-9-1}{RGB}{228,26,28}
\definecolor{Set1-9-A}{RGB}{228,26,28}
\definecolor{Set1-9-2}{RGB}{55,126,184}
\definecolor{Set1-9-B}{RGB}{55,126,184}
\definecolor{Set1-9-3}{RGB}{77,175,74}
\definecolor{Set1-9-C}{RGB}{77,175,74}
\definecolor{Set1-9-4}{RGB}{152,78,163}
\definecolor{Set1-9-D}{RGB}{152,78,163}
\definecolor{Set1-9-5}{RGB}{255,127,0}
\definecolor{Set1-9-E}{RGB}{255,127,0}
\definecolor{Set1-9-6}{RGB}{255,255,51}
\definecolor{Set1-9-F}{RGB}{255,255,51}
\definecolor{Set1-9-7}{RGB}{166,86,40}
\definecolor{Set1-9-G}{RGB}{166,86,40}
\definecolor{Set1-9-8}{RGB}{247,129,191}
\definecolor{Set1-9-H}{RGB}{247,129,191}
\definecolor{Set1-9-9}{RGB}{153,153,153}
\definecolor{Set1-9-I}{RGB}{153,153,153}
\definecolor{Pastel2-3-1}{RGB}{179,226,205}
\definecolor{Pastel2-3-A}{RGB}{179,226,205}
\definecolor{Pastel2-3-2}{RGB}{253,205,172}
\definecolor{Pastel2-3-B}{RGB}{253,205,172}
\definecolor{Pastel2-3-3}{RGB}{203,213,232}
\definecolor{Pastel2-3-C}{RGB}{203,213,232}
\definecolor{Pastel2-4-1}{RGB}{179,226,205}
\definecolor{Pastel2-4-A}{RGB}{179,226,205}
\definecolor{Pastel2-4-2}{RGB}{253,205,172}
\definecolor{Pastel2-4-B}{RGB}{253,205,172}
\definecolor{Pastel2-4-3}{RGB}{203,213,232}
\definecolor{Pastel2-4-C}{RGB}{203,213,232}
\definecolor{Pastel2-4-4}{RGB}{244,202,228}
\definecolor{Pastel2-4-D}{RGB}{244,202,228}
\definecolor{Pastel2-5-1}{RGB}{179,226,205}
\definecolor{Pastel2-5-A}{RGB}{179,226,205}
\definecolor{Pastel2-5-2}{RGB}{253,205,172}
\definecolor{Pastel2-5-B}{RGB}{253,205,172}
\definecolor{Pastel2-5-3}{RGB}{203,213,232}
\definecolor{Pastel2-5-C}{RGB}{203,213,232}
\definecolor{Pastel2-5-4}{RGB}{244,202,228}
\definecolor{Pastel2-5-D}{RGB}{244,202,228}
\definecolor{Pastel2-5-5}{RGB}{230,245,201}
\definecolor{Pastel2-5-E}{RGB}{230,245,201}
\definecolor{Pastel2-6-1}{RGB}{179,226,205}
\definecolor{Pastel2-6-A}{RGB}{179,226,205}
\definecolor{Pastel2-6-2}{RGB}{253,205,172}
\definecolor{Pastel2-6-B}{RGB}{253,205,172}
\definecolor{Pastel2-6-3}{RGB}{203,213,232}
\definecolor{Pastel2-6-C}{RGB}{203,213,232}
\definecolor{Pastel2-6-4}{RGB}{244,202,228}
\definecolor{Pastel2-6-D}{RGB}{244,202,228}
\definecolor{Pastel2-6-5}{RGB}{230,245,201}
\definecolor{Pastel2-6-E}{RGB}{230,245,201}
\definecolor{Pastel2-6-6}{RGB}{255,242,174}
\definecolor{Pastel2-6-F}{RGB}{255,242,174}
\definecolor{Pastel2-7-1}{RGB}{179,226,205}
\definecolor{Pastel2-7-A}{RGB}{179,226,205}
\definecolor{Pastel2-7-2}{RGB}{253,205,172}
\definecolor{Pastel2-7-B}{RGB}{253,205,172}
\definecolor{Pastel2-7-3}{RGB}{203,213,232}
\definecolor{Pastel2-7-C}{RGB}{203,213,232}
\definecolor{Pastel2-7-4}{RGB}{244,202,228}
\definecolor{Pastel2-7-D}{RGB}{244,202,228}
\definecolor{Pastel2-7-5}{RGB}{230,245,201}
\definecolor{Pastel2-7-E}{RGB}{230,245,201}
\definecolor{Pastel2-7-6}{RGB}{255,242,174}
\definecolor{Pastel2-7-F}{RGB}{255,242,174}
\definecolor{Pastel2-7-7}{RGB}{241,226,204}
\definecolor{Pastel2-7-G}{RGB}{241,226,204}
\definecolor{Pastel2-8-1}{RGB}{179,226,205}
\definecolor{Pastel2-8-A}{RGB}{179,226,205}
\definecolor{Pastel2-8-2}{RGB}{253,205,172}
\definecolor{Pastel2-8-B}{RGB}{253,205,172}
\definecolor{Pastel2-8-3}{RGB}{203,213,232}
\definecolor{Pastel2-8-C}{RGB}{203,213,232}
\definecolor{Pastel2-8-4}{RGB}{244,202,228}
\definecolor{Pastel2-8-D}{RGB}{244,202,228}
\definecolor{Pastel2-8-5}{RGB}{230,245,201}
\definecolor{Pastel2-8-E}{RGB}{230,245,201}
\definecolor{Pastel2-8-6}{RGB}{255,242,174}
\definecolor{Pastel2-8-F}{RGB}{255,242,174}
\definecolor{Pastel2-8-7}{RGB}{241,226,204}
\definecolor{Pastel2-8-G}{RGB}{241,226,204}
\definecolor{Pastel2-8-8}{RGB}{204,204,204}
\definecolor{Pastel2-8-H}{RGB}{204,204,204}
\definecolor{Set2-3-1}{RGB}{102,194,165}
\definecolor{Set2-3-A}{RGB}{102,194,165}
\definecolor{Set2-3-2}{RGB}{252,141,98}
\definecolor{Set2-3-B}{RGB}{252,141,98}
\definecolor{Set2-3-3}{RGB}{141,160,203}
\definecolor{Set2-3-C}{RGB}{141,160,203}
\definecolor{Set2-4-1}{RGB}{102,194,165}
\definecolor{Set2-4-A}{RGB}{102,194,165}
\definecolor{Set2-4-2}{RGB}{252,141,98}
\definecolor{Set2-4-B}{RGB}{252,141,98}
\definecolor{Set2-4-3}{RGB}{141,160,203}
\definecolor{Set2-4-C}{RGB}{141,160,203}
\definecolor{Set2-4-4}{RGB}{231,138,195}
\definecolor{Set2-4-D}{RGB}{231,138,195}
\definecolor{Set2-5-1}{RGB}{102,194,165}
\definecolor{Set2-5-A}{RGB}{102,194,165}
\definecolor{Set2-5-2}{RGB}{252,141,98}
\definecolor{Set2-5-B}{RGB}{252,141,98}
\definecolor{Set2-5-3}{RGB}{141,160,203}
\definecolor{Set2-5-C}{RGB}{141,160,203}
\definecolor{Set2-5-4}{RGB}{231,138,195}
\definecolor{Set2-5-D}{RGB}{231,138,195}
\definecolor{Set2-5-5}{RGB}{166,216,84}
\definecolor{Set2-5-E}{RGB}{166,216,84}
\definecolor{Set2-6-1}{RGB}{102,194,165}
\definecolor{Set2-6-A}{RGB}{102,194,165}
\definecolor{Set2-6-2}{RGB}{252,141,98}
\definecolor{Set2-6-B}{RGB}{252,141,98}
\definecolor{Set2-6-3}{RGB}{141,160,203}
\definecolor{Set2-6-C}{RGB}{141,160,203}
\definecolor{Set2-6-4}{RGB}{231,138,195}
\definecolor{Set2-6-D}{RGB}{231,138,195}
\definecolor{Set2-6-5}{RGB}{166,216,84}
\definecolor{Set2-6-E}{RGB}{166,216,84}
\definecolor{Set2-6-6}{RGB}{255,217,47}
\definecolor{Set2-6-F}{RGB}{255,217,47}
\definecolor{Set2-7-1}{RGB}{102,194,165}
\definecolor{Set2-7-A}{RGB}{102,194,165}
\definecolor{Set2-7-2}{RGB}{252,141,98}
\definecolor{Set2-7-B}{RGB}{252,141,98}
\definecolor{Set2-7-3}{RGB}{141,160,203}
\definecolor{Set2-7-C}{RGB}{141,160,203}
\definecolor{Set2-7-4}{RGB}{231,138,195}
\definecolor{Set2-7-D}{RGB}{231,138,195}
\definecolor{Set2-7-5}{RGB}{166,216,84}
\definecolor{Set2-7-E}{RGB}{166,216,84}
\definecolor{Set2-7-6}{RGB}{255,217,47}
\definecolor{Set2-7-F}{RGB}{255,217,47}
\definecolor{Set2-7-7}{RGB}{229,196,148}
\definecolor{Set2-7-G}{RGB}{229,196,148}
\definecolor{Set2-8-1}{RGB}{102,194,165}
\definecolor{Set2-8-A}{RGB}{102,194,165}
\definecolor{Set2-8-2}{RGB}{252,141,98}
\definecolor{Set2-8-B}{RGB}{252,141,98}
\definecolor{Set2-8-3}{RGB}{141,160,203}
\definecolor{Set2-8-C}{RGB}{141,160,203}
\definecolor{Set2-8-4}{RGB}{231,138,195}
\definecolor{Set2-8-D}{RGB}{231,138,195}
\definecolor{Set2-8-5}{RGB}{166,216,84}
\definecolor{Set2-8-E}{RGB}{166,216,84}
\definecolor{Set2-8-6}{RGB}{255,217,47}
\definecolor{Set2-8-F}{RGB}{255,217,47}
\definecolor{Set2-8-7}{RGB}{229,196,148}
\definecolor{Set2-8-G}{RGB}{229,196,148}
\definecolor{Set2-8-8}{RGB}{179,179,179}
\definecolor{Set2-8-H}{RGB}{179,179,179}
\definecolor{Dark2-3-1}{RGB}{27,158,119}
\definecolor{Dark2-3-A}{RGB}{27,158,119}
\definecolor{Dark2-3-2}{RGB}{217,95,2}
\definecolor{Dark2-3-B}{RGB}{217,95,2}
\definecolor{Dark2-3-3}{RGB}{117,112,179}
\definecolor{Dark2-3-C}{RGB}{117,112,179}
\definecolor{Dark2-4-1}{RGB}{27,158,119}
\definecolor{Dark2-4-A}{RGB}{27,158,119}
\definecolor{Dark2-4-2}{RGB}{217,95,2}
\definecolor{Dark2-4-B}{RGB}{217,95,2}
\definecolor{Dark2-4-3}{RGB}{117,112,179}
\definecolor{Dark2-4-C}{RGB}{117,112,179}
\definecolor{Dark2-4-4}{RGB}{231,41,138}
\definecolor{Dark2-4-D}{RGB}{231,41,138}
\definecolor{Dark2-5-1}{RGB}{27,158,119}
\definecolor{Dark2-5-A}{RGB}{27,158,119}
\definecolor{Dark2-5-2}{RGB}{217,95,2}
\definecolor{Dark2-5-B}{RGB}{217,95,2}
\definecolor{Dark2-5-3}{RGB}{117,112,179}
\definecolor{Dark2-5-C}{RGB}{117,112,179}
\definecolor{Dark2-5-4}{RGB}{231,41,138}
\definecolor{Dark2-5-D}{RGB}{231,41,138}
\definecolor{Dark2-5-5}{RGB}{102,166,30}
\definecolor{Dark2-5-E}{RGB}{102,166,30}
\definecolor{Dark2-6-1}{RGB}{27,158,119}
\definecolor{Dark2-6-A}{RGB}{27,158,119}
\definecolor{Dark2-6-2}{RGB}{217,95,2}
\definecolor{Dark2-6-B}{RGB}{217,95,2}
\definecolor{Dark2-6-3}{RGB}{117,112,179}
\definecolor{Dark2-6-C}{RGB}{117,112,179}
\definecolor{Dark2-6-4}{RGB}{231,41,138}
\definecolor{Dark2-6-D}{RGB}{231,41,138}
\definecolor{Dark2-6-5}{RGB}{102,166,30}
\definecolor{Dark2-6-E}{RGB}{102,166,30}
\definecolor{Dark2-6-6}{RGB}{230,171,2}
\definecolor{Dark2-6-F}{RGB}{230,171,2}
\definecolor{Dark2-7-1}{RGB}{27,158,119}
\definecolor{Dark2-7-A}{RGB}{27,158,119}
\definecolor{Dark2-7-2}{RGB}{217,95,2}
\definecolor{Dark2-7-B}{RGB}{217,95,2}
\definecolor{Dark2-7-3}{RGB}{117,112,179}
\definecolor{Dark2-7-C}{RGB}{117,112,179}
\definecolor{Dark2-7-4}{RGB}{231,41,138}
\definecolor{Dark2-7-D}{RGB}{231,41,138}
\definecolor{Dark2-7-5}{RGB}{102,166,30}
\definecolor{Dark2-7-E}{RGB}{102,166,30}
\definecolor{Dark2-7-6}{RGB}{230,171,2}
\definecolor{Dark2-7-F}{RGB}{230,171,2}
\definecolor{Dark2-7-7}{RGB}{166,118,29}
\definecolor{Dark2-7-G}{RGB}{166,118,29}
\definecolor{Dark2-8-1}{RGB}{27,158,119}
\definecolor{Dark2-8-A}{RGB}{27,158,119}
\definecolor{Dark2-8-2}{RGB}{217,95,2}
\definecolor{Dark2-8-B}{RGB}{217,95,2}
\definecolor{Dark2-8-3}{RGB}{117,112,179}
\definecolor{Dark2-8-C}{RGB}{117,112,179}
\definecolor{Dark2-8-4}{RGB}{231,41,138}
\definecolor{Dark2-8-D}{RGB}{231,41,138}
\definecolor{Dark2-8-5}{RGB}{102,166,30}
\definecolor{Dark2-8-E}{RGB}{102,166,30}
\definecolor{Dark2-8-6}{RGB}{230,171,2}
\definecolor{Dark2-8-F}{RGB}{230,171,2}
\definecolor{Dark2-8-7}{RGB}{166,118,29}
\definecolor{Dark2-8-G}{RGB}{166,118,29}
\definecolor{Dark2-8-8}{RGB}{102,102,102}
\definecolor{Dark2-8-H}{RGB}{102,102,102}
\definecolor{Paired-3-1}{RGB}{166,206,227}
\definecolor{Paired-3-A}{RGB}{166,206,227}
\definecolor{Paired-3-2}{RGB}{31,120,180}
\definecolor{Paired-3-B}{RGB}{31,120,180}
\definecolor{Paired-3-3}{RGB}{178,223,138}
\definecolor{Paired-3-C}{RGB}{178,223,138}
\definecolor{Paired-4-1}{RGB}{166,206,227}
\definecolor{Paired-4-A}{RGB}{166,206,227}
\definecolor{Paired-4-2}{RGB}{31,120,180}
\definecolor{Paired-4-B}{RGB}{31,120,180}
\definecolor{Paired-4-3}{RGB}{178,223,138}
\definecolor{Paired-4-C}{RGB}{178,223,138}
\definecolor{Paired-4-4}{RGB}{51,160,44}
\definecolor{Paired-4-D}{RGB}{51,160,44}
\definecolor{Paired-5-1}{RGB}{166,206,227}
\definecolor{Paired-5-A}{RGB}{166,206,227}
\definecolor{Paired-5-2}{RGB}{31,120,180}
\definecolor{Paired-5-B}{RGB}{31,120,180}
\definecolor{Paired-5-3}{RGB}{178,223,138}
\definecolor{Paired-5-C}{RGB}{178,223,138}
\definecolor{Paired-5-4}{RGB}{51,160,44}
\definecolor{Paired-5-D}{RGB}{51,160,44}
\definecolor{Paired-5-5}{RGB}{251,154,153}
\definecolor{Paired-5-E}{RGB}{251,154,153}
\definecolor{Paired-6-1}{RGB}{166,206,227}
\definecolor{Paired-6-A}{RGB}{166,206,227}
\definecolor{Paired-6-2}{RGB}{31,120,180}
\definecolor{Paired-6-B}{RGB}{31,120,180}
\definecolor{Paired-6-3}{RGB}{178,223,138}
\definecolor{Paired-6-C}{RGB}{178,223,138}
\definecolor{Paired-6-4}{RGB}{51,160,44}
\definecolor{Paired-6-D}{RGB}{51,160,44}
\definecolor{Paired-6-5}{RGB}{251,154,153}
\definecolor{Paired-6-E}{RGB}{251,154,153}
\definecolor{Paired-6-6}{RGB}{227,26,28}
\definecolor{Paired-6-F}{RGB}{227,26,28}
\definecolor{Paired-7-1}{RGB}{166,206,227}
\definecolor{Paired-7-A}{RGB}{166,206,227}
\definecolor{Paired-7-2}{RGB}{31,120,180}
\definecolor{Paired-7-B}{RGB}{31,120,180}
\definecolor{Paired-7-3}{RGB}{178,223,138}
\definecolor{Paired-7-C}{RGB}{178,223,138}
\definecolor{Paired-7-4}{RGB}{51,160,44}
\definecolor{Paired-7-D}{RGB}{51,160,44}
\definecolor{Paired-7-5}{RGB}{251,154,153}
\definecolor{Paired-7-E}{RGB}{251,154,153}
\definecolor{Paired-7-6}{RGB}{227,26,28}
\definecolor{Paired-7-F}{RGB}{227,26,28}
\definecolor{Paired-7-7}{RGB}{253,191,111}
\definecolor{Paired-7-G}{RGB}{253,191,111}
\definecolor{Paired-8-1}{RGB}{166,206,227}
\definecolor{Paired-8-A}{RGB}{166,206,227}
\definecolor{Paired-8-2}{RGB}{31,120,180}
\definecolor{Paired-8-B}{RGB}{31,120,180}
\definecolor{Paired-8-3}{RGB}{178,223,138}
\definecolor{Paired-8-C}{RGB}{178,223,138}
\definecolor{Paired-8-4}{RGB}{51,160,44}
\definecolor{Paired-8-D}{RGB}{51,160,44}
\definecolor{Paired-8-5}{RGB}{251,154,153}
\definecolor{Paired-8-E}{RGB}{251,154,153}
\definecolor{Paired-8-6}{RGB}{227,26,28}
\definecolor{Paired-8-F}{RGB}{227,26,28}
\definecolor{Paired-8-7}{RGB}{253,191,111}
\definecolor{Paired-8-G}{RGB}{253,191,111}
\definecolor{Paired-8-8}{RGB}{255,127,0}
\definecolor{Paired-8-H}{RGB}{255,127,0}
\definecolor{Paired-9-1}{RGB}{166,206,227}
\definecolor{Paired-9-A}{RGB}{166,206,227}
\definecolor{Paired-9-2}{RGB}{31,120,180}
\definecolor{Paired-9-B}{RGB}{31,120,180}
\definecolor{Paired-9-3}{RGB}{178,223,138}
\definecolor{Paired-9-C}{RGB}{178,223,138}
\definecolor{Paired-9-4}{RGB}{51,160,44}
\definecolor{Paired-9-D}{RGB}{51,160,44}
\definecolor{Paired-9-5}{RGB}{251,154,153}
\definecolor{Paired-9-E}{RGB}{251,154,153}
\definecolor{Paired-9-6}{RGB}{227,26,28}
\definecolor{Paired-9-F}{RGB}{227,26,28}
\definecolor{Paired-9-7}{RGB}{253,191,111}
\definecolor{Paired-9-G}{RGB}{253,191,111}
\definecolor{Paired-9-8}{RGB}{255,127,0}
\definecolor{Paired-9-H}{RGB}{255,127,0}
\definecolor{Paired-9-9}{RGB}{202,178,214}
\definecolor{Paired-9-I}{RGB}{202,178,214}
\definecolor{Paired-10-1}{RGB}{166,206,227}
\definecolor{Paired-10-A}{RGB}{166,206,227}
\definecolor{Paired-10-2}{RGB}{31,120,180}
\definecolor{Paired-10-B}{RGB}{31,120,180}
\definecolor{Paired-10-3}{RGB}{178,223,138}
\definecolor{Paired-10-C}{RGB}{178,223,138}
\definecolor{Paired-10-4}{RGB}{51,160,44}
\definecolor{Paired-10-D}{RGB}{51,160,44}
\definecolor{Paired-10-5}{RGB}{251,154,153}
\definecolor{Paired-10-E}{RGB}{251,154,153}
\definecolor{Paired-10-6}{RGB}{227,26,28}
\definecolor{Paired-10-F}{RGB}{227,26,28}
\definecolor{Paired-10-7}{RGB}{253,191,111}
\definecolor{Paired-10-G}{RGB}{253,191,111}
\definecolor{Paired-10-8}{RGB}{255,127,0}
\definecolor{Paired-10-H}{RGB}{255,127,0}
\definecolor{Paired-10-9}{RGB}{202,178,214}
\definecolor{Paired-10-I}{RGB}{202,178,214}
\definecolor{Paired-10-10}{RGB}{106,61,154}
\definecolor{Paired-10-J}{RGB}{106,61,154}
\definecolor{Paired-11-1}{RGB}{166,206,227}
\definecolor{Paired-11-A}{RGB}{166,206,227}
\definecolor{Paired-11-2}{RGB}{31,120,180}
\definecolor{Paired-11-B}{RGB}{31,120,180}
\definecolor{Paired-11-3}{RGB}{178,223,138}
\definecolor{Paired-11-C}{RGB}{178,223,138}
\definecolor{Paired-11-4}{RGB}{51,160,44}
\definecolor{Paired-11-D}{RGB}{51,160,44}
\definecolor{Paired-11-5}{RGB}{251,154,153}
\definecolor{Paired-11-E}{RGB}{251,154,153}
\definecolor{Paired-11-6}{RGB}{227,26,28}
\definecolor{Paired-11-F}{RGB}{227,26,28}
\definecolor{Paired-11-7}{RGB}{253,191,111}
\definecolor{Paired-11-G}{RGB}{253,191,111}
\definecolor{Paired-11-8}{RGB}{255,127,0}
\definecolor{Paired-11-H}{RGB}{255,127,0}
\definecolor{Paired-11-9}{RGB}{202,178,214}
\definecolor{Paired-11-I}{RGB}{202,178,214}
\definecolor{Paired-11-10}{RGB}{106,61,154}
\definecolor{Paired-11-J}{RGB}{106,61,154}
\definecolor{Paired-11-11}{RGB}{255,255,153}
\definecolor{Paired-11-K}{RGB}{255,255,153}
\definecolor{Paired-12-1}{RGB}{166,206,227}
\definecolor{Paired-12-A}{RGB}{166,206,227}
\definecolor{Paired-12-2}{RGB}{31,120,180}
\definecolor{Paired-12-B}{RGB}{31,120,180}
\definecolor{Paired-12-3}{RGB}{178,223,138}
\definecolor{Paired-12-C}{RGB}{178,223,138}
\definecolor{Paired-12-4}{RGB}{51,160,44}
\definecolor{Paired-12-D}{RGB}{51,160,44}
\definecolor{Paired-12-5}{RGB}{251,154,153}
\definecolor{Paired-12-E}{RGB}{251,154,153}
\definecolor{Paired-12-6}{RGB}{227,26,28}
\definecolor{Paired-12-F}{RGB}{227,26,28}
\definecolor{Paired-12-7}{RGB}{253,191,111}
\definecolor{Paired-12-G}{RGB}{253,191,111}
\definecolor{Paired-12-8}{RGB}{255,127,0}
\definecolor{Paired-12-H}{RGB}{255,127,0}
\definecolor{Paired-12-9}{RGB}{202,178,214}
\definecolor{Paired-12-I}{RGB}{202,178,214}
\definecolor{Paired-12-10}{RGB}{106,61,154}
\definecolor{Paired-12-J}{RGB}{106,61,154}
\definecolor{Paired-12-11}{RGB}{255,255,153}
\definecolor{Paired-12-K}{RGB}{255,255,153}
\definecolor{Paired-12-12}{RGB}{177,89,40}
\definecolor{Paired-12-L}{RGB}{177,89,40}
\definecolor{Accent-3-1}{RGB}{127,201,127}
\definecolor{Accent-3-A}{RGB}{127,201,127}
\definecolor{Accent-3-2}{RGB}{190,174,212}
\definecolor{Accent-3-B}{RGB}{190,174,212}
\definecolor{Accent-3-3}{RGB}{253,192,134}
\definecolor{Accent-3-C}{RGB}{253,192,134}
\definecolor{Accent-4-1}{RGB}{127,201,127}
\definecolor{Accent-4-A}{RGB}{127,201,127}
\definecolor{Accent-4-2}{RGB}{190,174,212}
\definecolor{Accent-4-B}{RGB}{190,174,212}
\definecolor{Accent-4-3}{RGB}{253,192,134}
\definecolor{Accent-4-C}{RGB}{253,192,134}
\definecolor{Accent-4-4}{RGB}{255,255,153}
\definecolor{Accent-4-D}{RGB}{255,255,153}
\definecolor{Accent-5-1}{RGB}{127,201,127}
\definecolor{Accent-5-A}{RGB}{127,201,127}
\definecolor{Accent-5-2}{RGB}{190,174,212}
\definecolor{Accent-5-B}{RGB}{190,174,212}
\definecolor{Accent-5-3}{RGB}{253,192,134}
\definecolor{Accent-5-C}{RGB}{253,192,134}
\definecolor{Accent-5-4}{RGB}{255,255,153}
\definecolor{Accent-5-D}{RGB}{255,255,153}
\definecolor{Accent-5-5}{RGB}{56,108,176}
\definecolor{Accent-5-E}{RGB}{56,108,176}
\definecolor{Accent-6-1}{RGB}{127,201,127}
\definecolor{Accent-6-A}{RGB}{127,201,127}
\definecolor{Accent-6-2}{RGB}{190,174,212}
\definecolor{Accent-6-B}{RGB}{190,174,212}
\definecolor{Accent-6-3}{RGB}{253,192,134}
\definecolor{Accent-6-C}{RGB}{253,192,134}
\definecolor{Accent-6-4}{RGB}{255,255,153}
\definecolor{Accent-6-D}{RGB}{255,255,153}
\definecolor{Accent-6-5}{RGB}{56,108,176}
\definecolor{Accent-6-E}{RGB}{56,108,176}
\definecolor{Accent-6-6}{RGB}{240,2,127}
\definecolor{Accent-6-F}{RGB}{240,2,127}
\definecolor{Accent-7-1}{RGB}{127,201,127}
\definecolor{Accent-7-A}{RGB}{127,201,127}
\definecolor{Accent-7-2}{RGB}{190,174,212}
\definecolor{Accent-7-B}{RGB}{190,174,212}
\definecolor{Accent-7-3}{RGB}{253,192,134}
\definecolor{Accent-7-C}{RGB}{253,192,134}
\definecolor{Accent-7-4}{RGB}{255,255,153}
\definecolor{Accent-7-D}{RGB}{255,255,153}
\definecolor{Accent-7-5}{RGB}{56,108,176}
\definecolor{Accent-7-E}{RGB}{56,108,176}
\definecolor{Accent-7-6}{RGB}{240,2,127}
\definecolor{Accent-7-F}{RGB}{240,2,127}
\definecolor{Accent-7-7}{RGB}{191,91,23}
\definecolor{Accent-7-G}{RGB}{191,91,23}
\definecolor{Accent-8-1}{RGB}{127,201,127}
\definecolor{Accent-8-A}{RGB}{127,201,127}
\definecolor{Accent-8-2}{RGB}{190,174,212}
\definecolor{Accent-8-B}{RGB}{190,174,212}
\definecolor{Accent-8-3}{RGB}{253,192,134}
\definecolor{Accent-8-C}{RGB}{253,192,134}
\definecolor{Accent-8-4}{RGB}{255,255,153}
\definecolor{Accent-8-D}{RGB}{255,255,153}
\definecolor{Accent-8-5}{RGB}{56,108,176}
\definecolor{Accent-8-E}{RGB}{56,108,176}
\definecolor{Accent-8-6}{RGB}{240,2,127}
\definecolor{Accent-8-F}{RGB}{240,2,127}
\definecolor{Accent-8-7}{RGB}{191,91,23}
\definecolor{Accent-8-G}{RGB}{191,91,23}
\definecolor{Accent-8-8}{RGB}{102,102,102}
\definecolor{Accent-8-H}{RGB}{102,102,102}

\pgfplotscreateplotcyclelist{YlGn-3}{%
  YlGn-3-1\\%
  YlGn-3-2\\%
  YlGn-3-3\\%
}
\pgfplotscreateplotcyclelist{YlGn-4}{%
  YlGn-4-1\\%
  YlGn-4-2\\%
  YlGn-4-3\\%
  YlGn-4-4\\%
}
\pgfplotscreateplotcyclelist{YlGn-5}{%
  YlGn-5-1\\%
  YlGn-5-2\\%
  YlGn-5-3\\%
  YlGn-5-4\\%
  YlGn-5-5\\%
}
\pgfplotscreateplotcyclelist{YlGn-6}{%
  YlGn-6-1\\%
  YlGn-6-2\\%
  YlGn-6-3\\%
  YlGn-6-4\\%
  YlGn-6-5\\%
  YlGn-6-6\\%
}
\pgfplotscreateplotcyclelist{YlGn-7}{%
  YlGn-7-1\\%
  YlGn-7-2\\%
  YlGn-7-3\\%
  YlGn-7-4\\%
  YlGn-7-5\\%
  YlGn-7-6\\%
  YlGn-7-7\\%
}
\pgfplotscreateplotcyclelist{YlGn-8}{%
  YlGn-8-1\\%
  YlGn-8-2\\%
  YlGn-8-3\\%
  YlGn-8-4\\%
  YlGn-8-5\\%
  YlGn-8-6\\%
  YlGn-8-7\\%
  YlGn-8-8\\%
}
\pgfplotscreateplotcyclelist{YlGn-9}{%
  YlGn-9-1\\%
  YlGn-9-2\\%
  YlGn-9-3\\%
  YlGn-9-4\\%
  YlGn-9-5\\%
  YlGn-9-6\\%
  YlGn-9-7\\%
  YlGn-9-8\\%
  YlGn-9-9\\%
}
\pgfplotscreateplotcyclelist{YlGnBu-3}{%
  YlGnBu-3-1\\%
  YlGnBu-3-2\\%
  YlGnBu-3-3\\%
}
\pgfplotscreateplotcyclelist{YlGnBu-4}{%
  YlGnBu-4-1\\%
  YlGnBu-4-2\\%
  YlGnBu-4-3\\%
  YlGnBu-4-4\\%
}
\pgfplotscreateplotcyclelist{YlGnBu-5}{%
  YlGnBu-5-1\\%
  YlGnBu-5-2\\%
  YlGnBu-5-3\\%
  YlGnBu-5-4\\%
  YlGnBu-5-5\\%
}
\pgfplotscreateplotcyclelist{YlGnBu-6}{%
  YlGnBu-6-1\\%
  YlGnBu-6-2\\%
  YlGnBu-6-3\\%
  YlGnBu-6-4\\%
  YlGnBu-6-5\\%
  YlGnBu-6-6\\%
}
\pgfplotscreateplotcyclelist{YlGnBu-7}{%
  YlGnBu-7-1\\%
  YlGnBu-7-2\\%
  YlGnBu-7-3\\%
  YlGnBu-7-4\\%
  YlGnBu-7-5\\%
  YlGnBu-7-6\\%
  YlGnBu-7-7\\%
}
\pgfplotscreateplotcyclelist{YlGnBu-8}{%
  YlGnBu-8-1\\%
  YlGnBu-8-2\\%
  YlGnBu-8-3\\%
  YlGnBu-8-4\\%
  YlGnBu-8-5\\%
  YlGnBu-8-6\\%
  YlGnBu-8-7\\%
  YlGnBu-8-8\\%
}
\pgfplotscreateplotcyclelist{YlGnBu-9}{%
  YlGnBu-9-1\\%
  YlGnBu-9-2\\%
  YlGnBu-9-3\\%
  YlGnBu-9-4\\%
  YlGnBu-9-5\\%
  YlGnBu-9-6\\%
  YlGnBu-9-7\\%
  YlGnBu-9-8\\%
  YlGnBu-9-9\\%
}
\pgfplotscreateplotcyclelist{GnBu-3}{%
  GnBu-3-1\\%
  GnBu-3-2\\%
  GnBu-3-3\\%
}
\pgfplotscreateplotcyclelist{GnBu-4}{%
  GnBu-4-1\\%
  GnBu-4-2\\%
  GnBu-4-3\\%
  GnBu-4-4\\%
}
\pgfplotscreateplotcyclelist{GnBu-5}{%
  GnBu-5-1\\%
  GnBu-5-2\\%
  GnBu-5-3\\%
  GnBu-5-4\\%
  GnBu-5-5\\%
}
\pgfplotscreateplotcyclelist{GnBu-6}{%
  GnBu-6-1\\%
  GnBu-6-2\\%
  GnBu-6-3\\%
  GnBu-6-4\\%
  GnBu-6-5\\%
  GnBu-6-6\\%
}
\pgfplotscreateplotcyclelist{GnBu-7}{%
  GnBu-7-1\\%
  GnBu-7-2\\%
  GnBu-7-3\\%
  GnBu-7-4\\%
  GnBu-7-5\\%
  GnBu-7-6\\%
  GnBu-7-7\\%
}
\pgfplotscreateplotcyclelist{GnBu-8}{%
  GnBu-8-1\\%
  GnBu-8-2\\%
  GnBu-8-3\\%
  GnBu-8-4\\%
  GnBu-8-5\\%
  GnBu-8-6\\%
  GnBu-8-7\\%
  GnBu-8-8\\%
}
\pgfplotscreateplotcyclelist{GnBu-9}{%
  GnBu-9-1\\%
  GnBu-9-2\\%
  GnBu-9-3\\%
  GnBu-9-4\\%
  GnBu-9-5\\%
  GnBu-9-6\\%
  GnBu-9-7\\%
  GnBu-9-8\\%
  GnBu-9-9\\%
}
\pgfplotscreateplotcyclelist{BuGn-3}{%
  BuGn-3-1\\%
  BuGn-3-2\\%
  BuGn-3-3\\%
}
\pgfplotscreateplotcyclelist{BuGn-4}{%
  BuGn-4-1\\%
  BuGn-4-2\\%
  BuGn-4-3\\%
  BuGn-4-4\\%
}
\pgfplotscreateplotcyclelist{BuGn-5}{%
  BuGn-5-1\\%
  BuGn-5-2\\%
  BuGn-5-3\\%
  BuGn-5-4\\%
  BuGn-5-5\\%
}
\pgfplotscreateplotcyclelist{BuGn-6}{%
  BuGn-6-1\\%
  BuGn-6-2\\%
  BuGn-6-3\\%
  BuGn-6-4\\%
  BuGn-6-5\\%
  BuGn-6-6\\%
}
\pgfplotscreateplotcyclelist{BuGn-7}{%
  BuGn-7-1\\%
  BuGn-7-2\\%
  BuGn-7-3\\%
  BuGn-7-4\\%
  BuGn-7-5\\%
  BuGn-7-6\\%
  BuGn-7-7\\%
}
\pgfplotscreateplotcyclelist{BuGn-8}{%
  BuGn-8-1\\%
  BuGn-8-2\\%
  BuGn-8-3\\%
  BuGn-8-4\\%
  BuGn-8-5\\%
  BuGn-8-6\\%
  BuGn-8-7\\%
  BuGn-8-8\\%
}
\pgfplotscreateplotcyclelist{BuGn-9}{%
  BuGn-9-1\\%
  BuGn-9-2\\%
  BuGn-9-3\\%
  BuGn-9-4\\%
  BuGn-9-5\\%
  BuGn-9-6\\%
  BuGn-9-7\\%
  BuGn-9-8\\%
  BuGn-9-9\\%
}
\pgfplotscreateplotcyclelist{PuBuGn-3}{%
  PuBuGn-3-1\\%
  PuBuGn-3-2\\%
  PuBuGn-3-3\\%
}
\pgfplotscreateplotcyclelist{PuBuGn-4}{%
  PuBuGn-4-1\\%
  PuBuGn-4-2\\%
  PuBuGn-4-3\\%
  PuBuGn-4-4\\%
}
\pgfplotscreateplotcyclelist{PuBuGn-5}{%
  PuBuGn-5-1\\%
  PuBuGn-5-2\\%
  PuBuGn-5-3\\%
  PuBuGn-5-4\\%
  PuBuGn-5-5\\%
}
\pgfplotscreateplotcyclelist{PuBuGn-6}{%
  PuBuGn-6-1\\%
  PuBuGn-6-2\\%
  PuBuGn-6-3\\%
  PuBuGn-6-4\\%
  PuBuGn-6-5\\%
  PuBuGn-6-6\\%
}
\pgfplotscreateplotcyclelist{PuBuGn-7}{%
  PuBuGn-7-1\\%
  PuBuGn-7-2\\%
  PuBuGn-7-3\\%
  PuBuGn-7-4\\%
  PuBuGn-7-5\\%
  PuBuGn-7-6\\%
  PuBuGn-7-7\\%
}
\pgfplotscreateplotcyclelist{PuBuGn-8}{%
  PuBuGn-8-1\\%
  PuBuGn-8-2\\%
  PuBuGn-8-3\\%
  PuBuGn-8-4\\%
  PuBuGn-8-5\\%
  PuBuGn-8-6\\%
  PuBuGn-8-7\\%
  PuBuGn-8-8\\%
}
\pgfplotscreateplotcyclelist{PuBuGn-9}{%
  PuBuGn-9-1\\%
  PuBuGn-9-2\\%
  PuBuGn-9-3\\%
  PuBuGn-9-4\\%
  PuBuGn-9-5\\%
  PuBuGn-9-6\\%
  PuBuGn-9-7\\%
  PuBuGn-9-8\\%
  PuBuGn-9-9\\%
}
\pgfplotscreateplotcyclelist{PuBu-3}{%
  PuBu-3-1\\%
  PuBu-3-2\\%
  PuBu-3-3\\%
}
\pgfplotscreateplotcyclelist{PuBu-4}{%
  PuBu-4-1\\%
  PuBu-4-2\\%
  PuBu-4-3\\%
  PuBu-4-4\\%
}
\pgfplotscreateplotcyclelist{PuBu-5}{%
  PuBu-5-1\\%
  PuBu-5-2\\%
  PuBu-5-3\\%
  PuBu-5-4\\%
  PuBu-5-5\\%
}
\pgfplotscreateplotcyclelist{PuBu-6}{%
  PuBu-6-1\\%
  PuBu-6-2\\%
  PuBu-6-3\\%
  PuBu-6-4\\%
  PuBu-6-5\\%
  PuBu-6-6\\%
}
\pgfplotscreateplotcyclelist{PuBu-7}{%
  PuBu-7-1\\%
  PuBu-7-2\\%
  PuBu-7-3\\%
  PuBu-7-4\\%
  PuBu-7-5\\%
  PuBu-7-6\\%
  PuBu-7-7\\%
}
\pgfplotscreateplotcyclelist{PuBu-8}{%
  PuBu-8-1\\%
  PuBu-8-2\\%
  PuBu-8-3\\%
  PuBu-8-4\\%
  PuBu-8-5\\%
  PuBu-8-6\\%
  PuBu-8-7\\%
  PuBu-8-8\\%
}
\pgfplotscreateplotcyclelist{PuBu-9}{%
  PuBu-9-1\\%
  PuBu-9-2\\%
  PuBu-9-3\\%
  PuBu-9-4\\%
  PuBu-9-5\\%
  PuBu-9-6\\%
  PuBu-9-7\\%
  PuBu-9-8\\%
  PuBu-9-9\\%
}
\pgfplotscreateplotcyclelist{BuPu-3}{%
  BuPu-3-1\\%
  BuPu-3-2\\%
  BuPu-3-3\\%
}
\pgfplotscreateplotcyclelist{BuPu-4}{%
  BuPu-4-1\\%
  BuPu-4-2\\%
  BuPu-4-3\\%
  BuPu-4-4\\%
}
\pgfplotscreateplotcyclelist{BuPu-5}{%
  BuPu-5-1\\%
  BuPu-5-2\\%
  BuPu-5-3\\%
  BuPu-5-4\\%
  BuPu-5-5\\%
}
\pgfplotscreateplotcyclelist{BuPu-6}{%
  BuPu-6-1\\%
  BuPu-6-2\\%
  BuPu-6-3\\%
  BuPu-6-4\\%
  BuPu-6-5\\%
  BuPu-6-6\\%
}
\pgfplotscreateplotcyclelist{BuPu-7}{%
  BuPu-7-1\\%
  BuPu-7-2\\%
  BuPu-7-3\\%
  BuPu-7-4\\%
  BuPu-7-5\\%
  BuPu-7-6\\%
  BuPu-7-7\\%
}
\pgfplotscreateplotcyclelist{BuPu-8}{%
  BuPu-8-1\\%
  BuPu-8-2\\%
  BuPu-8-3\\%
  BuPu-8-4\\%
  BuPu-8-5\\%
  BuPu-8-6\\%
  BuPu-8-7\\%
  BuPu-8-8\\%
}
\pgfplotscreateplotcyclelist{BuPu-9}{%
  BuPu-9-1\\%
  BuPu-9-2\\%
  BuPu-9-3\\%
  BuPu-9-4\\%
  BuPu-9-5\\%
  BuPu-9-6\\%
  BuPu-9-7\\%
  BuPu-9-8\\%
  BuPu-9-9\\%
}
\pgfplotscreateplotcyclelist{RdPu-3}{%
  RdPu-3-1\\%
  RdPu-3-2\\%
  RdPu-3-3\\%
}
\pgfplotscreateplotcyclelist{RdPu-4}{%
  RdPu-4-1\\%
  RdPu-4-2\\%
  RdPu-4-3\\%
  RdPu-4-4\\%
}
\pgfplotscreateplotcyclelist{RdPu-5}{%
  RdPu-5-1\\%
  RdPu-5-2\\%
  RdPu-5-3\\%
  RdPu-5-4\\%
  RdPu-5-5\\%
}
\pgfplotscreateplotcyclelist{RdPu-6}{%
  RdPu-6-1\\%
  RdPu-6-2\\%
  RdPu-6-3\\%
  RdPu-6-4\\%
  RdPu-6-5\\%
  RdPu-6-6\\%
}
\pgfplotscreateplotcyclelist{RdPu-7}{%
  RdPu-7-1\\%
  RdPu-7-2\\%
  RdPu-7-3\\%
  RdPu-7-4\\%
  RdPu-7-5\\%
  RdPu-7-6\\%
  RdPu-7-7\\%
}
\pgfplotscreateplotcyclelist{RdPu-8}{%
  RdPu-8-1\\%
  RdPu-8-2\\%
  RdPu-8-3\\%
  RdPu-8-4\\%
  RdPu-8-5\\%
  RdPu-8-6\\%
  RdPu-8-7\\%
  RdPu-8-8\\%
}
\pgfplotscreateplotcyclelist{RdPu-9}{%
  RdPu-9-1\\%
  RdPu-9-2\\%
  RdPu-9-3\\%
  RdPu-9-4\\%
  RdPu-9-5\\%
  RdPu-9-6\\%
  RdPu-9-7\\%
  RdPu-9-8\\%
  RdPu-9-9\\%
}
\pgfplotscreateplotcyclelist{PuRd-3}{%
  PuRd-3-1\\%
  PuRd-3-2\\%
  PuRd-3-3\\%
}
\pgfplotscreateplotcyclelist{PuRd-4}{%
  PuRd-4-1\\%
  PuRd-4-2\\%
  PuRd-4-3\\%
  PuRd-4-4\\%
}
\pgfplotscreateplotcyclelist{PuRd-5}{%
  PuRd-5-1\\%
  PuRd-5-2\\%
  PuRd-5-3\\%
  PuRd-5-4\\%
  PuRd-5-5\\%
}
\pgfplotscreateplotcyclelist{PuRd-6}{%
  PuRd-6-1\\%
  PuRd-6-2\\%
  PuRd-6-3\\%
  PuRd-6-4\\%
  PuRd-6-5\\%
  PuRd-6-6\\%
}
\pgfplotscreateplotcyclelist{PuRd-7}{%
  PuRd-7-1\\%
  PuRd-7-2\\%
  PuRd-7-3\\%
  PuRd-7-4\\%
  PuRd-7-5\\%
  PuRd-7-6\\%
  PuRd-7-7\\%
}
\pgfplotscreateplotcyclelist{PuRd-8}{%
  PuRd-8-1\\%
  PuRd-8-2\\%
  PuRd-8-3\\%
  PuRd-8-4\\%
  PuRd-8-5\\%
  PuRd-8-6\\%
  PuRd-8-7\\%
  PuRd-8-8\\%
}
\pgfplotscreateplotcyclelist{PuRd-9}{%
  PuRd-9-1\\%
  PuRd-9-2\\%
  PuRd-9-3\\%
  PuRd-9-4\\%
  PuRd-9-5\\%
  PuRd-9-6\\%
  PuRd-9-7\\%
  PuRd-9-8\\%
  PuRd-9-9\\%
}
\pgfplotscreateplotcyclelist{OrRd-3}{%
  OrRd-3-1\\%
  OrRd-3-2\\%
  OrRd-3-3\\%
}
\pgfplotscreateplotcyclelist{OrRd-4}{%
  OrRd-4-1\\%
  OrRd-4-2\\%
  OrRd-4-3\\%
  OrRd-4-4\\%
}
\pgfplotscreateplotcyclelist{OrRd-5}{%
  OrRd-5-1\\%
  OrRd-5-2\\%
  OrRd-5-3\\%
  OrRd-5-4\\%
  OrRd-5-5\\%
}
\pgfplotscreateplotcyclelist{OrRd-6}{%
  OrRd-6-1\\%
  OrRd-6-2\\%
  OrRd-6-3\\%
  OrRd-6-4\\%
  OrRd-6-5\\%
  OrRd-6-6\\%
}
\pgfplotscreateplotcyclelist{OrRd-7}{%
  OrRd-7-1\\%
  OrRd-7-2\\%
  OrRd-7-3\\%
  OrRd-7-4\\%
  OrRd-7-5\\%
  OrRd-7-6\\%
  OrRd-7-7\\%
}
\pgfplotscreateplotcyclelist{OrRd-8}{%
  OrRd-8-1\\%
  OrRd-8-2\\%
  OrRd-8-3\\%
  OrRd-8-4\\%
  OrRd-8-5\\%
  OrRd-8-6\\%
  OrRd-8-7\\%
  OrRd-8-8\\%
}
\pgfplotscreateplotcyclelist{OrRd-9}{%
  OrRd-9-1\\%
  OrRd-9-2\\%
  OrRd-9-3\\%
  OrRd-9-4\\%
  OrRd-9-5\\%
  OrRd-9-6\\%
  OrRd-9-7\\%
  OrRd-9-8\\%
  OrRd-9-9\\%
}
\pgfplotscreateplotcyclelist{YlOrRd-3}{%
  YlOrRd-3-1\\%
  YlOrRd-3-2\\%
  YlOrRd-3-3\\%
}
\pgfplotscreateplotcyclelist{YlOrRd-4}{%
  YlOrRd-4-1\\%
  YlOrRd-4-2\\%
  YlOrRd-4-3\\%
  YlOrRd-4-4\\%
}
\pgfplotscreateplotcyclelist{YlOrRd-5}{%
  YlOrRd-5-1\\%
  YlOrRd-5-2\\%
  YlOrRd-5-3\\%
  YlOrRd-5-4\\%
  YlOrRd-5-5\\%
}
\pgfplotscreateplotcyclelist{YlOrRd-6}{%
  YlOrRd-6-1\\%
  YlOrRd-6-2\\%
  YlOrRd-6-3\\%
  YlOrRd-6-4\\%
  YlOrRd-6-5\\%
  YlOrRd-6-6\\%
}
\pgfplotscreateplotcyclelist{YlOrRd-7}{%
  YlOrRd-7-1\\%
  YlOrRd-7-2\\%
  YlOrRd-7-3\\%
  YlOrRd-7-4\\%
  YlOrRd-7-5\\%
  YlOrRd-7-6\\%
  YlOrRd-7-7\\%
}
\pgfplotscreateplotcyclelist{YlOrRd-8}{%
  YlOrRd-8-1\\%
  YlOrRd-8-2\\%
  YlOrRd-8-3\\%
  YlOrRd-8-4\\%
  YlOrRd-8-5\\%
  YlOrRd-8-6\\%
  YlOrRd-8-7\\%
  YlOrRd-8-8\\%
}
\pgfplotscreateplotcyclelist{YlOrRd-9}{%
  YlOrRd-9-1\\%
  YlOrRd-9-2\\%
  YlOrRd-9-3\\%
  YlOrRd-9-4\\%
  YlOrRd-9-5\\%
  YlOrRd-9-6\\%
  YlOrRd-9-7\\%
  YlOrRd-9-8\\%
  YlOrRd-9-9\\%
}
\pgfplotscreateplotcyclelist{YlOrBr-3}{%
  YlOrBr-3-1\\%
  YlOrBr-3-2\\%
  YlOrBr-3-3\\%
}
\pgfplotscreateplotcyclelist{YlOrBr-4}{%
  YlOrBr-4-1\\%
  YlOrBr-4-2\\%
  YlOrBr-4-3\\%
  YlOrBr-4-4\\%
}
\pgfplotscreateplotcyclelist{YlOrBr-5}{%
  YlOrBr-5-1\\%
  YlOrBr-5-2\\%
  YlOrBr-5-3\\%
  YlOrBr-5-4\\%
  YlOrBr-5-5\\%
}
\pgfplotscreateplotcyclelist{YlOrBr-6}{%
  YlOrBr-6-1\\%
  YlOrBr-6-2\\%
  YlOrBr-6-3\\%
  YlOrBr-6-4\\%
  YlOrBr-6-5\\%
  YlOrBr-6-6\\%
}
\pgfplotscreateplotcyclelist{YlOrBr-7}{%
  YlOrBr-7-1\\%
  YlOrBr-7-2\\%
  YlOrBr-7-3\\%
  YlOrBr-7-4\\%
  YlOrBr-7-5\\%
  YlOrBr-7-6\\%
  YlOrBr-7-7\\%
}
\pgfplotscreateplotcyclelist{YlOrBr-8}{%
  YlOrBr-8-1\\%
  YlOrBr-8-2\\%
  YlOrBr-8-3\\%
  YlOrBr-8-4\\%
  YlOrBr-8-5\\%
  YlOrBr-8-6\\%
  YlOrBr-8-7\\%
  YlOrBr-8-8\\%
}
\pgfplotscreateplotcyclelist{YlOrBr-9}{%
  YlOrBr-9-1\\%
  YlOrBr-9-2\\%
  YlOrBr-9-3\\%
  YlOrBr-9-4\\%
  YlOrBr-9-5\\%
  YlOrBr-9-6\\%
  YlOrBr-9-7\\%
  YlOrBr-9-8\\%
  YlOrBr-9-9\\%
}
\pgfplotscreateplotcyclelist{Purples-3}{%
  Purples-3-1\\%
  Purples-3-2\\%
  Purples-3-3\\%
}
\pgfplotscreateplotcyclelist{Purples-4}{%
  Purples-4-1\\%
  Purples-4-2\\%
  Purples-4-3\\%
  Purples-4-4\\%
}
\pgfplotscreateplotcyclelist{Purples-5}{%
  Purples-5-1\\%
  Purples-5-2\\%
  Purples-5-3\\%
  Purples-5-4\\%
  Purples-5-5\\%
}
\pgfplotscreateplotcyclelist{Purples-6}{%
  Purples-6-1\\%
  Purples-6-2\\%
  Purples-6-3\\%
  Purples-6-4\\%
  Purples-6-5\\%
  Purples-6-6\\%
}
\pgfplotscreateplotcyclelist{Purples-7}{%
  Purples-7-1\\%
  Purples-7-2\\%
  Purples-7-3\\%
  Purples-7-4\\%
  Purples-7-5\\%
  Purples-7-6\\%
  Purples-7-7\\%
}
\pgfplotscreateplotcyclelist{Purples-8}{%
  Purples-8-1\\%
  Purples-8-2\\%
  Purples-8-3\\%
  Purples-8-4\\%
  Purples-8-5\\%
  Purples-8-6\\%
  Purples-8-7\\%
  Purples-8-8\\%
}
\pgfplotscreateplotcyclelist{Purples-9}{%
  Purples-9-1\\%
  Purples-9-2\\%
  Purples-9-3\\%
  Purples-9-4\\%
  Purples-9-5\\%
  Purples-9-6\\%
  Purples-9-7\\%
  Purples-9-8\\%
  Purples-9-9\\%
}
\pgfplotscreateplotcyclelist{Blues-3}{%
  Blues-3-1\\%
  Blues-3-2\\%
  Blues-3-3\\%
}
\pgfplotscreateplotcyclelist{Blues-4}{%
  Blues-4-1\\%
  Blues-4-2\\%
  Blues-4-3\\%
  Blues-4-4\\%
}
\pgfplotscreateplotcyclelist{Blues-5}{%
  Blues-5-1\\%
  Blues-5-2\\%
  Blues-5-3\\%
  Blues-5-4\\%
  Blues-5-5\\%
}
\pgfplotscreateplotcyclelist{Blues-6}{%
  Blues-6-1\\%
  Blues-6-2\\%
  Blues-6-3\\%
  Blues-6-4\\%
  Blues-6-5\\%
  Blues-6-6\\%
}
\pgfplotscreateplotcyclelist{Blues-7}{%
  Blues-7-1\\%
  Blues-7-2\\%
  Blues-7-3\\%
  Blues-7-4\\%
  Blues-7-5\\%
  Blues-7-6\\%
  Blues-7-7\\%
}
\pgfplotscreateplotcyclelist{Blues-8}{%
  Blues-8-1\\%
  Blues-8-2\\%
  Blues-8-3\\%
  Blues-8-4\\%
  Blues-8-5\\%
  Blues-8-6\\%
  Blues-8-7\\%
  Blues-8-8\\%
}
\pgfplotscreateplotcyclelist{Blues-9}{%
  Blues-9-1\\%
  Blues-9-2\\%
  Blues-9-3\\%
  Blues-9-4\\%
  Blues-9-5\\%
  Blues-9-6\\%
  Blues-9-7\\%
  Blues-9-8\\%
  Blues-9-9\\%
}
\pgfplotscreateplotcyclelist{Greens-3}{%
  Greens-3-1\\%
  Greens-3-2\\%
  Greens-3-3\\%
}
\pgfplotscreateplotcyclelist{Greens-4}{%
  Greens-4-1\\%
  Greens-4-2\\%
  Greens-4-3\\%
  Greens-4-4\\%
}
\pgfplotscreateplotcyclelist{Greens-5}{%
  Greens-5-1\\%
  Greens-5-2\\%
  Greens-5-3\\%
  Greens-5-4\\%
  Greens-5-5\\%
}
\pgfplotscreateplotcyclelist{Greens-6}{%
  Greens-6-1\\%
  Greens-6-2\\%
  Greens-6-3\\%
  Greens-6-4\\%
  Greens-6-5\\%
  Greens-6-6\\%
}
\pgfplotscreateplotcyclelist{Greens-7}{%
  Greens-7-1\\%
  Greens-7-2\\%
  Greens-7-3\\%
  Greens-7-4\\%
  Greens-7-5\\%
  Greens-7-6\\%
  Greens-7-7\\%
}
\pgfplotscreateplotcyclelist{Greens-8}{%
  Greens-8-1\\%
  Greens-8-2\\%
  Greens-8-3\\%
  Greens-8-4\\%
  Greens-8-5\\%
  Greens-8-6\\%
  Greens-8-7\\%
  Greens-8-8\\%
}
\pgfplotscreateplotcyclelist{Greens-9}{%
  Greens-9-1\\%
  Greens-9-2\\%
  Greens-9-3\\%
  Greens-9-4\\%
  Greens-9-5\\%
  Greens-9-6\\%
  Greens-9-7\\%
  Greens-9-8\\%
  Greens-9-9\\%
}
\pgfplotscreateplotcyclelist{Oranges-3}{%
  Oranges-3-1\\%
  Oranges-3-2\\%
  Oranges-3-3\\%
}
\pgfplotscreateplotcyclelist{Oranges-4}{%
  Oranges-4-1\\%
  Oranges-4-2\\%
  Oranges-4-3\\%
  Oranges-4-4\\%
}
\pgfplotscreateplotcyclelist{Oranges-5}{%
  Oranges-5-1\\%
  Oranges-5-2\\%
  Oranges-5-3\\%
  Oranges-5-4\\%
  Oranges-5-5\\%
}
\pgfplotscreateplotcyclelist{Oranges-6}{%
  Oranges-6-1\\%
  Oranges-6-2\\%
  Oranges-6-3\\%
  Oranges-6-4\\%
  Oranges-6-5\\%
  Oranges-6-6\\%
}
\pgfplotscreateplotcyclelist{Oranges-7}{%
  Oranges-7-1\\%
  Oranges-7-2\\%
  Oranges-7-3\\%
  Oranges-7-4\\%
  Oranges-7-5\\%
  Oranges-7-6\\%
  Oranges-7-7\\%
}
\pgfplotscreateplotcyclelist{Oranges-8}{%
  Oranges-8-1\\%
  Oranges-8-2\\%
  Oranges-8-3\\%
  Oranges-8-4\\%
  Oranges-8-5\\%
  Oranges-8-6\\%
  Oranges-8-7\\%
  Oranges-8-8\\%
}
\pgfplotscreateplotcyclelist{Oranges-9}{%
  Oranges-9-1\\%
  Oranges-9-2\\%
  Oranges-9-3\\%
  Oranges-9-4\\%
  Oranges-9-5\\%
  Oranges-9-6\\%
  Oranges-9-7\\%
  Oranges-9-8\\%
  Oranges-9-9\\%
}
\pgfplotscreateplotcyclelist{Reds-3}{%
  Reds-3-1\\%
  Reds-3-2\\%
  Reds-3-3\\%
}
\pgfplotscreateplotcyclelist{Reds-4}{%
  Reds-4-1\\%
  Reds-4-2\\%
  Reds-4-3\\%
  Reds-4-4\\%
}
\pgfplotscreateplotcyclelist{Reds-5}{%
  Reds-5-1\\%
  Reds-5-2\\%
  Reds-5-3\\%
  Reds-5-4\\%
  Reds-5-5\\%
}
\pgfplotscreateplotcyclelist{Reds-6}{%
  Reds-6-1\\%
  Reds-6-2\\%
  Reds-6-3\\%
  Reds-6-4\\%
  Reds-6-5\\%
  Reds-6-6\\%
}
\pgfplotscreateplotcyclelist{Reds-7}{%
  Reds-7-1\\%
  Reds-7-2\\%
  Reds-7-3\\%
  Reds-7-4\\%
  Reds-7-5\\%
  Reds-7-6\\%
  Reds-7-7\\%
}
\pgfplotscreateplotcyclelist{Reds-8}{%
  Reds-8-1\\%
  Reds-8-2\\%
  Reds-8-3\\%
  Reds-8-4\\%
  Reds-8-5\\%
  Reds-8-6\\%
  Reds-8-7\\%
  Reds-8-8\\%
}
\pgfplotscreateplotcyclelist{Reds-9}{%
  Reds-9-1\\%
  Reds-9-2\\%
  Reds-9-3\\%
  Reds-9-4\\%
  Reds-9-5\\%
  Reds-9-6\\%
  Reds-9-7\\%
  Reds-9-8\\%
  Reds-9-9\\%
}
\pgfplotscreateplotcyclelist{Greys-3}{%
  Greys-3-1\\%
  Greys-3-2\\%
  Greys-3-3\\%
}
\pgfplotscreateplotcyclelist{Greys-4}{%
  Greys-4-1\\%
  Greys-4-2\\%
  Greys-4-3\\%
  Greys-4-4\\%
}
\pgfplotscreateplotcyclelist{Greys-5}{%
  Greys-5-1\\%
  Greys-5-2\\%
  Greys-5-3\\%
  Greys-5-4\\%
  Greys-5-5\\%
}
\pgfplotscreateplotcyclelist{Greys-6}{%
  Greys-6-1\\%
  Greys-6-2\\%
  Greys-6-3\\%
  Greys-6-4\\%
  Greys-6-5\\%
  Greys-6-6\\%
}
\pgfplotscreateplotcyclelist{Greys-7}{%
  Greys-7-1\\%
  Greys-7-2\\%
  Greys-7-3\\%
  Greys-7-4\\%
  Greys-7-5\\%
  Greys-7-6\\%
  Greys-7-7\\%
}
\pgfplotscreateplotcyclelist{Greys-8}{%
  Greys-8-1\\%
  Greys-8-2\\%
  Greys-8-3\\%
  Greys-8-4\\%
  Greys-8-5\\%
  Greys-8-6\\%
  Greys-8-7\\%
  Greys-8-8\\%
}
\pgfplotscreateplotcyclelist{Greys-9}{%
  Greys-9-1\\%
  Greys-9-2\\%
  Greys-9-3\\%
  Greys-9-4\\%
  Greys-9-5\\%
  Greys-9-6\\%
  Greys-9-7\\%
  Greys-9-8\\%
  Greys-9-9\\%
}
\pgfplotscreateplotcyclelist{PuOr-3}{%
  PuOr-3-1\\%
  PuOr-3-2\\%
  PuOr-3-3\\%
}
\pgfplotscreateplotcyclelist{PuOr-4}{%
  PuOr-4-1\\%
  PuOr-4-2\\%
  PuOr-4-3\\%
  PuOr-4-4\\%
}
\pgfplotscreateplotcyclelist{PuOr-5}{%
  PuOr-5-1\\%
  PuOr-5-2\\%
  PuOr-5-3\\%
  PuOr-5-4\\%
  PuOr-5-5\\%
}
\pgfplotscreateplotcyclelist{PuOr-6}{%
  PuOr-6-1\\%
  PuOr-6-2\\%
  PuOr-6-3\\%
  PuOr-6-4\\%
  PuOr-6-5\\%
  PuOr-6-6\\%
}
\pgfplotscreateplotcyclelist{PuOr-7}{%
  PuOr-7-1\\%
  PuOr-7-2\\%
  PuOr-7-3\\%
  PuOr-7-4\\%
  PuOr-7-5\\%
  PuOr-7-6\\%
  PuOr-7-7\\%
}
\pgfplotscreateplotcyclelist{PuOr-8}{%
  PuOr-8-1\\%
  PuOr-8-2\\%
  PuOr-8-3\\%
  PuOr-8-4\\%
  PuOr-8-5\\%
  PuOr-8-6\\%
  PuOr-8-7\\%
  PuOr-8-8\\%
}
\pgfplotscreateplotcyclelist{PuOr-9}{%
  PuOr-9-1\\%
  PuOr-9-2\\%
  PuOr-9-3\\%
  PuOr-9-4\\%
  PuOr-9-5\\%
  PuOr-9-6\\%
  PuOr-9-7\\%
  PuOr-9-8\\%
  PuOr-9-9\\%
}
\pgfplotscreateplotcyclelist{PuOr-10}{%
  PuOr-10-1\\%
  PuOr-10-2\\%
  PuOr-10-3\\%
  PuOr-10-4\\%
  PuOr-10-5\\%
  PuOr-10-6\\%
  PuOr-10-7\\%
  PuOr-10-8\\%
  PuOr-10-9\\%
  PuOr-10-10\\%
}
\pgfplotscreateplotcyclelist{PuOr-11}{%
  PuOr-11-1\\%
  PuOr-11-2\\%
  PuOr-11-3\\%
  PuOr-11-4\\%
  PuOr-11-5\\%
  PuOr-11-6\\%
  PuOr-11-7\\%
  PuOr-11-8\\%
  PuOr-11-9\\%
  PuOr-11-10\\%
  PuOr-11-11\\%
}
\pgfplotscreateplotcyclelist{BrBG-3}{%
  BrBG-3-1\\%
  BrBG-3-2\\%
  BrBG-3-3\\%
}
\pgfplotscreateplotcyclelist{BrBG-4}{%
  BrBG-4-1\\%
  BrBG-4-2\\%
  BrBG-4-3\\%
  BrBG-4-4\\%
}
\pgfplotscreateplotcyclelist{BrBG-5}{%
  BrBG-5-1\\%
  BrBG-5-2\\%
  BrBG-5-3\\%
  BrBG-5-4\\%
  BrBG-5-5\\%
}
\pgfplotscreateplotcyclelist{BrBG-6}{%
  BrBG-6-1\\%
  BrBG-6-2\\%
  BrBG-6-3\\%
  BrBG-6-4\\%
  BrBG-6-5\\%
  BrBG-6-6\\%
}
\pgfplotscreateplotcyclelist{BrBG-7}{%
  BrBG-7-1\\%
  BrBG-7-2\\%
  BrBG-7-3\\%
  BrBG-7-4\\%
  BrBG-7-5\\%
  BrBG-7-6\\%
  BrBG-7-7\\%
}
\pgfplotscreateplotcyclelist{BrBG-8}{%
  BrBG-8-1\\%
  BrBG-8-2\\%
  BrBG-8-3\\%
  BrBG-8-4\\%
  BrBG-8-5\\%
  BrBG-8-6\\%
  BrBG-8-7\\%
  BrBG-8-8\\%
}
\pgfplotscreateplotcyclelist{BrBG-9}{%
  BrBG-9-1\\%
  BrBG-9-2\\%
  BrBG-9-3\\%
  BrBG-9-4\\%
  BrBG-9-5\\%
  BrBG-9-6\\%
  BrBG-9-7\\%
  BrBG-9-8\\%
  BrBG-9-9\\%
}
\pgfplotscreateplotcyclelist{BrBG-10}{%
  BrBG-10-1\\%
  BrBG-10-2\\%
  BrBG-10-3\\%
  BrBG-10-4\\%
  BrBG-10-5\\%
  BrBG-10-6\\%
  BrBG-10-7\\%
  BrBG-10-8\\%
  BrBG-10-9\\%
  BrBG-10-10\\%
}
\pgfplotscreateplotcyclelist{BrBG-11}{%
  BrBG-11-1\\%
  BrBG-11-2\\%
  BrBG-11-3\\%
  BrBG-11-4\\%
  BrBG-11-5\\%
  BrBG-11-6\\%
  BrBG-11-7\\%
  BrBG-11-8\\%
  BrBG-11-9\\%
  BrBG-11-10\\%
  BrBG-11-11\\%
}
\pgfplotscreateplotcyclelist{PRGn-3}{%
  PRGn-3-1\\%
  PRGn-3-2\\%
  PRGn-3-3\\%
}
\pgfplotscreateplotcyclelist{PRGn-4}{%
  PRGn-4-1\\%
  PRGn-4-2\\%
  PRGn-4-3\\%
  PRGn-4-4\\%
}
\pgfplotscreateplotcyclelist{PRGn-5}{%
  PRGn-5-1\\%
  PRGn-5-2\\%
  PRGn-5-3\\%
  PRGn-5-4\\%
  PRGn-5-5\\%
}
\pgfplotscreateplotcyclelist{PRGn-6}{%
  PRGn-6-1\\%
  PRGn-6-2\\%
  PRGn-6-3\\%
  PRGn-6-4\\%
  PRGn-6-5\\%
  PRGn-6-6\\%
}
\pgfplotscreateplotcyclelist{PRGn-7}{%
  PRGn-7-1\\%
  PRGn-7-2\\%
  PRGn-7-3\\%
  PRGn-7-4\\%
  PRGn-7-5\\%
  PRGn-7-6\\%
  PRGn-7-7\\%
}
\pgfplotscreateplotcyclelist{PRGn-8}{%
  PRGn-8-1\\%
  PRGn-8-2\\%
  PRGn-8-3\\%
  PRGn-8-4\\%
  PRGn-8-5\\%
  PRGn-8-6\\%
  PRGn-8-7\\%
  PRGn-8-8\\%
}
\pgfplotscreateplotcyclelist{PRGn-9}{%
  PRGn-9-1\\%
  PRGn-9-2\\%
  PRGn-9-3\\%
  PRGn-9-4\\%
  PRGn-9-5\\%
  PRGn-9-6\\%
  PRGn-9-7\\%
  PRGn-9-8\\%
  PRGn-9-9\\%
}
\pgfplotscreateplotcyclelist{PRGn-10}{%
  PRGn-10-1\\%
  PRGn-10-2\\%
  PRGn-10-3\\%
  PRGn-10-4\\%
  PRGn-10-5\\%
  PRGn-10-6\\%
  PRGn-10-7\\%
  PRGn-10-8\\%
  PRGn-10-9\\%
  PRGn-10-10\\%
}
\pgfplotscreateplotcyclelist{PRGn-11}{%
  PRGn-11-1\\%
  PRGn-11-2\\%
  PRGn-11-3\\%
  PRGn-11-4\\%
  PRGn-11-5\\%
  PRGn-11-6\\%
  PRGn-11-7\\%
  PRGn-11-8\\%
  PRGn-11-9\\%
  PRGn-11-10\\%
  PRGn-11-11\\%
}
\pgfplotscreateplotcyclelist{PiYG-3}{%
  PiYG-3-1\\%
  PiYG-3-2\\%
  PiYG-3-3\\%
}
\pgfplotscreateplotcyclelist{PiYG-4}{%
  PiYG-4-1\\%
  PiYG-4-2\\%
  PiYG-4-3\\%
  PiYG-4-4\\%
}
\pgfplotscreateplotcyclelist{PiYG-5}{%
  PiYG-5-1\\%
  PiYG-5-2\\%
  PiYG-5-3\\%
  PiYG-5-4\\%
  PiYG-5-5\\%
}
\pgfplotscreateplotcyclelist{PiYG-6}{%
  PiYG-6-1\\%
  PiYG-6-2\\%
  PiYG-6-3\\%
  PiYG-6-4\\%
  PiYG-6-5\\%
  PiYG-6-6\\%
}
\pgfplotscreateplotcyclelist{PiYG-7}{%
  PiYG-7-1\\%
  PiYG-7-2\\%
  PiYG-7-3\\%
  PiYG-7-4\\%
  PiYG-7-5\\%
  PiYG-7-6\\%
  PiYG-7-7\\%
}
\pgfplotscreateplotcyclelist{PiYG-8}{%
  PiYG-8-1\\%
  PiYG-8-2\\%
  PiYG-8-3\\%
  PiYG-8-4\\%
  PiYG-8-5\\%
  PiYG-8-6\\%
  PiYG-8-7\\%
  PiYG-8-8\\%
}
\pgfplotscreateplotcyclelist{PiYG-9}{%
  PiYG-9-1\\%
  PiYG-9-2\\%
  PiYG-9-3\\%
  PiYG-9-4\\%
  PiYG-9-5\\%
  PiYG-9-6\\%
  PiYG-9-7\\%
  PiYG-9-8\\%
  PiYG-9-9\\%
}
\pgfplotscreateplotcyclelist{PiYG-10}{%
  PiYG-10-1\\%
  PiYG-10-2\\%
  PiYG-10-3\\%
  PiYG-10-4\\%
  PiYG-10-5\\%
  PiYG-10-6\\%
  PiYG-10-7\\%
  PiYG-10-8\\%
  PiYG-10-9\\%
  PiYG-10-10\\%
}
\pgfplotscreateplotcyclelist{PiYG-11}{%
  PiYG-11-1\\%
  PiYG-11-2\\%
  PiYG-11-3\\%
  PiYG-11-4\\%
  PiYG-11-5\\%
  PiYG-11-6\\%
  PiYG-11-7\\%
  PiYG-11-8\\%
  PiYG-11-9\\%
  PiYG-11-10\\%
  PiYG-11-11\\%
}
\pgfplotscreateplotcyclelist{RdBu-3}{%
  RdBu-3-1\\%
  RdBu-3-2\\%
  RdBu-3-3\\%
}
\pgfplotscreateplotcyclelist{RdBu-4}{%
  RdBu-4-1\\%
  RdBu-4-2\\%
  RdBu-4-3\\%
  RdBu-4-4\\%
}
\pgfplotscreateplotcyclelist{RdBu-5}{%
  RdBu-5-1\\%
  RdBu-5-2\\%
  RdBu-5-3\\%
  RdBu-5-4\\%
  RdBu-5-5\\%
}
\pgfplotscreateplotcyclelist{RdBu-6}{%
  RdBu-6-1\\%
  RdBu-6-2\\%
  RdBu-6-3\\%
  RdBu-6-4\\%
  RdBu-6-5\\%
  RdBu-6-6\\%
}
\pgfplotscreateplotcyclelist{RdBu-7}{%
  RdBu-7-1\\%
  RdBu-7-2\\%
  RdBu-7-3\\%
  RdBu-7-4\\%
  RdBu-7-5\\%
  RdBu-7-6\\%
  RdBu-7-7\\%
}
\pgfplotscreateplotcyclelist{RdBu-8}{%
  RdBu-8-1\\%
  RdBu-8-2\\%
  RdBu-8-3\\%
  RdBu-8-4\\%
  RdBu-8-5\\%
  RdBu-8-6\\%
  RdBu-8-7\\%
  RdBu-8-8\\%
}
\pgfplotscreateplotcyclelist{RdBu-9}{%
  RdBu-9-1\\%
  RdBu-9-2\\%
  RdBu-9-3\\%
  RdBu-9-4\\%
  RdBu-9-5\\%
  RdBu-9-6\\%
  RdBu-9-7\\%
  RdBu-9-8\\%
  RdBu-9-9\\%
}
\pgfplotscreateplotcyclelist{RdBu-10}{%
  RdBu-10-1\\%
  RdBu-10-2\\%
  RdBu-10-3\\%
  RdBu-10-4\\%
  RdBu-10-5\\%
  RdBu-10-6\\%
  RdBu-10-7\\%
  RdBu-10-8\\%
  RdBu-10-9\\%
  RdBu-10-10\\%
}
\pgfplotscreateplotcyclelist{RdBu-11}{%
  RdBu-11-1\\%
  RdBu-11-2\\%
  RdBu-11-3\\%
  RdBu-11-4\\%
  RdBu-11-5\\%
  RdBu-11-6\\%
  RdBu-11-7\\%
  RdBu-11-8\\%
  RdBu-11-9\\%
  RdBu-11-10\\%
  RdBu-11-11\\%
}
\pgfplotscreateplotcyclelist{RdGy-3}{%
  RdGy-3-1\\%
  RdGy-3-2\\%
  RdGy-3-3\\%
}
\pgfplotscreateplotcyclelist{RdGy-4}{%
  RdGy-4-1\\%
  RdGy-4-2\\%
  RdGy-4-3\\%
  RdGy-4-4\\%
}
\pgfplotscreateplotcyclelist{RdGy-5}{%
  RdGy-5-1\\%
  RdGy-5-2\\%
  RdGy-5-3\\%
  RdGy-5-4\\%
  RdGy-5-5\\%
}
\pgfplotscreateplotcyclelist{RdGy-6}{%
  RdGy-6-1\\%
  RdGy-6-2\\%
  RdGy-6-3\\%
  RdGy-6-4\\%
  RdGy-6-5\\%
  RdGy-6-6\\%
}
\pgfplotscreateplotcyclelist{RdGy-7}{%
  RdGy-7-1\\%
  RdGy-7-2\\%
  RdGy-7-3\\%
  RdGy-7-4\\%
  RdGy-7-5\\%
  RdGy-7-6\\%
  RdGy-7-7\\%
}
\pgfplotscreateplotcyclelist{RdGy-8}{%
  RdGy-8-1\\%
  RdGy-8-2\\%
  RdGy-8-3\\%
  RdGy-8-4\\%
  RdGy-8-5\\%
  RdGy-8-6\\%
  RdGy-8-7\\%
  RdGy-8-8\\%
}
\pgfplotscreateplotcyclelist{RdGy-9}{%
  RdGy-9-1\\%
  RdGy-9-2\\%
  RdGy-9-3\\%
  RdGy-9-4\\%
  RdGy-9-5\\%
  RdGy-9-6\\%
  RdGy-9-7\\%
  RdGy-9-8\\%
  RdGy-9-9\\%
}
\pgfplotscreateplotcyclelist{RdGy-10}{%
  RdGy-10-1\\%
  RdGy-10-2\\%
  RdGy-10-3\\%
  RdGy-10-4\\%
  RdGy-10-5\\%
  RdGy-10-6\\%
  RdGy-10-7\\%
  RdGy-10-8\\%
  RdGy-10-9\\%
  RdGy-10-10\\%
}
\pgfplotscreateplotcyclelist{RdGy-11}{%
  RdGy-11-1\\%
  RdGy-11-2\\%
  RdGy-11-3\\%
  RdGy-11-4\\%
  RdGy-11-5\\%
  RdGy-11-6\\%
  RdGy-11-7\\%
  RdGy-11-8\\%
  RdGy-11-9\\%
  RdGy-11-10\\%
  RdGy-11-11\\%
}
\pgfplotscreateplotcyclelist{RdYlBu-3}{%
  RdYlBu-3-1\\%
  RdYlBu-3-2\\%
  RdYlBu-3-3\\%
}
\pgfplotscreateplotcyclelist{RdYlBu-4}{%
  RdYlBu-4-1\\%
  RdYlBu-4-2\\%
  RdYlBu-4-3\\%
  RdYlBu-4-4\\%
}
\pgfplotscreateplotcyclelist{RdYlBu-5}{%
  RdYlBu-5-1\\%
  RdYlBu-5-2\\%
  RdYlBu-5-3\\%
  RdYlBu-5-4\\%
  RdYlBu-5-5\\%
}
\pgfplotscreateplotcyclelist{RdYlBu-6}{%
  RdYlBu-6-1\\%
  RdYlBu-6-2\\%
  RdYlBu-6-3\\%
  RdYlBu-6-4\\%
  RdYlBu-6-5\\%
  RdYlBu-6-6\\%
}
\pgfplotscreateplotcyclelist{RdYlBu-7}{%
  RdYlBu-7-1\\%
  RdYlBu-7-2\\%
  RdYlBu-7-3\\%
  RdYlBu-7-4\\%
  RdYlBu-7-5\\%
  RdYlBu-7-6\\%
  RdYlBu-7-7\\%
}
\pgfplotscreateplotcyclelist{RdYlBu-8}{%
  RdYlBu-8-1\\%
  RdYlBu-8-2\\%
  RdYlBu-8-3\\%
  RdYlBu-8-4\\%
  RdYlBu-8-5\\%
  RdYlBu-8-6\\%
  RdYlBu-8-7\\%
  RdYlBu-8-8\\%
}
\pgfplotscreateplotcyclelist{RdYlBu-9}{%
  RdYlBu-9-1\\%
  RdYlBu-9-2\\%
  RdYlBu-9-3\\%
  RdYlBu-9-4\\%
  RdYlBu-9-5\\%
  RdYlBu-9-6\\%
  RdYlBu-9-7\\%
  RdYlBu-9-8\\%
  RdYlBu-9-9\\%
}
\pgfplotscreateplotcyclelist{RdYlBu-10}{%
  RdYlBu-10-1\\%
  RdYlBu-10-2\\%
  RdYlBu-10-3\\%
  RdYlBu-10-4\\%
  RdYlBu-10-5\\%
  RdYlBu-10-6\\%
  RdYlBu-10-7\\%
  RdYlBu-10-8\\%
  RdYlBu-10-9\\%
  RdYlBu-10-10\\%
}
\pgfplotscreateplotcyclelist{RdYlBu-11}{%
  RdYlBu-11-1\\%
  RdYlBu-11-2\\%
  RdYlBu-11-3\\%
  RdYlBu-11-4\\%
  RdYlBu-11-5\\%
  RdYlBu-11-6\\%
  RdYlBu-11-7\\%
  RdYlBu-11-8\\%
  RdYlBu-11-9\\%
  RdYlBu-11-10\\%
  RdYlBu-11-11\\%
}
\pgfplotscreateplotcyclelist{Spectral-3}{%
  Spectral-3-1\\%
  Spectral-3-2\\%
  Spectral-3-3\\%
}
\pgfplotscreateplotcyclelist{Spectral-4}{%
  Spectral-4-1\\%
  Spectral-4-2\\%
  Spectral-4-3\\%
  Spectral-4-4\\%
}
\pgfplotscreateplotcyclelist{Spectral-5}{%
  Spectral-5-1\\%
  Spectral-5-2\\%
  Spectral-5-3\\%
  Spectral-5-4\\%
  Spectral-5-5\\%
}
\pgfplotscreateplotcyclelist{Spectral-6}{%
  Spectral-6-1\\%
  Spectral-6-2\\%
  Spectral-6-3\\%
  Spectral-6-4\\%
  Spectral-6-5\\%
  Spectral-6-6\\%
}
\pgfplotscreateplotcyclelist{Spectral-7}{%
  Spectral-7-1\\%
  Spectral-7-2\\%
  Spectral-7-3\\%
  Spectral-7-4\\%
  Spectral-7-5\\%
  Spectral-7-6\\%
  Spectral-7-7\\%
}
\pgfplotscreateplotcyclelist{Spectral-8}{%
  Spectral-8-1\\%
  Spectral-8-2\\%
  Spectral-8-3\\%
  Spectral-8-4\\%
  Spectral-8-5\\%
  Spectral-8-6\\%
  Spectral-8-7\\%
  Spectral-8-8\\%
}
\pgfplotscreateplotcyclelist{Spectral-9}{%
  Spectral-9-1\\%
  Spectral-9-2\\%
  Spectral-9-3\\%
  Spectral-9-4\\%
  Spectral-9-5\\%
  Spectral-9-6\\%
  Spectral-9-7\\%
  Spectral-9-8\\%
  Spectral-9-9\\%
}
\pgfplotscreateplotcyclelist{Spectral-10}{%
  Spectral-10-1\\%
  Spectral-10-2\\%
  Spectral-10-3\\%
  Spectral-10-4\\%
  Spectral-10-5\\%
  Spectral-10-6\\%
  Spectral-10-7\\%
  Spectral-10-8\\%
  Spectral-10-9\\%
  Spectral-10-10\\%
}
\pgfplotscreateplotcyclelist{Spectral-11}{%
  Spectral-11-1\\%
  Spectral-11-2\\%
  Spectral-11-3\\%
  Spectral-11-4\\%
  Spectral-11-5\\%
  Spectral-11-6\\%
  Spectral-11-7\\%
  Spectral-11-8\\%
  Spectral-11-9\\%
  Spectral-11-10\\%
  Spectral-11-11\\%
}
\pgfplotscreateplotcyclelist{RdYlGn-3}{%
  RdYlGn-3-1\\%
  RdYlGn-3-2\\%
  RdYlGn-3-3\\%
}
\pgfplotscreateplotcyclelist{RdYlGn-4}{%
  RdYlGn-4-1\\%
  RdYlGn-4-2\\%
  RdYlGn-4-3\\%
  RdYlGn-4-4\\%
}
\pgfplotscreateplotcyclelist{RdYlGn-5}{%
  RdYlGn-5-1\\%
  RdYlGn-5-2\\%
  RdYlGn-5-3\\%
  RdYlGn-5-4\\%
  RdYlGn-5-5\\%
}
\pgfplotscreateplotcyclelist{RdYlGn-6}{%
  RdYlGn-6-1\\%
  RdYlGn-6-2\\%
  RdYlGn-6-3\\%
  RdYlGn-6-4\\%
  RdYlGn-6-5\\%
  RdYlGn-6-6\\%
}
\pgfplotscreateplotcyclelist{RdYlGn-7}{%
  RdYlGn-7-1\\%
  RdYlGn-7-2\\%
  RdYlGn-7-3\\%
  RdYlGn-7-4\\%
  RdYlGn-7-5\\%
  RdYlGn-7-6\\%
  RdYlGn-7-7\\%
}
\pgfplotscreateplotcyclelist{RdYlGn-8}{%
  RdYlGn-8-1\\%
  RdYlGn-8-2\\%
  RdYlGn-8-3\\%
  RdYlGn-8-4\\%
  RdYlGn-8-5\\%
  RdYlGn-8-6\\%
  RdYlGn-8-7\\%
  RdYlGn-8-8\\%
}
\pgfplotscreateplotcyclelist{RdYlGn-9}{%
  RdYlGn-9-1\\%
  RdYlGn-9-2\\%
  RdYlGn-9-3\\%
  RdYlGn-9-4\\%
  RdYlGn-9-5\\%
  RdYlGn-9-6\\%
  RdYlGn-9-7\\%
  RdYlGn-9-8\\%
  RdYlGn-9-9\\%
}
\pgfplotscreateplotcyclelist{RdYlGn-10}{%
  RdYlGn-10-1\\%
  RdYlGn-10-2\\%
  RdYlGn-10-3\\%
  RdYlGn-10-4\\%
  RdYlGn-10-5\\%
  RdYlGn-10-6\\%
  RdYlGn-10-7\\%
  RdYlGn-10-8\\%
  RdYlGn-10-9\\%
  RdYlGn-10-10\\%
}
\pgfplotscreateplotcyclelist{RdYlGn-11}{%
  RdYlGn-11-1\\%
  RdYlGn-11-2\\%
  RdYlGn-11-3\\%
  RdYlGn-11-4\\%
  RdYlGn-11-5\\%
  RdYlGn-11-6\\%
  RdYlGn-11-7\\%
  RdYlGn-11-8\\%
  RdYlGn-11-9\\%
  RdYlGn-11-10\\%
  RdYlGn-11-11\\%
}
\pgfplotscreateplotcyclelist{Set3-3}{%
  Set3-3-1\\%
  Set3-3-2\\%
  Set3-3-3\\%
}
\pgfplotscreateplotcyclelist{Set3-4}{%
  Set3-4-1\\%
  Set3-4-2\\%
  Set3-4-3\\%
  Set3-4-4\\%
}
\pgfplotscreateplotcyclelist{Set3-5}{%
  Set3-5-1\\%
  Set3-5-2\\%
  Set3-5-3\\%
  Set3-5-4\\%
  Set3-5-5\\%
}
\pgfplotscreateplotcyclelist{Set3-6}{%
  Set3-6-1\\%
  Set3-6-2\\%
  Set3-6-3\\%
  Set3-6-4\\%
  Set3-6-5\\%
  Set3-6-6\\%
}
\pgfplotscreateplotcyclelist{Set3-7}{%
  Set3-7-1\\%
  Set3-7-2\\%
  Set3-7-3\\%
  Set3-7-4\\%
  Set3-7-5\\%
  Set3-7-6\\%
  Set3-7-7\\%
}
\pgfplotscreateplotcyclelist{Set3-8}{%
  Set3-8-1\\%
  Set3-8-2\\%
  Set3-8-3\\%
  Set3-8-4\\%
  Set3-8-5\\%
  Set3-8-6\\%
  Set3-8-7\\%
  Set3-8-8\\%
}
\pgfplotscreateplotcyclelist{Set3-9}{%
  Set3-9-1\\%
  Set3-9-2\\%
  Set3-9-3\\%
  Set3-9-4\\%
  Set3-9-5\\%
  Set3-9-6\\%
  Set3-9-7\\%
  Set3-9-8\\%
  Set3-9-9\\%
}
\pgfplotscreateplotcyclelist{Set3-10}{%
  Set3-10-1\\%
  Set3-10-2\\%
  Set3-10-3\\%
  Set3-10-4\\%
  Set3-10-5\\%
  Set3-10-6\\%
  Set3-10-7\\%
  Set3-10-8\\%
  Set3-10-9\\%
  Set3-10-10\\%
}
\pgfplotscreateplotcyclelist{Set3-11}{%
  Set3-11-1\\%
  Set3-11-2\\%
  Set3-11-3\\%
  Set3-11-4\\%
  Set3-11-5\\%
  Set3-11-6\\%
  Set3-11-7\\%
  Set3-11-8\\%
  Set3-11-9\\%
  Set3-11-10\\%
  Set3-11-11\\%
}
\pgfplotscreateplotcyclelist{Set3-12}{%
  Set3-12-1\\%
  Set3-12-2\\%
  Set3-12-3\\%
  Set3-12-4\\%
  Set3-12-5\\%
  Set3-12-6\\%
  Set3-12-7\\%
  Set3-12-8\\%
  Set3-12-9\\%
  Set3-12-10\\%
  Set3-12-11\\%
  Set3-12-12\\%
}
\pgfplotscreateplotcyclelist{Pastel1-3}{%
  Pastel1-3-1\\%
  Pastel1-3-2\\%
  Pastel1-3-3\\%
}
\pgfplotscreateplotcyclelist{Pastel1-4}{%
  Pastel1-4-1\\%
  Pastel1-4-2\\%
  Pastel1-4-3\\%
  Pastel1-4-4\\%
}
\pgfplotscreateplotcyclelist{Pastel1-5}{%
  Pastel1-5-1\\%
  Pastel1-5-2\\%
  Pastel1-5-3\\%
  Pastel1-5-4\\%
  Pastel1-5-5\\%
}
\pgfplotscreateplotcyclelist{Pastel1-6}{%
  Pastel1-6-1\\%
  Pastel1-6-2\\%
  Pastel1-6-3\\%
  Pastel1-6-4\\%
  Pastel1-6-5\\%
  Pastel1-6-6\\%
}
\pgfplotscreateplotcyclelist{Pastel1-7}{%
  Pastel1-7-1\\%
  Pastel1-7-2\\%
  Pastel1-7-3\\%
  Pastel1-7-4\\%
  Pastel1-7-5\\%
  Pastel1-7-6\\%
  Pastel1-7-7\\%
}
\pgfplotscreateplotcyclelist{Pastel1-8}{%
  Pastel1-8-1\\%
  Pastel1-8-2\\%
  Pastel1-8-3\\%
  Pastel1-8-4\\%
  Pastel1-8-5\\%
  Pastel1-8-6\\%
  Pastel1-8-7\\%
  Pastel1-8-8\\%
}
\pgfplotscreateplotcyclelist{Pastel1-9}{%
  Pastel1-9-1\\%
  Pastel1-9-2\\%
  Pastel1-9-3\\%
  Pastel1-9-4\\%
  Pastel1-9-5\\%
  Pastel1-9-6\\%
  Pastel1-9-7\\%
  Pastel1-9-8\\%
  Pastel1-9-9\\%
}
\pgfplotscreateplotcyclelist{Set1-3}{%
  Set1-3-1\\%
  Set1-3-2\\%
  Set1-3-3\\%
}
\pgfplotscreateplotcyclelist{Set1-4}{%
  Set1-4-1\\%
  Set1-4-2\\%
  Set1-4-3\\%
  Set1-4-4\\%
}
\pgfplotscreateplotcyclelist{Set1-5}{%
  Set1-5-1\\%
  Set1-5-2\\%
  Set1-5-3\\%
  Set1-5-4\\%
  Set1-5-5\\%
}
\pgfplotscreateplotcyclelist{Set1-6}{%
  Set1-6-1\\%
  Set1-6-2\\%
  Set1-6-3\\%
  Set1-6-4\\%
  Set1-6-5\\%
  Set1-6-6\\%
}
\pgfplotscreateplotcyclelist{Set1-7}{%
  Set1-7-1\\%
  Set1-7-2\\%
  Set1-7-3\\%
  Set1-7-4\\%
  Set1-7-5\\%
  Set1-7-6\\%
  Set1-7-7\\%
}
\pgfplotscreateplotcyclelist{Set1-8}{%
  Set1-8-1\\%
  Set1-8-2\\%
  Set1-8-3\\%
  Set1-8-4\\%
  Set1-8-5\\%
  Set1-8-6\\%
  Set1-8-7\\%
  Set1-8-8\\%
}
\pgfplotscreateplotcyclelist{Set1-9}{%
  Set1-9-1\\%
  Set1-9-2\\%
  Set1-9-3\\%
  Set1-9-4\\%
  Set1-9-5\\%
  Set1-9-6\\%
  Set1-9-7\\%
  Set1-9-8\\%
  Set1-9-9\\%
}
\pgfplotscreateplotcyclelist{Pastel2-3}{%
  Pastel2-3-1\\%
  Pastel2-3-2\\%
  Pastel2-3-3\\%
}
\pgfplotscreateplotcyclelist{Pastel2-4}{%
  Pastel2-4-1\\%
  Pastel2-4-2\\%
  Pastel2-4-3\\%
  Pastel2-4-4\\%
}
\pgfplotscreateplotcyclelist{Pastel2-5}{%
  Pastel2-5-1\\%
  Pastel2-5-2\\%
  Pastel2-5-3\\%
  Pastel2-5-4\\%
  Pastel2-5-5\\%
}
\pgfplotscreateplotcyclelist{Pastel2-6}{%
  Pastel2-6-1\\%
  Pastel2-6-2\\%
  Pastel2-6-3\\%
  Pastel2-6-4\\%
  Pastel2-6-5\\%
  Pastel2-6-6\\%
}
\pgfplotscreateplotcyclelist{Pastel2-7}{%
  Pastel2-7-1\\%
  Pastel2-7-2\\%
  Pastel2-7-3\\%
  Pastel2-7-4\\%
  Pastel2-7-5\\%
  Pastel2-7-6\\%
  Pastel2-7-7\\%
}
\pgfplotscreateplotcyclelist{Pastel2-8}{%
  Pastel2-8-1\\%
  Pastel2-8-2\\%
  Pastel2-8-3\\%
  Pastel2-8-4\\%
  Pastel2-8-5\\%
  Pastel2-8-6\\%
  Pastel2-8-7\\%
  Pastel2-8-8\\%
}
\pgfplotscreateplotcyclelist{Set2-3}{%
  Set2-3-1\\%
  Set2-3-2\\%
  Set2-3-3\\%
}
\pgfplotscreateplotcyclelist{Set2-4}{%
  Set2-4-1\\%
  Set2-4-2\\%
  Set2-4-3\\%
  Set2-4-4\\%
}
\pgfplotscreateplotcyclelist{Set2-5}{%
  Set2-5-1\\%
  Set2-5-2\\%
  Set2-5-3\\%
  Set2-5-4\\%
  Set2-5-5\\%
}
\pgfplotscreateplotcyclelist{Set2-6}{%
  Set2-6-1\\%
  Set2-6-2\\%
  Set2-6-3\\%
  Set2-6-4\\%
  Set2-6-5\\%
  Set2-6-6\\%
}
\pgfplotscreateplotcyclelist{Set2-7}{%
  Set2-7-1\\%
  Set2-7-2\\%
  Set2-7-3\\%
  Set2-7-4\\%
  Set2-7-5\\%
  Set2-7-6\\%
  Set2-7-7\\%
}
\pgfplotscreateplotcyclelist{Set2-8}{%
  Set2-8-1\\%
  Set2-8-2\\%
  Set2-8-3\\%
  Set2-8-4\\%
  Set2-8-5\\%
  Set2-8-6\\%
  Set2-8-7\\%
  Set2-8-8\\%
}
\pgfplotscreateplotcyclelist{Dark2-3}{%
  Dark2-3-1\\%
  Dark2-3-2\\%
  Dark2-3-3\\%
}
\pgfplotscreateplotcyclelist{Dark2-4}{%
  Dark2-4-1\\%
  Dark2-4-2\\%
  Dark2-4-3\\%
  Dark2-4-4\\%
}
\pgfplotscreateplotcyclelist{Dark2-5}{%
  Dark2-5-1\\%
  Dark2-5-2\\%
  Dark2-5-3\\%
  Dark2-5-4\\%
  Dark2-5-5\\%
}
\pgfplotscreateplotcyclelist{Dark2-6}{%
  Dark2-6-1\\%
  Dark2-6-2\\%
  Dark2-6-3\\%
  Dark2-6-4\\%
  Dark2-6-5\\%
  Dark2-6-6\\%
}
\pgfplotscreateplotcyclelist{Dark2-7}{%
  Dark2-7-1\\%
  Dark2-7-2\\%
  Dark2-7-3\\%
  Dark2-7-4\\%
  Dark2-7-5\\%
  Dark2-7-6\\%
  Dark2-7-7\\%
}
\pgfplotscreateplotcyclelist{Dark2-8}{%
  Dark2-8-1\\%
  Dark2-8-2\\%
  Dark2-8-3\\%
  Dark2-8-4\\%
  Dark2-8-5\\%
  Dark2-8-6\\%
  Dark2-8-7\\%
  Dark2-8-8\\%
}
\pgfplotscreateplotcyclelist{Paired-3}{%
  Paired-3-1\\%
  Paired-3-2\\%
  Paired-3-3\\%
}
\pgfplotscreateplotcyclelist{Paired-4}{%
  Paired-4-1\\%
  Paired-4-2\\%
  Paired-4-3\\%
  Paired-4-4\\%
}
\pgfplotscreateplotcyclelist{Paired-5}{%
  Paired-5-1\\%
  Paired-5-2\\%
  Paired-5-3\\%
  Paired-5-4\\%
  Paired-5-5\\%
}
\pgfplotscreateplotcyclelist{Paired-6}{%
  Paired-6-1\\%
  Paired-6-2\\%
  Paired-6-3\\%
  Paired-6-4\\%
  Paired-6-5\\%
  Paired-6-6\\%
}
\pgfplotscreateplotcyclelist{Paired-7}{%
  Paired-7-1\\%
  Paired-7-2\\%
  Paired-7-3\\%
  Paired-7-4\\%
  Paired-7-5\\%
  Paired-7-6\\%
  Paired-7-7\\%
}
\pgfplotscreateplotcyclelist{Paired-8}{%
  Paired-8-1\\%
  Paired-8-2\\%
  Paired-8-3\\%
  Paired-8-4\\%
  Paired-8-5\\%
  Paired-8-6\\%
  Paired-8-7\\%
  Paired-8-8\\%
}
\pgfplotscreateplotcyclelist{Paired-9}{%
  Paired-9-1\\%
  Paired-9-2\\%
  Paired-9-3\\%
  Paired-9-4\\%
  Paired-9-5\\%
  Paired-9-6\\%
  Paired-9-7\\%
  Paired-9-8\\%
  Paired-9-9\\%
}
\pgfplotscreateplotcyclelist{Paired-10}{%
  Paired-10-1\\%
  Paired-10-2\\%
  Paired-10-3\\%
  Paired-10-4\\%
  Paired-10-5\\%
  Paired-10-6\\%
  Paired-10-7\\%
  Paired-10-8\\%
  Paired-10-9\\%
  Paired-10-10\\%
}
\pgfplotscreateplotcyclelist{Paired-11}{%
  Paired-11-1\\%
  Paired-11-2\\%
  Paired-11-3\\%
  Paired-11-4\\%
  Paired-11-5\\%
  Paired-11-6\\%
  Paired-11-7\\%
  Paired-11-8\\%
  Paired-11-9\\%
  Paired-11-10\\%
  Paired-11-11\\%
}
\pgfplotscreateplotcyclelist{Paired-12}{%
  Paired-12-1\\%
  Paired-12-2\\%
  Paired-12-3\\%
  Paired-12-4\\%
  Paired-12-5\\%
  Paired-12-6\\%
  Paired-12-7\\%
  Paired-12-8\\%
  Paired-12-9\\%
  Paired-12-10\\%
  Paired-12-11\\%
  Paired-12-12\\%
}
\pgfplotscreateplotcyclelist{Accent-3}{%
  Accent-3-1\\%
  Accent-3-2\\%
  Accent-3-3\\%
}
\pgfplotscreateplotcyclelist{Accent-4}{%
  Accent-4-1\\%
  Accent-4-2\\%
  Accent-4-3\\%
  Accent-4-4\\%
}
\pgfplotscreateplotcyclelist{Accent-5}{%
  Accent-5-1\\%
  Accent-5-2\\%
  Accent-5-3\\%
  Accent-5-4\\%
  Accent-5-5\\%
}
\pgfplotscreateplotcyclelist{Accent-6}{%
  Accent-6-1\\%
  Accent-6-2\\%
  Accent-6-3\\%
  Accent-6-4\\%
  Accent-6-5\\%
  Accent-6-6\\%
}
\pgfplotscreateplotcyclelist{Accent-7}{%
  Accent-7-1\\%
  Accent-7-2\\%
  Accent-7-3\\%
  Accent-7-4\\%
  Accent-7-5\\%
  Accent-7-6\\%
  Accent-7-7\\%
}
\pgfplotscreateplotcyclelist{Accent-8}{%
  Accent-8-1\\%
  Accent-8-2\\%
  Accent-8-3\\%
  Accent-8-4\\%
  Accent-8-5\\%
  Accent-8-6\\%
  Accent-8-7\\%
  Accent-8-8\\%
}
\pgfplotscreateplotcyclelist{Mark-YlGn-3}{%
 YlGn-3-1!70!black, every mark/.append style={solid,fill=YlGn-3-1},mark=*\\%
 YlGn-3-2!70!black, every mark/.append style={solid,fill=YlGn-3-2},mark=square*\\%
 YlGn-3-3!70!black, every mark/.append style={solid,fill=YlGn-3-3},mark=triangle*\\%
}
\pgfplotscreateplotcyclelist{Mark-YlGn-4}{%
 YlGn-4-1!70!black, every mark/.append style={solid,fill=YlGn-4-1},mark=*\\%
 YlGn-4-2!70!black, every mark/.append style={solid,fill=YlGn-4-2},mark=square*\\%
 YlGn-4-3!70!black, every mark/.append style={solid,fill=YlGn-4-3},mark=triangle*\\%
 YlGn-4-4!70!black, every mark/.append style={solid,fill=YlGn-4-4},mark=diamond*\\%
}
\pgfplotscreateplotcyclelist{Mark-YlGn-5}{%
 YlGn-5-1!70!black, every mark/.append style={solid,fill=YlGn-5-1},mark=*\\%
 YlGn-5-2!70!black, every mark/.append style={solid,fill=YlGn-5-2},mark=square*\\%
 YlGn-5-3!70!black, every mark/.append style={solid,fill=YlGn-5-3},mark=triangle*\\%
 YlGn-5-4!70!black, every mark/.append style={solid,fill=YlGn-5-4},mark=diamond*\\%
 YlGn-5-5!70!black, every mark/.append style={solid,fill=YlGn-5-5},mark=pentagon*\\%
}
\pgfplotscreateplotcyclelist{Mark-YlGn-6}{%
 YlGn-6-1!70!black, every mark/.append style={solid,fill=YlGn-6-1},mark=*\\%
 YlGn-6-2!70!black, every mark/.append style={solid,fill=YlGn-6-2},mark=square*\\%
 YlGn-6-3!70!black, every mark/.append style={solid,fill=YlGn-6-3},mark=triangle*\\%
 YlGn-6-4!70!black, every mark/.append style={solid,fill=YlGn-6-4},mark=diamond*\\%
 YlGn-6-5!70!black, every mark/.append style={solid,fill=YlGn-6-5},mark=pentagon*\\%
 YlGn-6-6!70!black, every mark/.append style={solid,fill=YlGn-6-6},mark=x\\%
}
\pgfplotscreateplotcyclelist{Mark-YlGn-7}{%
 YlGn-7-1!70!black, every mark/.append style={solid,fill=YlGn-7-1},mark=*\\%
 YlGn-7-2!70!black, every mark/.append style={solid,fill=YlGn-7-2},mark=square*\\%
 YlGn-7-3!70!black, every mark/.append style={solid,fill=YlGn-7-3},mark=triangle*\\%
 YlGn-7-4!70!black, every mark/.append style={solid,fill=YlGn-7-4},mark=diamond*\\%
 YlGn-7-5!70!black, every mark/.append style={solid,fill=YlGn-7-5},mark=pentagon*\\%
 YlGn-7-6!70!black, every mark/.append style={solid,fill=YlGn-7-6},mark=x\\%
 YlGn-7-7!70!black, every mark/.append style={solid,fill=YlGn-7-7},mark=+\\%
}
\pgfplotscreateplotcyclelist{Mark-YlGn-8}{%
 YlGn-8-1!70!black, every mark/.append style={solid,fill=YlGn-8-1},mark=*\\%
 YlGn-8-2!70!black, every mark/.append style={solid,fill=YlGn-8-2},mark=square*\\%
 YlGn-8-3!70!black, every mark/.append style={solid,fill=YlGn-8-3},mark=triangle*\\%
 YlGn-8-4!70!black, every mark/.append style={solid,fill=YlGn-8-4},mark=diamond*\\%
 YlGn-8-5!70!black, every mark/.append style={solid,fill=YlGn-8-5},mark=pentagon*\\%
 YlGn-8-6!70!black, every mark/.append style={solid,fill=YlGn-8-6},mark=x\\%
 YlGn-8-7!70!black, every mark/.append style={solid,fill=YlGn-8-7},mark=+\\%
 YlGn-8-8!70!black, every mark/.append style={solid,fill=YlGn-8-8},mark=star\\%
}
\pgfplotscreateplotcyclelist{Mark-YlGn-9}{%
 YlGn-9-1!70!black, every mark/.append style={solid,fill=YlGn-9-1},mark=*\\%
 YlGn-9-2!70!black, every mark/.append style={solid,fill=YlGn-9-2},mark=square*\\%
 YlGn-9-3!70!black, every mark/.append style={solid,fill=YlGn-9-3},mark=triangle*\\%
 YlGn-9-4!70!black, every mark/.append style={solid,fill=YlGn-9-4},mark=diamond*\\%
 YlGn-9-5!70!black, every mark/.append style={solid,fill=YlGn-9-5},mark=pentagon*\\%
 YlGn-9-6!70!black, every mark/.append style={solid,fill=YlGn-9-6},mark=x\\%
 YlGn-9-7!70!black, every mark/.append style={solid,fill=YlGn-9-7},mark=+\\%
 YlGn-9-8!70!black, every mark/.append style={solid,fill=YlGn-9-8},mark=star\\%
 YlGn-9-9!70!black, every mark/.append style={solid,fill=YlGn-9-9},mark=oplus*\\%
}
\pgfplotscreateplotcyclelist{Mark-YlGnBu-3}{%
 YlGnBu-3-1!70!black, every mark/.append style={solid,fill=YlGnBu-3-1},mark=*\\%
 YlGnBu-3-2!70!black, every mark/.append style={solid,fill=YlGnBu-3-2},mark=square*\\%
 YlGnBu-3-3!70!black, every mark/.append style={solid,fill=YlGnBu-3-3},mark=triangle*\\%
}
\pgfplotscreateplotcyclelist{Mark-YlGnBu-4}{%
 YlGnBu-4-1!70!black, every mark/.append style={solid,fill=YlGnBu-4-1},mark=*\\%
 YlGnBu-4-2!70!black, every mark/.append style={solid,fill=YlGnBu-4-2},mark=square*\\%
 YlGnBu-4-3!70!black, every mark/.append style={solid,fill=YlGnBu-4-3},mark=triangle*\\%
 YlGnBu-4-4!70!black, every mark/.append style={solid,fill=YlGnBu-4-4},mark=diamond*\\%
}
\pgfplotscreateplotcyclelist{Mark-YlGnBu-5}{%
 YlGnBu-5-1!70!black, every mark/.append style={solid,fill=YlGnBu-5-1},mark=*\\%
 YlGnBu-5-2!70!black, every mark/.append style={solid,fill=YlGnBu-5-2},mark=square*\\%
 YlGnBu-5-3!70!black, every mark/.append style={solid,fill=YlGnBu-5-3},mark=triangle*\\%
 YlGnBu-5-4!70!black, every mark/.append style={solid,fill=YlGnBu-5-4},mark=diamond*\\%
 YlGnBu-5-5!70!black, every mark/.append style={solid,fill=YlGnBu-5-5},mark=pentagon*\\%
}
\pgfplotscreateplotcyclelist{Mark-YlGnBu-6}{%
 YlGnBu-6-1!70!black, every mark/.append style={solid,fill=YlGnBu-6-1},mark=*\\%
 YlGnBu-6-2!70!black, every mark/.append style={solid,fill=YlGnBu-6-2},mark=square*\\%
 YlGnBu-6-3!70!black, every mark/.append style={solid,fill=YlGnBu-6-3},mark=triangle*\\%
 YlGnBu-6-4!70!black, every mark/.append style={solid,fill=YlGnBu-6-4},mark=diamond*\\%
 YlGnBu-6-5!70!black, every mark/.append style={solid,fill=YlGnBu-6-5},mark=pentagon*\\%
 YlGnBu-6-6!70!black, every mark/.append style={solid,fill=YlGnBu-6-6},mark=x\\%
}
\pgfplotscreateplotcyclelist{Mark-YlGnBu-7}{%
 YlGnBu-7-1!70!black, every mark/.append style={solid,fill=YlGnBu-7-1},mark=*\\%
 YlGnBu-7-2!70!black, every mark/.append style={solid,fill=YlGnBu-7-2},mark=square*\\%
 YlGnBu-7-3!70!black, every mark/.append style={solid,fill=YlGnBu-7-3},mark=triangle*\\%
 YlGnBu-7-4!70!black, every mark/.append style={solid,fill=YlGnBu-7-4},mark=diamond*\\%
 YlGnBu-7-5!70!black, every mark/.append style={solid,fill=YlGnBu-7-5},mark=pentagon*\\%
 YlGnBu-7-6!70!black, every mark/.append style={solid,fill=YlGnBu-7-6},mark=x\\%
 YlGnBu-7-7!70!black, every mark/.append style={solid,fill=YlGnBu-7-7},mark=+\\%
}
\pgfplotscreateplotcyclelist{Mark-YlGnBu-8}{%
 YlGnBu-8-1!70!black, every mark/.append style={solid,fill=YlGnBu-8-1},mark=*\\%
 YlGnBu-8-2!70!black, every mark/.append style={solid,fill=YlGnBu-8-2},mark=square*\\%
 YlGnBu-8-3!70!black, every mark/.append style={solid,fill=YlGnBu-8-3},mark=triangle*\\%
 YlGnBu-8-4!70!black, every mark/.append style={solid,fill=YlGnBu-8-4},mark=diamond*\\%
 YlGnBu-8-5!70!black, every mark/.append style={solid,fill=YlGnBu-8-5},mark=pentagon*\\%
 YlGnBu-8-6!70!black, every mark/.append style={solid,fill=YlGnBu-8-6},mark=x\\%
 YlGnBu-8-7!70!black, every mark/.append style={solid,fill=YlGnBu-8-7},mark=+\\%
 YlGnBu-8-8!70!black, every mark/.append style={solid,fill=YlGnBu-8-8},mark=star\\%
}
\pgfplotscreateplotcyclelist{Mark-YlGnBu-9}{%
 YlGnBu-9-1!70!black, every mark/.append style={solid,fill=YlGnBu-9-1},mark=*\\%
 YlGnBu-9-2!70!black, every mark/.append style={solid,fill=YlGnBu-9-2},mark=square*\\%
 YlGnBu-9-3!70!black, every mark/.append style={solid,fill=YlGnBu-9-3},mark=triangle*\\%
 YlGnBu-9-4!70!black, every mark/.append style={solid,fill=YlGnBu-9-4},mark=diamond*\\%
 YlGnBu-9-5!70!black, every mark/.append style={solid,fill=YlGnBu-9-5},mark=pentagon*\\%
 YlGnBu-9-6!70!black, every mark/.append style={solid,fill=YlGnBu-9-6},mark=x\\%
 YlGnBu-9-7!70!black, every mark/.append style={solid,fill=YlGnBu-9-7},mark=+\\%
 YlGnBu-9-8!70!black, every mark/.append style={solid,fill=YlGnBu-9-8},mark=star\\%
 YlGnBu-9-9!70!black, every mark/.append style={solid,fill=YlGnBu-9-9},mark=oplus*\\%
}
\pgfplotscreateplotcyclelist{Mark-GnBu-3}{%
 GnBu-3-1!70!black, every mark/.append style={solid,fill=GnBu-3-1},mark=*\\%
 GnBu-3-2!70!black, every mark/.append style={solid,fill=GnBu-3-2},mark=square*\\%
 GnBu-3-3!70!black, every mark/.append style={solid,fill=GnBu-3-3},mark=triangle*\\%
}
\pgfplotscreateplotcyclelist{Mark-GnBu-4}{%
 GnBu-4-1!70!black, every mark/.append style={solid,fill=GnBu-4-1},mark=*\\%
 GnBu-4-2!70!black, every mark/.append style={solid,fill=GnBu-4-2},mark=square*\\%
 GnBu-4-3!70!black, every mark/.append style={solid,fill=GnBu-4-3},mark=triangle*\\%
 GnBu-4-4!70!black, every mark/.append style={solid,fill=GnBu-4-4},mark=diamond*\\%
}
\pgfplotscreateplotcyclelist{Mark-GnBu-5}{%
 GnBu-5-1!70!black, every mark/.append style={solid,fill=GnBu-5-1},mark=*\\%
 GnBu-5-2!70!black, every mark/.append style={solid,fill=GnBu-5-2},mark=square*\\%
 GnBu-5-3!70!black, every mark/.append style={solid,fill=GnBu-5-3},mark=triangle*\\%
 GnBu-5-4!70!black, every mark/.append style={solid,fill=GnBu-5-4},mark=diamond*\\%
 GnBu-5-5!70!black, every mark/.append style={solid,fill=GnBu-5-5},mark=pentagon*\\%
}
\pgfplotscreateplotcyclelist{Mark-GnBu-6}{%
 GnBu-6-1!70!black, every mark/.append style={solid,fill=GnBu-6-1},mark=*\\%
 GnBu-6-2!70!black, every mark/.append style={solid,fill=GnBu-6-2},mark=square*\\%
 GnBu-6-3!70!black, every mark/.append style={solid,fill=GnBu-6-3},mark=triangle*\\%
 GnBu-6-4!70!black, every mark/.append style={solid,fill=GnBu-6-4},mark=diamond*\\%
 GnBu-6-5!70!black, every mark/.append style={solid,fill=GnBu-6-5},mark=pentagon*\\%
 GnBu-6-6!70!black, every mark/.append style={solid,fill=GnBu-6-6},mark=x\\%
}
\pgfplotscreateplotcyclelist{Mark-GnBu-7}{%
 GnBu-7-1!70!black, every mark/.append style={solid,fill=GnBu-7-1},mark=*\\%
 GnBu-7-2!70!black, every mark/.append style={solid,fill=GnBu-7-2},mark=square*\\%
 GnBu-7-3!70!black, every mark/.append style={solid,fill=GnBu-7-3},mark=triangle*\\%
 GnBu-7-4!70!black, every mark/.append style={solid,fill=GnBu-7-4},mark=diamond*\\%
 GnBu-7-5!70!black, every mark/.append style={solid,fill=GnBu-7-5},mark=pentagon*\\%
 GnBu-7-6!70!black, every mark/.append style={solid,fill=GnBu-7-6},mark=x\\%
 GnBu-7-7!70!black, every mark/.append style={solid,fill=GnBu-7-7},mark=+\\%
}
\pgfplotscreateplotcyclelist{Mark-GnBu-8}{%
 GnBu-8-1!70!black, every mark/.append style={solid,fill=GnBu-8-1},mark=*\\%
 GnBu-8-2!70!black, every mark/.append style={solid,fill=GnBu-8-2},mark=square*\\%
 GnBu-8-3!70!black, every mark/.append style={solid,fill=GnBu-8-3},mark=triangle*\\%
 GnBu-8-4!70!black, every mark/.append style={solid,fill=GnBu-8-4},mark=diamond*\\%
 GnBu-8-5!70!black, every mark/.append style={solid,fill=GnBu-8-5},mark=pentagon*\\%
 GnBu-8-6!70!black, every mark/.append style={solid,fill=GnBu-8-6},mark=x\\%
 GnBu-8-7!70!black, every mark/.append style={solid,fill=GnBu-8-7},mark=+\\%
 GnBu-8-8!70!black, every mark/.append style={solid,fill=GnBu-8-8},mark=star\\%
}
\pgfplotscreateplotcyclelist{Mark-GnBu-9}{%
 GnBu-9-1!70!black, every mark/.append style={solid,fill=GnBu-9-1},mark=*\\%
 GnBu-9-2!70!black, every mark/.append style={solid,fill=GnBu-9-2},mark=square*\\%
 GnBu-9-3!70!black, every mark/.append style={solid,fill=GnBu-9-3},mark=triangle*\\%
 GnBu-9-4!70!black, every mark/.append style={solid,fill=GnBu-9-4},mark=diamond*\\%
 GnBu-9-5!70!black, every mark/.append style={solid,fill=GnBu-9-5},mark=pentagon*\\%
 GnBu-9-6!70!black, every mark/.append style={solid,fill=GnBu-9-6},mark=x\\%
 GnBu-9-7!70!black, every mark/.append style={solid,fill=GnBu-9-7},mark=+\\%
 GnBu-9-8!70!black, every mark/.append style={solid,fill=GnBu-9-8},mark=star\\%
 GnBu-9-9!70!black, every mark/.append style={solid,fill=GnBu-9-9},mark=oplus*\\%
}
\pgfplotscreateplotcyclelist{Mark-BuGn-3}{%
 BuGn-3-1!70!black, every mark/.append style={solid,fill=BuGn-3-1},mark=*\\%
 BuGn-3-2!70!black, every mark/.append style={solid,fill=BuGn-3-2},mark=square*\\%
 BuGn-3-3!70!black, every mark/.append style={solid,fill=BuGn-3-3},mark=triangle*\\%
}
\pgfplotscreateplotcyclelist{Mark-BuGn-4}{%
 BuGn-4-1!70!black, every mark/.append style={solid,fill=BuGn-4-1},mark=*\\%
 BuGn-4-2!70!black, every mark/.append style={solid,fill=BuGn-4-2},mark=square*\\%
 BuGn-4-3!70!black, every mark/.append style={solid,fill=BuGn-4-3},mark=triangle*\\%
 BuGn-4-4!70!black, every mark/.append style={solid,fill=BuGn-4-4},mark=diamond*\\%
}
\pgfplotscreateplotcyclelist{Mark-BuGn-5}{%
 BuGn-5-1!70!black, every mark/.append style={solid,fill=BuGn-5-1},mark=*\\%
 BuGn-5-2!70!black, every mark/.append style={solid,fill=BuGn-5-2},mark=square*\\%
 BuGn-5-3!70!black, every mark/.append style={solid,fill=BuGn-5-3},mark=triangle*\\%
 BuGn-5-4!70!black, every mark/.append style={solid,fill=BuGn-5-4},mark=diamond*\\%
 BuGn-5-5!70!black, every mark/.append style={solid,fill=BuGn-5-5},mark=pentagon*\\%
}
\pgfplotscreateplotcyclelist{Mark-BuGn-6}{%
 BuGn-6-1!70!black, every mark/.append style={solid,fill=BuGn-6-1},mark=*\\%
 BuGn-6-2!70!black, every mark/.append style={solid,fill=BuGn-6-2},mark=square*\\%
 BuGn-6-3!70!black, every mark/.append style={solid,fill=BuGn-6-3},mark=triangle*\\%
 BuGn-6-4!70!black, every mark/.append style={solid,fill=BuGn-6-4},mark=diamond*\\%
 BuGn-6-5!70!black, every mark/.append style={solid,fill=BuGn-6-5},mark=pentagon*\\%
 BuGn-6-6!70!black, every mark/.append style={solid,fill=BuGn-6-6},mark=x\\%
}
\pgfplotscreateplotcyclelist{Mark-BuGn-7}{%
 BuGn-7-1!70!black, every mark/.append style={solid,fill=BuGn-7-1},mark=*\\%
 BuGn-7-2!70!black, every mark/.append style={solid,fill=BuGn-7-2},mark=square*\\%
 BuGn-7-3!70!black, every mark/.append style={solid,fill=BuGn-7-3},mark=triangle*\\%
 BuGn-7-4!70!black, every mark/.append style={solid,fill=BuGn-7-4},mark=diamond*\\%
 BuGn-7-5!70!black, every mark/.append style={solid,fill=BuGn-7-5},mark=pentagon*\\%
 BuGn-7-6!70!black, every mark/.append style={solid,fill=BuGn-7-6},mark=x\\%
 BuGn-7-7!70!black, every mark/.append style={solid,fill=BuGn-7-7},mark=+\\%
}
\pgfplotscreateplotcyclelist{Mark-BuGn-8}{%
 BuGn-8-1!70!black, every mark/.append style={solid,fill=BuGn-8-1},mark=*\\%
 BuGn-8-2!70!black, every mark/.append style={solid,fill=BuGn-8-2},mark=square*\\%
 BuGn-8-3!70!black, every mark/.append style={solid,fill=BuGn-8-3},mark=triangle*\\%
 BuGn-8-4!70!black, every mark/.append style={solid,fill=BuGn-8-4},mark=diamond*\\%
 BuGn-8-5!70!black, every mark/.append style={solid,fill=BuGn-8-5},mark=pentagon*\\%
 BuGn-8-6!70!black, every mark/.append style={solid,fill=BuGn-8-6},mark=x\\%
 BuGn-8-7!70!black, every mark/.append style={solid,fill=BuGn-8-7},mark=+\\%
 BuGn-8-8!70!black, every mark/.append style={solid,fill=BuGn-8-8},mark=star\\%
}
\pgfplotscreateplotcyclelist{Mark-BuGn-9}{%
 BuGn-9-1!70!black, every mark/.append style={solid,fill=BuGn-9-1},mark=*\\%
 BuGn-9-2!70!black, every mark/.append style={solid,fill=BuGn-9-2},mark=square*\\%
 BuGn-9-3!70!black, every mark/.append style={solid,fill=BuGn-9-3},mark=triangle*\\%
 BuGn-9-4!70!black, every mark/.append style={solid,fill=BuGn-9-4},mark=diamond*\\%
 BuGn-9-5!70!black, every mark/.append style={solid,fill=BuGn-9-5},mark=pentagon*\\%
 BuGn-9-6!70!black, every mark/.append style={solid,fill=BuGn-9-6},mark=x\\%
 BuGn-9-7!70!black, every mark/.append style={solid,fill=BuGn-9-7},mark=+\\%
 BuGn-9-8!70!black, every mark/.append style={solid,fill=BuGn-9-8},mark=star\\%
 BuGn-9-9!70!black, every mark/.append style={solid,fill=BuGn-9-9},mark=oplus*\\%
}
\pgfplotscreateplotcyclelist{Mark-PuBuGn-3}{%
 PuBuGn-3-1!70!black, every mark/.append style={solid,fill=PuBuGn-3-1},mark=*\\%
 PuBuGn-3-2!70!black, every mark/.append style={solid,fill=PuBuGn-3-2},mark=square*\\%
 PuBuGn-3-3!70!black, every mark/.append style={solid,fill=PuBuGn-3-3},mark=triangle*\\%
}
\pgfplotscreateplotcyclelist{Mark-PuBuGn-4}{%
 PuBuGn-4-1!70!black, every mark/.append style={solid,fill=PuBuGn-4-1},mark=*\\%
 PuBuGn-4-2!70!black, every mark/.append style={solid,fill=PuBuGn-4-2},mark=square*\\%
 PuBuGn-4-3!70!black, every mark/.append style={solid,fill=PuBuGn-4-3},mark=triangle*\\%
 PuBuGn-4-4!70!black, every mark/.append style={solid,fill=PuBuGn-4-4},mark=diamond*\\%
}
\pgfplotscreateplotcyclelist{Mark-PuBuGn-5}{%
 PuBuGn-5-1!70!black, every mark/.append style={solid,fill=PuBuGn-5-1},mark=*\\%
 PuBuGn-5-2!70!black, every mark/.append style={solid,fill=PuBuGn-5-2},mark=square*\\%
 PuBuGn-5-3!70!black, every mark/.append style={solid,fill=PuBuGn-5-3},mark=triangle*\\%
 PuBuGn-5-4!70!black, every mark/.append style={solid,fill=PuBuGn-5-4},mark=diamond*\\%
 PuBuGn-5-5!70!black, every mark/.append style={solid,fill=PuBuGn-5-5},mark=pentagon*\\%
}
\pgfplotscreateplotcyclelist{Mark-PuBuGn-6}{%
 PuBuGn-6-1!70!black, every mark/.append style={solid,fill=PuBuGn-6-1},mark=*\\%
 PuBuGn-6-2!70!black, every mark/.append style={solid,fill=PuBuGn-6-2},mark=square*\\%
 PuBuGn-6-3!70!black, every mark/.append style={solid,fill=PuBuGn-6-3},mark=triangle*\\%
 PuBuGn-6-4!70!black, every mark/.append style={solid,fill=PuBuGn-6-4},mark=diamond*\\%
 PuBuGn-6-5!70!black, every mark/.append style={solid,fill=PuBuGn-6-5},mark=pentagon*\\%
 PuBuGn-6-6!70!black, every mark/.append style={solid,fill=PuBuGn-6-6},mark=x\\%
}
\pgfplotscreateplotcyclelist{Mark-PuBuGn-7}{%
 PuBuGn-7-1!70!black, every mark/.append style={solid,fill=PuBuGn-7-1},mark=*\\%
 PuBuGn-7-2!70!black, every mark/.append style={solid,fill=PuBuGn-7-2},mark=square*\\%
 PuBuGn-7-3!70!black, every mark/.append style={solid,fill=PuBuGn-7-3},mark=triangle*\\%
 PuBuGn-7-4!70!black, every mark/.append style={solid,fill=PuBuGn-7-4},mark=diamond*\\%
 PuBuGn-7-5!70!black, every mark/.append style={solid,fill=PuBuGn-7-5},mark=pentagon*\\%
 PuBuGn-7-6!70!black, every mark/.append style={solid,fill=PuBuGn-7-6},mark=x\\%
 PuBuGn-7-7!70!black, every mark/.append style={solid,fill=PuBuGn-7-7},mark=+\\%
}
\pgfplotscreateplotcyclelist{Mark-PuBuGn-8}{%
 PuBuGn-8-1!70!black, every mark/.append style={solid,fill=PuBuGn-8-1},mark=*\\%
 PuBuGn-8-2!70!black, every mark/.append style={solid,fill=PuBuGn-8-2},mark=square*\\%
 PuBuGn-8-3!70!black, every mark/.append style={solid,fill=PuBuGn-8-3},mark=triangle*\\%
 PuBuGn-8-4!70!black, every mark/.append style={solid,fill=PuBuGn-8-4},mark=diamond*\\%
 PuBuGn-8-5!70!black, every mark/.append style={solid,fill=PuBuGn-8-5},mark=pentagon*\\%
 PuBuGn-8-6!70!black, every mark/.append style={solid,fill=PuBuGn-8-6},mark=x\\%
 PuBuGn-8-7!70!black, every mark/.append style={solid,fill=PuBuGn-8-7},mark=+\\%
 PuBuGn-8-8!70!black, every mark/.append style={solid,fill=PuBuGn-8-8},mark=star\\%
}
\pgfplotscreateplotcyclelist{Mark-PuBuGn-9}{%
 PuBuGn-9-1!70!black, every mark/.append style={solid,fill=PuBuGn-9-1},mark=*\\%
 PuBuGn-9-2!70!black, every mark/.append style={solid,fill=PuBuGn-9-2},mark=square*\\%
 PuBuGn-9-3!70!black, every mark/.append style={solid,fill=PuBuGn-9-3},mark=triangle*\\%
 PuBuGn-9-4!70!black, every mark/.append style={solid,fill=PuBuGn-9-4},mark=diamond*\\%
 PuBuGn-9-5!70!black, every mark/.append style={solid,fill=PuBuGn-9-5},mark=pentagon*\\%
 PuBuGn-9-6!70!black, every mark/.append style={solid,fill=PuBuGn-9-6},mark=x\\%
 PuBuGn-9-7!70!black, every mark/.append style={solid,fill=PuBuGn-9-7},mark=+\\%
 PuBuGn-9-8!70!black, every mark/.append style={solid,fill=PuBuGn-9-8},mark=star\\%
 PuBuGn-9-9!70!black, every mark/.append style={solid,fill=PuBuGn-9-9},mark=oplus*\\%
}
\pgfplotscreateplotcyclelist{Mark-PuBu-3}{%
 PuBu-3-1!70!black, every mark/.append style={solid,fill=PuBu-3-1},mark=*\\%
 PuBu-3-2!70!black, every mark/.append style={solid,fill=PuBu-3-2},mark=square*\\%
 PuBu-3-3!70!black, every mark/.append style={solid,fill=PuBu-3-3},mark=triangle*\\%
}
\pgfplotscreateplotcyclelist{Mark-PuBu-4}{%
 PuBu-4-1!70!black, every mark/.append style={solid,fill=PuBu-4-1},mark=*\\%
 PuBu-4-2!70!black, every mark/.append style={solid,fill=PuBu-4-2},mark=square*\\%
 PuBu-4-3!70!black, every mark/.append style={solid,fill=PuBu-4-3},mark=triangle*\\%
 PuBu-4-4!70!black, every mark/.append style={solid,fill=PuBu-4-4},mark=diamond*\\%
}
\pgfplotscreateplotcyclelist{Mark-PuBu-5}{%
 PuBu-5-1!70!black, every mark/.append style={solid,fill=PuBu-5-1},mark=*\\%
 PuBu-5-2!70!black, every mark/.append style={solid,fill=PuBu-5-2},mark=square*\\%
 PuBu-5-3!70!black, every mark/.append style={solid,fill=PuBu-5-3},mark=triangle*\\%
 PuBu-5-4!70!black, every mark/.append style={solid,fill=PuBu-5-4},mark=diamond*\\%
 PuBu-5-5!70!black, every mark/.append style={solid,fill=PuBu-5-5},mark=pentagon*\\%
}
\pgfplotscreateplotcyclelist{Mark-PuBu-6}{%
 PuBu-6-1!70!black, every mark/.append style={solid,fill=PuBu-6-1},mark=*\\%
 PuBu-6-2!70!black, every mark/.append style={solid,fill=PuBu-6-2},mark=square*\\%
 PuBu-6-3!70!black, every mark/.append style={solid,fill=PuBu-6-3},mark=triangle*\\%
 PuBu-6-4!70!black, every mark/.append style={solid,fill=PuBu-6-4},mark=diamond*\\%
 PuBu-6-5!70!black, every mark/.append style={solid,fill=PuBu-6-5},mark=pentagon*\\%
 PuBu-6-6!70!black, every mark/.append style={solid,fill=PuBu-6-6},mark=x\\%
}
\pgfplotscreateplotcyclelist{Mark-PuBu-7}{%
 PuBu-7-1!70!black, every mark/.append style={solid,fill=PuBu-7-1},mark=*\\%
 PuBu-7-2!70!black, every mark/.append style={solid,fill=PuBu-7-2},mark=square*\\%
 PuBu-7-3!70!black, every mark/.append style={solid,fill=PuBu-7-3},mark=triangle*\\%
 PuBu-7-4!70!black, every mark/.append style={solid,fill=PuBu-7-4},mark=diamond*\\%
 PuBu-7-5!70!black, every mark/.append style={solid,fill=PuBu-7-5},mark=pentagon*\\%
 PuBu-7-6!70!black, every mark/.append style={solid,fill=PuBu-7-6},mark=x\\%
 PuBu-7-7!70!black, every mark/.append style={solid,fill=PuBu-7-7},mark=+\\%
}
\pgfplotscreateplotcyclelist{Mark-PuBu-8}{%
 PuBu-8-1!70!black, every mark/.append style={solid,fill=PuBu-8-1},mark=*\\%
 PuBu-8-2!70!black, every mark/.append style={solid,fill=PuBu-8-2},mark=square*\\%
 PuBu-8-3!70!black, every mark/.append style={solid,fill=PuBu-8-3},mark=triangle*\\%
 PuBu-8-4!70!black, every mark/.append style={solid,fill=PuBu-8-4},mark=diamond*\\%
 PuBu-8-5!70!black, every mark/.append style={solid,fill=PuBu-8-5},mark=pentagon*\\%
 PuBu-8-6!70!black, every mark/.append style={solid,fill=PuBu-8-6},mark=x\\%
 PuBu-8-7!70!black, every mark/.append style={solid,fill=PuBu-8-7},mark=+\\%
 PuBu-8-8!70!black, every mark/.append style={solid,fill=PuBu-8-8},mark=star\\%
}
\pgfplotscreateplotcyclelist{Mark-PuBu-9}{%
 PuBu-9-1!70!black, every mark/.append style={solid,fill=PuBu-9-1},mark=*\\%
 PuBu-9-2!70!black, every mark/.append style={solid,fill=PuBu-9-2},mark=square*\\%
 PuBu-9-3!70!black, every mark/.append style={solid,fill=PuBu-9-3},mark=triangle*\\%
 PuBu-9-4!70!black, every mark/.append style={solid,fill=PuBu-9-4},mark=diamond*\\%
 PuBu-9-5!70!black, every mark/.append style={solid,fill=PuBu-9-5},mark=pentagon*\\%
 PuBu-9-6!70!black, every mark/.append style={solid,fill=PuBu-9-6},mark=x\\%
 PuBu-9-7!70!black, every mark/.append style={solid,fill=PuBu-9-7},mark=+\\%
 PuBu-9-8!70!black, every mark/.append style={solid,fill=PuBu-9-8},mark=star\\%
 PuBu-9-9!70!black, every mark/.append style={solid,fill=PuBu-9-9},mark=oplus*\\%
}
\pgfplotscreateplotcyclelist{Mark-BuPu-3}{%
 BuPu-3-1!70!black, every mark/.append style={solid,fill=BuPu-3-1},mark=*\\%
 BuPu-3-2!70!black, every mark/.append style={solid,fill=BuPu-3-2},mark=square*\\%
 BuPu-3-3!70!black, every mark/.append style={solid,fill=BuPu-3-3},mark=triangle*\\%
}
\pgfplotscreateplotcyclelist{Mark-BuPu-4}{%
 BuPu-4-1!70!black, every mark/.append style={solid,fill=BuPu-4-1},mark=*\\%
 BuPu-4-2!70!black, every mark/.append style={solid,fill=BuPu-4-2},mark=square*\\%
 BuPu-4-3!70!black, every mark/.append style={solid,fill=BuPu-4-3},mark=triangle*\\%
 BuPu-4-4!70!black, every mark/.append style={solid,fill=BuPu-4-4},mark=diamond*\\%
}
\pgfplotscreateplotcyclelist{Mark-BuPu-5}{%
 BuPu-5-1!70!black, every mark/.append style={solid,fill=BuPu-5-1},mark=*\\%
 BuPu-5-2!70!black, every mark/.append style={solid,fill=BuPu-5-2},mark=square*\\%
 BuPu-5-3!70!black, every mark/.append style={solid,fill=BuPu-5-3},mark=triangle*\\%
 BuPu-5-4!70!black, every mark/.append style={solid,fill=BuPu-5-4},mark=diamond*\\%
 BuPu-5-5!70!black, every mark/.append style={solid,fill=BuPu-5-5},mark=pentagon*\\%
}
\pgfplotscreateplotcyclelist{Mark-BuPu-6}{%
 BuPu-6-1!70!black, every mark/.append style={solid,fill=BuPu-6-1},mark=*\\%
 BuPu-6-2!70!black, every mark/.append style={solid,fill=BuPu-6-2},mark=square*\\%
 BuPu-6-3!70!black, every mark/.append style={solid,fill=BuPu-6-3},mark=triangle*\\%
 BuPu-6-4!70!black, every mark/.append style={solid,fill=BuPu-6-4},mark=diamond*\\%
 BuPu-6-5!70!black, every mark/.append style={solid,fill=BuPu-6-5},mark=pentagon*\\%
 BuPu-6-6!70!black, every mark/.append style={solid,fill=BuPu-6-6},mark=x\\%
}
\pgfplotscreateplotcyclelist{Mark-BuPu-7}{%
 BuPu-7-1!70!black, every mark/.append style={solid,fill=BuPu-7-1},mark=*\\%
 BuPu-7-2!70!black, every mark/.append style={solid,fill=BuPu-7-2},mark=square*\\%
 BuPu-7-3!70!black, every mark/.append style={solid,fill=BuPu-7-3},mark=triangle*\\%
 BuPu-7-4!70!black, every mark/.append style={solid,fill=BuPu-7-4},mark=diamond*\\%
 BuPu-7-5!70!black, every mark/.append style={solid,fill=BuPu-7-5},mark=pentagon*\\%
 BuPu-7-6!70!black, every mark/.append style={solid,fill=BuPu-7-6},mark=x\\%
 BuPu-7-7!70!black, every mark/.append style={solid,fill=BuPu-7-7},mark=+\\%
}
\pgfplotscreateplotcyclelist{Mark-BuPu-8}{%
 BuPu-8-1!70!black, every mark/.append style={solid,fill=BuPu-8-1},mark=*\\%
 BuPu-8-2!70!black, every mark/.append style={solid,fill=BuPu-8-2},mark=square*\\%
 BuPu-8-3!70!black, every mark/.append style={solid,fill=BuPu-8-3},mark=triangle*\\%
 BuPu-8-4!70!black, every mark/.append style={solid,fill=BuPu-8-4},mark=diamond*\\%
 BuPu-8-5!70!black, every mark/.append style={solid,fill=BuPu-8-5},mark=pentagon*\\%
 BuPu-8-6!70!black, every mark/.append style={solid,fill=BuPu-8-6},mark=x\\%
 BuPu-8-7!70!black, every mark/.append style={solid,fill=BuPu-8-7},mark=+\\%
 BuPu-8-8!70!black, every mark/.append style={solid,fill=BuPu-8-8},mark=star\\%
}
\pgfplotscreateplotcyclelist{Mark-BuPu-9}{%
 BuPu-9-1!70!black, every mark/.append style={solid,fill=BuPu-9-1},mark=*\\%
 BuPu-9-2!70!black, every mark/.append style={solid,fill=BuPu-9-2},mark=square*\\%
 BuPu-9-3!70!black, every mark/.append style={solid,fill=BuPu-9-3},mark=triangle*\\%
 BuPu-9-4!70!black, every mark/.append style={solid,fill=BuPu-9-4},mark=diamond*\\%
 BuPu-9-5!70!black, every mark/.append style={solid,fill=BuPu-9-5},mark=pentagon*\\%
 BuPu-9-6!70!black, every mark/.append style={solid,fill=BuPu-9-6},mark=x\\%
 BuPu-9-7!70!black, every mark/.append style={solid,fill=BuPu-9-7},mark=+\\%
 BuPu-9-8!70!black, every mark/.append style={solid,fill=BuPu-9-8},mark=star\\%
 BuPu-9-9!70!black, every mark/.append style={solid,fill=BuPu-9-9},mark=oplus*\\%
}
\pgfplotscreateplotcyclelist{Mark-RdPu-3}{%
 RdPu-3-1!70!black, every mark/.append style={solid,fill=RdPu-3-1},mark=*\\%
 RdPu-3-2!70!black, every mark/.append style={solid,fill=RdPu-3-2},mark=square*\\%
 RdPu-3-3!70!black, every mark/.append style={solid,fill=RdPu-3-3},mark=triangle*\\%
}
\pgfplotscreateplotcyclelist{Mark-RdPu-4}{%
 RdPu-4-1!70!black, every mark/.append style={solid,fill=RdPu-4-1},mark=*\\%
 RdPu-4-2!70!black, every mark/.append style={solid,fill=RdPu-4-2},mark=square*\\%
 RdPu-4-3!70!black, every mark/.append style={solid,fill=RdPu-4-3},mark=triangle*\\%
 RdPu-4-4!70!black, every mark/.append style={solid,fill=RdPu-4-4},mark=diamond*\\%
}
\pgfplotscreateplotcyclelist{Mark-RdPu-5}{%
 RdPu-5-1!70!black, every mark/.append style={solid,fill=RdPu-5-1},mark=*\\%
 RdPu-5-2!70!black, every mark/.append style={solid,fill=RdPu-5-2},mark=square*\\%
 RdPu-5-3!70!black, every mark/.append style={solid,fill=RdPu-5-3},mark=triangle*\\%
 RdPu-5-4!70!black, every mark/.append style={solid,fill=RdPu-5-4},mark=diamond*\\%
 RdPu-5-5!70!black, every mark/.append style={solid,fill=RdPu-5-5},mark=pentagon*\\%
}
\pgfplotscreateplotcyclelist{Mark-RdPu-6}{%
 RdPu-6-1!70!black, every mark/.append style={solid,fill=RdPu-6-1},mark=*\\%
 RdPu-6-2!70!black, every mark/.append style={solid,fill=RdPu-6-2},mark=square*\\%
 RdPu-6-3!70!black, every mark/.append style={solid,fill=RdPu-6-3},mark=triangle*\\%
 RdPu-6-4!70!black, every mark/.append style={solid,fill=RdPu-6-4},mark=diamond*\\%
 RdPu-6-5!70!black, every mark/.append style={solid,fill=RdPu-6-5},mark=pentagon*\\%
 RdPu-6-6!70!black, every mark/.append style={solid,fill=RdPu-6-6},mark=x\\%
}
\pgfplotscreateplotcyclelist{Mark-RdPu-7}{%
 RdPu-7-1!70!black, every mark/.append style={solid,fill=RdPu-7-1},mark=*\\%
 RdPu-7-2!70!black, every mark/.append style={solid,fill=RdPu-7-2},mark=square*\\%
 RdPu-7-3!70!black, every mark/.append style={solid,fill=RdPu-7-3},mark=triangle*\\%
 RdPu-7-4!70!black, every mark/.append style={solid,fill=RdPu-7-4},mark=diamond*\\%
 RdPu-7-5!70!black, every mark/.append style={solid,fill=RdPu-7-5},mark=pentagon*\\%
 RdPu-7-6!70!black, every mark/.append style={solid,fill=RdPu-7-6},mark=x\\%
 RdPu-7-7!70!black, every mark/.append style={solid,fill=RdPu-7-7},mark=+\\%
}
\pgfplotscreateplotcyclelist{Mark-RdPu-8}{%
 RdPu-8-1!70!black, every mark/.append style={solid,fill=RdPu-8-1},mark=*\\%
 RdPu-8-2!70!black, every mark/.append style={solid,fill=RdPu-8-2},mark=square*\\%
 RdPu-8-3!70!black, every mark/.append style={solid,fill=RdPu-8-3},mark=triangle*\\%
 RdPu-8-4!70!black, every mark/.append style={solid,fill=RdPu-8-4},mark=diamond*\\%
 RdPu-8-5!70!black, every mark/.append style={solid,fill=RdPu-8-5},mark=pentagon*\\%
 RdPu-8-6!70!black, every mark/.append style={solid,fill=RdPu-8-6},mark=x\\%
 RdPu-8-7!70!black, every mark/.append style={solid,fill=RdPu-8-7},mark=+\\%
 RdPu-8-8!70!black, every mark/.append style={solid,fill=RdPu-8-8},mark=star\\%
}
\pgfplotscreateplotcyclelist{Mark-RdPu-9}{%
 RdPu-9-1!70!black, every mark/.append style={solid,fill=RdPu-9-1},mark=*\\%
 RdPu-9-2!70!black, every mark/.append style={solid,fill=RdPu-9-2},mark=square*\\%
 RdPu-9-3!70!black, every mark/.append style={solid,fill=RdPu-9-3},mark=triangle*\\%
 RdPu-9-4!70!black, every mark/.append style={solid,fill=RdPu-9-4},mark=diamond*\\%
 RdPu-9-5!70!black, every mark/.append style={solid,fill=RdPu-9-5},mark=pentagon*\\%
 RdPu-9-6!70!black, every mark/.append style={solid,fill=RdPu-9-6},mark=x\\%
 RdPu-9-7!70!black, every mark/.append style={solid,fill=RdPu-9-7},mark=+\\%
 RdPu-9-8!70!black, every mark/.append style={solid,fill=RdPu-9-8},mark=star\\%
 RdPu-9-9!70!black, every mark/.append style={solid,fill=RdPu-9-9},mark=oplus*\\%
}
\pgfplotscreateplotcyclelist{Mark-PuRd-3}{%
 PuRd-3-1!70!black, every mark/.append style={solid,fill=PuRd-3-1},mark=*\\%
 PuRd-3-2!70!black, every mark/.append style={solid,fill=PuRd-3-2},mark=square*\\%
 PuRd-3-3!70!black, every mark/.append style={solid,fill=PuRd-3-3},mark=triangle*\\%
}
\pgfplotscreateplotcyclelist{Mark-PuRd-4}{%
 PuRd-4-1!70!black, every mark/.append style={solid,fill=PuRd-4-1},mark=*\\%
 PuRd-4-2!70!black, every mark/.append style={solid,fill=PuRd-4-2},mark=square*\\%
 PuRd-4-3!70!black, every mark/.append style={solid,fill=PuRd-4-3},mark=triangle*\\%
 PuRd-4-4!70!black, every mark/.append style={solid,fill=PuRd-4-4},mark=diamond*\\%
}
\pgfplotscreateplotcyclelist{Mark-PuRd-5}{%
 PuRd-5-1!70!black, every mark/.append style={solid,fill=PuRd-5-1},mark=*\\%
 PuRd-5-2!70!black, every mark/.append style={solid,fill=PuRd-5-2},mark=square*\\%
 PuRd-5-3!70!black, every mark/.append style={solid,fill=PuRd-5-3},mark=triangle*\\%
 PuRd-5-4!70!black, every mark/.append style={solid,fill=PuRd-5-4},mark=diamond*\\%
 PuRd-5-5!70!black, every mark/.append style={solid,fill=PuRd-5-5},mark=pentagon*\\%
}
\pgfplotscreateplotcyclelist{Mark-PuRd-6}{%
 PuRd-6-1!70!black, every mark/.append style={solid,fill=PuRd-6-1},mark=*\\%
 PuRd-6-2!70!black, every mark/.append style={solid,fill=PuRd-6-2},mark=square*\\%
 PuRd-6-3!70!black, every mark/.append style={solid,fill=PuRd-6-3},mark=triangle*\\%
 PuRd-6-4!70!black, every mark/.append style={solid,fill=PuRd-6-4},mark=diamond*\\%
 PuRd-6-5!70!black, every mark/.append style={solid,fill=PuRd-6-5},mark=pentagon*\\%
 PuRd-6-6!70!black, every mark/.append style={solid,fill=PuRd-6-6},mark=x\\%
}
\pgfplotscreateplotcyclelist{Mark-PuRd-7}{%
 PuRd-7-1!70!black, every mark/.append style={solid,fill=PuRd-7-1},mark=*\\%
 PuRd-7-2!70!black, every mark/.append style={solid,fill=PuRd-7-2},mark=square*\\%
 PuRd-7-3!70!black, every mark/.append style={solid,fill=PuRd-7-3},mark=triangle*\\%
 PuRd-7-4!70!black, every mark/.append style={solid,fill=PuRd-7-4},mark=diamond*\\%
 PuRd-7-5!70!black, every mark/.append style={solid,fill=PuRd-7-5},mark=pentagon*\\%
 PuRd-7-6!70!black, every mark/.append style={solid,fill=PuRd-7-6},mark=x\\%
 PuRd-7-7!70!black, every mark/.append style={solid,fill=PuRd-7-7},mark=+\\%
}
\pgfplotscreateplotcyclelist{Mark-PuRd-8}{%
 PuRd-8-1!70!black, every mark/.append style={solid,fill=PuRd-8-1},mark=*\\%
 PuRd-8-2!70!black, every mark/.append style={solid,fill=PuRd-8-2},mark=square*\\%
 PuRd-8-3!70!black, every mark/.append style={solid,fill=PuRd-8-3},mark=triangle*\\%
 PuRd-8-4!70!black, every mark/.append style={solid,fill=PuRd-8-4},mark=diamond*\\%
 PuRd-8-5!70!black, every mark/.append style={solid,fill=PuRd-8-5},mark=pentagon*\\%
 PuRd-8-6!70!black, every mark/.append style={solid,fill=PuRd-8-6},mark=x\\%
 PuRd-8-7!70!black, every mark/.append style={solid,fill=PuRd-8-7},mark=+\\%
 PuRd-8-8!70!black, every mark/.append style={solid,fill=PuRd-8-8},mark=star\\%
}
\pgfplotscreateplotcyclelist{Mark-PuRd-9}{%
 PuRd-9-1!70!black, every mark/.append style={solid,fill=PuRd-9-1},mark=*\\%
 PuRd-9-2!70!black, every mark/.append style={solid,fill=PuRd-9-2},mark=square*\\%
 PuRd-9-3!70!black, every mark/.append style={solid,fill=PuRd-9-3},mark=triangle*\\%
 PuRd-9-4!70!black, every mark/.append style={solid,fill=PuRd-9-4},mark=diamond*\\%
 PuRd-9-5!70!black, every mark/.append style={solid,fill=PuRd-9-5},mark=pentagon*\\%
 PuRd-9-6!70!black, every mark/.append style={solid,fill=PuRd-9-6},mark=x\\%
 PuRd-9-7!70!black, every mark/.append style={solid,fill=PuRd-9-7},mark=+\\%
 PuRd-9-8!70!black, every mark/.append style={solid,fill=PuRd-9-8},mark=star\\%
 PuRd-9-9!70!black, every mark/.append style={solid,fill=PuRd-9-9},mark=oplus*\\%
}
\pgfplotscreateplotcyclelist{Mark-OrRd-3}{%
 OrRd-3-1!70!black, every mark/.append style={solid,fill=OrRd-3-1},mark=*\\%
 OrRd-3-2!70!black, every mark/.append style={solid,fill=OrRd-3-2},mark=square*\\%
 OrRd-3-3!70!black, every mark/.append style={solid,fill=OrRd-3-3},mark=triangle*\\%
}
\pgfplotscreateplotcyclelist{Mark-OrRd-4}{%
 OrRd-4-1!70!black, every mark/.append style={solid,fill=OrRd-4-1},mark=*\\%
 OrRd-4-2!70!black, every mark/.append style={solid,fill=OrRd-4-2},mark=square*\\%
 OrRd-4-3!70!black, every mark/.append style={solid,fill=OrRd-4-3},mark=triangle*\\%
 OrRd-4-4!70!black, every mark/.append style={solid,fill=OrRd-4-4},mark=diamond*\\%
}
\pgfplotscreateplotcyclelist{Mark-OrRd-5}{%
 OrRd-5-1!70!black, every mark/.append style={solid,fill=OrRd-5-1},mark=*\\%
 OrRd-5-2!70!black, every mark/.append style={solid,fill=OrRd-5-2},mark=square*\\%
 OrRd-5-3!70!black, every mark/.append style={solid,fill=OrRd-5-3},mark=triangle*\\%
 OrRd-5-4!70!black, every mark/.append style={solid,fill=OrRd-5-4},mark=diamond*\\%
 OrRd-5-5!70!black, every mark/.append style={solid,fill=OrRd-5-5},mark=pentagon*\\%
}
\pgfplotscreateplotcyclelist{Mark-OrRd-6}{%
 OrRd-6-1!70!black, every mark/.append style={solid,fill=OrRd-6-1},mark=*\\%
 OrRd-6-2!70!black, every mark/.append style={solid,fill=OrRd-6-2},mark=square*\\%
 OrRd-6-3!70!black, every mark/.append style={solid,fill=OrRd-6-3},mark=triangle*\\%
 OrRd-6-4!70!black, every mark/.append style={solid,fill=OrRd-6-4},mark=diamond*\\%
 OrRd-6-5!70!black, every mark/.append style={solid,fill=OrRd-6-5},mark=pentagon*\\%
 OrRd-6-6!70!black, every mark/.append style={solid,fill=OrRd-6-6},mark=x\\%
}
\pgfplotscreateplotcyclelist{Mark-OrRd-7}{%
 OrRd-7-1!70!black, every mark/.append style={solid,fill=OrRd-7-1},mark=*\\%
 OrRd-7-2!70!black, every mark/.append style={solid,fill=OrRd-7-2},mark=square*\\%
 OrRd-7-3!70!black, every mark/.append style={solid,fill=OrRd-7-3},mark=triangle*\\%
 OrRd-7-4!70!black, every mark/.append style={solid,fill=OrRd-7-4},mark=diamond*\\%
 OrRd-7-5!70!black, every mark/.append style={solid,fill=OrRd-7-5},mark=pentagon*\\%
 OrRd-7-6!70!black, every mark/.append style={solid,fill=OrRd-7-6},mark=x\\%
 OrRd-7-7!70!black, every mark/.append style={solid,fill=OrRd-7-7},mark=+\\%
}
\pgfplotscreateplotcyclelist{Mark-OrRd-8}{%
 OrRd-8-1!70!black, every mark/.append style={solid,fill=OrRd-8-1},mark=*\\%
 OrRd-8-2!70!black, every mark/.append style={solid,fill=OrRd-8-2},mark=square*\\%
 OrRd-8-3!70!black, every mark/.append style={solid,fill=OrRd-8-3},mark=triangle*\\%
 OrRd-8-4!70!black, every mark/.append style={solid,fill=OrRd-8-4},mark=diamond*\\%
 OrRd-8-5!70!black, every mark/.append style={solid,fill=OrRd-8-5},mark=pentagon*\\%
 OrRd-8-6!70!black, every mark/.append style={solid,fill=OrRd-8-6},mark=x\\%
 OrRd-8-7!70!black, every mark/.append style={solid,fill=OrRd-8-7},mark=+\\%
 OrRd-8-8!70!black, every mark/.append style={solid,fill=OrRd-8-8},mark=star\\%
}
\pgfplotscreateplotcyclelist{Mark-OrRd-9}{%
 OrRd-9-1!70!black, every mark/.append style={solid,fill=OrRd-9-1},mark=*\\%
 OrRd-9-2!70!black, every mark/.append style={solid,fill=OrRd-9-2},mark=square*\\%
 OrRd-9-3!70!black, every mark/.append style={solid,fill=OrRd-9-3},mark=triangle*\\%
 OrRd-9-4!70!black, every mark/.append style={solid,fill=OrRd-9-4},mark=diamond*\\%
 OrRd-9-5!70!black, every mark/.append style={solid,fill=OrRd-9-5},mark=pentagon*\\%
 OrRd-9-6!70!black, every mark/.append style={solid,fill=OrRd-9-6},mark=x\\%
 OrRd-9-7!70!black, every mark/.append style={solid,fill=OrRd-9-7},mark=+\\%
 OrRd-9-8!70!black, every mark/.append style={solid,fill=OrRd-9-8},mark=star\\%
 OrRd-9-9!70!black, every mark/.append style={solid,fill=OrRd-9-9},mark=oplus*\\%
}
\pgfplotscreateplotcyclelist{Mark-YlOrRd-3}{%
 YlOrRd-3-1!70!black, every mark/.append style={solid,fill=YlOrRd-3-1},mark=*\\%
 YlOrRd-3-2!70!black, every mark/.append style={solid,fill=YlOrRd-3-2},mark=square*\\%
 YlOrRd-3-3!70!black, every mark/.append style={solid,fill=YlOrRd-3-3},mark=triangle*\\%
}
\pgfplotscreateplotcyclelist{Mark-YlOrRd-4}{%
 YlOrRd-4-1!70!black, every mark/.append style={solid,fill=YlOrRd-4-1},mark=*\\%
 YlOrRd-4-2!70!black, every mark/.append style={solid,fill=YlOrRd-4-2},mark=square*\\%
 YlOrRd-4-3!70!black, every mark/.append style={solid,fill=YlOrRd-4-3},mark=triangle*\\%
 YlOrRd-4-4!70!black, every mark/.append style={solid,fill=YlOrRd-4-4},mark=diamond*\\%
}
\pgfplotscreateplotcyclelist{Mark-YlOrRd-5}{%
 YlOrRd-5-1!70!black, every mark/.append style={solid,fill=YlOrRd-5-1},mark=*\\%
 YlOrRd-5-2!70!black, every mark/.append style={solid,fill=YlOrRd-5-2},mark=square*\\%
 YlOrRd-5-3!70!black, every mark/.append style={solid,fill=YlOrRd-5-3},mark=triangle*\\%
 YlOrRd-5-4!70!black, every mark/.append style={solid,fill=YlOrRd-5-4},mark=diamond*\\%
 YlOrRd-5-5!70!black, every mark/.append style={solid,fill=YlOrRd-5-5},mark=pentagon*\\%
}
\pgfplotscreateplotcyclelist{Mark-YlOrRd-6}{%
 YlOrRd-6-1!70!black, every mark/.append style={solid,fill=YlOrRd-6-1},mark=*\\%
 YlOrRd-6-2!70!black, every mark/.append style={solid,fill=YlOrRd-6-2},mark=square*\\%
 YlOrRd-6-3!70!black, every mark/.append style={solid,fill=YlOrRd-6-3},mark=triangle*\\%
 YlOrRd-6-4!70!black, every mark/.append style={solid,fill=YlOrRd-6-4},mark=diamond*\\%
 YlOrRd-6-5!70!black, every mark/.append style={solid,fill=YlOrRd-6-5},mark=pentagon*\\%
 YlOrRd-6-6!70!black, every mark/.append style={solid,fill=YlOrRd-6-6},mark=x\\%
}
\pgfplotscreateplotcyclelist{Mark-YlOrRd-7}{%
 YlOrRd-7-1!70!black, every mark/.append style={solid,fill=YlOrRd-7-1},mark=*\\%
 YlOrRd-7-2!70!black, every mark/.append style={solid,fill=YlOrRd-7-2},mark=square*\\%
 YlOrRd-7-3!70!black, every mark/.append style={solid,fill=YlOrRd-7-3},mark=triangle*\\%
 YlOrRd-7-4!70!black, every mark/.append style={solid,fill=YlOrRd-7-4},mark=diamond*\\%
 YlOrRd-7-5!70!black, every mark/.append style={solid,fill=YlOrRd-7-5},mark=pentagon*\\%
 YlOrRd-7-6!70!black, every mark/.append style={solid,fill=YlOrRd-7-6},mark=x\\%
 YlOrRd-7-7!70!black, every mark/.append style={solid,fill=YlOrRd-7-7},mark=+\\%
}
\pgfplotscreateplotcyclelist{Mark-YlOrRd-8}{%
 YlOrRd-8-1!70!black, every mark/.append style={solid,fill=YlOrRd-8-1},mark=*\\%
 YlOrRd-8-2!70!black, every mark/.append style={solid,fill=YlOrRd-8-2},mark=square*\\%
 YlOrRd-8-3!70!black, every mark/.append style={solid,fill=YlOrRd-8-3},mark=triangle*\\%
 YlOrRd-8-4!70!black, every mark/.append style={solid,fill=YlOrRd-8-4},mark=diamond*\\%
 YlOrRd-8-5!70!black, every mark/.append style={solid,fill=YlOrRd-8-5},mark=pentagon*\\%
 YlOrRd-8-6!70!black, every mark/.append style={solid,fill=YlOrRd-8-6},mark=x\\%
 YlOrRd-8-7!70!black, every mark/.append style={solid,fill=YlOrRd-8-7},mark=+\\%
 YlOrRd-8-8!70!black, every mark/.append style={solid,fill=YlOrRd-8-8},mark=star\\%
}
\pgfplotscreateplotcyclelist{Mark-YlOrRd-9}{%
 YlOrRd-9-1!70!black, every mark/.append style={solid,fill=YlOrRd-9-1},mark=*\\%
 YlOrRd-9-2!70!black, every mark/.append style={solid,fill=YlOrRd-9-2},mark=square*\\%
 YlOrRd-9-3!70!black, every mark/.append style={solid,fill=YlOrRd-9-3},mark=triangle*\\%
 YlOrRd-9-4!70!black, every mark/.append style={solid,fill=YlOrRd-9-4},mark=diamond*\\%
 YlOrRd-9-5!70!black, every mark/.append style={solid,fill=YlOrRd-9-5},mark=pentagon*\\%
 YlOrRd-9-6!70!black, every mark/.append style={solid,fill=YlOrRd-9-6},mark=x\\%
 YlOrRd-9-7!70!black, every mark/.append style={solid,fill=YlOrRd-9-7},mark=+\\%
 YlOrRd-9-8!70!black, every mark/.append style={solid,fill=YlOrRd-9-8},mark=star\\%
 YlOrRd-9-9!70!black, every mark/.append style={solid,fill=YlOrRd-9-9},mark=oplus*\\%
}
\pgfplotscreateplotcyclelist{Mark-YlOrBr-3}{%
 YlOrBr-3-1!70!black, every mark/.append style={solid,fill=YlOrBr-3-1},mark=*\\%
 YlOrBr-3-2!70!black, every mark/.append style={solid,fill=YlOrBr-3-2},mark=square*\\%
 YlOrBr-3-3!70!black, every mark/.append style={solid,fill=YlOrBr-3-3},mark=triangle*\\%
}
\pgfplotscreateplotcyclelist{Mark-YlOrBr-4}{%
 YlOrBr-4-1!70!black, every mark/.append style={solid,fill=YlOrBr-4-1},mark=*\\%
 YlOrBr-4-2!70!black, every mark/.append style={solid,fill=YlOrBr-4-2},mark=square*\\%
 YlOrBr-4-3!70!black, every mark/.append style={solid,fill=YlOrBr-4-3},mark=triangle*\\%
 YlOrBr-4-4!70!black, every mark/.append style={solid,fill=YlOrBr-4-4},mark=diamond*\\%
}
\pgfplotscreateplotcyclelist{Mark-YlOrBr-5}{%
 YlOrBr-5-1!70!black, every mark/.append style={solid,fill=YlOrBr-5-1},mark=*\\%
 YlOrBr-5-2!70!black, every mark/.append style={solid,fill=YlOrBr-5-2},mark=square*\\%
 YlOrBr-5-3!70!black, every mark/.append style={solid,fill=YlOrBr-5-3},mark=triangle*\\%
 YlOrBr-5-4!70!black, every mark/.append style={solid,fill=YlOrBr-5-4},mark=diamond*\\%
 YlOrBr-5-5!70!black, every mark/.append style={solid,fill=YlOrBr-5-5},mark=pentagon*\\%
}
\pgfplotscreateplotcyclelist{Mark-YlOrBr-6}{%
 YlOrBr-6-1!70!black, every mark/.append style={solid,fill=YlOrBr-6-1},mark=*\\%
 YlOrBr-6-2!70!black, every mark/.append style={solid,fill=YlOrBr-6-2},mark=square*\\%
 YlOrBr-6-3!70!black, every mark/.append style={solid,fill=YlOrBr-6-3},mark=triangle*\\%
 YlOrBr-6-4!70!black, every mark/.append style={solid,fill=YlOrBr-6-4},mark=diamond*\\%
 YlOrBr-6-5!70!black, every mark/.append style={solid,fill=YlOrBr-6-5},mark=pentagon*\\%
 YlOrBr-6-6!70!black, every mark/.append style={solid,fill=YlOrBr-6-6},mark=x\\%
}
\pgfplotscreateplotcyclelist{Mark-YlOrBr-7}{%
 YlOrBr-7-1!70!black, every mark/.append style={solid,fill=YlOrBr-7-1},mark=*\\%
 YlOrBr-7-2!70!black, every mark/.append style={solid,fill=YlOrBr-7-2},mark=square*\\%
 YlOrBr-7-3!70!black, every mark/.append style={solid,fill=YlOrBr-7-3},mark=triangle*\\%
 YlOrBr-7-4!70!black, every mark/.append style={solid,fill=YlOrBr-7-4},mark=diamond*\\%
 YlOrBr-7-5!70!black, every mark/.append style={solid,fill=YlOrBr-7-5},mark=pentagon*\\%
 YlOrBr-7-6!70!black, every mark/.append style={solid,fill=YlOrBr-7-6},mark=x\\%
 YlOrBr-7-7!70!black, every mark/.append style={solid,fill=YlOrBr-7-7},mark=+\\%
}
\pgfplotscreateplotcyclelist{Mark-YlOrBr-8}{%
 YlOrBr-8-1!70!black, every mark/.append style={solid,fill=YlOrBr-8-1},mark=*\\%
 YlOrBr-8-2!70!black, every mark/.append style={solid,fill=YlOrBr-8-2},mark=square*\\%
 YlOrBr-8-3!70!black, every mark/.append style={solid,fill=YlOrBr-8-3},mark=triangle*\\%
 YlOrBr-8-4!70!black, every mark/.append style={solid,fill=YlOrBr-8-4},mark=diamond*\\%
 YlOrBr-8-5!70!black, every mark/.append style={solid,fill=YlOrBr-8-5},mark=pentagon*\\%
 YlOrBr-8-6!70!black, every mark/.append style={solid,fill=YlOrBr-8-6},mark=x\\%
 YlOrBr-8-7!70!black, every mark/.append style={solid,fill=YlOrBr-8-7},mark=+\\%
 YlOrBr-8-8!70!black, every mark/.append style={solid,fill=YlOrBr-8-8},mark=star\\%
}
\pgfplotscreateplotcyclelist{Mark-YlOrBr-9}{%
 YlOrBr-9-1!70!black, every mark/.append style={solid,fill=YlOrBr-9-1},mark=*\\%
 YlOrBr-9-2!70!black, every mark/.append style={solid,fill=YlOrBr-9-2},mark=square*\\%
 YlOrBr-9-3!70!black, every mark/.append style={solid,fill=YlOrBr-9-3},mark=triangle*\\%
 YlOrBr-9-4!70!black, every mark/.append style={solid,fill=YlOrBr-9-4},mark=diamond*\\%
 YlOrBr-9-5!70!black, every mark/.append style={solid,fill=YlOrBr-9-5},mark=pentagon*\\%
 YlOrBr-9-6!70!black, every mark/.append style={solid,fill=YlOrBr-9-6},mark=x\\%
 YlOrBr-9-7!70!black, every mark/.append style={solid,fill=YlOrBr-9-7},mark=+\\%
 YlOrBr-9-8!70!black, every mark/.append style={solid,fill=YlOrBr-9-8},mark=star\\%
 YlOrBr-9-9!70!black, every mark/.append style={solid,fill=YlOrBr-9-9},mark=oplus*\\%
}
\pgfplotscreateplotcyclelist{Mark-Purples-3}{%
 Purples-3-1!70!black, every mark/.append style={solid,fill=Purples-3-1},mark=*\\%
 Purples-3-2!70!black, every mark/.append style={solid,fill=Purples-3-2},mark=square*\\%
 Purples-3-3!70!black, every mark/.append style={solid,fill=Purples-3-3},mark=triangle*\\%
}
\pgfplotscreateplotcyclelist{Mark-Purples-4}{%
 Purples-4-1!70!black, every mark/.append style={solid,fill=Purples-4-1},mark=*\\%
 Purples-4-2!70!black, every mark/.append style={solid,fill=Purples-4-2},mark=square*\\%
 Purples-4-3!70!black, every mark/.append style={solid,fill=Purples-4-3},mark=triangle*\\%
 Purples-4-4!70!black, every mark/.append style={solid,fill=Purples-4-4},mark=diamond*\\%
}
\pgfplotscreateplotcyclelist{Mark-Purples-5}{%
 Purples-5-1!70!black, every mark/.append style={solid,fill=Purples-5-1},mark=*\\%
 Purples-5-2!70!black, every mark/.append style={solid,fill=Purples-5-2},mark=square*\\%
 Purples-5-3!70!black, every mark/.append style={solid,fill=Purples-5-3},mark=triangle*\\%
 Purples-5-4!70!black, every mark/.append style={solid,fill=Purples-5-4},mark=diamond*\\%
 Purples-5-5!70!black, every mark/.append style={solid,fill=Purples-5-5},mark=pentagon*\\%
}
\pgfplotscreateplotcyclelist{Mark-Purples-6}{%
 Purples-6-1!70!black, every mark/.append style={solid,fill=Purples-6-1},mark=*\\%
 Purples-6-2!70!black, every mark/.append style={solid,fill=Purples-6-2},mark=square*\\%
 Purples-6-3!70!black, every mark/.append style={solid,fill=Purples-6-3},mark=triangle*\\%
 Purples-6-4!70!black, every mark/.append style={solid,fill=Purples-6-4},mark=diamond*\\%
 Purples-6-5!70!black, every mark/.append style={solid,fill=Purples-6-5},mark=pentagon*\\%
 Purples-6-6!70!black, every mark/.append style={solid,fill=Purples-6-6},mark=x\\%
}
\pgfplotscreateplotcyclelist{Mark-Purples-7}{%
 Purples-7-1!70!black, every mark/.append style={solid,fill=Purples-7-1},mark=*\\%
 Purples-7-2!70!black, every mark/.append style={solid,fill=Purples-7-2},mark=square*\\%
 Purples-7-3!70!black, every mark/.append style={solid,fill=Purples-7-3},mark=triangle*\\%
 Purples-7-4!70!black, every mark/.append style={solid,fill=Purples-7-4},mark=diamond*\\%
 Purples-7-5!70!black, every mark/.append style={solid,fill=Purples-7-5},mark=pentagon*\\%
 Purples-7-6!70!black, every mark/.append style={solid,fill=Purples-7-6},mark=x\\%
 Purples-7-7!70!black, every mark/.append style={solid,fill=Purples-7-7},mark=+\\%
}
\pgfplotscreateplotcyclelist{Mark-Purples-8}{%
 Purples-8-1!70!black, every mark/.append style={solid,fill=Purples-8-1},mark=*\\%
 Purples-8-2!70!black, every mark/.append style={solid,fill=Purples-8-2},mark=square*\\%
 Purples-8-3!70!black, every mark/.append style={solid,fill=Purples-8-3},mark=triangle*\\%
 Purples-8-4!70!black, every mark/.append style={solid,fill=Purples-8-4},mark=diamond*\\%
 Purples-8-5!70!black, every mark/.append style={solid,fill=Purples-8-5},mark=pentagon*\\%
 Purples-8-6!70!black, every mark/.append style={solid,fill=Purples-8-6},mark=x\\%
 Purples-8-7!70!black, every mark/.append style={solid,fill=Purples-8-7},mark=+\\%
 Purples-8-8!70!black, every mark/.append style={solid,fill=Purples-8-8},mark=star\\%
}
\pgfplotscreateplotcyclelist{Mark-Purples-9}{%
 Purples-9-1!70!black, every mark/.append style={solid,fill=Purples-9-1},mark=*\\%
 Purples-9-2!70!black, every mark/.append style={solid,fill=Purples-9-2},mark=square*\\%
 Purples-9-3!70!black, every mark/.append style={solid,fill=Purples-9-3},mark=triangle*\\%
 Purples-9-4!70!black, every mark/.append style={solid,fill=Purples-9-4},mark=diamond*\\%
 Purples-9-5!70!black, every mark/.append style={solid,fill=Purples-9-5},mark=pentagon*\\%
 Purples-9-6!70!black, every mark/.append style={solid,fill=Purples-9-6},mark=x\\%
 Purples-9-7!70!black, every mark/.append style={solid,fill=Purples-9-7},mark=+\\%
 Purples-9-8!70!black, every mark/.append style={solid,fill=Purples-9-8},mark=star\\%
 Purples-9-9!70!black, every mark/.append style={solid,fill=Purples-9-9},mark=oplus*\\%
}
\pgfplotscreateplotcyclelist{Mark-Blues-3}{%
 Blues-3-1!70!black, every mark/.append style={solid,fill=Blues-3-1},mark=*\\%
 Blues-3-2!70!black, every mark/.append style={solid,fill=Blues-3-2},mark=square*\\%
 Blues-3-3!70!black, every mark/.append style={solid,fill=Blues-3-3},mark=triangle*\\%
}
\pgfplotscreateplotcyclelist{Mark-Blues-4}{%
 Blues-4-1!70!black, every mark/.append style={solid,fill=Blues-4-1},mark=*\\%
 Blues-4-2!70!black, every mark/.append style={solid,fill=Blues-4-2},mark=square*\\%
 Blues-4-3!70!black, every mark/.append style={solid,fill=Blues-4-3},mark=triangle*\\%
 Blues-4-4!70!black, every mark/.append style={solid,fill=Blues-4-4},mark=diamond*\\%
}
\pgfplotscreateplotcyclelist{Mark-Blues-5}{%
 Blues-5-1!70!black, every mark/.append style={solid,fill=Blues-5-1},mark=*\\%
 Blues-5-2!70!black, every mark/.append style={solid,fill=Blues-5-2},mark=square*\\%
 Blues-5-3!70!black, every mark/.append style={solid,fill=Blues-5-3},mark=triangle*\\%
 Blues-5-4!70!black, every mark/.append style={solid,fill=Blues-5-4},mark=diamond*\\%
 Blues-5-5!70!black, every mark/.append style={solid,fill=Blues-5-5},mark=pentagon*\\%
}
\pgfplotscreateplotcyclelist{Mark-Blues-6}{%
 Blues-6-1!70!black, every mark/.append style={solid,fill=Blues-6-1},mark=*\\%
 Blues-6-2!70!black, every mark/.append style={solid,fill=Blues-6-2},mark=square*\\%
 Blues-6-3!70!black, every mark/.append style={solid,fill=Blues-6-3},mark=triangle*\\%
 Blues-6-4!70!black, every mark/.append style={solid,fill=Blues-6-4},mark=diamond*\\%
 Blues-6-5!70!black, every mark/.append style={solid,fill=Blues-6-5},mark=pentagon*\\%
 Blues-6-6!70!black, every mark/.append style={solid,fill=Blues-6-6},mark=x\\%
}
\pgfplotscreateplotcyclelist{Mark-Blues-7}{%
 Blues-7-1!70!black, every mark/.append style={solid,fill=Blues-7-1},mark=*\\%
 Blues-7-2!70!black, every mark/.append style={solid,fill=Blues-7-2},mark=square*\\%
 Blues-7-3!70!black, every mark/.append style={solid,fill=Blues-7-3},mark=triangle*\\%
 Blues-7-4!70!black, every mark/.append style={solid,fill=Blues-7-4},mark=diamond*\\%
 Blues-7-5!70!black, every mark/.append style={solid,fill=Blues-7-5},mark=pentagon*\\%
 Blues-7-6!70!black, every mark/.append style={solid,fill=Blues-7-6},mark=x\\%
 Blues-7-7!70!black, every mark/.append style={solid,fill=Blues-7-7},mark=+\\%
}
\pgfplotscreateplotcyclelist{Mark-Blues-8}{%
 Blues-8-1!70!black, every mark/.append style={solid,fill=Blues-8-1},mark=*\\%
 Blues-8-2!70!black, every mark/.append style={solid,fill=Blues-8-2},mark=square*\\%
 Blues-8-3!70!black, every mark/.append style={solid,fill=Blues-8-3},mark=triangle*\\%
 Blues-8-4!70!black, every mark/.append style={solid,fill=Blues-8-4},mark=diamond*\\%
 Blues-8-5!70!black, every mark/.append style={solid,fill=Blues-8-5},mark=pentagon*\\%
 Blues-8-6!70!black, every mark/.append style={solid,fill=Blues-8-6},mark=x\\%
 Blues-8-7!70!black, every mark/.append style={solid,fill=Blues-8-7},mark=+\\%
 Blues-8-8!70!black, every mark/.append style={solid,fill=Blues-8-8},mark=star\\%
}
\pgfplotscreateplotcyclelist{Mark-Blues-9}{%
 Blues-9-1!70!black, every mark/.append style={solid,fill=Blues-9-1},mark=*\\%
 Blues-9-2!70!black, every mark/.append style={solid,fill=Blues-9-2},mark=square*\\%
 Blues-9-3!70!black, every mark/.append style={solid,fill=Blues-9-3},mark=triangle*\\%
 Blues-9-4!70!black, every mark/.append style={solid,fill=Blues-9-4},mark=diamond*\\%
 Blues-9-5!70!black, every mark/.append style={solid,fill=Blues-9-5},mark=pentagon*\\%
 Blues-9-6!70!black, every mark/.append style={solid,fill=Blues-9-6},mark=x\\%
 Blues-9-7!70!black, every mark/.append style={solid,fill=Blues-9-7},mark=+\\%
 Blues-9-8!70!black, every mark/.append style={solid,fill=Blues-9-8},mark=star\\%
 Blues-9-9!70!black, every mark/.append style={solid,fill=Blues-9-9},mark=oplus*\\%
}
\pgfplotscreateplotcyclelist{Mark-Greens-3}{%
 Greens-3-1!70!black, every mark/.append style={solid,fill=Greens-3-1},mark=*\\%
 Greens-3-2!70!black, every mark/.append style={solid,fill=Greens-3-2},mark=square*\\%
 Greens-3-3!70!black, every mark/.append style={solid,fill=Greens-3-3},mark=triangle*\\%
}
\pgfplotscreateplotcyclelist{Mark-Greens-4}{%
 Greens-4-1!70!black, every mark/.append style={solid,fill=Greens-4-1},mark=*\\%
 Greens-4-2!70!black, every mark/.append style={solid,fill=Greens-4-2},mark=square*\\%
 Greens-4-3!70!black, every mark/.append style={solid,fill=Greens-4-3},mark=triangle*\\%
 Greens-4-4!70!black, every mark/.append style={solid,fill=Greens-4-4},mark=diamond*\\%
}
\pgfplotscreateplotcyclelist{Mark-Greens-5}{%
 Greens-5-1!70!black, every mark/.append style={solid,fill=Greens-5-1},mark=*\\%
 Greens-5-2!70!black, every mark/.append style={solid,fill=Greens-5-2},mark=square*\\%
 Greens-5-3!70!black, every mark/.append style={solid,fill=Greens-5-3},mark=triangle*\\%
 Greens-5-4!70!black, every mark/.append style={solid,fill=Greens-5-4},mark=diamond*\\%
 Greens-5-5!70!black, every mark/.append style={solid,fill=Greens-5-5},mark=pentagon*\\%
}
\pgfplotscreateplotcyclelist{Mark-Greens-6}{%
 Greens-6-1!70!black, every mark/.append style={solid,fill=Greens-6-1},mark=*\\%
 Greens-6-2!70!black, every mark/.append style={solid,fill=Greens-6-2},mark=square*\\%
 Greens-6-3!70!black, every mark/.append style={solid,fill=Greens-6-3},mark=triangle*\\%
 Greens-6-4!70!black, every mark/.append style={solid,fill=Greens-6-4},mark=diamond*\\%
 Greens-6-5!70!black, every mark/.append style={solid,fill=Greens-6-5},mark=pentagon*\\%
 Greens-6-6!70!black, every mark/.append style={solid,fill=Greens-6-6},mark=x\\%
}
\pgfplotscreateplotcyclelist{Mark-Greens-7}{%
 Greens-7-1!70!black, every mark/.append style={solid,fill=Greens-7-1},mark=*\\%
 Greens-7-2!70!black, every mark/.append style={solid,fill=Greens-7-2},mark=square*\\%
 Greens-7-3!70!black, every mark/.append style={solid,fill=Greens-7-3},mark=triangle*\\%
 Greens-7-4!70!black, every mark/.append style={solid,fill=Greens-7-4},mark=diamond*\\%
 Greens-7-5!70!black, every mark/.append style={solid,fill=Greens-7-5},mark=pentagon*\\%
 Greens-7-6!70!black, every mark/.append style={solid,fill=Greens-7-6},mark=x\\%
 Greens-7-7!70!black, every mark/.append style={solid,fill=Greens-7-7},mark=+\\%
}
\pgfplotscreateplotcyclelist{Mark-Greens-8}{%
 Greens-8-1!70!black, every mark/.append style={solid,fill=Greens-8-1},mark=*\\%
 Greens-8-2!70!black, every mark/.append style={solid,fill=Greens-8-2},mark=square*\\%
 Greens-8-3!70!black, every mark/.append style={solid,fill=Greens-8-3},mark=triangle*\\%
 Greens-8-4!70!black, every mark/.append style={solid,fill=Greens-8-4},mark=diamond*\\%
 Greens-8-5!70!black, every mark/.append style={solid,fill=Greens-8-5},mark=pentagon*\\%
 Greens-8-6!70!black, every mark/.append style={solid,fill=Greens-8-6},mark=x\\%
 Greens-8-7!70!black, every mark/.append style={solid,fill=Greens-8-7},mark=+\\%
 Greens-8-8!70!black, every mark/.append style={solid,fill=Greens-8-8},mark=star\\%
}
\pgfplotscreateplotcyclelist{Mark-Greens-9}{%
 Greens-9-1!70!black, every mark/.append style={solid,fill=Greens-9-1},mark=*\\%
 Greens-9-2!70!black, every mark/.append style={solid,fill=Greens-9-2},mark=square*\\%
 Greens-9-3!70!black, every mark/.append style={solid,fill=Greens-9-3},mark=triangle*\\%
 Greens-9-4!70!black, every mark/.append style={solid,fill=Greens-9-4},mark=diamond*\\%
 Greens-9-5!70!black, every mark/.append style={solid,fill=Greens-9-5},mark=pentagon*\\%
 Greens-9-6!70!black, every mark/.append style={solid,fill=Greens-9-6},mark=x\\%
 Greens-9-7!70!black, every mark/.append style={solid,fill=Greens-9-7},mark=+\\%
 Greens-9-8!70!black, every mark/.append style={solid,fill=Greens-9-8},mark=star\\%
 Greens-9-9!70!black, every mark/.append style={solid,fill=Greens-9-9},mark=oplus*\\%
}
\pgfplotscreateplotcyclelist{Mark-Oranges-3}{%
 Oranges-3-1!70!black, every mark/.append style={solid,fill=Oranges-3-1},mark=*\\%
 Oranges-3-2!70!black, every mark/.append style={solid,fill=Oranges-3-2},mark=square*\\%
 Oranges-3-3!70!black, every mark/.append style={solid,fill=Oranges-3-3},mark=triangle*\\%
}
\pgfplotscreateplotcyclelist{Mark-Oranges-4}{%
 Oranges-4-1!70!black, every mark/.append style={solid,fill=Oranges-4-1},mark=*\\%
 Oranges-4-2!70!black, every mark/.append style={solid,fill=Oranges-4-2},mark=square*\\%
 Oranges-4-3!70!black, every mark/.append style={solid,fill=Oranges-4-3},mark=triangle*\\%
 Oranges-4-4!70!black, every mark/.append style={solid,fill=Oranges-4-4},mark=diamond*\\%
}
\pgfplotscreateplotcyclelist{Mark-Oranges-5}{%
 Oranges-5-1!70!black, every mark/.append style={solid,fill=Oranges-5-1},mark=*\\%
 Oranges-5-2!70!black, every mark/.append style={solid,fill=Oranges-5-2},mark=square*\\%
 Oranges-5-3!70!black, every mark/.append style={solid,fill=Oranges-5-3},mark=triangle*\\%
 Oranges-5-4!70!black, every mark/.append style={solid,fill=Oranges-5-4},mark=diamond*\\%
 Oranges-5-5!70!black, every mark/.append style={solid,fill=Oranges-5-5},mark=pentagon*\\%
}
\pgfplotscreateplotcyclelist{Mark-Oranges-6}{%
 Oranges-6-1!70!black, every mark/.append style={solid,fill=Oranges-6-1},mark=*\\%
 Oranges-6-2!70!black, every mark/.append style={solid,fill=Oranges-6-2},mark=square*\\%
 Oranges-6-3!70!black, every mark/.append style={solid,fill=Oranges-6-3},mark=triangle*\\%
 Oranges-6-4!70!black, every mark/.append style={solid,fill=Oranges-6-4},mark=diamond*\\%
 Oranges-6-5!70!black, every mark/.append style={solid,fill=Oranges-6-5},mark=pentagon*\\%
 Oranges-6-6!70!black, every mark/.append style={solid,fill=Oranges-6-6},mark=x\\%
}
\pgfplotscreateplotcyclelist{Mark-Oranges-7}{%
 Oranges-7-1!70!black, every mark/.append style={solid,fill=Oranges-7-1},mark=*\\%
 Oranges-7-2!70!black, every mark/.append style={solid,fill=Oranges-7-2},mark=square*\\%
 Oranges-7-3!70!black, every mark/.append style={solid,fill=Oranges-7-3},mark=triangle*\\%
 Oranges-7-4!70!black, every mark/.append style={solid,fill=Oranges-7-4},mark=diamond*\\%
 Oranges-7-5!70!black, every mark/.append style={solid,fill=Oranges-7-5},mark=pentagon*\\%
 Oranges-7-6!70!black, every mark/.append style={solid,fill=Oranges-7-6},mark=x\\%
 Oranges-7-7!70!black, every mark/.append style={solid,fill=Oranges-7-7},mark=+\\%
}
\pgfplotscreateplotcyclelist{Mark-Oranges-8}{%
 Oranges-8-1!70!black, every mark/.append style={solid,fill=Oranges-8-1},mark=*\\%
 Oranges-8-2!70!black, every mark/.append style={solid,fill=Oranges-8-2},mark=square*\\%
 Oranges-8-3!70!black, every mark/.append style={solid,fill=Oranges-8-3},mark=triangle*\\%
 Oranges-8-4!70!black, every mark/.append style={solid,fill=Oranges-8-4},mark=diamond*\\%
 Oranges-8-5!70!black, every mark/.append style={solid,fill=Oranges-8-5},mark=pentagon*\\%
 Oranges-8-6!70!black, every mark/.append style={solid,fill=Oranges-8-6},mark=x\\%
 Oranges-8-7!70!black, every mark/.append style={solid,fill=Oranges-8-7},mark=+\\%
 Oranges-8-8!70!black, every mark/.append style={solid,fill=Oranges-8-8},mark=star\\%
}
\pgfplotscreateplotcyclelist{Mark-Oranges-9}{%
 Oranges-9-1!70!black, every mark/.append style={solid,fill=Oranges-9-1},mark=*\\%
 Oranges-9-2!70!black, every mark/.append style={solid,fill=Oranges-9-2},mark=square*\\%
 Oranges-9-3!70!black, every mark/.append style={solid,fill=Oranges-9-3},mark=triangle*\\%
 Oranges-9-4!70!black, every mark/.append style={solid,fill=Oranges-9-4},mark=diamond*\\%
 Oranges-9-5!70!black, every mark/.append style={solid,fill=Oranges-9-5},mark=pentagon*\\%
 Oranges-9-6!70!black, every mark/.append style={solid,fill=Oranges-9-6},mark=x\\%
 Oranges-9-7!70!black, every mark/.append style={solid,fill=Oranges-9-7},mark=+\\%
 Oranges-9-8!70!black, every mark/.append style={solid,fill=Oranges-9-8},mark=star\\%
 Oranges-9-9!70!black, every mark/.append style={solid,fill=Oranges-9-9},mark=oplus*\\%
}
\pgfplotscreateplotcyclelist{Mark-Reds-3}{%
 Reds-3-1!70!black, every mark/.append style={solid,fill=Reds-3-1},mark=*\\%
 Reds-3-2!70!black, every mark/.append style={solid,fill=Reds-3-2},mark=square*\\%
 Reds-3-3!70!black, every mark/.append style={solid,fill=Reds-3-3},mark=triangle*\\%
}
\pgfplotscreateplotcyclelist{Mark-Reds-4}{%
 Reds-4-1!70!black, every mark/.append style={solid,fill=Reds-4-1},mark=*\\%
 Reds-4-2!70!black, every mark/.append style={solid,fill=Reds-4-2},mark=square*\\%
 Reds-4-3!70!black, every mark/.append style={solid,fill=Reds-4-3},mark=triangle*\\%
 Reds-4-4!70!black, every mark/.append style={solid,fill=Reds-4-4},mark=diamond*\\%
}
\pgfplotscreateplotcyclelist{Mark-Reds-5}{%
 Reds-5-1!70!black, every mark/.append style={solid,fill=Reds-5-1},mark=*\\%
 Reds-5-2!70!black, every mark/.append style={solid,fill=Reds-5-2},mark=square*\\%
 Reds-5-3!70!black, every mark/.append style={solid,fill=Reds-5-3},mark=triangle*\\%
 Reds-5-4!70!black, every mark/.append style={solid,fill=Reds-5-4},mark=diamond*\\%
 Reds-5-5!70!black, every mark/.append style={solid,fill=Reds-5-5},mark=pentagon*\\%
}
\pgfplotscreateplotcyclelist{Mark-Reds-6}{%
 Reds-6-1!70!black, every mark/.append style={solid,fill=Reds-6-1},mark=*\\%
 Reds-6-2!70!black, every mark/.append style={solid,fill=Reds-6-2},mark=square*\\%
 Reds-6-3!70!black, every mark/.append style={solid,fill=Reds-6-3},mark=triangle*\\%
 Reds-6-4!70!black, every mark/.append style={solid,fill=Reds-6-4},mark=diamond*\\%
 Reds-6-5!70!black, every mark/.append style={solid,fill=Reds-6-5},mark=pentagon*\\%
 Reds-6-6!70!black, every mark/.append style={solid,fill=Reds-6-6},mark=x\\%
}
\pgfplotscreateplotcyclelist{Mark-Reds-7}{%
 Reds-7-1!70!black, every mark/.append style={solid,fill=Reds-7-1},mark=*\\%
 Reds-7-2!70!black, every mark/.append style={solid,fill=Reds-7-2},mark=square*\\%
 Reds-7-3!70!black, every mark/.append style={solid,fill=Reds-7-3},mark=triangle*\\%
 Reds-7-4!70!black, every mark/.append style={solid,fill=Reds-7-4},mark=diamond*\\%
 Reds-7-5!70!black, every mark/.append style={solid,fill=Reds-7-5},mark=pentagon*\\%
 Reds-7-6!70!black, every mark/.append style={solid,fill=Reds-7-6},mark=x\\%
 Reds-7-7!70!black, every mark/.append style={solid,fill=Reds-7-7},mark=+\\%
}
\pgfplotscreateplotcyclelist{Mark-Reds-8}{%
 Reds-8-1!70!black, every mark/.append style={solid,fill=Reds-8-1},mark=*\\%
 Reds-8-2!70!black, every mark/.append style={solid,fill=Reds-8-2},mark=square*\\%
 Reds-8-3!70!black, every mark/.append style={solid,fill=Reds-8-3},mark=triangle*\\%
 Reds-8-4!70!black, every mark/.append style={solid,fill=Reds-8-4},mark=diamond*\\%
 Reds-8-5!70!black, every mark/.append style={solid,fill=Reds-8-5},mark=pentagon*\\%
 Reds-8-6!70!black, every mark/.append style={solid,fill=Reds-8-6},mark=x\\%
 Reds-8-7!70!black, every mark/.append style={solid,fill=Reds-8-7},mark=+\\%
 Reds-8-8!70!black, every mark/.append style={solid,fill=Reds-8-8},mark=star\\%
}
\pgfplotscreateplotcyclelist{Mark-Reds-9}{%
 Reds-9-1!70!black, every mark/.append style={solid,fill=Reds-9-1},mark=*\\%
 Reds-9-2!70!black, every mark/.append style={solid,fill=Reds-9-2},mark=square*\\%
 Reds-9-3!70!black, every mark/.append style={solid,fill=Reds-9-3},mark=triangle*\\%
 Reds-9-4!70!black, every mark/.append style={solid,fill=Reds-9-4},mark=diamond*\\%
 Reds-9-5!70!black, every mark/.append style={solid,fill=Reds-9-5},mark=pentagon*\\%
 Reds-9-6!70!black, every mark/.append style={solid,fill=Reds-9-6},mark=x\\%
 Reds-9-7!70!black, every mark/.append style={solid,fill=Reds-9-7},mark=+\\%
 Reds-9-8!70!black, every mark/.append style={solid,fill=Reds-9-8},mark=star\\%
 Reds-9-9!70!black, every mark/.append style={solid,fill=Reds-9-9},mark=oplus*\\%
}
\pgfplotscreateplotcyclelist{Mark-Greys-3}{%
 Greys-3-1!70!black, every mark/.append style={solid,fill=Greys-3-1},mark=*\\%
 Greys-3-2!70!black, every mark/.append style={solid,fill=Greys-3-2},mark=square*\\%
 Greys-3-3!70!black, every mark/.append style={solid,fill=Greys-3-3},mark=triangle*\\%
}
\pgfplotscreateplotcyclelist{Mark-Greys-4}{%
 Greys-4-1!70!black, every mark/.append style={solid,fill=Greys-4-1},mark=*\\%
 Greys-4-2!70!black, every mark/.append style={solid,fill=Greys-4-2},mark=square*\\%
 Greys-4-3!70!black, every mark/.append style={solid,fill=Greys-4-3},mark=triangle*\\%
 Greys-4-4!70!black, every mark/.append style={solid,fill=Greys-4-4},mark=diamond*\\%
}
\pgfplotscreateplotcyclelist{Mark-Greys-5}{%
 Greys-5-1!70!black, every mark/.append style={solid,fill=Greys-5-1},mark=*\\%
 Greys-5-2!70!black, every mark/.append style={solid,fill=Greys-5-2},mark=square*\\%
 Greys-5-3!70!black, every mark/.append style={solid,fill=Greys-5-3},mark=triangle*\\%
 Greys-5-4!70!black, every mark/.append style={solid,fill=Greys-5-4},mark=diamond*\\%
 Greys-5-5!70!black, every mark/.append style={solid,fill=Greys-5-5},mark=pentagon*\\%
}
\pgfplotscreateplotcyclelist{Mark-Greys-6}{%
 Greys-6-1!70!black, every mark/.append style={solid,fill=Greys-6-1},mark=*\\%
 Greys-6-2!70!black, every mark/.append style={solid,fill=Greys-6-2},mark=square*\\%
 Greys-6-3!70!black, every mark/.append style={solid,fill=Greys-6-3},mark=triangle*\\%
 Greys-6-4!70!black, every mark/.append style={solid,fill=Greys-6-4},mark=diamond*\\%
 Greys-6-5!70!black, every mark/.append style={solid,fill=Greys-6-5},mark=pentagon*\\%
 Greys-6-6!70!black, every mark/.append style={solid,fill=Greys-6-6},mark=x\\%
}
\pgfplotscreateplotcyclelist{Mark-Greys-7}{%
 Greys-7-1!70!black, every mark/.append style={solid,fill=Greys-7-1},mark=*\\%
 Greys-7-2!70!black, every mark/.append style={solid,fill=Greys-7-2},mark=square*\\%
 Greys-7-3!70!black, every mark/.append style={solid,fill=Greys-7-3},mark=triangle*\\%
 Greys-7-4!70!black, every mark/.append style={solid,fill=Greys-7-4},mark=diamond*\\%
 Greys-7-5!70!black, every mark/.append style={solid,fill=Greys-7-5},mark=pentagon*\\%
 Greys-7-6!70!black, every mark/.append style={solid,fill=Greys-7-6},mark=x\\%
 Greys-7-7!70!black, every mark/.append style={solid,fill=Greys-7-7},mark=+\\%
}
\pgfplotscreateplotcyclelist{Mark-Greys-8}{%
 Greys-8-1!70!black, every mark/.append style={solid,fill=Greys-8-1},mark=*\\%
 Greys-8-2!70!black, every mark/.append style={solid,fill=Greys-8-2},mark=square*\\%
 Greys-8-3!70!black, every mark/.append style={solid,fill=Greys-8-3},mark=triangle*\\%
 Greys-8-4!70!black, every mark/.append style={solid,fill=Greys-8-4},mark=diamond*\\%
 Greys-8-5!70!black, every mark/.append style={solid,fill=Greys-8-5},mark=pentagon*\\%
 Greys-8-6!70!black, every mark/.append style={solid,fill=Greys-8-6},mark=x\\%
 Greys-8-7!70!black, every mark/.append style={solid,fill=Greys-8-7},mark=+\\%
 Greys-8-8!70!black, every mark/.append style={solid,fill=Greys-8-8},mark=star\\%
}
\pgfplotscreateplotcyclelist{Mark-Greys-9}{%
 Greys-9-1!70!black, every mark/.append style={solid,fill=Greys-9-1},mark=*\\%
 Greys-9-2!70!black, every mark/.append style={solid,fill=Greys-9-2},mark=square*\\%
 Greys-9-3!70!black, every mark/.append style={solid,fill=Greys-9-3},mark=triangle*\\%
 Greys-9-4!70!black, every mark/.append style={solid,fill=Greys-9-4},mark=diamond*\\%
 Greys-9-5!70!black, every mark/.append style={solid,fill=Greys-9-5},mark=pentagon*\\%
 Greys-9-6!70!black, every mark/.append style={solid,fill=Greys-9-6},mark=x\\%
 Greys-9-7!70!black, every mark/.append style={solid,fill=Greys-9-7},mark=+\\%
 Greys-9-8!70!black, every mark/.append style={solid,fill=Greys-9-8},mark=star\\%
 Greys-9-9!70!black, every mark/.append style={solid,fill=Greys-9-9},mark=oplus*\\%
}
\pgfplotscreateplotcyclelist{Mark-PuOr-3}{%
 PuOr-3-1!70!black, every mark/.append style={solid,fill=PuOr-3-1},mark=*\\%
 PuOr-3-2!70!black, every mark/.append style={solid,fill=PuOr-3-2},mark=square*\\%
 PuOr-3-3!70!black, every mark/.append style={solid,fill=PuOr-3-3},mark=triangle*\\%
}
\pgfplotscreateplotcyclelist{Mark-PuOr-4}{%
 PuOr-4-1!70!black, every mark/.append style={solid,fill=PuOr-4-1},mark=*\\%
 PuOr-4-2!70!black, every mark/.append style={solid,fill=PuOr-4-2},mark=square*\\%
 PuOr-4-3!70!black, every mark/.append style={solid,fill=PuOr-4-3},mark=triangle*\\%
 PuOr-4-4!70!black, every mark/.append style={solid,fill=PuOr-4-4},mark=diamond*\\%
}
\pgfplotscreateplotcyclelist{Mark-PuOr-5}{%
 PuOr-5-1!70!black, every mark/.append style={solid,fill=PuOr-5-1},mark=*\\%
 PuOr-5-2!70!black, every mark/.append style={solid,fill=PuOr-5-2},mark=square*\\%
 PuOr-5-3!70!black, every mark/.append style={solid,fill=PuOr-5-3},mark=triangle*\\%
 PuOr-5-4!70!black, every mark/.append style={solid,fill=PuOr-5-4},mark=diamond*\\%
 PuOr-5-5!70!black, every mark/.append style={solid,fill=PuOr-5-5},mark=pentagon*\\%
}
\pgfplotscreateplotcyclelist{Mark-PuOr-6}{%
 PuOr-6-1!70!black, every mark/.append style={solid,fill=PuOr-6-1},mark=*\\%
 PuOr-6-2!70!black, every mark/.append style={solid,fill=PuOr-6-2},mark=square*\\%
 PuOr-6-3!70!black, every mark/.append style={solid,fill=PuOr-6-3},mark=triangle*\\%
 PuOr-6-4!70!black, every mark/.append style={solid,fill=PuOr-6-4},mark=diamond*\\%
 PuOr-6-5!70!black, every mark/.append style={solid,fill=PuOr-6-5},mark=pentagon*\\%
 PuOr-6-6!70!black, every mark/.append style={solid,fill=PuOr-6-6},mark=x\\%
}
\pgfplotscreateplotcyclelist{Mark-PuOr-7}{%
 PuOr-7-1!70!black, every mark/.append style={solid,fill=PuOr-7-1},mark=*\\%
 PuOr-7-2!70!black, every mark/.append style={solid,fill=PuOr-7-2},mark=square*\\%
 PuOr-7-3!70!black, every mark/.append style={solid,fill=PuOr-7-3},mark=triangle*\\%
 PuOr-7-4!70!black, every mark/.append style={solid,fill=PuOr-7-4},mark=diamond*\\%
 PuOr-7-5!70!black, every mark/.append style={solid,fill=PuOr-7-5},mark=pentagon*\\%
 PuOr-7-6!70!black, every mark/.append style={solid,fill=PuOr-7-6},mark=x\\%
 PuOr-7-7!70!black, every mark/.append style={solid,fill=PuOr-7-7},mark=+\\%
}
\pgfplotscreateplotcyclelist{Mark-PuOr-8}{%
 PuOr-8-1!70!black, every mark/.append style={solid,fill=PuOr-8-1},mark=*\\%
 PuOr-8-2!70!black, every mark/.append style={solid,fill=PuOr-8-2},mark=square*\\%
 PuOr-8-3!70!black, every mark/.append style={solid,fill=PuOr-8-3},mark=triangle*\\%
 PuOr-8-4!70!black, every mark/.append style={solid,fill=PuOr-8-4},mark=diamond*\\%
 PuOr-8-5!70!black, every mark/.append style={solid,fill=PuOr-8-5},mark=pentagon*\\%
 PuOr-8-6!70!black, every mark/.append style={solid,fill=PuOr-8-6},mark=x\\%
 PuOr-8-7!70!black, every mark/.append style={solid,fill=PuOr-8-7},mark=+\\%
 PuOr-8-8!70!black, every mark/.append style={solid,fill=PuOr-8-8},mark=star\\%
}
\pgfplotscreateplotcyclelist{Mark-PuOr-9}{%
 PuOr-9-1!70!black, every mark/.append style={solid,fill=PuOr-9-1},mark=*\\%
 PuOr-9-2!70!black, every mark/.append style={solid,fill=PuOr-9-2},mark=square*\\%
 PuOr-9-3!70!black, every mark/.append style={solid,fill=PuOr-9-3},mark=triangle*\\%
 PuOr-9-4!70!black, every mark/.append style={solid,fill=PuOr-9-4},mark=diamond*\\%
 PuOr-9-5!70!black, every mark/.append style={solid,fill=PuOr-9-5},mark=pentagon*\\%
 PuOr-9-6!70!black, every mark/.append style={solid,fill=PuOr-9-6},mark=x\\%
 PuOr-9-7!70!black, every mark/.append style={solid,fill=PuOr-9-7},mark=+\\%
 PuOr-9-8!70!black, every mark/.append style={solid,fill=PuOr-9-8},mark=star\\%
 PuOr-9-9!70!black, every mark/.append style={solid,fill=PuOr-9-9},mark=oplus*\\%
}
\pgfplotscreateplotcyclelist{Mark-PuOr-10}{%
 PuOr-10-1!70!black, every mark/.append style={solid,fill=PuOr-10-1},mark=*\\%
 PuOr-10-2!70!black, every mark/.append style={solid,fill=PuOr-10-2},mark=square*\\%
 PuOr-10-3!70!black, every mark/.append style={solid,fill=PuOr-10-3},mark=triangle*\\%
 PuOr-10-4!70!black, every mark/.append style={solid,fill=PuOr-10-4},mark=diamond*\\%
 PuOr-10-5!70!black, every mark/.append style={solid,fill=PuOr-10-5},mark=pentagon*\\%
 PuOr-10-6!70!black, every mark/.append style={solid,fill=PuOr-10-6},mark=x\\%
 PuOr-10-7!70!black, every mark/.append style={solid,fill=PuOr-10-7},mark=+\\%
 PuOr-10-8!70!black, every mark/.append style={solid,fill=PuOr-10-8},mark=star\\%
 PuOr-10-9!70!black, every mark/.append style={solid,fill=PuOr-10-9},mark=oplus*\\%
 PuOr-10-10!70!black, every mark/.append style={solid,fill=PuOr-10-10},mark=otimes*\\%
}
\pgfplotscreateplotcyclelist{Mark-PuOr-11}{%
 PuOr-11-1!70!black, every mark/.append style={solid,fill=PuOr-11-1},mark=*\\%
 PuOr-11-2!70!black, every mark/.append style={solid,fill=PuOr-11-2},mark=square*\\%
 PuOr-11-3!70!black, every mark/.append style={solid,fill=PuOr-11-3},mark=triangle*\\%
 PuOr-11-4!70!black, every mark/.append style={solid,fill=PuOr-11-4},mark=diamond*\\%
 PuOr-11-5!70!black, every mark/.append style={solid,fill=PuOr-11-5},mark=pentagon*\\%
 PuOr-11-6!70!black, every mark/.append style={solid,fill=PuOr-11-6},mark=x\\%
 PuOr-11-7!70!black, every mark/.append style={solid,fill=PuOr-11-7},mark=+\\%
 PuOr-11-8!70!black, every mark/.append style={solid,fill=PuOr-11-8},mark=star\\%
 PuOr-11-9!70!black, every mark/.append style={solid,fill=PuOr-11-9},mark=oplus*\\%
 PuOr-11-10!70!black, every mark/.append style={solid,fill=PuOr-11-10},mark=otimes*\\%
}
\pgfplotscreateplotcyclelist{Mark-BrBG-3}{%
 BrBG-3-1!70!black, every mark/.append style={solid,fill=BrBG-3-1},mark=*\\%
 BrBG-3-2!70!black, every mark/.append style={solid,fill=BrBG-3-2},mark=square*\\%
 BrBG-3-3!70!black, every mark/.append style={solid,fill=BrBG-3-3},mark=triangle*\\%
}
\pgfplotscreateplotcyclelist{Mark-BrBG-4}{%
 BrBG-4-1!70!black, every mark/.append style={solid,fill=BrBG-4-1},mark=*\\%
 BrBG-4-2!70!black, every mark/.append style={solid,fill=BrBG-4-2},mark=square*\\%
 BrBG-4-3!70!black, every mark/.append style={solid,fill=BrBG-4-3},mark=triangle*\\%
 BrBG-4-4!70!black, every mark/.append style={solid,fill=BrBG-4-4},mark=diamond*\\%
}
\pgfplotscreateplotcyclelist{Mark-BrBG-5}{%
 BrBG-5-1!70!black, every mark/.append style={solid,fill=BrBG-5-1},mark=*\\%
 BrBG-5-2!70!black, every mark/.append style={solid,fill=BrBG-5-2},mark=square*\\%
 BrBG-5-3!70!black, every mark/.append style={solid,fill=BrBG-5-3},mark=triangle*\\%
 BrBG-5-4!70!black, every mark/.append style={solid,fill=BrBG-5-4},mark=diamond*\\%
 BrBG-5-5!70!black, every mark/.append style={solid,fill=BrBG-5-5},mark=pentagon*\\%
}
\pgfplotscreateplotcyclelist{Mark-BrBG-6}{%
 BrBG-6-1!70!black, every mark/.append style={solid,fill=BrBG-6-1},mark=*\\%
 BrBG-6-2!70!black, every mark/.append style={solid,fill=BrBG-6-2},mark=square*\\%
 BrBG-6-3!70!black, every mark/.append style={solid,fill=BrBG-6-3},mark=triangle*\\%
 BrBG-6-4!70!black, every mark/.append style={solid,fill=BrBG-6-4},mark=diamond*\\%
 BrBG-6-5!70!black, every mark/.append style={solid,fill=BrBG-6-5},mark=pentagon*\\%
 BrBG-6-6!70!black, every mark/.append style={solid,fill=BrBG-6-6},mark=x\\%
}
\pgfplotscreateplotcyclelist{Mark-BrBG-7}{%
 BrBG-7-1!70!black, every mark/.append style={solid,fill=BrBG-7-1},mark=*\\%
 BrBG-7-2!70!black, every mark/.append style={solid,fill=BrBG-7-2},mark=square*\\%
 BrBG-7-3!70!black, every mark/.append style={solid,fill=BrBG-7-3},mark=triangle*\\%
 BrBG-7-4!70!black, every mark/.append style={solid,fill=BrBG-7-4},mark=diamond*\\%
 BrBG-7-5!70!black, every mark/.append style={solid,fill=BrBG-7-5},mark=pentagon*\\%
 BrBG-7-6!70!black, every mark/.append style={solid,fill=BrBG-7-6},mark=x\\%
 BrBG-7-7!70!black, every mark/.append style={solid,fill=BrBG-7-7},mark=+\\%
}
\pgfplotscreateplotcyclelist{Mark-BrBG-8}{%
 BrBG-8-1!70!black, every mark/.append style={solid,fill=BrBG-8-1},mark=*\\%
 BrBG-8-2!70!black, every mark/.append style={solid,fill=BrBG-8-2},mark=square*\\%
 BrBG-8-3!70!black, every mark/.append style={solid,fill=BrBG-8-3},mark=triangle*\\%
 BrBG-8-4!70!black, every mark/.append style={solid,fill=BrBG-8-4},mark=diamond*\\%
 BrBG-8-5!70!black, every mark/.append style={solid,fill=BrBG-8-5},mark=pentagon*\\%
 BrBG-8-6!70!black, every mark/.append style={solid,fill=BrBG-8-6},mark=x\\%
 BrBG-8-7!70!black, every mark/.append style={solid,fill=BrBG-8-7},mark=+\\%
 BrBG-8-8!70!black, every mark/.append style={solid,fill=BrBG-8-8},mark=star\\%
}
\pgfplotscreateplotcyclelist{Mark-BrBG-9}{%
 BrBG-9-1!70!black, every mark/.append style={solid,fill=BrBG-9-1},mark=*\\%
 BrBG-9-2!70!black, every mark/.append style={solid,fill=BrBG-9-2},mark=square*\\%
 BrBG-9-3!70!black, every mark/.append style={solid,fill=BrBG-9-3},mark=triangle*\\%
 BrBG-9-4!70!black, every mark/.append style={solid,fill=BrBG-9-4},mark=diamond*\\%
 BrBG-9-5!70!black, every mark/.append style={solid,fill=BrBG-9-5},mark=pentagon*\\%
 BrBG-9-6!70!black, every mark/.append style={solid,fill=BrBG-9-6},mark=x\\%
 BrBG-9-7!70!black, every mark/.append style={solid,fill=BrBG-9-7},mark=+\\%
 BrBG-9-8!70!black, every mark/.append style={solid,fill=BrBG-9-8},mark=star\\%
 BrBG-9-9!70!black, every mark/.append style={solid,fill=BrBG-9-9},mark=oplus*\\%
}
\pgfplotscreateplotcyclelist{Mark-BrBG-10}{%
 BrBG-10-1!70!black, every mark/.append style={solid,fill=BrBG-10-1},mark=*\\%
 BrBG-10-2!70!black, every mark/.append style={solid,fill=BrBG-10-2},mark=square*\\%
 BrBG-10-3!70!black, every mark/.append style={solid,fill=BrBG-10-3},mark=triangle*\\%
 BrBG-10-4!70!black, every mark/.append style={solid,fill=BrBG-10-4},mark=diamond*\\%
 BrBG-10-5!70!black, every mark/.append style={solid,fill=BrBG-10-5},mark=pentagon*\\%
 BrBG-10-6!70!black, every mark/.append style={solid,fill=BrBG-10-6},mark=x\\%
 BrBG-10-7!70!black, every mark/.append style={solid,fill=BrBG-10-7},mark=+\\%
 BrBG-10-8!70!black, every mark/.append style={solid,fill=BrBG-10-8},mark=star\\%
 BrBG-10-9!70!black, every mark/.append style={solid,fill=BrBG-10-9},mark=oplus*\\%
 BrBG-10-10!70!black, every mark/.append style={solid,fill=BrBG-10-10},mark=otimes*\\%
}
\pgfplotscreateplotcyclelist{Mark-BrBG-11}{%
 BrBG-11-1!70!black, every mark/.append style={solid,fill=BrBG-11-1},mark=*\\%
 BrBG-11-2!70!black, every mark/.append style={solid,fill=BrBG-11-2},mark=square*\\%
 BrBG-11-3!70!black, every mark/.append style={solid,fill=BrBG-11-3},mark=triangle*\\%
 BrBG-11-4!70!black, every mark/.append style={solid,fill=BrBG-11-4},mark=diamond*\\%
 BrBG-11-5!70!black, every mark/.append style={solid,fill=BrBG-11-5},mark=pentagon*\\%
 BrBG-11-6!70!black, every mark/.append style={solid,fill=BrBG-11-6},mark=x\\%
 BrBG-11-7!70!black, every mark/.append style={solid,fill=BrBG-11-7},mark=+\\%
 BrBG-11-8!70!black, every mark/.append style={solid,fill=BrBG-11-8},mark=star\\%
 BrBG-11-9!70!black, every mark/.append style={solid,fill=BrBG-11-9},mark=oplus*\\%
 BrBG-11-10!70!black, every mark/.append style={solid,fill=BrBG-11-10},mark=otimes*\\%
}
\pgfplotscreateplotcyclelist{Mark-PRGn-3}{%
 PRGn-3-1!70!black, every mark/.append style={solid,fill=PRGn-3-1},mark=*\\%
 PRGn-3-2!70!black, every mark/.append style={solid,fill=PRGn-3-2},mark=square*\\%
 PRGn-3-3!70!black, every mark/.append style={solid,fill=PRGn-3-3},mark=triangle*\\%
}
\pgfplotscreateplotcyclelist{Mark-PRGn-4}{%
 PRGn-4-1!70!black, every mark/.append style={solid,fill=PRGn-4-1},mark=*\\%
 PRGn-4-2!70!black, every mark/.append style={solid,fill=PRGn-4-2},mark=square*\\%
 PRGn-4-3!70!black, every mark/.append style={solid,fill=PRGn-4-3},mark=triangle*\\%
 PRGn-4-4!70!black, every mark/.append style={solid,fill=PRGn-4-4},mark=diamond*\\%
}
\pgfplotscreateplotcyclelist{Mark-PRGn-5}{%
 PRGn-5-1!70!black, every mark/.append style={solid,fill=PRGn-5-1},mark=*\\%
 PRGn-5-2!70!black, every mark/.append style={solid,fill=PRGn-5-2},mark=square*\\%
 PRGn-5-3!70!black, every mark/.append style={solid,fill=PRGn-5-3},mark=triangle*\\%
 PRGn-5-4!70!black, every mark/.append style={solid,fill=PRGn-5-4},mark=diamond*\\%
 PRGn-5-5!70!black, every mark/.append style={solid,fill=PRGn-5-5},mark=pentagon*\\%
}
\pgfplotscreateplotcyclelist{Mark-PRGn-6}{%
 PRGn-6-1!70!black, every mark/.append style={solid,fill=PRGn-6-1},mark=*\\%
 PRGn-6-2!70!black, every mark/.append style={solid,fill=PRGn-6-2},mark=square*\\%
 PRGn-6-3!70!black, every mark/.append style={solid,fill=PRGn-6-3},mark=triangle*\\%
 PRGn-6-4!70!black, every mark/.append style={solid,fill=PRGn-6-4},mark=diamond*\\%
 PRGn-6-5!70!black, every mark/.append style={solid,fill=PRGn-6-5},mark=pentagon*\\%
 PRGn-6-6!70!black, every mark/.append style={solid,fill=PRGn-6-6},mark=x\\%
}
\pgfplotscreateplotcyclelist{Mark-PRGn-7}{%
 PRGn-7-1!70!black, every mark/.append style={solid,fill=PRGn-7-1},mark=*\\%
 PRGn-7-2!70!black, every mark/.append style={solid,fill=PRGn-7-2},mark=square*\\%
 PRGn-7-3!70!black, every mark/.append style={solid,fill=PRGn-7-3},mark=triangle*\\%
 PRGn-7-4!70!black, every mark/.append style={solid,fill=PRGn-7-4},mark=diamond*\\%
 PRGn-7-5!70!black, every mark/.append style={solid,fill=PRGn-7-5},mark=pentagon*\\%
 PRGn-7-6!70!black, every mark/.append style={solid,fill=PRGn-7-6},mark=x\\%
 PRGn-7-7!70!black, every mark/.append style={solid,fill=PRGn-7-7},mark=+\\%
}
\pgfplotscreateplotcyclelist{Mark-PRGn-8}{%
 PRGn-8-1!70!black, every mark/.append style={solid,fill=PRGn-8-1},mark=*\\%
 PRGn-8-2!70!black, every mark/.append style={solid,fill=PRGn-8-2},mark=square*\\%
 PRGn-8-3!70!black, every mark/.append style={solid,fill=PRGn-8-3},mark=triangle*\\%
 PRGn-8-4!70!black, every mark/.append style={solid,fill=PRGn-8-4},mark=diamond*\\%
 PRGn-8-5!70!black, every mark/.append style={solid,fill=PRGn-8-5},mark=pentagon*\\%
 PRGn-8-6!70!black, every mark/.append style={solid,fill=PRGn-8-6},mark=x\\%
 PRGn-8-7!70!black, every mark/.append style={solid,fill=PRGn-8-7},mark=+\\%
 PRGn-8-8!70!black, every mark/.append style={solid,fill=PRGn-8-8},mark=star\\%
}
\pgfplotscreateplotcyclelist{Mark-PRGn-9}{%
 PRGn-9-1!70!black, every mark/.append style={solid,fill=PRGn-9-1},mark=*\\%
 PRGn-9-2!70!black, every mark/.append style={solid,fill=PRGn-9-2},mark=square*\\%
 PRGn-9-3!70!black, every mark/.append style={solid,fill=PRGn-9-3},mark=triangle*\\%
 PRGn-9-4!70!black, every mark/.append style={solid,fill=PRGn-9-4},mark=diamond*\\%
 PRGn-9-5!70!black, every mark/.append style={solid,fill=PRGn-9-5},mark=pentagon*\\%
 PRGn-9-6!70!black, every mark/.append style={solid,fill=PRGn-9-6},mark=x\\%
 PRGn-9-7!70!black, every mark/.append style={solid,fill=PRGn-9-7},mark=+\\%
 PRGn-9-8!70!black, every mark/.append style={solid,fill=PRGn-9-8},mark=star\\%
 PRGn-9-9!70!black, every mark/.append style={solid,fill=PRGn-9-9},mark=oplus*\\%
}
\pgfplotscreateplotcyclelist{Mark-PRGn-10}{%
 PRGn-10-1!70!black, every mark/.append style={solid,fill=PRGn-10-1},mark=*\\%
 PRGn-10-2!70!black, every mark/.append style={solid,fill=PRGn-10-2},mark=square*\\%
 PRGn-10-3!70!black, every mark/.append style={solid,fill=PRGn-10-3},mark=triangle*\\%
 PRGn-10-4!70!black, every mark/.append style={solid,fill=PRGn-10-4},mark=diamond*\\%
 PRGn-10-5!70!black, every mark/.append style={solid,fill=PRGn-10-5},mark=pentagon*\\%
 PRGn-10-6!70!black, every mark/.append style={solid,fill=PRGn-10-6},mark=x\\%
 PRGn-10-7!70!black, every mark/.append style={solid,fill=PRGn-10-7},mark=+\\%
 PRGn-10-8!70!black, every mark/.append style={solid,fill=PRGn-10-8},mark=star\\%
 PRGn-10-9!70!black, every mark/.append style={solid,fill=PRGn-10-9},mark=oplus*\\%
 PRGn-10-10!70!black, every mark/.append style={solid,fill=PRGn-10-10},mark=otimes*\\%
}
\pgfplotscreateplotcyclelist{Mark-PRGn-11}{%
 PRGn-11-1!70!black, every mark/.append style={solid,fill=PRGn-11-1},mark=*\\%
 PRGn-11-2!70!black, every mark/.append style={solid,fill=PRGn-11-2},mark=square*\\%
 PRGn-11-3!70!black, every mark/.append style={solid,fill=PRGn-11-3},mark=triangle*\\%
 PRGn-11-4!70!black, every mark/.append style={solid,fill=PRGn-11-4},mark=diamond*\\%
 PRGn-11-5!70!black, every mark/.append style={solid,fill=PRGn-11-5},mark=pentagon*\\%
 PRGn-11-6!70!black, every mark/.append style={solid,fill=PRGn-11-6},mark=x\\%
 PRGn-11-7!70!black, every mark/.append style={solid,fill=PRGn-11-7},mark=+\\%
 PRGn-11-8!70!black, every mark/.append style={solid,fill=PRGn-11-8},mark=star\\%
 PRGn-11-9!70!black, every mark/.append style={solid,fill=PRGn-11-9},mark=oplus*\\%
 PRGn-11-10!70!black, every mark/.append style={solid,fill=PRGn-11-10},mark=otimes*\\%
}
\pgfplotscreateplotcyclelist{Mark-PiYG-3}{%
 PiYG-3-1!70!black, every mark/.append style={solid,fill=PiYG-3-1},mark=*\\%
 PiYG-3-2!70!black, every mark/.append style={solid,fill=PiYG-3-2},mark=square*\\%
 PiYG-3-3!70!black, every mark/.append style={solid,fill=PiYG-3-3},mark=triangle*\\%
}
\pgfplotscreateplotcyclelist{Mark-PiYG-4}{%
 PiYG-4-1!70!black, every mark/.append style={solid,fill=PiYG-4-1},mark=*\\%
 PiYG-4-2!70!black, every mark/.append style={solid,fill=PiYG-4-2},mark=square*\\%
 PiYG-4-3!70!black, every mark/.append style={solid,fill=PiYG-4-3},mark=triangle*\\%
 PiYG-4-4!70!black, every mark/.append style={solid,fill=PiYG-4-4},mark=diamond*\\%
}
\pgfplotscreateplotcyclelist{Mark-PiYG-5}{%
 PiYG-5-1!70!black, every mark/.append style={solid,fill=PiYG-5-1},mark=*\\%
 PiYG-5-2!70!black, every mark/.append style={solid,fill=PiYG-5-2},mark=square*\\%
 PiYG-5-3!70!black, every mark/.append style={solid,fill=PiYG-5-3},mark=triangle*\\%
 PiYG-5-4!70!black, every mark/.append style={solid,fill=PiYG-5-4},mark=diamond*\\%
 PiYG-5-5!70!black, every mark/.append style={solid,fill=PiYG-5-5},mark=pentagon*\\%
}
\pgfplotscreateplotcyclelist{Mark-PiYG-6}{%
 PiYG-6-1!70!black, every mark/.append style={solid,fill=PiYG-6-1},mark=*\\%
 PiYG-6-2!70!black, every mark/.append style={solid,fill=PiYG-6-2},mark=square*\\%
 PiYG-6-3!70!black, every mark/.append style={solid,fill=PiYG-6-3},mark=triangle*\\%
 PiYG-6-4!70!black, every mark/.append style={solid,fill=PiYG-6-4},mark=diamond*\\%
 PiYG-6-5!70!black, every mark/.append style={solid,fill=PiYG-6-5},mark=pentagon*\\%
 PiYG-6-6!70!black, every mark/.append style={solid,fill=PiYG-6-6},mark=x\\%
}
\pgfplotscreateplotcyclelist{Mark-PiYG-7}{%
 PiYG-7-1!70!black, every mark/.append style={solid,fill=PiYG-7-1},mark=*\\%
 PiYG-7-2!70!black, every mark/.append style={solid,fill=PiYG-7-2},mark=square*\\%
 PiYG-7-3!70!black, every mark/.append style={solid,fill=PiYG-7-3},mark=triangle*\\%
 PiYG-7-4!70!black, every mark/.append style={solid,fill=PiYG-7-4},mark=diamond*\\%
 PiYG-7-5!70!black, every mark/.append style={solid,fill=PiYG-7-5},mark=pentagon*\\%
 PiYG-7-6!70!black, every mark/.append style={solid,fill=PiYG-7-6},mark=x\\%
 PiYG-7-7!70!black, every mark/.append style={solid,fill=PiYG-7-7},mark=+\\%
}
\pgfplotscreateplotcyclelist{Mark-PiYG-8}{%
 PiYG-8-1!70!black, every mark/.append style={solid,fill=PiYG-8-1},mark=*\\%
 PiYG-8-2!70!black, every mark/.append style={solid,fill=PiYG-8-2},mark=square*\\%
 PiYG-8-3!70!black, every mark/.append style={solid,fill=PiYG-8-3},mark=triangle*\\%
 PiYG-8-4!70!black, every mark/.append style={solid,fill=PiYG-8-4},mark=diamond*\\%
 PiYG-8-5!70!black, every mark/.append style={solid,fill=PiYG-8-5},mark=pentagon*\\%
 PiYG-8-6!70!black, every mark/.append style={solid,fill=PiYG-8-6},mark=x\\%
 PiYG-8-7!70!black, every mark/.append style={solid,fill=PiYG-8-7},mark=+\\%
 PiYG-8-8!70!black, every mark/.append style={solid,fill=PiYG-8-8},mark=star\\%
}
\pgfplotscreateplotcyclelist{Mark-PiYG-9}{%
 PiYG-9-1!70!black, every mark/.append style={solid,fill=PiYG-9-1},mark=*\\%
 PiYG-9-2!70!black, every mark/.append style={solid,fill=PiYG-9-2},mark=square*\\%
 PiYG-9-3!70!black, every mark/.append style={solid,fill=PiYG-9-3},mark=triangle*\\%
 PiYG-9-4!70!black, every mark/.append style={solid,fill=PiYG-9-4},mark=diamond*\\%
 PiYG-9-5!70!black, every mark/.append style={solid,fill=PiYG-9-5},mark=pentagon*\\%
 PiYG-9-6!70!black, every mark/.append style={solid,fill=PiYG-9-6},mark=x\\%
 PiYG-9-7!70!black, every mark/.append style={solid,fill=PiYG-9-7},mark=+\\%
 PiYG-9-8!70!black, every mark/.append style={solid,fill=PiYG-9-8},mark=star\\%
 PiYG-9-9!70!black, every mark/.append style={solid,fill=PiYG-9-9},mark=oplus*\\%
}
\pgfplotscreateplotcyclelist{Mark-PiYG-10}{%
 PiYG-10-1!70!black, every mark/.append style={solid,fill=PiYG-10-1},mark=*\\%
 PiYG-10-2!70!black, every mark/.append style={solid,fill=PiYG-10-2},mark=square*\\%
 PiYG-10-3!70!black, every mark/.append style={solid,fill=PiYG-10-3},mark=triangle*\\%
 PiYG-10-4!70!black, every mark/.append style={solid,fill=PiYG-10-4},mark=diamond*\\%
 PiYG-10-5!70!black, every mark/.append style={solid,fill=PiYG-10-5},mark=pentagon*\\%
 PiYG-10-6!70!black, every mark/.append style={solid,fill=PiYG-10-6},mark=x\\%
 PiYG-10-7!70!black, every mark/.append style={solid,fill=PiYG-10-7},mark=+\\%
 PiYG-10-8!70!black, every mark/.append style={solid,fill=PiYG-10-8},mark=star\\%
 PiYG-10-9!70!black, every mark/.append style={solid,fill=PiYG-10-9},mark=oplus*\\%
 PiYG-10-10!70!black, every mark/.append style={solid,fill=PiYG-10-10},mark=otimes*\\%
}
\pgfplotscreateplotcyclelist{Mark-PiYG-11}{%
 PiYG-11-1!70!black, every mark/.append style={solid,fill=PiYG-11-1},mark=*\\%
 PiYG-11-2!70!black, every mark/.append style={solid,fill=PiYG-11-2},mark=square*\\%
 PiYG-11-3!70!black, every mark/.append style={solid,fill=PiYG-11-3},mark=triangle*\\%
 PiYG-11-4!70!black, every mark/.append style={solid,fill=PiYG-11-4},mark=diamond*\\%
 PiYG-11-5!70!black, every mark/.append style={solid,fill=PiYG-11-5},mark=pentagon*\\%
 PiYG-11-6!70!black, every mark/.append style={solid,fill=PiYG-11-6},mark=x\\%
 PiYG-11-7!70!black, every mark/.append style={solid,fill=PiYG-11-7},mark=+\\%
 PiYG-11-8!70!black, every mark/.append style={solid,fill=PiYG-11-8},mark=star\\%
 PiYG-11-9!70!black, every mark/.append style={solid,fill=PiYG-11-9},mark=oplus*\\%
 PiYG-11-10!70!black, every mark/.append style={solid,fill=PiYG-11-10},mark=otimes*\\%
}
\pgfplotscreateplotcyclelist{Mark-RdBu-3}{%
 RdBu-3-1!70!black, every mark/.append style={solid,fill=RdBu-3-1},mark=*\\%
 RdBu-3-2!70!black, every mark/.append style={solid,fill=RdBu-3-2},mark=square*\\%
 RdBu-3-3!70!black, every mark/.append style={solid,fill=RdBu-3-3},mark=triangle*\\%
}
\pgfplotscreateplotcyclelist{Mark-RdBu-4}{%
 RdBu-4-1!70!black, every mark/.append style={solid,fill=RdBu-4-1},mark=*\\%
 RdBu-4-2!70!black, every mark/.append style={solid,fill=RdBu-4-2},mark=square*\\%
 RdBu-4-3!70!black, every mark/.append style={solid,fill=RdBu-4-3},mark=triangle*\\%
 RdBu-4-4!70!black, every mark/.append style={solid,fill=RdBu-4-4},mark=diamond*\\%
}
\pgfplotscreateplotcyclelist{Mark-RdBu-5}{%
 RdBu-5-1!70!black, every mark/.append style={solid,fill=RdBu-5-1},mark=*\\%
 RdBu-5-2!70!black, every mark/.append style={solid,fill=RdBu-5-2},mark=square*\\%
 RdBu-5-3!70!black, every mark/.append style={solid,fill=RdBu-5-3},mark=triangle*\\%
 RdBu-5-4!70!black, every mark/.append style={solid,fill=RdBu-5-4},mark=diamond*\\%
 RdBu-5-5!70!black, every mark/.append style={solid,fill=RdBu-5-5},mark=pentagon*\\%
}
\pgfplotscreateplotcyclelist{Mark-RdBu-6}{%
 RdBu-6-1!70!black, every mark/.append style={solid,fill=RdBu-6-1},mark=*\\%
 RdBu-6-2!70!black, every mark/.append style={solid,fill=RdBu-6-2},mark=square*\\%
 RdBu-6-3!70!black, every mark/.append style={solid,fill=RdBu-6-3},mark=triangle*\\%
 RdBu-6-4!70!black, every mark/.append style={solid,fill=RdBu-6-4},mark=diamond*\\%
 RdBu-6-5!70!black, every mark/.append style={solid,fill=RdBu-6-5},mark=pentagon*\\%
 RdBu-6-6!70!black, every mark/.append style={solid,fill=RdBu-6-6},mark=x\\%
}
\pgfplotscreateplotcyclelist{Mark-RdBu-7}{%
 RdBu-7-1!70!black, every mark/.append style={solid,fill=RdBu-7-1},mark=*\\%
 RdBu-7-2!70!black, every mark/.append style={solid,fill=RdBu-7-2},mark=square*\\%
 RdBu-7-3!70!black, every mark/.append style={solid,fill=RdBu-7-3},mark=triangle*\\%
 RdBu-7-4!70!black, every mark/.append style={solid,fill=RdBu-7-4},mark=diamond*\\%
 RdBu-7-5!70!black, every mark/.append style={solid,fill=RdBu-7-5},mark=pentagon*\\%
 RdBu-7-6!70!black, every mark/.append style={solid,fill=RdBu-7-6},mark=x\\%
 RdBu-7-7!70!black, every mark/.append style={solid,fill=RdBu-7-7},mark=+\\%
}
\pgfplotscreateplotcyclelist{Mark-RdBu-8}{%
 RdBu-8-1!70!black, every mark/.append style={solid,fill=RdBu-8-1},mark=*\\%
 RdBu-8-2!70!black, every mark/.append style={solid,fill=RdBu-8-2},mark=square*\\%
 RdBu-8-3!70!black, every mark/.append style={solid,fill=RdBu-8-3},mark=triangle*\\%
 RdBu-8-4!70!black, every mark/.append style={solid,fill=RdBu-8-4},mark=diamond*\\%
 RdBu-8-5!70!black, every mark/.append style={solid,fill=RdBu-8-5},mark=pentagon*\\%
 RdBu-8-6!70!black, every mark/.append style={solid,fill=RdBu-8-6},mark=x\\%
 RdBu-8-7!70!black, every mark/.append style={solid,fill=RdBu-8-7},mark=+\\%
 RdBu-8-8!70!black, every mark/.append style={solid,fill=RdBu-8-8},mark=star\\%
}
\pgfplotscreateplotcyclelist{Mark-RdBu-9}{%
 RdBu-9-1!70!black, every mark/.append style={solid,fill=RdBu-9-1},mark=*\\%
 RdBu-9-2!70!black, every mark/.append style={solid,fill=RdBu-9-2},mark=square*\\%
 RdBu-9-3!70!black, every mark/.append style={solid,fill=RdBu-9-3},mark=triangle*\\%
 RdBu-9-4!70!black, every mark/.append style={solid,fill=RdBu-9-4},mark=diamond*\\%
 RdBu-9-5!70!black, every mark/.append style={solid,fill=RdBu-9-5},mark=pentagon*\\%
 RdBu-9-6!70!black, every mark/.append style={solid,fill=RdBu-9-6},mark=x\\%
 RdBu-9-7!70!black, every mark/.append style={solid,fill=RdBu-9-7},mark=+\\%
 RdBu-9-8!70!black, every mark/.append style={solid,fill=RdBu-9-8},mark=star\\%
 RdBu-9-9!70!black, every mark/.append style={solid,fill=RdBu-9-9},mark=oplus*\\%
}
\pgfplotscreateplotcyclelist{Mark-RdBu-10}{%
 RdBu-10-1!70!black, every mark/.append style={solid,fill=RdBu-10-1},mark=*\\%
 RdBu-10-2!70!black, every mark/.append style={solid,fill=RdBu-10-2},mark=square*\\%
 RdBu-10-3!70!black, every mark/.append style={solid,fill=RdBu-10-3},mark=triangle*\\%
 RdBu-10-4!70!black, every mark/.append style={solid,fill=RdBu-10-4},mark=diamond*\\%
 RdBu-10-5!70!black, every mark/.append style={solid,fill=RdBu-10-5},mark=pentagon*\\%
 RdBu-10-6!70!black, every mark/.append style={solid,fill=RdBu-10-6},mark=x\\%
 RdBu-10-7!70!black, every mark/.append style={solid,fill=RdBu-10-7},mark=+\\%
 RdBu-10-8!70!black, every mark/.append style={solid,fill=RdBu-10-8},mark=star\\%
 RdBu-10-9!70!black, every mark/.append style={solid,fill=RdBu-10-9},mark=oplus*\\%
 RdBu-10-10!70!black, every mark/.append style={solid,fill=RdBu-10-10},mark=otimes*\\%
}
\pgfplotscreateplotcyclelist{Mark-RdBu-11}{%
 RdBu-11-1!70!black, every mark/.append style={solid,fill=RdBu-11-1},mark=*\\%
 RdBu-11-2!70!black, every mark/.append style={solid,fill=RdBu-11-2},mark=square*\\%
 RdBu-11-3!70!black, every mark/.append style={solid,fill=RdBu-11-3},mark=triangle*\\%
 RdBu-11-4!70!black, every mark/.append style={solid,fill=RdBu-11-4},mark=diamond*\\%
 RdBu-11-5!70!black, every mark/.append style={solid,fill=RdBu-11-5},mark=pentagon*\\%
 RdBu-11-6!70!black, every mark/.append style={solid,fill=RdBu-11-6},mark=x\\%
 RdBu-11-7!70!black, every mark/.append style={solid,fill=RdBu-11-7},mark=+\\%
 RdBu-11-8!70!black, every mark/.append style={solid,fill=RdBu-11-8},mark=star\\%
 RdBu-11-9!70!black, every mark/.append style={solid,fill=RdBu-11-9},mark=oplus*\\%
 RdBu-11-10!70!black, every mark/.append style={solid,fill=RdBu-11-10},mark=otimes*\\%
}
\pgfplotscreateplotcyclelist{Mark-RdGy-3}{%
 RdGy-3-1!70!black, every mark/.append style={solid,fill=RdGy-3-1},mark=*\\%
 RdGy-3-2!70!black, every mark/.append style={solid,fill=RdGy-3-2},mark=square*\\%
 RdGy-3-3!70!black, every mark/.append style={solid,fill=RdGy-3-3},mark=triangle*\\%
}
\pgfplotscreateplotcyclelist{Mark-RdGy-4}{%
 RdGy-4-1!70!black, every mark/.append style={solid,fill=RdGy-4-1},mark=*\\%
 RdGy-4-2!70!black, every mark/.append style={solid,fill=RdGy-4-2},mark=square*\\%
 RdGy-4-3!70!black, every mark/.append style={solid,fill=RdGy-4-3},mark=triangle*\\%
 RdGy-4-4!70!black, every mark/.append style={solid,fill=RdGy-4-4},mark=diamond*\\%
}
\pgfplotscreateplotcyclelist{Mark-RdGy-5}{%
 RdGy-5-1!70!black, every mark/.append style={solid,fill=RdGy-5-1},mark=*\\%
 RdGy-5-2!70!black, every mark/.append style={solid,fill=RdGy-5-2},mark=square*\\%
 RdGy-5-3!70!black, every mark/.append style={solid,fill=RdGy-5-3},mark=triangle*\\%
 RdGy-5-4!70!black, every mark/.append style={solid,fill=RdGy-5-4},mark=diamond*\\%
 RdGy-5-5!70!black, every mark/.append style={solid,fill=RdGy-5-5},mark=pentagon*\\%
}
\pgfplotscreateplotcyclelist{Mark-RdGy-6}{%
 RdGy-6-1!70!black, every mark/.append style={solid,fill=RdGy-6-1},mark=*\\%
 RdGy-6-2!70!black, every mark/.append style={solid,fill=RdGy-6-2},mark=square*\\%
 RdGy-6-3!70!black, every mark/.append style={solid,fill=RdGy-6-3},mark=triangle*\\%
 RdGy-6-4!70!black, every mark/.append style={solid,fill=RdGy-6-4},mark=diamond*\\%
 RdGy-6-5!70!black, every mark/.append style={solid,fill=RdGy-6-5},mark=pentagon*\\%
 RdGy-6-6!70!black, every mark/.append style={solid,fill=RdGy-6-6},mark=x\\%
}
\pgfplotscreateplotcyclelist{Mark-RdGy-7}{%
 RdGy-7-1!70!black, every mark/.append style={solid,fill=RdGy-7-1},mark=*\\%
 RdGy-7-2!70!black, every mark/.append style={solid,fill=RdGy-7-2},mark=square*\\%
 RdGy-7-3!70!black, every mark/.append style={solid,fill=RdGy-7-3},mark=triangle*\\%
 RdGy-7-4!70!black, every mark/.append style={solid,fill=RdGy-7-4},mark=diamond*\\%
 RdGy-7-5!70!black, every mark/.append style={solid,fill=RdGy-7-5},mark=pentagon*\\%
 RdGy-7-6!70!black, every mark/.append style={solid,fill=RdGy-7-6},mark=x\\%
 RdGy-7-7!70!black, every mark/.append style={solid,fill=RdGy-7-7},mark=+\\%
}
\pgfplotscreateplotcyclelist{Mark-RdGy-8}{%
 RdGy-8-1!70!black, every mark/.append style={solid,fill=RdGy-8-1},mark=*\\%
 RdGy-8-2!70!black, every mark/.append style={solid,fill=RdGy-8-2},mark=square*\\%
 RdGy-8-3!70!black, every mark/.append style={solid,fill=RdGy-8-3},mark=triangle*\\%
 RdGy-8-4!70!black, every mark/.append style={solid,fill=RdGy-8-4},mark=diamond*\\%
 RdGy-8-5!70!black, every mark/.append style={solid,fill=RdGy-8-5},mark=pentagon*\\%
 RdGy-8-6!70!black, every mark/.append style={solid,fill=RdGy-8-6},mark=x\\%
 RdGy-8-7!70!black, every mark/.append style={solid,fill=RdGy-8-7},mark=+\\%
 RdGy-8-8!70!black, every mark/.append style={solid,fill=RdGy-8-8},mark=star\\%
}
\pgfplotscreateplotcyclelist{Mark-RdGy-9}{%
 RdGy-9-1!70!black, every mark/.append style={solid,fill=RdGy-9-1},mark=*\\%
 RdGy-9-2!70!black, every mark/.append style={solid,fill=RdGy-9-2},mark=square*\\%
 RdGy-9-3!70!black, every mark/.append style={solid,fill=RdGy-9-3},mark=triangle*\\%
 RdGy-9-4!70!black, every mark/.append style={solid,fill=RdGy-9-4},mark=diamond*\\%
 RdGy-9-5!70!black, every mark/.append style={solid,fill=RdGy-9-5},mark=pentagon*\\%
 RdGy-9-6!70!black, every mark/.append style={solid,fill=RdGy-9-6},mark=x\\%
 RdGy-9-7!70!black, every mark/.append style={solid,fill=RdGy-9-7},mark=+\\%
 RdGy-9-8!70!black, every mark/.append style={solid,fill=RdGy-9-8},mark=star\\%
 RdGy-9-9!70!black, every mark/.append style={solid,fill=RdGy-9-9},mark=oplus*\\%
}
\pgfplotscreateplotcyclelist{Mark-RdGy-10}{%
 RdGy-10-1!70!black, every mark/.append style={solid,fill=RdGy-10-1},mark=*\\%
 RdGy-10-2!70!black, every mark/.append style={solid,fill=RdGy-10-2},mark=square*\\%
 RdGy-10-3!70!black, every mark/.append style={solid,fill=RdGy-10-3},mark=triangle*\\%
 RdGy-10-4!70!black, every mark/.append style={solid,fill=RdGy-10-4},mark=diamond*\\%
 RdGy-10-5!70!black, every mark/.append style={solid,fill=RdGy-10-5},mark=pentagon*\\%
 RdGy-10-6!70!black, every mark/.append style={solid,fill=RdGy-10-6},mark=x\\%
 RdGy-10-7!70!black, every mark/.append style={solid,fill=RdGy-10-7},mark=+\\%
 RdGy-10-8!70!black, every mark/.append style={solid,fill=RdGy-10-8},mark=star\\%
 RdGy-10-9!70!black, every mark/.append style={solid,fill=RdGy-10-9},mark=oplus*\\%
 RdGy-10-10!70!black, every mark/.append style={solid,fill=RdGy-10-10},mark=otimes*\\%
}
\pgfplotscreateplotcyclelist{Mark-RdGy-11}{%
 RdGy-11-1!70!black, every mark/.append style={solid,fill=RdGy-11-1},mark=*\\%
 RdGy-11-2!70!black, every mark/.append style={solid,fill=RdGy-11-2},mark=square*\\%
 RdGy-11-3!70!black, every mark/.append style={solid,fill=RdGy-11-3},mark=triangle*\\%
 RdGy-11-4!70!black, every mark/.append style={solid,fill=RdGy-11-4},mark=diamond*\\%
 RdGy-11-5!70!black, every mark/.append style={solid,fill=RdGy-11-5},mark=pentagon*\\%
 RdGy-11-6!70!black, every mark/.append style={solid,fill=RdGy-11-6},mark=x\\%
 RdGy-11-7!70!black, every mark/.append style={solid,fill=RdGy-11-7},mark=+\\%
 RdGy-11-8!70!black, every mark/.append style={solid,fill=RdGy-11-8},mark=star\\%
 RdGy-11-9!70!black, every mark/.append style={solid,fill=RdGy-11-9},mark=oplus*\\%
 RdGy-11-10!70!black, every mark/.append style={solid,fill=RdGy-11-10},mark=otimes*\\%
}
\pgfplotscreateplotcyclelist{Mark-RdYlBu-3}{%
 RdYlBu-3-1!70!black, every mark/.append style={solid,fill=RdYlBu-3-1},mark=*\\%
 RdYlBu-3-2!70!black, every mark/.append style={solid,fill=RdYlBu-3-2},mark=square*\\%
 RdYlBu-3-3!70!black, every mark/.append style={solid,fill=RdYlBu-3-3},mark=triangle*\\%
}
\pgfplotscreateplotcyclelist{Mark-RdYlBu-4}{%
 RdYlBu-4-1!70!black, every mark/.append style={solid,fill=RdYlBu-4-1},mark=*\\%
 RdYlBu-4-2!70!black, every mark/.append style={solid,fill=RdYlBu-4-2},mark=square*\\%
 RdYlBu-4-3!70!black, every mark/.append style={solid,fill=RdYlBu-4-3},mark=triangle*\\%
 RdYlBu-4-4!70!black, every mark/.append style={solid,fill=RdYlBu-4-4},mark=diamond*\\%
}
\pgfplotscreateplotcyclelist{Mark-RdYlBu-5}{%
 RdYlBu-5-1!70!black, every mark/.append style={solid,fill=RdYlBu-5-1},mark=*\\%
 RdYlBu-5-2!70!black, every mark/.append style={solid,fill=RdYlBu-5-2},mark=square*\\%
 RdYlBu-5-3!70!black, every mark/.append style={solid,fill=RdYlBu-5-3},mark=triangle*\\%
 RdYlBu-5-4!70!black, every mark/.append style={solid,fill=RdYlBu-5-4},mark=diamond*\\%
 RdYlBu-5-5!70!black, every mark/.append style={solid,fill=RdYlBu-5-5},mark=pentagon*\\%
}
\pgfplotscreateplotcyclelist{Mark-RdYlBu-6}{%
 RdYlBu-6-1!70!black, every mark/.append style={solid,fill=RdYlBu-6-1},mark=*\\%
 RdYlBu-6-2!70!black, every mark/.append style={solid,fill=RdYlBu-6-2},mark=square*\\%
 RdYlBu-6-3!70!black, every mark/.append style={solid,fill=RdYlBu-6-3},mark=triangle*\\%
 RdYlBu-6-4!70!black, every mark/.append style={solid,fill=RdYlBu-6-4},mark=diamond*\\%
 RdYlBu-6-5!70!black, every mark/.append style={solid,fill=RdYlBu-6-5},mark=pentagon*\\%
 RdYlBu-6-6!70!black, every mark/.append style={solid,fill=RdYlBu-6-6},mark=x\\%
}
\pgfplotscreateplotcyclelist{Mark-RdYlBu-7}{%
 RdYlBu-7-1!70!black, every mark/.append style={solid,fill=RdYlBu-7-1},mark=*\\%
 RdYlBu-7-2!70!black, every mark/.append style={solid,fill=RdYlBu-7-2},mark=square*\\%
 RdYlBu-7-3!70!black, every mark/.append style={solid,fill=RdYlBu-7-3},mark=triangle*\\%
 RdYlBu-7-4!70!black, every mark/.append style={solid,fill=RdYlBu-7-4},mark=diamond*\\%
 RdYlBu-7-5!70!black, every mark/.append style={solid,fill=RdYlBu-7-5},mark=pentagon*\\%
 RdYlBu-7-6!70!black, every mark/.append style={solid,fill=RdYlBu-7-6},mark=x\\%
 RdYlBu-7-7!70!black, every mark/.append style={solid,fill=RdYlBu-7-7},mark=+\\%
}
\pgfplotscreateplotcyclelist{Mark-RdYlBu-8}{%
 RdYlBu-8-1!70!black, every mark/.append style={solid,fill=RdYlBu-8-1},mark=*\\%
 RdYlBu-8-2!70!black, every mark/.append style={solid,fill=RdYlBu-8-2},mark=square*\\%
 RdYlBu-8-3!70!black, every mark/.append style={solid,fill=RdYlBu-8-3},mark=triangle*\\%
 RdYlBu-8-4!70!black, every mark/.append style={solid,fill=RdYlBu-8-4},mark=diamond*\\%
 RdYlBu-8-5!70!black, every mark/.append style={solid,fill=RdYlBu-8-5},mark=pentagon*\\%
 RdYlBu-8-6!70!black, every mark/.append style={solid,fill=RdYlBu-8-6},mark=x\\%
 RdYlBu-8-7!70!black, every mark/.append style={solid,fill=RdYlBu-8-7},mark=+\\%
 RdYlBu-8-8!70!black, every mark/.append style={solid,fill=RdYlBu-8-8},mark=star\\%
}
\pgfplotscreateplotcyclelist{Mark-RdYlBu-9}{%
 RdYlBu-9-1!70!black, every mark/.append style={solid,fill=RdYlBu-9-1},mark=*\\%
 RdYlBu-9-2!70!black, every mark/.append style={solid,fill=RdYlBu-9-2},mark=square*\\%
 RdYlBu-9-3!70!black, every mark/.append style={solid,fill=RdYlBu-9-3},mark=triangle*\\%
 RdYlBu-9-4!70!black, every mark/.append style={solid,fill=RdYlBu-9-4},mark=diamond*\\%
 RdYlBu-9-5!70!black, every mark/.append style={solid,fill=RdYlBu-9-5},mark=pentagon*\\%
 RdYlBu-9-6!70!black, every mark/.append style={solid,fill=RdYlBu-9-6},mark=x\\%
 RdYlBu-9-7!70!black, every mark/.append style={solid,fill=RdYlBu-9-7},mark=+\\%
 RdYlBu-9-8!70!black, every mark/.append style={solid,fill=RdYlBu-9-8},mark=star\\%
 RdYlBu-9-9!70!black, every mark/.append style={solid,fill=RdYlBu-9-9},mark=oplus*\\%
}
\pgfplotscreateplotcyclelist{Mark-RdYlBu-10}{%
 RdYlBu-10-1!70!black, every mark/.append style={solid,fill=RdYlBu-10-1},mark=*\\%
 RdYlBu-10-2!70!black, every mark/.append style={solid,fill=RdYlBu-10-2},mark=square*\\%
 RdYlBu-10-3!70!black, every mark/.append style={solid,fill=RdYlBu-10-3},mark=triangle*\\%
 RdYlBu-10-4!70!black, every mark/.append style={solid,fill=RdYlBu-10-4},mark=diamond*\\%
 RdYlBu-10-5!70!black, every mark/.append style={solid,fill=RdYlBu-10-5},mark=pentagon*\\%
 RdYlBu-10-6!70!black, every mark/.append style={solid,fill=RdYlBu-10-6},mark=x\\%
 RdYlBu-10-7!70!black, every mark/.append style={solid,fill=RdYlBu-10-7},mark=+\\%
 RdYlBu-10-8!70!black, every mark/.append style={solid,fill=RdYlBu-10-8},mark=star\\%
 RdYlBu-10-9!70!black, every mark/.append style={solid,fill=RdYlBu-10-9},mark=oplus*\\%
 RdYlBu-10-10!70!black, every mark/.append style={solid,fill=RdYlBu-10-10},mark=otimes*\\%
}
\pgfplotscreateplotcyclelist{Mark-RdYlBu-11}{%
 RdYlBu-11-1!70!black, every mark/.append style={solid,fill=RdYlBu-11-1},mark=*\\%
 RdYlBu-11-2!70!black, every mark/.append style={solid,fill=RdYlBu-11-2},mark=square*\\%
 RdYlBu-11-3!70!black, every mark/.append style={solid,fill=RdYlBu-11-3},mark=triangle*\\%
 RdYlBu-11-4!70!black, every mark/.append style={solid,fill=RdYlBu-11-4},mark=diamond*\\%
 RdYlBu-11-5!70!black, every mark/.append style={solid,fill=RdYlBu-11-5},mark=pentagon*\\%
 RdYlBu-11-6!70!black, every mark/.append style={solid,fill=RdYlBu-11-6},mark=x\\%
 RdYlBu-11-7!70!black, every mark/.append style={solid,fill=RdYlBu-11-7},mark=+\\%
 RdYlBu-11-8!70!black, every mark/.append style={solid,fill=RdYlBu-11-8},mark=star\\%
 RdYlBu-11-9!70!black, every mark/.append style={solid,fill=RdYlBu-11-9},mark=oplus*\\%
 RdYlBu-11-10!70!black, every mark/.append style={solid,fill=RdYlBu-11-10},mark=otimes*\\%
}
\pgfplotscreateplotcyclelist{Mark-Spectral-3}{%
 Spectral-3-1!70!black, every mark/.append style={solid,fill=Spectral-3-1},mark=*\\%
 Spectral-3-2!70!black, every mark/.append style={solid,fill=Spectral-3-2},mark=square*\\%
 Spectral-3-3!70!black, every mark/.append style={solid,fill=Spectral-3-3},mark=triangle*\\%
}
\pgfplotscreateplotcyclelist{Mark-Spectral-4}{%
 Spectral-4-1!70!black, every mark/.append style={solid,fill=Spectral-4-1},mark=*\\%
 Spectral-4-2!70!black, every mark/.append style={solid,fill=Spectral-4-2},mark=square*\\%
 Spectral-4-3!70!black, every mark/.append style={solid,fill=Spectral-4-3},mark=triangle*\\%
 Spectral-4-4!70!black, every mark/.append style={solid,fill=Spectral-4-4},mark=diamond*\\%
}
\pgfplotscreateplotcyclelist{Mark-Spectral-5}{%
 Spectral-5-1!70!black, every mark/.append style={solid,fill=Spectral-5-1},mark=*\\%
 Spectral-5-2!70!black, every mark/.append style={solid,fill=Spectral-5-2},mark=square*\\%
 Spectral-5-3!70!black, every mark/.append style={solid,fill=Spectral-5-3},mark=triangle*\\%
 Spectral-5-4!70!black, every mark/.append style={solid,fill=Spectral-5-4},mark=diamond*\\%
 Spectral-5-5!70!black, every mark/.append style={solid,fill=Spectral-5-5},mark=pentagon*\\%
}
\pgfplotscreateplotcyclelist{Mark-Spectral-6}{%
 Spectral-6-1!70!black, every mark/.append style={solid,fill=Spectral-6-1},mark=*\\%
 Spectral-6-2!70!black, every mark/.append style={solid,fill=Spectral-6-2},mark=square*\\%
 Spectral-6-3!70!black, every mark/.append style={solid,fill=Spectral-6-3},mark=triangle*\\%
 Spectral-6-4!70!black, every mark/.append style={solid,fill=Spectral-6-4},mark=diamond*\\%
 Spectral-6-5!70!black, every mark/.append style={solid,fill=Spectral-6-5},mark=pentagon*\\%
 Spectral-6-6!70!black, every mark/.append style={solid,fill=Spectral-6-6},mark=x\\%
}
\pgfplotscreateplotcyclelist{Mark-Spectral-7}{%
 Spectral-7-1!70!black, every mark/.append style={solid,fill=Spectral-7-1},mark=*\\%
 Spectral-7-2!70!black, every mark/.append style={solid,fill=Spectral-7-2},mark=square*\\%
 Spectral-7-3!70!black, every mark/.append style={solid,fill=Spectral-7-3},mark=triangle*\\%
 Spectral-7-4!70!black, every mark/.append style={solid,fill=Spectral-7-4},mark=diamond*\\%
 Spectral-7-5!70!black, every mark/.append style={solid,fill=Spectral-7-5},mark=pentagon*\\%
 Spectral-7-6!70!black, every mark/.append style={solid,fill=Spectral-7-6},mark=x\\%
 Spectral-7-7!70!black, every mark/.append style={solid,fill=Spectral-7-7},mark=+\\%
}
\pgfplotscreateplotcyclelist{Mark-Spectral-8}{%
 Spectral-8-1!70!black, every mark/.append style={solid,fill=Spectral-8-1},mark=*\\%
 Spectral-8-2!70!black, every mark/.append style={solid,fill=Spectral-8-2},mark=square*\\%
 Spectral-8-3!70!black, every mark/.append style={solid,fill=Spectral-8-3},mark=triangle*\\%
 Spectral-8-4!70!black, every mark/.append style={solid,fill=Spectral-8-4},mark=diamond*\\%
 Spectral-8-5!70!black, every mark/.append style={solid,fill=Spectral-8-5},mark=pentagon*\\%
 Spectral-8-6!70!black, every mark/.append style={solid,fill=Spectral-8-6},mark=x\\%
 Spectral-8-7!70!black, every mark/.append style={solid,fill=Spectral-8-7},mark=+\\%
 Spectral-8-8!70!black, every mark/.append style={solid,fill=Spectral-8-8},mark=star\\%
}
\pgfplotscreateplotcyclelist{Mark-Spectral-9}{%
 Spectral-9-1!70!black, every mark/.append style={solid,fill=Spectral-9-1},mark=*\\%
 Spectral-9-2!70!black, every mark/.append style={solid,fill=Spectral-9-2},mark=square*\\%
 Spectral-9-3!70!black, every mark/.append style={solid,fill=Spectral-9-3},mark=triangle*\\%
 Spectral-9-4!70!black, every mark/.append style={solid,fill=Spectral-9-4},mark=diamond*\\%
 Spectral-9-5!70!black, every mark/.append style={solid,fill=Spectral-9-5},mark=pentagon*\\%
 Spectral-9-6!70!black, every mark/.append style={solid,fill=Spectral-9-6},mark=x\\%
 Spectral-9-7!70!black, every mark/.append style={solid,fill=Spectral-9-7},mark=+\\%
 Spectral-9-8!70!black, every mark/.append style={solid,fill=Spectral-9-8},mark=star\\%
 Spectral-9-9!70!black, every mark/.append style={solid,fill=Spectral-9-9},mark=oplus*\\%
}
\pgfplotscreateplotcyclelist{Mark-Spectral-10}{%
 Spectral-10-1!70!black, every mark/.append style={solid,fill=Spectral-10-1},mark=*\\%
 Spectral-10-2!70!black, every mark/.append style={solid,fill=Spectral-10-2},mark=square*\\%
 Spectral-10-3!70!black, every mark/.append style={solid,fill=Spectral-10-3},mark=triangle*\\%
 Spectral-10-4!70!black, every mark/.append style={solid,fill=Spectral-10-4},mark=diamond*\\%
 Spectral-10-5!70!black, every mark/.append style={solid,fill=Spectral-10-5},mark=pentagon*\\%
 Spectral-10-6!70!black, every mark/.append style={solid,fill=Spectral-10-6},mark=x\\%
 Spectral-10-7!70!black, every mark/.append style={solid,fill=Spectral-10-7},mark=+\\%
 Spectral-10-8!70!black, every mark/.append style={solid,fill=Spectral-10-8},mark=star\\%
 Spectral-10-9!70!black, every mark/.append style={solid,fill=Spectral-10-9},mark=oplus*\\%
 Spectral-10-10!70!black, every mark/.append style={solid,fill=Spectral-10-10},mark=otimes*\\%
}
\pgfplotscreateplotcyclelist{Mark-Spectral-11}{%
 Spectral-11-1!70!black, every mark/.append style={solid,fill=Spectral-11-1},mark=*\\%
 Spectral-11-2!70!black, every mark/.append style={solid,fill=Spectral-11-2},mark=square*\\%
 Spectral-11-3!70!black, every mark/.append style={solid,fill=Spectral-11-3},mark=triangle*\\%
 Spectral-11-4!70!black, every mark/.append style={solid,fill=Spectral-11-4},mark=diamond*\\%
 Spectral-11-5!70!black, every mark/.append style={solid,fill=Spectral-11-5},mark=pentagon*\\%
 Spectral-11-6!70!black, every mark/.append style={solid,fill=Spectral-11-6},mark=x\\%
 Spectral-11-7!70!black, every mark/.append style={solid,fill=Spectral-11-7},mark=+\\%
 Spectral-11-8!70!black, every mark/.append style={solid,fill=Spectral-11-8},mark=star\\%
 Spectral-11-9!70!black, every mark/.append style={solid,fill=Spectral-11-9},mark=oplus*\\%
 Spectral-11-10!70!black, every mark/.append style={solid,fill=Spectral-11-10},mark=otimes*\\%
}
\pgfplotscreateplotcyclelist{Mark-RdYlGn-3}{%
 RdYlGn-3-1!70!black, every mark/.append style={solid,fill=RdYlGn-3-1},mark=*\\%
 RdYlGn-3-2!70!black, every mark/.append style={solid,fill=RdYlGn-3-2},mark=square*\\%
 RdYlGn-3-3!70!black, every mark/.append style={solid,fill=RdYlGn-3-3},mark=triangle*\\%
}
\pgfplotscreateplotcyclelist{Mark-RdYlGn-4}{%
 RdYlGn-4-1!70!black, every mark/.append style={solid,fill=RdYlGn-4-1},mark=*\\%
 RdYlGn-4-2!70!black, every mark/.append style={solid,fill=RdYlGn-4-2},mark=square*\\%
 RdYlGn-4-3!70!black, every mark/.append style={solid,fill=RdYlGn-4-3},mark=triangle*\\%
 RdYlGn-4-4!70!black, every mark/.append style={solid,fill=RdYlGn-4-4},mark=diamond*\\%
}
\pgfplotscreateplotcyclelist{Mark-RdYlGn-5}{%
 RdYlGn-5-1!70!black, every mark/.append style={solid,fill=RdYlGn-5-1},mark=*\\%
 RdYlGn-5-2!70!black, every mark/.append style={solid,fill=RdYlGn-5-2},mark=square*\\%
 RdYlGn-5-3!70!black, every mark/.append style={solid,fill=RdYlGn-5-3},mark=triangle*\\%
 RdYlGn-5-4!70!black, every mark/.append style={solid,fill=RdYlGn-5-4},mark=diamond*\\%
 RdYlGn-5-5!70!black, every mark/.append style={solid,fill=RdYlGn-5-5},mark=pentagon*\\%
}
\pgfplotscreateplotcyclelist{Mark-RdYlGn-6}{%
 RdYlGn-6-1!70!black, every mark/.append style={solid,fill=RdYlGn-6-1},mark=*\\%
 RdYlGn-6-2!70!black, every mark/.append style={solid,fill=RdYlGn-6-2},mark=square*\\%
 RdYlGn-6-3!70!black, every mark/.append style={solid,fill=RdYlGn-6-3},mark=triangle*\\%
 RdYlGn-6-4!70!black, every mark/.append style={solid,fill=RdYlGn-6-4},mark=diamond*\\%
 RdYlGn-6-5!70!black, every mark/.append style={solid,fill=RdYlGn-6-5},mark=pentagon*\\%
 RdYlGn-6-6!70!black, every mark/.append style={solid,fill=RdYlGn-6-6},mark=x\\%
}
\pgfplotscreateplotcyclelist{Mark-RdYlGn-7}{%
 RdYlGn-7-1!70!black, every mark/.append style={solid,fill=RdYlGn-7-1},mark=*\\%
 RdYlGn-7-2!70!black, every mark/.append style={solid,fill=RdYlGn-7-2},mark=square*\\%
 RdYlGn-7-3!70!black, every mark/.append style={solid,fill=RdYlGn-7-3},mark=triangle*\\%
 RdYlGn-7-4!70!black, every mark/.append style={solid,fill=RdYlGn-7-4},mark=diamond*\\%
 RdYlGn-7-5!70!black, every mark/.append style={solid,fill=RdYlGn-7-5},mark=pentagon*\\%
 RdYlGn-7-6!70!black, every mark/.append style={solid,fill=RdYlGn-7-6},mark=x\\%
 RdYlGn-7-7!70!black, every mark/.append style={solid,fill=RdYlGn-7-7},mark=+\\%
}
\pgfplotscreateplotcyclelist{Mark-RdYlGn-8}{%
 RdYlGn-8-1!70!black, every mark/.append style={solid,fill=RdYlGn-8-1},mark=*\\%
 RdYlGn-8-2!70!black, every mark/.append style={solid,fill=RdYlGn-8-2},mark=square*\\%
 RdYlGn-8-3!70!black, every mark/.append style={solid,fill=RdYlGn-8-3},mark=triangle*\\%
 RdYlGn-8-4!70!black, every mark/.append style={solid,fill=RdYlGn-8-4},mark=diamond*\\%
 RdYlGn-8-5!70!black, every mark/.append style={solid,fill=RdYlGn-8-5},mark=pentagon*\\%
 RdYlGn-8-6!70!black, every mark/.append style={solid,fill=RdYlGn-8-6},mark=x\\%
 RdYlGn-8-7!70!black, every mark/.append style={solid,fill=RdYlGn-8-7},mark=+\\%
 RdYlGn-8-8!70!black, every mark/.append style={solid,fill=RdYlGn-8-8},mark=star\\%
}
\pgfplotscreateplotcyclelist{Mark-RdYlGn-9}{%
 RdYlGn-9-1!70!black, every mark/.append style={solid,fill=RdYlGn-9-1},mark=*\\%
 RdYlGn-9-2!70!black, every mark/.append style={solid,fill=RdYlGn-9-2},mark=square*\\%
 RdYlGn-9-3!70!black, every mark/.append style={solid,fill=RdYlGn-9-3},mark=triangle*\\%
 RdYlGn-9-4!70!black, every mark/.append style={solid,fill=RdYlGn-9-4},mark=diamond*\\%
 RdYlGn-9-5!70!black, every mark/.append style={solid,fill=RdYlGn-9-5},mark=pentagon*\\%
 RdYlGn-9-6!70!black, every mark/.append style={solid,fill=RdYlGn-9-6},mark=x\\%
 RdYlGn-9-7!70!black, every mark/.append style={solid,fill=RdYlGn-9-7},mark=+\\%
 RdYlGn-9-8!70!black, every mark/.append style={solid,fill=RdYlGn-9-8},mark=star\\%
 RdYlGn-9-9!70!black, every mark/.append style={solid,fill=RdYlGn-9-9},mark=oplus*\\%
}
\pgfplotscreateplotcyclelist{Mark-RdYlGn-10}{%
 RdYlGn-10-1!70!black, every mark/.append style={solid,fill=RdYlGn-10-1},mark=*\\%
 RdYlGn-10-2!70!black, every mark/.append style={solid,fill=RdYlGn-10-2},mark=square*\\%
 RdYlGn-10-3!70!black, every mark/.append style={solid,fill=RdYlGn-10-3},mark=triangle*\\%
 RdYlGn-10-4!70!black, every mark/.append style={solid,fill=RdYlGn-10-4},mark=diamond*\\%
 RdYlGn-10-5!70!black, every mark/.append style={solid,fill=RdYlGn-10-5},mark=pentagon*\\%
 RdYlGn-10-6!70!black, every mark/.append style={solid,fill=RdYlGn-10-6},mark=x\\%
 RdYlGn-10-7!70!black, every mark/.append style={solid,fill=RdYlGn-10-7},mark=+\\%
 RdYlGn-10-8!70!black, every mark/.append style={solid,fill=RdYlGn-10-8},mark=star\\%
 RdYlGn-10-9!70!black, every mark/.append style={solid,fill=RdYlGn-10-9},mark=oplus*\\%
 RdYlGn-10-10!70!black, every mark/.append style={solid,fill=RdYlGn-10-10},mark=otimes*\\%
}
\pgfplotscreateplotcyclelist{Mark-RdYlGn-11}{%
 RdYlGn-11-1!70!black, every mark/.append style={solid,fill=RdYlGn-11-1},mark=*\\%
 RdYlGn-11-2!70!black, every mark/.append style={solid,fill=RdYlGn-11-2},mark=square*\\%
 RdYlGn-11-3!70!black, every mark/.append style={solid,fill=RdYlGn-11-3},mark=triangle*\\%
 RdYlGn-11-4!70!black, every mark/.append style={solid,fill=RdYlGn-11-4},mark=diamond*\\%
 RdYlGn-11-5!70!black, every mark/.append style={solid,fill=RdYlGn-11-5},mark=pentagon*\\%
 RdYlGn-11-6!70!black, every mark/.append style={solid,fill=RdYlGn-11-6},mark=x\\%
 RdYlGn-11-7!70!black, every mark/.append style={solid,fill=RdYlGn-11-7},mark=+\\%
 RdYlGn-11-8!70!black, every mark/.append style={solid,fill=RdYlGn-11-8},mark=star\\%
 RdYlGn-11-9!70!black, every mark/.append style={solid,fill=RdYlGn-11-9},mark=oplus*\\%
 RdYlGn-11-10!70!black, every mark/.append style={solid,fill=RdYlGn-11-10},mark=otimes*\\%
}
\pgfplotscreateplotcyclelist{Mark-Set3-3}{%
 Set3-3-1!70!black, every mark/.append style={solid,fill=Set3-3-1},mark=*\\%
 Set3-3-2!70!black, every mark/.append style={solid,fill=Set3-3-2},mark=square*\\%
 Set3-3-3!70!black, every mark/.append style={solid,fill=Set3-3-3},mark=triangle*\\%
}
\pgfplotscreateplotcyclelist{Mark-Set3-4}{%
 Set3-4-1!70!black, every mark/.append style={solid,fill=Set3-4-1},mark=*\\%
 Set3-4-2!70!black, every mark/.append style={solid,fill=Set3-4-2},mark=square*\\%
 Set3-4-3!70!black, every mark/.append style={solid,fill=Set3-4-3},mark=triangle*\\%
 Set3-4-4!70!black, every mark/.append style={solid,fill=Set3-4-4},mark=diamond*\\%
}
\pgfplotscreateplotcyclelist{Mark-Set3-5}{%
 Set3-5-1!70!black, every mark/.append style={solid,fill=Set3-5-1},mark=*\\%
 Set3-5-2!70!black, every mark/.append style={solid,fill=Set3-5-2},mark=square*\\%
 Set3-5-3!70!black, every mark/.append style={solid,fill=Set3-5-3},mark=triangle*\\%
 Set3-5-4!70!black, every mark/.append style={solid,fill=Set3-5-4},mark=diamond*\\%
 Set3-5-5!70!black, every mark/.append style={solid,fill=Set3-5-5},mark=pentagon*\\%
}
\pgfplotscreateplotcyclelist{Mark-Set3-6}{%
 Set3-6-1!70!black, every mark/.append style={solid,fill=Set3-6-1},mark=*\\%
 Set3-6-2!70!black, every mark/.append style={solid,fill=Set3-6-2},mark=square*\\%
 Set3-6-3!70!black, every mark/.append style={solid,fill=Set3-6-3},mark=triangle*\\%
 Set3-6-4!70!black, every mark/.append style={solid,fill=Set3-6-4},mark=diamond*\\%
 Set3-6-5!70!black, every mark/.append style={solid,fill=Set3-6-5},mark=pentagon*\\%
 Set3-6-6!70!black, every mark/.append style={solid,fill=Set3-6-6},mark=x\\%
}
\pgfplotscreateplotcyclelist{Mark-Set3-7}{%
 Set3-7-1!70!black, every mark/.append style={solid,fill=Set3-7-1},mark=*\\%
 Set3-7-2!70!black, every mark/.append style={solid,fill=Set3-7-2},mark=square*\\%
 Set3-7-3!70!black, every mark/.append style={solid,fill=Set3-7-3},mark=triangle*\\%
 Set3-7-4!70!black, every mark/.append style={solid,fill=Set3-7-4},mark=diamond*\\%
 Set3-7-5!70!black, every mark/.append style={solid,fill=Set3-7-5},mark=pentagon*\\%
 Set3-7-6!70!black, every mark/.append style={solid,fill=Set3-7-6},mark=x\\%
 Set3-7-7!70!black, every mark/.append style={solid,fill=Set3-7-7},mark=+\\%
}
\pgfplotscreateplotcyclelist{Mark-Set3-8}{%
 Set3-8-1!70!black, every mark/.append style={solid,fill=Set3-8-1},mark=*\\%
 Set3-8-2!70!black, every mark/.append style={solid,fill=Set3-8-2},mark=square*\\%
 Set3-8-3!70!black, every mark/.append style={solid,fill=Set3-8-3},mark=triangle*\\%
 Set3-8-4!70!black, every mark/.append style={solid,fill=Set3-8-4},mark=diamond*\\%
 Set3-8-5!70!black, every mark/.append style={solid,fill=Set3-8-5},mark=pentagon*\\%
 Set3-8-6!70!black, every mark/.append style={solid,fill=Set3-8-6},mark=x\\%
 Set3-8-7!70!black, every mark/.append style={solid,fill=Set3-8-7},mark=+\\%
 Set3-8-8!70!black, every mark/.append style={solid,fill=Set3-8-8},mark=star\\%
}
\pgfplotscreateplotcyclelist{Mark-Set3-9}{%
 Set3-9-1!70!black, every mark/.append style={solid,fill=Set3-9-1},mark=*\\%
 Set3-9-2!70!black, every mark/.append style={solid,fill=Set3-9-2},mark=square*\\%
 Set3-9-3!70!black, every mark/.append style={solid,fill=Set3-9-3},mark=triangle*\\%
 Set3-9-4!70!black, every mark/.append style={solid,fill=Set3-9-4},mark=diamond*\\%
 Set3-9-5!70!black, every mark/.append style={solid,fill=Set3-9-5},mark=pentagon*\\%
 Set3-9-6!70!black, every mark/.append style={solid,fill=Set3-9-6},mark=x\\%
 Set3-9-7!70!black, every mark/.append style={solid,fill=Set3-9-7},mark=+\\%
 Set3-9-8!70!black, every mark/.append style={solid,fill=Set3-9-8},mark=star\\%
 Set3-9-9!70!black, every mark/.append style={solid,fill=Set3-9-9},mark=oplus*\\%
}
\pgfplotscreateplotcyclelist{Mark-Set3-10}{%
 Set3-10-1!70!black, every mark/.append style={solid,fill=Set3-10-1},mark=*\\%
 Set3-10-2!70!black, every mark/.append style={solid,fill=Set3-10-2},mark=square*\\%
 Set3-10-3!70!black, every mark/.append style={solid,fill=Set3-10-3},mark=triangle*\\%
 Set3-10-4!70!black, every mark/.append style={solid,fill=Set3-10-4},mark=diamond*\\%
 Set3-10-5!70!black, every mark/.append style={solid,fill=Set3-10-5},mark=pentagon*\\%
 Set3-10-6!70!black, every mark/.append style={solid,fill=Set3-10-6},mark=x\\%
 Set3-10-7!70!black, every mark/.append style={solid,fill=Set3-10-7},mark=+\\%
 Set3-10-8!70!black, every mark/.append style={solid,fill=Set3-10-8},mark=star\\%
 Set3-10-9!70!black, every mark/.append style={solid,fill=Set3-10-9},mark=oplus*\\%
 Set3-10-10!70!black, every mark/.append style={solid,fill=Set3-10-10},mark=otimes*\\%
}
\pgfplotscreateplotcyclelist{Mark-Set3-11}{%
 Set3-11-1!70!black, every mark/.append style={solid,fill=Set3-11-1},mark=*\\%
 Set3-11-2!70!black, every mark/.append style={solid,fill=Set3-11-2},mark=square*\\%
 Set3-11-3!70!black, every mark/.append style={solid,fill=Set3-11-3},mark=triangle*\\%
 Set3-11-4!70!black, every mark/.append style={solid,fill=Set3-11-4},mark=diamond*\\%
 Set3-11-5!70!black, every mark/.append style={solid,fill=Set3-11-5},mark=pentagon*\\%
 Set3-11-6!70!black, every mark/.append style={solid,fill=Set3-11-6},mark=x\\%
 Set3-11-7!70!black, every mark/.append style={solid,fill=Set3-11-7},mark=+\\%
 Set3-11-8!70!black, every mark/.append style={solid,fill=Set3-11-8},mark=star\\%
 Set3-11-9!70!black, every mark/.append style={solid,fill=Set3-11-9},mark=oplus*\\%
 Set3-11-10!70!black, every mark/.append style={solid,fill=Set3-11-10},mark=otimes*\\%
}
\pgfplotscreateplotcyclelist{Mark-Set3-12}{%
 Set3-12-1!70!black, every mark/.append style={solid,fill=Set3-12-1},mark=*\\%
 Set3-12-2!70!black, every mark/.append style={solid,fill=Set3-12-2},mark=square*\\%
 Set3-12-3!70!black, every mark/.append style={solid,fill=Set3-12-3},mark=triangle*\\%
 Set3-12-4!70!black, every mark/.append style={solid,fill=Set3-12-4},mark=diamond*\\%
 Set3-12-5!70!black, every mark/.append style={solid,fill=Set3-12-5},mark=pentagon*\\%
 Set3-12-6!70!black, every mark/.append style={solid,fill=Set3-12-6},mark=x\\%
 Set3-12-7!70!black, every mark/.append style={solid,fill=Set3-12-7},mark=+\\%
 Set3-12-8!70!black, every mark/.append style={solid,fill=Set3-12-8},mark=star\\%
 Set3-12-9!70!black, every mark/.append style={solid,fill=Set3-12-9},mark=oplus*\\%
 Set3-12-10!70!black, every mark/.append style={solid,fill=Set3-12-10},mark=otimes*\\%
}
\pgfplotscreateplotcyclelist{Mark-Pastel1-3}{%
 Pastel1-3-1!70!black, every mark/.append style={solid,fill=Pastel1-3-1},mark=*\\%
 Pastel1-3-2!70!black, every mark/.append style={solid,fill=Pastel1-3-2},mark=square*\\%
 Pastel1-3-3!70!black, every mark/.append style={solid,fill=Pastel1-3-3},mark=triangle*\\%
}
\pgfplotscreateplotcyclelist{Mark-Pastel1-4}{%
 Pastel1-4-1!70!black, every mark/.append style={solid,fill=Pastel1-4-1},mark=*\\%
 Pastel1-4-2!70!black, every mark/.append style={solid,fill=Pastel1-4-2},mark=square*\\%
 Pastel1-4-3!70!black, every mark/.append style={solid,fill=Pastel1-4-3},mark=triangle*\\%
 Pastel1-4-4!70!black, every mark/.append style={solid,fill=Pastel1-4-4},mark=diamond*\\%
}
\pgfplotscreateplotcyclelist{Mark-Pastel1-5}{%
 Pastel1-5-1!70!black, every mark/.append style={solid,fill=Pastel1-5-1},mark=*\\%
 Pastel1-5-2!70!black, every mark/.append style={solid,fill=Pastel1-5-2},mark=square*\\%
 Pastel1-5-3!70!black, every mark/.append style={solid,fill=Pastel1-5-3},mark=triangle*\\%
 Pastel1-5-4!70!black, every mark/.append style={solid,fill=Pastel1-5-4},mark=diamond*\\%
 Pastel1-5-5!70!black, every mark/.append style={solid,fill=Pastel1-5-5},mark=pentagon*\\%
}
\pgfplotscreateplotcyclelist{Mark-Pastel1-6}{%
 Pastel1-6-1!70!black, every mark/.append style={solid,fill=Pastel1-6-1},mark=*\\%
 Pastel1-6-2!70!black, every mark/.append style={solid,fill=Pastel1-6-2},mark=square*\\%
 Pastel1-6-3!70!black, every mark/.append style={solid,fill=Pastel1-6-3},mark=triangle*\\%
 Pastel1-6-4!70!black, every mark/.append style={solid,fill=Pastel1-6-4},mark=diamond*\\%
 Pastel1-6-5!70!black, every mark/.append style={solid,fill=Pastel1-6-5},mark=pentagon*\\%
 Pastel1-6-6!70!black, every mark/.append style={solid,fill=Pastel1-6-6},mark=x\\%
}
\pgfplotscreateplotcyclelist{Mark-Pastel1-7}{%
 Pastel1-7-1!70!black, every mark/.append style={solid,fill=Pastel1-7-1},mark=*\\%
 Pastel1-7-2!70!black, every mark/.append style={solid,fill=Pastel1-7-2},mark=square*\\%
 Pastel1-7-3!70!black, every mark/.append style={solid,fill=Pastel1-7-3},mark=triangle*\\%
 Pastel1-7-4!70!black, every mark/.append style={solid,fill=Pastel1-7-4},mark=diamond*\\%
 Pastel1-7-5!70!black, every mark/.append style={solid,fill=Pastel1-7-5},mark=pentagon*\\%
 Pastel1-7-6!70!black, every mark/.append style={solid,fill=Pastel1-7-6},mark=x\\%
 Pastel1-7-7!70!black, every mark/.append style={solid,fill=Pastel1-7-7},mark=+\\%
}
\pgfplotscreateplotcyclelist{Mark-Pastel1-8}{%
 Pastel1-8-1!70!black, every mark/.append style={solid,fill=Pastel1-8-1},mark=*\\%
 Pastel1-8-2!70!black, every mark/.append style={solid,fill=Pastel1-8-2},mark=square*\\%
 Pastel1-8-3!70!black, every mark/.append style={solid,fill=Pastel1-8-3},mark=triangle*\\%
 Pastel1-8-4!70!black, every mark/.append style={solid,fill=Pastel1-8-4},mark=diamond*\\%
 Pastel1-8-5!70!black, every mark/.append style={solid,fill=Pastel1-8-5},mark=pentagon*\\%
 Pastel1-8-6!70!black, every mark/.append style={solid,fill=Pastel1-8-6},mark=x\\%
 Pastel1-8-7!70!black, every mark/.append style={solid,fill=Pastel1-8-7},mark=+\\%
 Pastel1-8-8!70!black, every mark/.append style={solid,fill=Pastel1-8-8},mark=star\\%
}
\pgfplotscreateplotcyclelist{Mark-Pastel1-9}{%
 Pastel1-9-1!70!black, every mark/.append style={solid,fill=Pastel1-9-1},mark=*\\%
 Pastel1-9-2!70!black, every mark/.append style={solid,fill=Pastel1-9-2},mark=square*\\%
 Pastel1-9-3!70!black, every mark/.append style={solid,fill=Pastel1-9-3},mark=triangle*\\%
 Pastel1-9-4!70!black, every mark/.append style={solid,fill=Pastel1-9-4},mark=diamond*\\%
 Pastel1-9-5!70!black, every mark/.append style={solid,fill=Pastel1-9-5},mark=pentagon*\\%
 Pastel1-9-6!70!black, every mark/.append style={solid,fill=Pastel1-9-6},mark=x\\%
 Pastel1-9-7!70!black, every mark/.append style={solid,fill=Pastel1-9-7},mark=+\\%
 Pastel1-9-8!70!black, every mark/.append style={solid,fill=Pastel1-9-8},mark=star\\%
 Pastel1-9-9!70!black, every mark/.append style={solid,fill=Pastel1-9-9},mark=oplus*\\%
}
\pgfplotscreateplotcyclelist{Mark-Set1-3}{%
 Set1-3-1!70!black, every mark/.append style={solid,fill=Set1-3-1},mark=*\\%
 Set1-3-2!70!black, every mark/.append style={solid,fill=Set1-3-2},mark=square*\\%
 Set1-3-3!70!black, every mark/.append style={solid,fill=Set1-3-3},mark=triangle*\\%
}
\pgfplotscreateplotcyclelist{Mark-Set1-4}{%
 Set1-4-1!70!black, every mark/.append style={solid,fill=Set1-4-1},mark=*\\%
 Set1-4-2!70!black, every mark/.append style={solid,fill=Set1-4-2},mark=square*\\%
 Set1-4-3!70!black, every mark/.append style={solid,fill=Set1-4-3},mark=triangle*\\%
 Set1-4-4!70!black, every mark/.append style={solid,fill=Set1-4-4},mark=diamond*\\%
}
\pgfplotscreateplotcyclelist{Mark-Set1-5}{%
 Set1-5-1!70!black, every mark/.append style={solid,fill=Set1-5-1},mark=*\\%
 Set1-5-2!70!black, every mark/.append style={solid,fill=Set1-5-2},mark=square*\\%
 Set1-5-3!70!black, every mark/.append style={solid,fill=Set1-5-3},mark=triangle*\\%
 Set1-5-4!70!black, every mark/.append style={solid,fill=Set1-5-4},mark=diamond*\\%
 Set1-5-5!70!black, every mark/.append style={solid,fill=Set1-5-5},mark=pentagon*\\%
}
\pgfplotscreateplotcyclelist{Mark-Set1-6}{%
 Set1-6-1!70!black, every mark/.append style={solid,fill=Set1-6-1},mark=*\\%
 Set1-6-2!70!black, every mark/.append style={solid,fill=Set1-6-2},mark=square*\\%
 Set1-6-3!70!black, every mark/.append style={solid,fill=Set1-6-3},mark=triangle*\\%
 Set1-6-4!70!black, every mark/.append style={solid,fill=Set1-6-4},mark=diamond*\\%
 Set1-6-5!70!black, every mark/.append style={solid,fill=Set1-6-5},mark=pentagon*\\%
 Set1-6-6!70!black, every mark/.append style={solid,fill=Set1-6-6},mark=x\\%
}
\pgfplotscreateplotcyclelist{Mark-Set1-7}{%
 Set1-7-1!70!black, every mark/.append style={solid,fill=Set1-7-1},mark=*\\%
 Set1-7-2!70!black, every mark/.append style={solid,fill=Set1-7-2},mark=square*\\%
 Set1-7-3!70!black, every mark/.append style={solid,fill=Set1-7-3},mark=triangle*\\%
 Set1-7-4!70!black, every mark/.append style={solid,fill=Set1-7-4},mark=diamond*\\%
 Set1-7-5!70!black, every mark/.append style={solid,fill=Set1-7-5},mark=pentagon*\\%
 Set1-7-6!70!black, every mark/.append style={solid,fill=Set1-7-6},mark=x\\%
 Set1-7-7!70!black, every mark/.append style={solid,fill=Set1-7-7},mark=+\\%
}
\pgfplotscreateplotcyclelist{Mark-Set1-8}{%
 Set1-8-1!70!black, every mark/.append style={solid,fill=Set1-8-1},mark=*\\%
 Set1-8-2!70!black, every mark/.append style={solid,fill=Set1-8-2},mark=square*\\%
 Set1-8-3!70!black, every mark/.append style={solid,fill=Set1-8-3},mark=triangle*\\%
 Set1-8-4!70!black, every mark/.append style={solid,fill=Set1-8-4},mark=diamond*\\%
 Set1-8-5!70!black, every mark/.append style={solid,fill=Set1-8-5},mark=pentagon*\\%
 Set1-8-6!70!black, every mark/.append style={solid,fill=Set1-8-6},mark=x\\%
 Set1-8-7!70!black, every mark/.append style={solid,fill=Set1-8-7},mark=+\\%
 Set1-8-8!70!black, every mark/.append style={solid,fill=Set1-8-8},mark=star\\%
}
\pgfplotscreateplotcyclelist{Mark-Set1-9}{%
 Set1-9-1!70!black, every mark/.append style={solid,fill=Set1-9-1},mark=*\\%
 Set1-9-2!70!black, every mark/.append style={solid,fill=Set1-9-2},mark=square*\\%
 Set1-9-3!70!black, every mark/.append style={solid,fill=Set1-9-3},mark=triangle*\\%
 Set1-9-4!70!black, every mark/.append style={solid,fill=Set1-9-4},mark=diamond*\\%
 Set1-9-5!70!black, every mark/.append style={solid,fill=Set1-9-5},mark=pentagon*\\%
 Set1-9-6!70!black, every mark/.append style={solid,fill=Set1-9-6},mark=x\\%
 Set1-9-7!70!black, every mark/.append style={solid,fill=Set1-9-7},mark=+\\%
 Set1-9-8!70!black, every mark/.append style={solid,fill=Set1-9-8},mark=star\\%
 Set1-9-9!70!black, every mark/.append style={solid,fill=Set1-9-9},mark=oplus*\\%
}
\pgfplotscreateplotcyclelist{Mark-Pastel2-3}{%
 Pastel2-3-1!70!black, every mark/.append style={solid,fill=Pastel2-3-1},mark=*\\%
 Pastel2-3-2!70!black, every mark/.append style={solid,fill=Pastel2-3-2},mark=square*\\%
 Pastel2-3-3!70!black, every mark/.append style={solid,fill=Pastel2-3-3},mark=triangle*\\%
}
\pgfplotscreateplotcyclelist{Mark-Pastel2-4}{%
 Pastel2-4-1!70!black, every mark/.append style={solid,fill=Pastel2-4-1},mark=*\\%
 Pastel2-4-2!70!black, every mark/.append style={solid,fill=Pastel2-4-2},mark=square*\\%
 Pastel2-4-3!70!black, every mark/.append style={solid,fill=Pastel2-4-3},mark=triangle*\\%
 Pastel2-4-4!70!black, every mark/.append style={solid,fill=Pastel2-4-4},mark=diamond*\\%
}
\pgfplotscreateplotcyclelist{Mark-Pastel2-5}{%
 Pastel2-5-1!70!black, every mark/.append style={solid,fill=Pastel2-5-1},mark=*\\%
 Pastel2-5-2!70!black, every mark/.append style={solid,fill=Pastel2-5-2},mark=square*\\%
 Pastel2-5-3!70!black, every mark/.append style={solid,fill=Pastel2-5-3},mark=triangle*\\%
 Pastel2-5-4!70!black, every mark/.append style={solid,fill=Pastel2-5-4},mark=diamond*\\%
 Pastel2-5-5!70!black, every mark/.append style={solid,fill=Pastel2-5-5},mark=pentagon*\\%
}
\pgfplotscreateplotcyclelist{Mark-Pastel2-6}{%
 Pastel2-6-1!70!black, every mark/.append style={solid,fill=Pastel2-6-1},mark=*\\%
 Pastel2-6-2!70!black, every mark/.append style={solid,fill=Pastel2-6-2},mark=square*\\%
 Pastel2-6-3!70!black, every mark/.append style={solid,fill=Pastel2-6-3},mark=triangle*\\%
 Pastel2-6-4!70!black, every mark/.append style={solid,fill=Pastel2-6-4},mark=diamond*\\%
 Pastel2-6-5!70!black, every mark/.append style={solid,fill=Pastel2-6-5},mark=pentagon*\\%
 Pastel2-6-6!70!black, every mark/.append style={solid,fill=Pastel2-6-6},mark=x\\%
}
\pgfplotscreateplotcyclelist{Mark-Pastel2-7}{%
 Pastel2-7-1!70!black, every mark/.append style={solid,fill=Pastel2-7-1},mark=*\\%
 Pastel2-7-2!70!black, every mark/.append style={solid,fill=Pastel2-7-2},mark=square*\\%
 Pastel2-7-3!70!black, every mark/.append style={solid,fill=Pastel2-7-3},mark=triangle*\\%
 Pastel2-7-4!70!black, every mark/.append style={solid,fill=Pastel2-7-4},mark=diamond*\\%
 Pastel2-7-5!70!black, every mark/.append style={solid,fill=Pastel2-7-5},mark=pentagon*\\%
 Pastel2-7-6!70!black, every mark/.append style={solid,fill=Pastel2-7-6},mark=x\\%
 Pastel2-7-7!70!black, every mark/.append style={solid,fill=Pastel2-7-7},mark=+\\%
}
\pgfplotscreateplotcyclelist{Mark-Pastel2-8}{%
 Pastel2-8-1!70!black, every mark/.append style={solid,fill=Pastel2-8-1},mark=*\\%
 Pastel2-8-2!70!black, every mark/.append style={solid,fill=Pastel2-8-2},mark=square*\\%
 Pastel2-8-3!70!black, every mark/.append style={solid,fill=Pastel2-8-3},mark=triangle*\\%
 Pastel2-8-4!70!black, every mark/.append style={solid,fill=Pastel2-8-4},mark=diamond*\\%
 Pastel2-8-5!70!black, every mark/.append style={solid,fill=Pastel2-8-5},mark=pentagon*\\%
 Pastel2-8-6!70!black, every mark/.append style={solid,fill=Pastel2-8-6},mark=x\\%
 Pastel2-8-7!70!black, every mark/.append style={solid,fill=Pastel2-8-7},mark=+\\%
 Pastel2-8-8!70!black, every mark/.append style={solid,fill=Pastel2-8-8},mark=star\\%
}
\pgfplotscreateplotcyclelist{Mark-Set2-3}{%
 Set2-3-1!70!black, every mark/.append style={solid,fill=Set2-3-1},mark=*\\%
 Set2-3-2!70!black, every mark/.append style={solid,fill=Set2-3-2},mark=square*\\%
 Set2-3-3!70!black, every mark/.append style={solid,fill=Set2-3-3},mark=triangle*\\%
}
\pgfplotscreateplotcyclelist{Mark-Set2-4}{%
 Set2-4-1!70!black, every mark/.append style={solid,fill=Set2-4-1},mark=*\\%
 Set2-4-2!70!black, every mark/.append style={solid,fill=Set2-4-2},mark=square*\\%
 Set2-4-3!70!black, every mark/.append style={solid,fill=Set2-4-3},mark=triangle*\\%
 Set2-4-4!70!black, every mark/.append style={solid,fill=Set2-4-4},mark=diamond*\\%
}
\pgfplotscreateplotcyclelist{Mark-Set2-5}{%
 Set2-5-1!70!black, every mark/.append style={solid,fill=Set2-5-1},mark=*\\%
 Set2-5-2!70!black, every mark/.append style={solid,fill=Set2-5-2},mark=square*\\%
 Set2-5-3!70!black, every mark/.append style={solid,fill=Set2-5-3},mark=triangle*\\%
 Set2-5-4!70!black, every mark/.append style={solid,fill=Set2-5-4},mark=diamond*\\%
 Set2-5-5!70!black, every mark/.append style={solid,fill=Set2-5-5},mark=pentagon*\\%
}
\pgfplotscreateplotcyclelist{Mark-Set2-6}{%
 Set2-6-1!70!black, every mark/.append style={solid,fill=Set2-6-1},mark=*\\%
 Set2-6-2!70!black, every mark/.append style={solid,fill=Set2-6-2},mark=square*\\%
 Set2-6-3!70!black, every mark/.append style={solid,fill=Set2-6-3},mark=triangle*\\%
 Set2-6-4!70!black, every mark/.append style={solid,fill=Set2-6-4},mark=diamond*\\%
 Set2-6-5!70!black, every mark/.append style={solid,fill=Set2-6-5},mark=pentagon*\\%
 Set2-6-6!70!black, every mark/.append style={solid,fill=Set2-6-6},mark=x\\%
}
\pgfplotscreateplotcyclelist{Mark-Set2-7}{%
 Set2-7-1!70!black, every mark/.append style={solid,fill=Set2-7-1},mark=*\\%
 Set2-7-2!70!black, every mark/.append style={solid,fill=Set2-7-2},mark=square*\\%
 Set2-7-3!70!black, every mark/.append style={solid,fill=Set2-7-3},mark=triangle*\\%
 Set2-7-4!70!black, every mark/.append style={solid,fill=Set2-7-4},mark=diamond*\\%
 Set2-7-5!70!black, every mark/.append style={solid,fill=Set2-7-5},mark=pentagon*\\%
 Set2-7-6!70!black, every mark/.append style={solid,fill=Set2-7-6},mark=x\\%
 Set2-7-7!70!black, every mark/.append style={solid,fill=Set2-7-7},mark=+\\%
}
\pgfplotscreateplotcyclelist{Mark-Set2-8}{%
 Set2-8-1!70!black, every mark/.append style={solid,fill=Set2-8-1},mark=*\\%
 Set2-8-2!70!black, every mark/.append style={solid,fill=Set2-8-2},mark=square*\\%
 Set2-8-3!70!black, every mark/.append style={solid,fill=Set2-8-3},mark=triangle*\\%
 Set2-8-4!70!black, every mark/.append style={solid,fill=Set2-8-4},mark=diamond*\\%
 Set2-8-5!70!black, every mark/.append style={solid,fill=Set2-8-5},mark=pentagon*\\%
 Set2-8-6!70!black, every mark/.append style={solid,fill=Set2-8-6},mark=x\\%
 Set2-8-7!70!black, every mark/.append style={solid,fill=Set2-8-7},mark=+\\%
 Set2-8-8!70!black, every mark/.append style={solid,fill=Set2-8-8},mark=star\\%
}
\pgfplotscreateplotcyclelist{Mark-Dark2-3}{%
 Dark2-3-1!70!black, every mark/.append style={solid,fill=Dark2-3-1},mark=*\\%
 Dark2-3-2!70!black, every mark/.append style={solid,fill=Dark2-3-2},mark=square*\\%
 Dark2-3-3!70!black, every mark/.append style={solid,fill=Dark2-3-3},mark=triangle*\\%
}
\pgfplotscreateplotcyclelist{Mark-Dark2-4}{%
 Dark2-4-1!70!black, every mark/.append style={solid,fill=Dark2-4-1},mark=*\\%
 Dark2-4-2!70!black, every mark/.append style={solid,fill=Dark2-4-2},mark=square*\\%
 Dark2-4-3!70!black, every mark/.append style={solid,fill=Dark2-4-3},mark=triangle*\\%
 Dark2-4-4!70!black, every mark/.append style={solid,fill=Dark2-4-4},mark=diamond*\\%
}
\pgfplotscreateplotcyclelist{Mark-Dark2-5}{%
 Dark2-5-1!70!black, every mark/.append style={solid,fill=Dark2-5-1},mark=*\\%
 Dark2-5-2!70!black, every mark/.append style={solid,fill=Dark2-5-2},mark=square*\\%
 Dark2-5-3!70!black, every mark/.append style={solid,fill=Dark2-5-3},mark=triangle*\\%
 Dark2-5-4!70!black, every mark/.append style={solid,fill=Dark2-5-4},mark=diamond*\\%
 Dark2-5-5!70!black, every mark/.append style={solid,fill=Dark2-5-5},mark=pentagon*\\%
}
\pgfplotscreateplotcyclelist{Mark-Dark2-6}{%
 Dark2-6-1!70!black, every mark/.append style={solid,fill=Dark2-6-1},mark=*\\%
 Dark2-6-2!70!black, every mark/.append style={solid,fill=Dark2-6-2},mark=square*\\%
 Dark2-6-3!70!black, every mark/.append style={solid,fill=Dark2-6-3},mark=triangle*\\%
 Dark2-6-4!70!black, every mark/.append style={solid,fill=Dark2-6-4},mark=diamond*\\%
 Dark2-6-5!70!black, every mark/.append style={solid,fill=Dark2-6-5},mark=pentagon*\\%
 Dark2-6-6!70!black, every mark/.append style={solid,fill=Dark2-6-6},mark=x\\%
}
\pgfplotscreateplotcyclelist{Mark-Dark2-7}{%
 Dark2-7-1!70!black, every mark/.append style={solid,fill=Dark2-7-1},mark=*\\%
 Dark2-7-2!70!black, every mark/.append style={solid,fill=Dark2-7-2},mark=square*\\%
 Dark2-7-3!70!black, every mark/.append style={solid,fill=Dark2-7-3},mark=triangle*\\%
 Dark2-7-4!70!black, every mark/.append style={solid,fill=Dark2-7-4},mark=diamond*\\%
 Dark2-7-5!70!black, every mark/.append style={solid,fill=Dark2-7-5},mark=pentagon*\\%
 Dark2-7-6!70!black, every mark/.append style={solid,fill=Dark2-7-6},mark=x\\%
 Dark2-7-7!70!black, every mark/.append style={solid,fill=Dark2-7-7},mark=+\\%
}
\pgfplotscreateplotcyclelist{Mark-Dark2-8}{%
 Dark2-8-1!70!black, every mark/.append style={solid,fill=Dark2-8-1},mark=*\\%
 Dark2-8-2!70!black, every mark/.append style={solid,fill=Dark2-8-2},mark=square*\\%
 Dark2-8-3!70!black, every mark/.append style={solid,fill=Dark2-8-3},mark=triangle*\\%
 Dark2-8-4!70!black, every mark/.append style={solid,fill=Dark2-8-4},mark=diamond*\\%
 Dark2-8-5!70!black, every mark/.append style={solid,fill=Dark2-8-5},mark=pentagon*\\%
 Dark2-8-6!70!black, every mark/.append style={solid,fill=Dark2-8-6},mark=x\\%
 Dark2-8-7!70!black, every mark/.append style={solid,fill=Dark2-8-7},mark=+\\%
 Dark2-8-8!70!black, every mark/.append style={solid,fill=Dark2-8-8},mark=star\\%
}
\pgfplotscreateplotcyclelist{Mark-Paired-3}{%
 Paired-3-1!70!black, every mark/.append style={solid,fill=Paired-3-1},mark=*\\%
 Paired-3-2!70!black, every mark/.append style={solid,fill=Paired-3-2},mark=square*\\%
 Paired-3-3!70!black, every mark/.append style={solid,fill=Paired-3-3},mark=triangle*\\%
}
\pgfplotscreateplotcyclelist{Mark-Paired-4}{%
 Paired-4-1!70!black, every mark/.append style={solid,fill=Paired-4-1},mark=*\\%
 Paired-4-2!70!black, every mark/.append style={solid,fill=Paired-4-2},mark=square*\\%
 Paired-4-3!70!black, every mark/.append style={solid,fill=Paired-4-3},mark=triangle*\\%
 Paired-4-4!70!black, every mark/.append style={solid,fill=Paired-4-4},mark=diamond*\\%
}
\pgfplotscreateplotcyclelist{Mark-Paired-5}{%
 Paired-5-1!70!black, every mark/.append style={solid,fill=Paired-5-1},mark=*\\%
 Paired-5-2!70!black, every mark/.append style={solid,fill=Paired-5-2},mark=square*\\%
 Paired-5-3!70!black, every mark/.append style={solid,fill=Paired-5-3},mark=triangle*\\%
 Paired-5-4!70!black, every mark/.append style={solid,fill=Paired-5-4},mark=diamond*\\%
 Paired-5-5!70!black, every mark/.append style={solid,fill=Paired-5-5},mark=pentagon*\\%
}
\pgfplotscreateplotcyclelist{Mark-Paired-6}{%
 Paired-6-1!70!black, every mark/.append style={solid,fill=Paired-6-1},mark=*\\%
 Paired-6-2!70!black, every mark/.append style={solid,fill=Paired-6-2},mark=square*\\%
 Paired-6-3!70!black, every mark/.append style={solid,fill=Paired-6-3},mark=triangle*\\%
 Paired-6-4!70!black, every mark/.append style={solid,fill=Paired-6-4},mark=diamond*\\%
 Paired-6-5!70!black, every mark/.append style={solid,fill=Paired-6-5},mark=pentagon*\\%
 Paired-6-6!70!black, every mark/.append style={solid,fill=Paired-6-6},mark=x\\%
}
\pgfplotscreateplotcyclelist{Mark-Paired-7}{%
 Paired-7-1!70!black, every mark/.append style={solid,fill=Paired-7-1},mark=*\\%
 Paired-7-2!70!black, every mark/.append style={solid,fill=Paired-7-2},mark=square*\\%
 Paired-7-3!70!black, every mark/.append style={solid,fill=Paired-7-3},mark=triangle*\\%
 Paired-7-4!70!black, every mark/.append style={solid,fill=Paired-7-4},mark=diamond*\\%
 Paired-7-5!70!black, every mark/.append style={solid,fill=Paired-7-5},mark=pentagon*\\%
 Paired-7-6!70!black, every mark/.append style={solid,fill=Paired-7-6},mark=x\\%
 Paired-7-7!70!black, every mark/.append style={solid,fill=Paired-7-7},mark=+\\%
}
\pgfplotscreateplotcyclelist{Mark-Paired-8}{%
 Paired-8-1!70!black, every mark/.append style={solid,fill=Paired-8-1},mark=*\\%
 Paired-8-2!70!black, every mark/.append style={solid,fill=Paired-8-2},mark=square*\\%
 Paired-8-3!70!black, every mark/.append style={solid,fill=Paired-8-3},mark=triangle*\\%
 Paired-8-4!70!black, every mark/.append style={solid,fill=Paired-8-4},mark=diamond*\\%
 Paired-8-5!70!black, every mark/.append style={solid,fill=Paired-8-5},mark=pentagon*\\%
 Paired-8-6!70!black, every mark/.append style={solid,fill=Paired-8-6},mark=x\\%
 Paired-8-7!70!black, every mark/.append style={solid,fill=Paired-8-7},mark=+\\%
 Paired-8-8!70!black, every mark/.append style={solid,fill=Paired-8-8},mark=star\\%
}
\pgfplotscreateplotcyclelist{Mark-Paired-9}{%
 Paired-9-1!70!black, every mark/.append style={solid,fill=Paired-9-1},mark=*\\%
 Paired-9-2!70!black, every mark/.append style={solid,fill=Paired-9-2},mark=square*\\%
 Paired-9-3!70!black, every mark/.append style={solid,fill=Paired-9-3},mark=triangle*\\%
 Paired-9-4!70!black, every mark/.append style={solid,fill=Paired-9-4},mark=diamond*\\%
 Paired-9-5!70!black, every mark/.append style={solid,fill=Paired-9-5},mark=pentagon*\\%
 Paired-9-6!70!black, every mark/.append style={solid,fill=Paired-9-6},mark=x\\%
 Paired-9-7!70!black, every mark/.append style={solid,fill=Paired-9-7},mark=+\\%
 Paired-9-8!70!black, every mark/.append style={solid,fill=Paired-9-8},mark=star\\%
 Paired-9-9!70!black, every mark/.append style={solid,fill=Paired-9-9},mark=oplus*\\%
}
\pgfplotscreateplotcyclelist{Mark-Paired-10}{%
 Paired-10-1!70!black, every mark/.append style={solid,fill=Paired-10-1},mark=*\\%
 Paired-10-2!70!black, every mark/.append style={solid,fill=Paired-10-2},mark=square*\\%
 Paired-10-3!70!black, every mark/.append style={solid,fill=Paired-10-3},mark=triangle*\\%
 Paired-10-4!70!black, every mark/.append style={solid,fill=Paired-10-4},mark=diamond*\\%
 Paired-10-5!70!black, every mark/.append style={solid,fill=Paired-10-5},mark=pentagon*\\%
 Paired-10-6!70!black, every mark/.append style={solid,fill=Paired-10-6},mark=x\\%
 Paired-10-7!70!black, every mark/.append style={solid,fill=Paired-10-7},mark=+\\%
 Paired-10-8!70!black, every mark/.append style={solid,fill=Paired-10-8},mark=star\\%
 Paired-10-9!70!black, every mark/.append style={solid,fill=Paired-10-9},mark=oplus*\\%
 Paired-10-10!70!black, every mark/.append style={solid,fill=Paired-10-10},mark=otimes*\\%
}
\pgfplotscreateplotcyclelist{Mark-Paired-11}{%
 Paired-11-1!70!black, every mark/.append style={solid,fill=Paired-11-1},mark=*\\%
 Paired-11-2!70!black, every mark/.append style={solid,fill=Paired-11-2},mark=square*\\%
 Paired-11-3!70!black, every mark/.append style={solid,fill=Paired-11-3},mark=triangle*\\%
 Paired-11-4!70!black, every mark/.append style={solid,fill=Paired-11-4},mark=diamond*\\%
 Paired-11-5!70!black, every mark/.append style={solid,fill=Paired-11-5},mark=pentagon*\\%
 Paired-11-6!70!black, every mark/.append style={solid,fill=Paired-11-6},mark=x\\%
 Paired-11-7!70!black, every mark/.append style={solid,fill=Paired-11-7},mark=+\\%
 Paired-11-8!70!black, every mark/.append style={solid,fill=Paired-11-8},mark=star\\%
 Paired-11-9!70!black, every mark/.append style={solid,fill=Paired-11-9},mark=oplus*\\%
 Paired-11-10!70!black, every mark/.append style={solid,fill=Paired-11-10},mark=otimes*\\%
}
\pgfplotscreateplotcyclelist{Mark-Paired-12}{%
 Paired-12-1!70!black, every mark/.append style={solid,fill=Paired-12-1},mark=*\\%
 Paired-12-2!70!black, every mark/.append style={solid,fill=Paired-12-2},mark=square*\\%
 Paired-12-3!70!black, every mark/.append style={solid,fill=Paired-12-3},mark=triangle*\\%
 Paired-12-4!70!black, every mark/.append style={solid,fill=Paired-12-4},mark=diamond*\\%
 Paired-12-5!70!black, every mark/.append style={solid,fill=Paired-12-5},mark=pentagon*\\%
 Paired-12-6!70!black, every mark/.append style={solid,fill=Paired-12-6},mark=x\\%
 Paired-12-7!70!black, every mark/.append style={solid,fill=Paired-12-7},mark=+\\%
 Paired-12-8!70!black, every mark/.append style={solid,fill=Paired-12-8},mark=star\\%
 Paired-12-9!70!black, every mark/.append style={solid,fill=Paired-12-9},mark=oplus*\\%
 Paired-12-10!70!black, every mark/.append style={solid,fill=Paired-12-10},mark=otimes*\\%
}
\pgfplotscreateplotcyclelist{Mark-Accent-3}{%
 Accent-3-1!70!black, every mark/.append style={solid,fill=Accent-3-1},mark=*\\%
 Accent-3-2!70!black, every mark/.append style={solid,fill=Accent-3-2},mark=square*\\%
 Accent-3-3!70!black, every mark/.append style={solid,fill=Accent-3-3},mark=triangle*\\%
}
\pgfplotscreateplotcyclelist{Mark-Accent-4}{%
 Accent-4-1!70!black, every mark/.append style={solid,fill=Accent-4-1},mark=*\\%
 Accent-4-2!70!black, every mark/.append style={solid,fill=Accent-4-2},mark=square*\\%
 Accent-4-3!70!black, every mark/.append style={solid,fill=Accent-4-3},mark=triangle*\\%
 Accent-4-4!70!black, every mark/.append style={solid,fill=Accent-4-4},mark=diamond*\\%
}
\pgfplotscreateplotcyclelist{Mark-Accent-5}{%
 Accent-5-1!70!black, every mark/.append style={solid,fill=Accent-5-1},mark=*\\%
 Accent-5-2!70!black, every mark/.append style={solid,fill=Accent-5-2},mark=square*\\%
 Accent-5-3!70!black, every mark/.append style={solid,fill=Accent-5-3},mark=triangle*\\%
 Accent-5-4!70!black, every mark/.append style={solid,fill=Accent-5-4},mark=diamond*\\%
 Accent-5-5!70!black, every mark/.append style={solid,fill=Accent-5-5},mark=pentagon*\\%
}
\pgfplotscreateplotcyclelist{Mark-Accent-6}{%
 Accent-6-1!70!black, every mark/.append style={solid,fill=Accent-6-1},mark=*\\%
 Accent-6-2!70!black, every mark/.append style={solid,fill=Accent-6-2},mark=square*\\%
 Accent-6-3!70!black, every mark/.append style={solid,fill=Accent-6-3},mark=triangle*\\%
 Accent-6-4!70!black, every mark/.append style={solid,fill=Accent-6-4},mark=diamond*\\%
 Accent-6-5!70!black, every mark/.append style={solid,fill=Accent-6-5},mark=pentagon*\\%
 Accent-6-6!70!black, every mark/.append style={solid,fill=Accent-6-6},mark=x\\%
}
\pgfplotscreateplotcyclelist{Mark-Accent-7}{%
 Accent-7-1!70!black, every mark/.append style={solid,fill=Accent-7-1},mark=*\\%
 Accent-7-2!70!black, every mark/.append style={solid,fill=Accent-7-2},mark=square*\\%
 Accent-7-3!70!black, every mark/.append style={solid,fill=Accent-7-3},mark=triangle*\\%
 Accent-7-4!70!black, every mark/.append style={solid,fill=Accent-7-4},mark=diamond*\\%
 Accent-7-5!70!black, every mark/.append style={solid,fill=Accent-7-5},mark=pentagon*\\%
 Accent-7-6!70!black, every mark/.append style={solid,fill=Accent-7-6},mark=x\\%
 Accent-7-7!70!black, every mark/.append style={solid,fill=Accent-7-7},mark=+\\%
}
\pgfplotscreateplotcyclelist{Mark-Accent-8}{%
 Accent-8-1!70!black, every mark/.append style={solid,fill=Accent-8-1},mark=*\\%
 Accent-8-2!70!black, every mark/.append style={solid,fill=Accent-8-2},mark=square*\\%
 Accent-8-3!70!black, every mark/.append style={solid,fill=Accent-8-3},mark=triangle*\\%
 Accent-8-4!70!black, every mark/.append style={solid,fill=Accent-8-4},mark=diamond*\\%
 Accent-8-5!70!black, every mark/.append style={solid,fill=Accent-8-5},mark=pentagon*\\%
 Accent-8-6!70!black, every mark/.append style={solid,fill=Accent-8-6},mark=x\\%
 Accent-8-7!70!black, every mark/.append style={solid,fill=Accent-8-7},mark=+\\%
 Accent-8-8!70!black, every mark/.append style={solid,fill=Accent-8-8},mark=star\\%
}
\pgfplotsset{
colormap={YlGn-3}{
  color=(YlGn-3-1);
  color=(YlGn-3-2);
  color=(YlGn-3-3);
},
colormap={YlGn-4}{
  color=(YlGn-4-1);
  color=(YlGn-4-2);
  color=(YlGn-4-3);
  color=(YlGn-4-4);
},
colormap={YlGn-5}{
  color=(YlGn-5-1);
  color=(YlGn-5-2);
  color=(YlGn-5-3);
  color=(YlGn-5-4);
  color=(YlGn-5-5);
},
colormap={YlGn-6}{
  color=(YlGn-6-1);
  color=(YlGn-6-2);
  color=(YlGn-6-3);
  color=(YlGn-6-4);
  color=(YlGn-6-5);
  color=(YlGn-6-6);
},
colormap={YlGn-7}{
  color=(YlGn-7-1);
  color=(YlGn-7-2);
  color=(YlGn-7-3);
  color=(YlGn-7-4);
  color=(YlGn-7-5);
  color=(YlGn-7-6);
  color=(YlGn-7-7);
},
colormap={YlGn-8}{
  color=(YlGn-8-1);
  color=(YlGn-8-2);
  color=(YlGn-8-3);
  color=(YlGn-8-4);
  color=(YlGn-8-5);
  color=(YlGn-8-6);
  color=(YlGn-8-7);
  color=(YlGn-8-8);
},
colormap={YlGn-9}{
  color=(YlGn-9-1);
  color=(YlGn-9-2);
  color=(YlGn-9-3);
  color=(YlGn-9-4);
  color=(YlGn-9-5);
  color=(YlGn-9-6);
  color=(YlGn-9-7);
  color=(YlGn-9-8);
  color=(YlGn-9-9);
},
colormap={YlGnBu-3}{
  color=(YlGnBu-3-1);
  color=(YlGnBu-3-2);
  color=(YlGnBu-3-3);
},
colormap={YlGnBu-4}{
  color=(YlGnBu-4-1);
  color=(YlGnBu-4-2);
  color=(YlGnBu-4-3);
  color=(YlGnBu-4-4);
},
colormap={YlGnBu-5}{
  color=(YlGnBu-5-1);
  color=(YlGnBu-5-2);
  color=(YlGnBu-5-3);
  color=(YlGnBu-5-4);
  color=(YlGnBu-5-5);
},
colormap={YlGnBu-6}{
  color=(YlGnBu-6-1);
  color=(YlGnBu-6-2);
  color=(YlGnBu-6-3);
  color=(YlGnBu-6-4);
  color=(YlGnBu-6-5);
  color=(YlGnBu-6-6);
},
colormap={YlGnBu-7}{
  color=(YlGnBu-7-1);
  color=(YlGnBu-7-2);
  color=(YlGnBu-7-3);
  color=(YlGnBu-7-4);
  color=(YlGnBu-7-5);
  color=(YlGnBu-7-6);
  color=(YlGnBu-7-7);
},
colormap={YlGnBu-8}{
  color=(YlGnBu-8-1);
  color=(YlGnBu-8-2);
  color=(YlGnBu-8-3);
  color=(YlGnBu-8-4);
  color=(YlGnBu-8-5);
  color=(YlGnBu-8-6);
  color=(YlGnBu-8-7);
  color=(YlGnBu-8-8);
},
colormap={YlGnBu-9}{
  color=(YlGnBu-9-1);
  color=(YlGnBu-9-2);
  color=(YlGnBu-9-3);
  color=(YlGnBu-9-4);
  color=(YlGnBu-9-5);
  color=(YlGnBu-9-6);
  color=(YlGnBu-9-7);
  color=(YlGnBu-9-8);
  color=(YlGnBu-9-9);
},
colormap={GnBu-3}{
  color=(GnBu-3-1);
  color=(GnBu-3-2);
  color=(GnBu-3-3);
},
colormap={GnBu-4}{
  color=(GnBu-4-1);
  color=(GnBu-4-2);
  color=(GnBu-4-3);
  color=(GnBu-4-4);
},
colormap={GnBu-5}{
  color=(GnBu-5-1);
  color=(GnBu-5-2);
  color=(GnBu-5-3);
  color=(GnBu-5-4);
  color=(GnBu-5-5);
},
colormap={GnBu-6}{
  color=(GnBu-6-1);
  color=(GnBu-6-2);
  color=(GnBu-6-3);
  color=(GnBu-6-4);
  color=(GnBu-6-5);
  color=(GnBu-6-6);
},
colormap={GnBu-7}{
  color=(GnBu-7-1);
  color=(GnBu-7-2);
  color=(GnBu-7-3);
  color=(GnBu-7-4);
  color=(GnBu-7-5);
  color=(GnBu-7-6);
  color=(GnBu-7-7);
},
colormap={GnBu-8}{
  color=(GnBu-8-1);
  color=(GnBu-8-2);
  color=(GnBu-8-3);
  color=(GnBu-8-4);
  color=(GnBu-8-5);
  color=(GnBu-8-6);
  color=(GnBu-8-7);
  color=(GnBu-8-8);
},
colormap={GnBu-9}{
  color=(GnBu-9-1);
  color=(GnBu-9-2);
  color=(GnBu-9-3);
  color=(GnBu-9-4);
  color=(GnBu-9-5);
  color=(GnBu-9-6);
  color=(GnBu-9-7);
  color=(GnBu-9-8);
  color=(GnBu-9-9);
},
colormap={BuGn-3}{
  color=(BuGn-3-1);
  color=(BuGn-3-2);
  color=(BuGn-3-3);
},
colormap={BuGn-4}{
  color=(BuGn-4-1);
  color=(BuGn-4-2);
  color=(BuGn-4-3);
  color=(BuGn-4-4);
},
colormap={BuGn-5}{
  color=(BuGn-5-1);
  color=(BuGn-5-2);
  color=(BuGn-5-3);
  color=(BuGn-5-4);
  color=(BuGn-5-5);
},
colormap={BuGn-6}{
  color=(BuGn-6-1);
  color=(BuGn-6-2);
  color=(BuGn-6-3);
  color=(BuGn-6-4);
  color=(BuGn-6-5);
  color=(BuGn-6-6);
},
colormap={BuGn-7}{
  color=(BuGn-7-1);
  color=(BuGn-7-2);
  color=(BuGn-7-3);
  color=(BuGn-7-4);
  color=(BuGn-7-5);
  color=(BuGn-7-6);
  color=(BuGn-7-7);
},
colormap={BuGn-8}{
  color=(BuGn-8-1);
  color=(BuGn-8-2);
  color=(BuGn-8-3);
  color=(BuGn-8-4);
  color=(BuGn-8-5);
  color=(BuGn-8-6);
  color=(BuGn-8-7);
  color=(BuGn-8-8);
},
colormap={BuGn-9}{
  color=(BuGn-9-1);
  color=(BuGn-9-2);
  color=(BuGn-9-3);
  color=(BuGn-9-4);
  color=(BuGn-9-5);
  color=(BuGn-9-6);
  color=(BuGn-9-7);
  color=(BuGn-9-8);
  color=(BuGn-9-9);
},
colormap={PuBuGn-3}{
  color=(PuBuGn-3-1);
  color=(PuBuGn-3-2);
  color=(PuBuGn-3-3);
},
colormap={PuBuGn-4}{
  color=(PuBuGn-4-1);
  color=(PuBuGn-4-2);
  color=(PuBuGn-4-3);
  color=(PuBuGn-4-4);
},
colormap={PuBuGn-5}{
  color=(PuBuGn-5-1);
  color=(PuBuGn-5-2);
  color=(PuBuGn-5-3);
  color=(PuBuGn-5-4);
  color=(PuBuGn-5-5);
},
colormap={PuBuGn-6}{
  color=(PuBuGn-6-1);
  color=(PuBuGn-6-2);
  color=(PuBuGn-6-3);
  color=(PuBuGn-6-4);
  color=(PuBuGn-6-5);
  color=(PuBuGn-6-6);
},
colormap={PuBuGn-7}{
  color=(PuBuGn-7-1);
  color=(PuBuGn-7-2);
  color=(PuBuGn-7-3);
  color=(PuBuGn-7-4);
  color=(PuBuGn-7-5);
  color=(PuBuGn-7-6);
  color=(PuBuGn-7-7);
},
colormap={PuBuGn-8}{
  color=(PuBuGn-8-1);
  color=(PuBuGn-8-2);
  color=(PuBuGn-8-3);
  color=(PuBuGn-8-4);
  color=(PuBuGn-8-5);
  color=(PuBuGn-8-6);
  color=(PuBuGn-8-7);
  color=(PuBuGn-8-8);
},
colormap={PuBuGn-9}{
  color=(PuBuGn-9-1);
  color=(PuBuGn-9-2);
  color=(PuBuGn-9-3);
  color=(PuBuGn-9-4);
  color=(PuBuGn-9-5);
  color=(PuBuGn-9-6);
  color=(PuBuGn-9-7);
  color=(PuBuGn-9-8);
  color=(PuBuGn-9-9);
},
colormap={PuBu-3}{
  color=(PuBu-3-1);
  color=(PuBu-3-2);
  color=(PuBu-3-3);
},
colormap={PuBu-4}{
  color=(PuBu-4-1);
  color=(PuBu-4-2);
  color=(PuBu-4-3);
  color=(PuBu-4-4);
},
colormap={PuBu-5}{
  color=(PuBu-5-1);
  color=(PuBu-5-2);
  color=(PuBu-5-3);
  color=(PuBu-5-4);
  color=(PuBu-5-5);
},
colormap={PuBu-6}{
  color=(PuBu-6-1);
  color=(PuBu-6-2);
  color=(PuBu-6-3);
  color=(PuBu-6-4);
  color=(PuBu-6-5);
  color=(PuBu-6-6);
},
colormap={PuBu-7}{
  color=(PuBu-7-1);
  color=(PuBu-7-2);
  color=(PuBu-7-3);
  color=(PuBu-7-4);
  color=(PuBu-7-5);
  color=(PuBu-7-6);
  color=(PuBu-7-7);
},
colormap={PuBu-8}{
  color=(PuBu-8-1);
  color=(PuBu-8-2);
  color=(PuBu-8-3);
  color=(PuBu-8-4);
  color=(PuBu-8-5);
  color=(PuBu-8-6);
  color=(PuBu-8-7);
  color=(PuBu-8-8);
},
colormap={PuBu-9}{
  color=(PuBu-9-1);
  color=(PuBu-9-2);
  color=(PuBu-9-3);
  color=(PuBu-9-4);
  color=(PuBu-9-5);
  color=(PuBu-9-6);
  color=(PuBu-9-7);
  color=(PuBu-9-8);
  color=(PuBu-9-9);
},
colormap={BuPu-3}{
  color=(BuPu-3-1);
  color=(BuPu-3-2);
  color=(BuPu-3-3);
},
colormap={BuPu-4}{
  color=(BuPu-4-1);
  color=(BuPu-4-2);
  color=(BuPu-4-3);
  color=(BuPu-4-4);
},
colormap={BuPu-5}{
  color=(BuPu-5-1);
  color=(BuPu-5-2);
  color=(BuPu-5-3);
  color=(BuPu-5-4);
  color=(BuPu-5-5);
},
colormap={BuPu-6}{
  color=(BuPu-6-1);
  color=(BuPu-6-2);
  color=(BuPu-6-3);
  color=(BuPu-6-4);
  color=(BuPu-6-5);
  color=(BuPu-6-6);
},
colormap={BuPu-7}{
  color=(BuPu-7-1);
  color=(BuPu-7-2);
  color=(BuPu-7-3);
  color=(BuPu-7-4);
  color=(BuPu-7-5);
  color=(BuPu-7-6);
  color=(BuPu-7-7);
},
colormap={BuPu-8}{
  color=(BuPu-8-1);
  color=(BuPu-8-2);
  color=(BuPu-8-3);
  color=(BuPu-8-4);
  color=(BuPu-8-5);
  color=(BuPu-8-6);
  color=(BuPu-8-7);
  color=(BuPu-8-8);
},
colormap={BuPu-9}{
  color=(BuPu-9-1);
  color=(BuPu-9-2);
  color=(BuPu-9-3);
  color=(BuPu-9-4);
  color=(BuPu-9-5);
  color=(BuPu-9-6);
  color=(BuPu-9-7);
  color=(BuPu-9-8);
  color=(BuPu-9-9);
},
colormap={RdPu-3}{
  color=(RdPu-3-1);
  color=(RdPu-3-2);
  color=(RdPu-3-3);
},
colormap={RdPu-4}{
  color=(RdPu-4-1);
  color=(RdPu-4-2);
  color=(RdPu-4-3);
  color=(RdPu-4-4);
},
colormap={RdPu-5}{
  color=(RdPu-5-1);
  color=(RdPu-5-2);
  color=(RdPu-5-3);
  color=(RdPu-5-4);
  color=(RdPu-5-5);
},
colormap={RdPu-6}{
  color=(RdPu-6-1);
  color=(RdPu-6-2);
  color=(RdPu-6-3);
  color=(RdPu-6-4);
  color=(RdPu-6-5);
  color=(RdPu-6-6);
},
colormap={RdPu-7}{
  color=(RdPu-7-1);
  color=(RdPu-7-2);
  color=(RdPu-7-3);
  color=(RdPu-7-4);
  color=(RdPu-7-5);
  color=(RdPu-7-6);
  color=(RdPu-7-7);
},
colormap={RdPu-8}{
  color=(RdPu-8-1);
  color=(RdPu-8-2);
  color=(RdPu-8-3);
  color=(RdPu-8-4);
  color=(RdPu-8-5);
  color=(RdPu-8-6);
  color=(RdPu-8-7);
  color=(RdPu-8-8);
},
colormap={RdPu-9}{
  color=(RdPu-9-1);
  color=(RdPu-9-2);
  color=(RdPu-9-3);
  color=(RdPu-9-4);
  color=(RdPu-9-5);
  color=(RdPu-9-6);
  color=(RdPu-9-7);
  color=(RdPu-9-8);
  color=(RdPu-9-9);
},
colormap={PuRd-3}{
  color=(PuRd-3-1);
  color=(PuRd-3-2);
  color=(PuRd-3-3);
},
colormap={PuRd-4}{
  color=(PuRd-4-1);
  color=(PuRd-4-2);
  color=(PuRd-4-3);
  color=(PuRd-4-4);
},
colormap={PuRd-5}{
  color=(PuRd-5-1);
  color=(PuRd-5-2);
  color=(PuRd-5-3);
  color=(PuRd-5-4);
  color=(PuRd-5-5);
},
colormap={PuRd-6}{
  color=(PuRd-6-1);
  color=(PuRd-6-2);
  color=(PuRd-6-3);
  color=(PuRd-6-4);
  color=(PuRd-6-5);
  color=(PuRd-6-6);
},
colormap={PuRd-7}{
  color=(PuRd-7-1);
  color=(PuRd-7-2);
  color=(PuRd-7-3);
  color=(PuRd-7-4);
  color=(PuRd-7-5);
  color=(PuRd-7-6);
  color=(PuRd-7-7);
},
colormap={PuRd-8}{
  color=(PuRd-8-1);
  color=(PuRd-8-2);
  color=(PuRd-8-3);
  color=(PuRd-8-4);
  color=(PuRd-8-5);
  color=(PuRd-8-6);
  color=(PuRd-8-7);
  color=(PuRd-8-8);
},
colormap={PuRd-9}{
  color=(PuRd-9-1);
  color=(PuRd-9-2);
  color=(PuRd-9-3);
  color=(PuRd-9-4);
  color=(PuRd-9-5);
  color=(PuRd-9-6);
  color=(PuRd-9-7);
  color=(PuRd-9-8);
  color=(PuRd-9-9);
},
colormap={OrRd-3}{
  color=(OrRd-3-1);
  color=(OrRd-3-2);
  color=(OrRd-3-3);
},
colormap={OrRd-4}{
  color=(OrRd-4-1);
  color=(OrRd-4-2);
  color=(OrRd-4-3);
  color=(OrRd-4-4);
},
colormap={OrRd-5}{
  color=(OrRd-5-1);
  color=(OrRd-5-2);
  color=(OrRd-5-3);
  color=(OrRd-5-4);
  color=(OrRd-5-5);
},
colormap={OrRd-6}{
  color=(OrRd-6-1);
  color=(OrRd-6-2);
  color=(OrRd-6-3);
  color=(OrRd-6-4);
  color=(OrRd-6-5);
  color=(OrRd-6-6);
},
colormap={OrRd-7}{
  color=(OrRd-7-1);
  color=(OrRd-7-2);
  color=(OrRd-7-3);
  color=(OrRd-7-4);
  color=(OrRd-7-5);
  color=(OrRd-7-6);
  color=(OrRd-7-7);
},
colormap={OrRd-8}{
  color=(OrRd-8-1);
  color=(OrRd-8-2);
  color=(OrRd-8-3);
  color=(OrRd-8-4);
  color=(OrRd-8-5);
  color=(OrRd-8-6);
  color=(OrRd-8-7);
  color=(OrRd-8-8);
},
colormap={OrRd-9}{
  color=(OrRd-9-1);
  color=(OrRd-9-2);
  color=(OrRd-9-3);
  color=(OrRd-9-4);
  color=(OrRd-9-5);
  color=(OrRd-9-6);
  color=(OrRd-9-7);
  color=(OrRd-9-8);
  color=(OrRd-9-9);
},
colormap={YlOrRd-3}{
  color=(YlOrRd-3-1);
  color=(YlOrRd-3-2);
  color=(YlOrRd-3-3);
},
colormap={YlOrRd-4}{
  color=(YlOrRd-4-1);
  color=(YlOrRd-4-2);
  color=(YlOrRd-4-3);
  color=(YlOrRd-4-4);
},
colormap={YlOrRd-5}{
  color=(YlOrRd-5-1);
  color=(YlOrRd-5-2);
  color=(YlOrRd-5-3);
  color=(YlOrRd-5-4);
  color=(YlOrRd-5-5);
},
colormap={YlOrRd-6}{
  color=(YlOrRd-6-1);
  color=(YlOrRd-6-2);
  color=(YlOrRd-6-3);
  color=(YlOrRd-6-4);
  color=(YlOrRd-6-5);
  color=(YlOrRd-6-6);
},
colormap={YlOrRd-7}{
  color=(YlOrRd-7-1);
  color=(YlOrRd-7-2);
  color=(YlOrRd-7-3);
  color=(YlOrRd-7-4);
  color=(YlOrRd-7-5);
  color=(YlOrRd-7-6);
  color=(YlOrRd-7-7);
},
colormap={YlOrRd-8}{
  color=(YlOrRd-8-1);
  color=(YlOrRd-8-2);
  color=(YlOrRd-8-3);
  color=(YlOrRd-8-4);
  color=(YlOrRd-8-5);
  color=(YlOrRd-8-6);
  color=(YlOrRd-8-7);
  color=(YlOrRd-8-8);
},
colormap={YlOrRd-9}{
  color=(YlOrRd-9-1);
  color=(YlOrRd-9-2);
  color=(YlOrRd-9-3);
  color=(YlOrRd-9-4);
  color=(YlOrRd-9-5);
  color=(YlOrRd-9-6);
  color=(YlOrRd-9-7);
  color=(YlOrRd-9-8);
  color=(YlOrRd-9-9);
},
colormap={YlOrBr-3}{
  color=(YlOrBr-3-1);
  color=(YlOrBr-3-2);
  color=(YlOrBr-3-3);
},
colormap={YlOrBr-4}{
  color=(YlOrBr-4-1);
  color=(YlOrBr-4-2);
  color=(YlOrBr-4-3);
  color=(YlOrBr-4-4);
},
colormap={YlOrBr-5}{
  color=(YlOrBr-5-1);
  color=(YlOrBr-5-2);
  color=(YlOrBr-5-3);
  color=(YlOrBr-5-4);
  color=(YlOrBr-5-5);
},
colormap={YlOrBr-6}{
  color=(YlOrBr-6-1);
  color=(YlOrBr-6-2);
  color=(YlOrBr-6-3);
  color=(YlOrBr-6-4);
  color=(YlOrBr-6-5);
  color=(YlOrBr-6-6);
},
colormap={YlOrBr-7}{
  color=(YlOrBr-7-1);
  color=(YlOrBr-7-2);
  color=(YlOrBr-7-3);
  color=(YlOrBr-7-4);
  color=(YlOrBr-7-5);
  color=(YlOrBr-7-6);
  color=(YlOrBr-7-7);
},
colormap={YlOrBr-8}{
  color=(YlOrBr-8-1);
  color=(YlOrBr-8-2);
  color=(YlOrBr-8-3);
  color=(YlOrBr-8-4);
  color=(YlOrBr-8-5);
  color=(YlOrBr-8-6);
  color=(YlOrBr-8-7);
  color=(YlOrBr-8-8);
},
colormap={YlOrBr-9}{
  color=(YlOrBr-9-1);
  color=(YlOrBr-9-2);
  color=(YlOrBr-9-3);
  color=(YlOrBr-9-4);
  color=(YlOrBr-9-5);
  color=(YlOrBr-9-6);
  color=(YlOrBr-9-7);
  color=(YlOrBr-9-8);
  color=(YlOrBr-9-9);
},
colormap={Purples-3}{
  color=(Purples-3-1);
  color=(Purples-3-2);
  color=(Purples-3-3);
},
colormap={Purples-4}{
  color=(Purples-4-1);
  color=(Purples-4-2);
  color=(Purples-4-3);
  color=(Purples-4-4);
},
colormap={Purples-5}{
  color=(Purples-5-1);
  color=(Purples-5-2);
  color=(Purples-5-3);
  color=(Purples-5-4);
  color=(Purples-5-5);
},
colormap={Purples-6}{
  color=(Purples-6-1);
  color=(Purples-6-2);
  color=(Purples-6-3);
  color=(Purples-6-4);
  color=(Purples-6-5);
  color=(Purples-6-6);
},
colormap={Purples-7}{
  color=(Purples-7-1);
  color=(Purples-7-2);
  color=(Purples-7-3);
  color=(Purples-7-4);
  color=(Purples-7-5);
  color=(Purples-7-6);
  color=(Purples-7-7);
},
colormap={Purples-8}{
  color=(Purples-8-1);
  color=(Purples-8-2);
  color=(Purples-8-3);
  color=(Purples-8-4);
  color=(Purples-8-5);
  color=(Purples-8-6);
  color=(Purples-8-7);
  color=(Purples-8-8);
},
colormap={Purples-9}{
  color=(Purples-9-1);
  color=(Purples-9-2);
  color=(Purples-9-3);
  color=(Purples-9-4);
  color=(Purples-9-5);
  color=(Purples-9-6);
  color=(Purples-9-7);
  color=(Purples-9-8);
  color=(Purples-9-9);
},
colormap={Blues-3}{
  color=(Blues-3-1);
  color=(Blues-3-2);
  color=(Blues-3-3);
},
colormap={Blues-4}{
  color=(Blues-4-1);
  color=(Blues-4-2);
  color=(Blues-4-3);
  color=(Blues-4-4);
},
colormap={Blues-5}{
  color=(Blues-5-1);
  color=(Blues-5-2);
  color=(Blues-5-3);
  color=(Blues-5-4);
  color=(Blues-5-5);
},
colormap={Blues-6}{
  color=(Blues-6-1);
  color=(Blues-6-2);
  color=(Blues-6-3);
  color=(Blues-6-4);
  color=(Blues-6-5);
  color=(Blues-6-6);
},
colormap={Blues-7}{
  color=(Blues-7-1);
  color=(Blues-7-2);
  color=(Blues-7-3);
  color=(Blues-7-4);
  color=(Blues-7-5);
  color=(Blues-7-6);
  color=(Blues-7-7);
},
colormap={Blues-8}{
  color=(Blues-8-1);
  color=(Blues-8-2);
  color=(Blues-8-3);
  color=(Blues-8-4);
  color=(Blues-8-5);
  color=(Blues-8-6);
  color=(Blues-8-7);
  color=(Blues-8-8);
},
colormap={Blues-9}{
  color=(Blues-9-1);
  color=(Blues-9-2);
  color=(Blues-9-3);
  color=(Blues-9-4);
  color=(Blues-9-5);
  color=(Blues-9-6);
  color=(Blues-9-7);
  color=(Blues-9-8);
  color=(Blues-9-9);
},
colormap={Greens-3}{
  color=(Greens-3-1);
  color=(Greens-3-2);
  color=(Greens-3-3);
},
colormap={Greens-4}{
  color=(Greens-4-1);
  color=(Greens-4-2);
  color=(Greens-4-3);
  color=(Greens-4-4);
},
colormap={Greens-5}{
  color=(Greens-5-1);
  color=(Greens-5-2);
  color=(Greens-5-3);
  color=(Greens-5-4);
  color=(Greens-5-5);
},
colormap={Greens-6}{
  color=(Greens-6-1);
  color=(Greens-6-2);
  color=(Greens-6-3);
  color=(Greens-6-4);
  color=(Greens-6-5);
  color=(Greens-6-6);
},
colormap={Greens-7}{
  color=(Greens-7-1);
  color=(Greens-7-2);
  color=(Greens-7-3);
  color=(Greens-7-4);
  color=(Greens-7-5);
  color=(Greens-7-6);
  color=(Greens-7-7);
},
colormap={Greens-8}{
  color=(Greens-8-1);
  color=(Greens-8-2);
  color=(Greens-8-3);
  color=(Greens-8-4);
  color=(Greens-8-5);
  color=(Greens-8-6);
  color=(Greens-8-7);
  color=(Greens-8-8);
},
colormap={Greens-9}{
  color=(Greens-9-1);
  color=(Greens-9-2);
  color=(Greens-9-3);
  color=(Greens-9-4);
  color=(Greens-9-5);
  color=(Greens-9-6);
  color=(Greens-9-7);
  color=(Greens-9-8);
  color=(Greens-9-9);
},
colormap={Oranges-3}{
  color=(Oranges-3-1);
  color=(Oranges-3-2);
  color=(Oranges-3-3);
},
colormap={Oranges-4}{
  color=(Oranges-4-1);
  color=(Oranges-4-2);
  color=(Oranges-4-3);
  color=(Oranges-4-4);
},
colormap={Oranges-5}{
  color=(Oranges-5-1);
  color=(Oranges-5-2);
  color=(Oranges-5-3);
  color=(Oranges-5-4);
  color=(Oranges-5-5);
},
colormap={Oranges-6}{
  color=(Oranges-6-1);
  color=(Oranges-6-2);
  color=(Oranges-6-3);
  color=(Oranges-6-4);
  color=(Oranges-6-5);
  color=(Oranges-6-6);
},
colormap={Oranges-7}{
  color=(Oranges-7-1);
  color=(Oranges-7-2);
  color=(Oranges-7-3);
  color=(Oranges-7-4);
  color=(Oranges-7-5);
  color=(Oranges-7-6);
  color=(Oranges-7-7);
},
colormap={Oranges-8}{
  color=(Oranges-8-1);
  color=(Oranges-8-2);
  color=(Oranges-8-3);
  color=(Oranges-8-4);
  color=(Oranges-8-5);
  color=(Oranges-8-6);
  color=(Oranges-8-7);
  color=(Oranges-8-8);
},
colormap={Oranges-9}{
  color=(Oranges-9-1);
  color=(Oranges-9-2);
  color=(Oranges-9-3);
  color=(Oranges-9-4);
  color=(Oranges-9-5);
  color=(Oranges-9-6);
  color=(Oranges-9-7);
  color=(Oranges-9-8);
  color=(Oranges-9-9);
},
colormap={Reds-3}{
  color=(Reds-3-1);
  color=(Reds-3-2);
  color=(Reds-3-3);
},
colormap={Reds-4}{
  color=(Reds-4-1);
  color=(Reds-4-2);
  color=(Reds-4-3);
  color=(Reds-4-4);
},
colormap={Reds-5}{
  color=(Reds-5-1);
  color=(Reds-5-2);
  color=(Reds-5-3);
  color=(Reds-5-4);
  color=(Reds-5-5);
},
colormap={Reds-6}{
  color=(Reds-6-1);
  color=(Reds-6-2);
  color=(Reds-6-3);
  color=(Reds-6-4);
  color=(Reds-6-5);
  color=(Reds-6-6);
},
colormap={Reds-7}{
  color=(Reds-7-1);
  color=(Reds-7-2);
  color=(Reds-7-3);
  color=(Reds-7-4);
  color=(Reds-7-5);
  color=(Reds-7-6);
  color=(Reds-7-7);
},
colormap={Reds-8}{
  color=(Reds-8-1);
  color=(Reds-8-2);
  color=(Reds-8-3);
  color=(Reds-8-4);
  color=(Reds-8-5);
  color=(Reds-8-6);
  color=(Reds-8-7);
  color=(Reds-8-8);
},
colormap={Reds-9}{
  color=(Reds-9-1);
  color=(Reds-9-2);
  color=(Reds-9-3);
  color=(Reds-9-4);
  color=(Reds-9-5);
  color=(Reds-9-6);
  color=(Reds-9-7);
  color=(Reds-9-8);
  color=(Reds-9-9);
},
colormap={Greys-3}{
  color=(Greys-3-1);
  color=(Greys-3-2);
  color=(Greys-3-3);
},
colormap={Greys-4}{
  color=(Greys-4-1);
  color=(Greys-4-2);
  color=(Greys-4-3);
  color=(Greys-4-4);
},
colormap={Greys-5}{
  color=(Greys-5-1);
  color=(Greys-5-2);
  color=(Greys-5-3);
  color=(Greys-5-4);
  color=(Greys-5-5);
},
colormap={Greys-6}{
  color=(Greys-6-1);
  color=(Greys-6-2);
  color=(Greys-6-3);
  color=(Greys-6-4);
  color=(Greys-6-5);
  color=(Greys-6-6);
},
colormap={Greys-7}{
  color=(Greys-7-1);
  color=(Greys-7-2);
  color=(Greys-7-3);
  color=(Greys-7-4);
  color=(Greys-7-5);
  color=(Greys-7-6);
  color=(Greys-7-7);
},
colormap={Greys-8}{
  color=(Greys-8-1);
  color=(Greys-8-2);
  color=(Greys-8-3);
  color=(Greys-8-4);
  color=(Greys-8-5);
  color=(Greys-8-6);
  color=(Greys-8-7);
  color=(Greys-8-8);
},
colormap={Greys-9}{
  color=(Greys-9-1);
  color=(Greys-9-2);
  color=(Greys-9-3);
  color=(Greys-9-4);
  color=(Greys-9-5);
  color=(Greys-9-6);
  color=(Greys-9-7);
  color=(Greys-9-8);
  color=(Greys-9-9);
},
colormap={PuOr-3}{
  color=(PuOr-3-1);
  color=(PuOr-3-2);
  color=(PuOr-3-3);
},
colormap={PuOr-4}{
  color=(PuOr-4-1);
  color=(PuOr-4-2);
  color=(PuOr-4-3);
  color=(PuOr-4-4);
},
colormap={PuOr-5}{
  color=(PuOr-5-1);
  color=(PuOr-5-2);
  color=(PuOr-5-3);
  color=(PuOr-5-4);
  color=(PuOr-5-5);
},
colormap={PuOr-6}{
  color=(PuOr-6-1);
  color=(PuOr-6-2);
  color=(PuOr-6-3);
  color=(PuOr-6-4);
  color=(PuOr-6-5);
  color=(PuOr-6-6);
},
colormap={PuOr-7}{
  color=(PuOr-7-1);
  color=(PuOr-7-2);
  color=(PuOr-7-3);
  color=(PuOr-7-4);
  color=(PuOr-7-5);
  color=(PuOr-7-6);
  color=(PuOr-7-7);
},
colormap={PuOr-8}{
  color=(PuOr-8-1);
  color=(PuOr-8-2);
  color=(PuOr-8-3);
  color=(PuOr-8-4);
  color=(PuOr-8-5);
  color=(PuOr-8-6);
  color=(PuOr-8-7);
  color=(PuOr-8-8);
},
colormap={PuOr-9}{
  color=(PuOr-9-1);
  color=(PuOr-9-2);
  color=(PuOr-9-3);
  color=(PuOr-9-4);
  color=(PuOr-9-5);
  color=(PuOr-9-6);
  color=(PuOr-9-7);
  color=(PuOr-9-8);
  color=(PuOr-9-9);
},
colormap={PuOr-10}{
  color=(PuOr-10-1);
  color=(PuOr-10-2);
  color=(PuOr-10-3);
  color=(PuOr-10-4);
  color=(PuOr-10-5);
  color=(PuOr-10-6);
  color=(PuOr-10-7);
  color=(PuOr-10-8);
  color=(PuOr-10-9);
  color=(PuOr-10-10);
},
colormap={PuOr-11}{
  color=(PuOr-11-1);
  color=(PuOr-11-2);
  color=(PuOr-11-3);
  color=(PuOr-11-4);
  color=(PuOr-11-5);
  color=(PuOr-11-6);
  color=(PuOr-11-7);
  color=(PuOr-11-8);
  color=(PuOr-11-9);
  color=(PuOr-11-10);
  color=(PuOr-11-11);
},
colormap={BrBG-3}{
  color=(BrBG-3-1);
  color=(BrBG-3-2);
  color=(BrBG-3-3);
},
colormap={BrBG-4}{
  color=(BrBG-4-1);
  color=(BrBG-4-2);
  color=(BrBG-4-3);
  color=(BrBG-4-4);
},
colormap={BrBG-5}{
  color=(BrBG-5-1);
  color=(BrBG-5-2);
  color=(BrBG-5-3);
  color=(BrBG-5-4);
  color=(BrBG-5-5);
},
colormap={BrBG-6}{
  color=(BrBG-6-1);
  color=(BrBG-6-2);
  color=(BrBG-6-3);
  color=(BrBG-6-4);
  color=(BrBG-6-5);
  color=(BrBG-6-6);
},
colormap={BrBG-7}{
  color=(BrBG-7-1);
  color=(BrBG-7-2);
  color=(BrBG-7-3);
  color=(BrBG-7-4);
  color=(BrBG-7-5);
  color=(BrBG-7-6);
  color=(BrBG-7-7);
},
colormap={BrBG-8}{
  color=(BrBG-8-1);
  color=(BrBG-8-2);
  color=(BrBG-8-3);
  color=(BrBG-8-4);
  color=(BrBG-8-5);
  color=(BrBG-8-6);
  color=(BrBG-8-7);
  color=(BrBG-8-8);
},
colormap={BrBG-9}{
  color=(BrBG-9-1);
  color=(BrBG-9-2);
  color=(BrBG-9-3);
  color=(BrBG-9-4);
  color=(BrBG-9-5);
  color=(BrBG-9-6);
  color=(BrBG-9-7);
  color=(BrBG-9-8);
  color=(BrBG-9-9);
},
colormap={BrBG-10}{
  color=(BrBG-10-1);
  color=(BrBG-10-2);
  color=(BrBG-10-3);
  color=(BrBG-10-4);
  color=(BrBG-10-5);
  color=(BrBG-10-6);
  color=(BrBG-10-7);
  color=(BrBG-10-8);
  color=(BrBG-10-9);
  color=(BrBG-10-10);
},
colormap={BrBG-11}{
  color=(BrBG-11-1);
  color=(BrBG-11-2);
  color=(BrBG-11-3);
  color=(BrBG-11-4);
  color=(BrBG-11-5);
  color=(BrBG-11-6);
  color=(BrBG-11-7);
  color=(BrBG-11-8);
  color=(BrBG-11-9);
  color=(BrBG-11-10);
  color=(BrBG-11-11);
},
colormap={PRGn-3}{
  color=(PRGn-3-1);
  color=(PRGn-3-2);
  color=(PRGn-3-3);
},
colormap={PRGn-4}{
  color=(PRGn-4-1);
  color=(PRGn-4-2);
  color=(PRGn-4-3);
  color=(PRGn-4-4);
},
colormap={PRGn-5}{
  color=(PRGn-5-1);
  color=(PRGn-5-2);
  color=(PRGn-5-3);
  color=(PRGn-5-4);
  color=(PRGn-5-5);
},
colormap={PRGn-6}{
  color=(PRGn-6-1);
  color=(PRGn-6-2);
  color=(PRGn-6-3);
  color=(PRGn-6-4);
  color=(PRGn-6-5);
  color=(PRGn-6-6);
},
colormap={PRGn-7}{
  color=(PRGn-7-1);
  color=(PRGn-7-2);
  color=(PRGn-7-3);
  color=(PRGn-7-4);
  color=(PRGn-7-5);
  color=(PRGn-7-6);
  color=(PRGn-7-7);
},
colormap={PRGn-8}{
  color=(PRGn-8-1);
  color=(PRGn-8-2);
  color=(PRGn-8-3);
  color=(PRGn-8-4);
  color=(PRGn-8-5);
  color=(PRGn-8-6);
  color=(PRGn-8-7);
  color=(PRGn-8-8);
},
colormap={PRGn-9}{
  color=(PRGn-9-1);
  color=(PRGn-9-2);
  color=(PRGn-9-3);
  color=(PRGn-9-4);
  color=(PRGn-9-5);
  color=(PRGn-9-6);
  color=(PRGn-9-7);
  color=(PRGn-9-8);
  color=(PRGn-9-9);
},
colormap={PRGn-10}{
  color=(PRGn-10-1);
  color=(PRGn-10-2);
  color=(PRGn-10-3);
  color=(PRGn-10-4);
  color=(PRGn-10-5);
  color=(PRGn-10-6);
  color=(PRGn-10-7);
  color=(PRGn-10-8);
  color=(PRGn-10-9);
  color=(PRGn-10-10);
},
colormap={PRGn-11}{
  color=(PRGn-11-1);
  color=(PRGn-11-2);
  color=(PRGn-11-3);
  color=(PRGn-11-4);
  color=(PRGn-11-5);
  color=(PRGn-11-6);
  color=(PRGn-11-7);
  color=(PRGn-11-8);
  color=(PRGn-11-9);
  color=(PRGn-11-10);
  color=(PRGn-11-11);
},
colormap={PiYG-3}{
  color=(PiYG-3-1);
  color=(PiYG-3-2);
  color=(PiYG-3-3);
},
colormap={PiYG-4}{
  color=(PiYG-4-1);
  color=(PiYG-4-2);
  color=(PiYG-4-3);
  color=(PiYG-4-4);
},
colormap={PiYG-5}{
  color=(PiYG-5-1);
  color=(PiYG-5-2);
  color=(PiYG-5-3);
  color=(PiYG-5-4);
  color=(PiYG-5-5);
},
colormap={PiYG-6}{
  color=(PiYG-6-1);
  color=(PiYG-6-2);
  color=(PiYG-6-3);
  color=(PiYG-6-4);
  color=(PiYG-6-5);
  color=(PiYG-6-6);
},
colormap={PiYG-7}{
  color=(PiYG-7-1);
  color=(PiYG-7-2);
  color=(PiYG-7-3);
  color=(PiYG-7-4);
  color=(PiYG-7-5);
  color=(PiYG-7-6);
  color=(PiYG-7-7);
},
colormap={PiYG-8}{
  color=(PiYG-8-1);
  color=(PiYG-8-2);
  color=(PiYG-8-3);
  color=(PiYG-8-4);
  color=(PiYG-8-5);
  color=(PiYG-8-6);
  color=(PiYG-8-7);
  color=(PiYG-8-8);
},
colormap={PiYG-9}{
  color=(PiYG-9-1);
  color=(PiYG-9-2);
  color=(PiYG-9-3);
  color=(PiYG-9-4);
  color=(PiYG-9-5);
  color=(PiYG-9-6);
  color=(PiYG-9-7);
  color=(PiYG-9-8);
  color=(PiYG-9-9);
},
colormap={PiYG-10}{
  color=(PiYG-10-1);
  color=(PiYG-10-2);
  color=(PiYG-10-3);
  color=(PiYG-10-4);
  color=(PiYG-10-5);
  color=(PiYG-10-6);
  color=(PiYG-10-7);
  color=(PiYG-10-8);
  color=(PiYG-10-9);
  color=(PiYG-10-10);
},
colormap={PiYG-11}{
  color=(PiYG-11-1);
  color=(PiYG-11-2);
  color=(PiYG-11-3);
  color=(PiYG-11-4);
  color=(PiYG-11-5);
  color=(PiYG-11-6);
  color=(PiYG-11-7);
  color=(PiYG-11-8);
  color=(PiYG-11-9);
  color=(PiYG-11-10);
  color=(PiYG-11-11);
},
colormap={RdBu-3}{
  color=(RdBu-3-1);
  color=(RdBu-3-2);
  color=(RdBu-3-3);
},
colormap={RdBu-4}{
  color=(RdBu-4-1);
  color=(RdBu-4-2);
  color=(RdBu-4-3);
  color=(RdBu-4-4);
},
colormap={RdBu-5}{
  color=(RdBu-5-1);
  color=(RdBu-5-2);
  color=(RdBu-5-3);
  color=(RdBu-5-4);
  color=(RdBu-5-5);
},
colormap={RdBu-6}{
  color=(RdBu-6-1);
  color=(RdBu-6-2);
  color=(RdBu-6-3);
  color=(RdBu-6-4);
  color=(RdBu-6-5);
  color=(RdBu-6-6);
},
colormap={RdBu-7}{
  color=(RdBu-7-1);
  color=(RdBu-7-2);
  color=(RdBu-7-3);
  color=(RdBu-7-4);
  color=(RdBu-7-5);
  color=(RdBu-7-6);
  color=(RdBu-7-7);
},
colormap={RdBu-8}{
  color=(RdBu-8-1);
  color=(RdBu-8-2);
  color=(RdBu-8-3);
  color=(RdBu-8-4);
  color=(RdBu-8-5);
  color=(RdBu-8-6);
  color=(RdBu-8-7);
  color=(RdBu-8-8);
},
colormap={RdBu-9}{
  color=(RdBu-9-1);
  color=(RdBu-9-2);
  color=(RdBu-9-3);
  color=(RdBu-9-4);
  color=(RdBu-9-5);
  color=(RdBu-9-6);
  color=(RdBu-9-7);
  color=(RdBu-9-8);
  color=(RdBu-9-9);
},
colormap={RdBu-10}{
  color=(RdBu-10-1);
  color=(RdBu-10-2);
  color=(RdBu-10-3);
  color=(RdBu-10-4);
  color=(RdBu-10-5);
  color=(RdBu-10-6);
  color=(RdBu-10-7);
  color=(RdBu-10-8);
  color=(RdBu-10-9);
  color=(RdBu-10-10);
},
colormap={RdBu-11}{
  color=(RdBu-11-1);
  color=(RdBu-11-2);
  color=(RdBu-11-3);
  color=(RdBu-11-4);
  color=(RdBu-11-5);
  color=(RdBu-11-6);
  color=(RdBu-11-7);
  color=(RdBu-11-8);
  color=(RdBu-11-9);
  color=(RdBu-11-10);
  color=(RdBu-11-11);
},
colormap={RdGy-3}{
  color=(RdGy-3-1);
  color=(RdGy-3-2);
  color=(RdGy-3-3);
},
colormap={RdGy-4}{
  color=(RdGy-4-1);
  color=(RdGy-4-2);
  color=(RdGy-4-3);
  color=(RdGy-4-4);
},
colormap={RdGy-5}{
  color=(RdGy-5-1);
  color=(RdGy-5-2);
  color=(RdGy-5-3);
  color=(RdGy-5-4);
  color=(RdGy-5-5);
},
colormap={RdGy-6}{
  color=(RdGy-6-1);
  color=(RdGy-6-2);
  color=(RdGy-6-3);
  color=(RdGy-6-4);
  color=(RdGy-6-5);
  color=(RdGy-6-6);
},
colormap={RdGy-7}{
  color=(RdGy-7-1);
  color=(RdGy-7-2);
  color=(RdGy-7-3);
  color=(RdGy-7-4);
  color=(RdGy-7-5);
  color=(RdGy-7-6);
  color=(RdGy-7-7);
},
colormap={RdGy-8}{
  color=(RdGy-8-1);
  color=(RdGy-8-2);
  color=(RdGy-8-3);
  color=(RdGy-8-4);
  color=(RdGy-8-5);
  color=(RdGy-8-6);
  color=(RdGy-8-7);
  color=(RdGy-8-8);
},
colormap={RdGy-9}{
  color=(RdGy-9-1);
  color=(RdGy-9-2);
  color=(RdGy-9-3);
  color=(RdGy-9-4);
  color=(RdGy-9-5);
  color=(RdGy-9-6);
  color=(RdGy-9-7);
  color=(RdGy-9-8);
  color=(RdGy-9-9);
},
colormap={RdGy-10}{
  color=(RdGy-10-1);
  color=(RdGy-10-2);
  color=(RdGy-10-3);
  color=(RdGy-10-4);
  color=(RdGy-10-5);
  color=(RdGy-10-6);
  color=(RdGy-10-7);
  color=(RdGy-10-8);
  color=(RdGy-10-9);
  color=(RdGy-10-10);
},
colormap={RdGy-11}{
  color=(RdGy-11-1);
  color=(RdGy-11-2);
  color=(RdGy-11-3);
  color=(RdGy-11-4);
  color=(RdGy-11-5);
  color=(RdGy-11-6);
  color=(RdGy-11-7);
  color=(RdGy-11-8);
  color=(RdGy-11-9);
  color=(RdGy-11-10);
  color=(RdGy-11-11);
},
colormap={RdYlBu-3}{
  color=(RdYlBu-3-1);
  color=(RdYlBu-3-2);
  color=(RdYlBu-3-3);
},
colormap={RdYlBu-4}{
  color=(RdYlBu-4-1);
  color=(RdYlBu-4-2);
  color=(RdYlBu-4-3);
  color=(RdYlBu-4-4);
},
colormap={RdYlBu-5}{
  color=(RdYlBu-5-1);
  color=(RdYlBu-5-2);
  color=(RdYlBu-5-3);
  color=(RdYlBu-5-4);
  color=(RdYlBu-5-5);
},
colormap={RdYlBu-6}{
  color=(RdYlBu-6-1);
  color=(RdYlBu-6-2);
  color=(RdYlBu-6-3);
  color=(RdYlBu-6-4);
  color=(RdYlBu-6-5);
  color=(RdYlBu-6-6);
},
colormap={RdYlBu-7}{
  color=(RdYlBu-7-1);
  color=(RdYlBu-7-2);
  color=(RdYlBu-7-3);
  color=(RdYlBu-7-4);
  color=(RdYlBu-7-5);
  color=(RdYlBu-7-6);
  color=(RdYlBu-7-7);
},
colormap={RdYlBu-8}{
  color=(RdYlBu-8-1);
  color=(RdYlBu-8-2);
  color=(RdYlBu-8-3);
  color=(RdYlBu-8-4);
  color=(RdYlBu-8-5);
  color=(RdYlBu-8-6);
  color=(RdYlBu-8-7);
  color=(RdYlBu-8-8);
},
colormap={RdYlBu-9}{
  color=(RdYlBu-9-1);
  color=(RdYlBu-9-2);
  color=(RdYlBu-9-3);
  color=(RdYlBu-9-4);
  color=(RdYlBu-9-5);
  color=(RdYlBu-9-6);
  color=(RdYlBu-9-7);
  color=(RdYlBu-9-8);
  color=(RdYlBu-9-9);
},
colormap={RdYlBu-10}{
  color=(RdYlBu-10-1);
  color=(RdYlBu-10-2);
  color=(RdYlBu-10-3);
  color=(RdYlBu-10-4);
  color=(RdYlBu-10-5);
  color=(RdYlBu-10-6);
  color=(RdYlBu-10-7);
  color=(RdYlBu-10-8);
  color=(RdYlBu-10-9);
  color=(RdYlBu-10-10);
},
colormap={RdYlBu-11}{
  color=(RdYlBu-11-1);
  color=(RdYlBu-11-2);
  color=(RdYlBu-11-3);
  color=(RdYlBu-11-4);
  color=(RdYlBu-11-5);
  color=(RdYlBu-11-6);
  color=(RdYlBu-11-7);
  color=(RdYlBu-11-8);
  color=(RdYlBu-11-9);
  color=(RdYlBu-11-10);
  color=(RdYlBu-11-11);
},
colormap={Spectral-3}{
  color=(Spectral-3-1);
  color=(Spectral-3-2);
  color=(Spectral-3-3);
},
colormap={Spectral-4}{
  color=(Spectral-4-1);
  color=(Spectral-4-2);
  color=(Spectral-4-3);
  color=(Spectral-4-4);
},
colormap={Spectral-5}{
  color=(Spectral-5-1);
  color=(Spectral-5-2);
  color=(Spectral-5-3);
  color=(Spectral-5-4);
  color=(Spectral-5-5);
},
colormap={Spectral-6}{
  color=(Spectral-6-1);
  color=(Spectral-6-2);
  color=(Spectral-6-3);
  color=(Spectral-6-4);
  color=(Spectral-6-5);
  color=(Spectral-6-6);
},
colormap={Spectral-7}{
  color=(Spectral-7-1);
  color=(Spectral-7-2);
  color=(Spectral-7-3);
  color=(Spectral-7-4);
  color=(Spectral-7-5);
  color=(Spectral-7-6);
  color=(Spectral-7-7);
},
colormap={Spectral-8}{
  color=(Spectral-8-1);
  color=(Spectral-8-2);
  color=(Spectral-8-3);
  color=(Spectral-8-4);
  color=(Spectral-8-5);
  color=(Spectral-8-6);
  color=(Spectral-8-7);
  color=(Spectral-8-8);
},
colormap={Spectral-9}{
  color=(Spectral-9-1);
  color=(Spectral-9-2);
  color=(Spectral-9-3);
  color=(Spectral-9-4);
  color=(Spectral-9-5);
  color=(Spectral-9-6);
  color=(Spectral-9-7);
  color=(Spectral-9-8);
  color=(Spectral-9-9);
},
colormap={Spectral-10}{
  color=(Spectral-10-1);
  color=(Spectral-10-2);
  color=(Spectral-10-3);
  color=(Spectral-10-4);
  color=(Spectral-10-5);
  color=(Spectral-10-6);
  color=(Spectral-10-7);
  color=(Spectral-10-8);
  color=(Spectral-10-9);
  color=(Spectral-10-10);
},
colormap={Spectral-11}{
  color=(Spectral-11-1);
  color=(Spectral-11-2);
  color=(Spectral-11-3);
  color=(Spectral-11-4);
  color=(Spectral-11-5);
  color=(Spectral-11-6);
  color=(Spectral-11-7);
  color=(Spectral-11-8);
  color=(Spectral-11-9);
  color=(Spectral-11-10);
  color=(Spectral-11-11);
},
colormap={RdYlGn-3}{
  color=(RdYlGn-3-1);
  color=(RdYlGn-3-2);
  color=(RdYlGn-3-3);
},
colormap={RdYlGn-4}{
  color=(RdYlGn-4-1);
  color=(RdYlGn-4-2);
  color=(RdYlGn-4-3);
  color=(RdYlGn-4-4);
},
colormap={RdYlGn-5}{
  color=(RdYlGn-5-1);
  color=(RdYlGn-5-2);
  color=(RdYlGn-5-3);
  color=(RdYlGn-5-4);
  color=(RdYlGn-5-5);
},
colormap={RdYlGn-6}{
  color=(RdYlGn-6-1);
  color=(RdYlGn-6-2);
  color=(RdYlGn-6-3);
  color=(RdYlGn-6-4);
  color=(RdYlGn-6-5);
  color=(RdYlGn-6-6);
},
colormap={RdYlGn-7}{
  color=(RdYlGn-7-1);
  color=(RdYlGn-7-2);
  color=(RdYlGn-7-3);
  color=(RdYlGn-7-4);
  color=(RdYlGn-7-5);
  color=(RdYlGn-7-6);
  color=(RdYlGn-7-7);
},
colormap={RdYlGn-8}{
  color=(RdYlGn-8-1);
  color=(RdYlGn-8-2);
  color=(RdYlGn-8-3);
  color=(RdYlGn-8-4);
  color=(RdYlGn-8-5);
  color=(RdYlGn-8-6);
  color=(RdYlGn-8-7);
  color=(RdYlGn-8-8);
},
colormap={RdYlGn-9}{
  color=(RdYlGn-9-1);
  color=(RdYlGn-9-2);
  color=(RdYlGn-9-3);
  color=(RdYlGn-9-4);
  color=(RdYlGn-9-5);
  color=(RdYlGn-9-6);
  color=(RdYlGn-9-7);
  color=(RdYlGn-9-8);
  color=(RdYlGn-9-9);
},
colormap={RdYlGn-10}{
  color=(RdYlGn-10-1);
  color=(RdYlGn-10-2);
  color=(RdYlGn-10-3);
  color=(RdYlGn-10-4);
  color=(RdYlGn-10-5);
  color=(RdYlGn-10-6);
  color=(RdYlGn-10-7);
  color=(RdYlGn-10-8);
  color=(RdYlGn-10-9);
  color=(RdYlGn-10-10);
},
colormap={RdYlGn-11}{
  color=(RdYlGn-11-1);
  color=(RdYlGn-11-2);
  color=(RdYlGn-11-3);
  color=(RdYlGn-11-4);
  color=(RdYlGn-11-5);
  color=(RdYlGn-11-6);
  color=(RdYlGn-11-7);
  color=(RdYlGn-11-8);
  color=(RdYlGn-11-9);
  color=(RdYlGn-11-10);
  color=(RdYlGn-11-11);
},
colormap={Set3-3}{
  color=(Set3-3-1);
  color=(Set3-3-2);
  color=(Set3-3-3);
},
colormap={Set3-4}{
  color=(Set3-4-1);
  color=(Set3-4-2);
  color=(Set3-4-3);
  color=(Set3-4-4);
},
colormap={Set3-5}{
  color=(Set3-5-1);
  color=(Set3-5-2);
  color=(Set3-5-3);
  color=(Set3-5-4);
  color=(Set3-5-5);
},
colormap={Set3-6}{
  color=(Set3-6-1);
  color=(Set3-6-2);
  color=(Set3-6-3);
  color=(Set3-6-4);
  color=(Set3-6-5);
  color=(Set3-6-6);
},
colormap={Set3-7}{
  color=(Set3-7-1);
  color=(Set3-7-2);
  color=(Set3-7-3);
  color=(Set3-7-4);
  color=(Set3-7-5);
  color=(Set3-7-6);
  color=(Set3-7-7);
},
colormap={Set3-8}{
  color=(Set3-8-1);
  color=(Set3-8-2);
  color=(Set3-8-3);
  color=(Set3-8-4);
  color=(Set3-8-5);
  color=(Set3-8-6);
  color=(Set3-8-7);
  color=(Set3-8-8);
},
colormap={Set3-9}{
  color=(Set3-9-1);
  color=(Set3-9-2);
  color=(Set3-9-3);
  color=(Set3-9-4);
  color=(Set3-9-5);
  color=(Set3-9-6);
  color=(Set3-9-7);
  color=(Set3-9-8);
  color=(Set3-9-9);
},
colormap={Set3-10}{
  color=(Set3-10-1);
  color=(Set3-10-2);
  color=(Set3-10-3);
  color=(Set3-10-4);
  color=(Set3-10-5);
  color=(Set3-10-6);
  color=(Set3-10-7);
  color=(Set3-10-8);
  color=(Set3-10-9);
  color=(Set3-10-10);
},
colormap={Set3-11}{
  color=(Set3-11-1);
  color=(Set3-11-2);
  color=(Set3-11-3);
  color=(Set3-11-4);
  color=(Set3-11-5);
  color=(Set3-11-6);
  color=(Set3-11-7);
  color=(Set3-11-8);
  color=(Set3-11-9);
  color=(Set3-11-10);
  color=(Set3-11-11);
},
colormap={Set3-12}{
  color=(Set3-12-1);
  color=(Set3-12-2);
  color=(Set3-12-3);
  color=(Set3-12-4);
  color=(Set3-12-5);
  color=(Set3-12-6);
  color=(Set3-12-7);
  color=(Set3-12-8);
  color=(Set3-12-9);
  color=(Set3-12-10);
  color=(Set3-12-11);
  color=(Set3-12-12);
},
colormap={Pastel1-3}{
  color=(Pastel1-3-1);
  color=(Pastel1-3-2);
  color=(Pastel1-3-3);
},
colormap={Pastel1-4}{
  color=(Pastel1-4-1);
  color=(Pastel1-4-2);
  color=(Pastel1-4-3);
  color=(Pastel1-4-4);
},
colormap={Pastel1-5}{
  color=(Pastel1-5-1);
  color=(Pastel1-5-2);
  color=(Pastel1-5-3);
  color=(Pastel1-5-4);
  color=(Pastel1-5-5);
},
colormap={Pastel1-6}{
  color=(Pastel1-6-1);
  color=(Pastel1-6-2);
  color=(Pastel1-6-3);
  color=(Pastel1-6-4);
  color=(Pastel1-6-5);
  color=(Pastel1-6-6);
},
colormap={Pastel1-7}{
  color=(Pastel1-7-1);
  color=(Pastel1-7-2);
  color=(Pastel1-7-3);
  color=(Pastel1-7-4);
  color=(Pastel1-7-5);
  color=(Pastel1-7-6);
  color=(Pastel1-7-7);
},
colormap={Pastel1-8}{
  color=(Pastel1-8-1);
  color=(Pastel1-8-2);
  color=(Pastel1-8-3);
  color=(Pastel1-8-4);
  color=(Pastel1-8-5);
  color=(Pastel1-8-6);
  color=(Pastel1-8-7);
  color=(Pastel1-8-8);
},
colormap={Pastel1-9}{
  color=(Pastel1-9-1);
  color=(Pastel1-9-2);
  color=(Pastel1-9-3);
  color=(Pastel1-9-4);
  color=(Pastel1-9-5);
  color=(Pastel1-9-6);
  color=(Pastel1-9-7);
  color=(Pastel1-9-8);
  color=(Pastel1-9-9);
},
colormap={Set1-3}{
  color=(Set1-3-1);
  color=(Set1-3-2);
  color=(Set1-3-3);
},
colormap={Set1-4}{
  color=(Set1-4-1);
  color=(Set1-4-2);
  color=(Set1-4-3);
  color=(Set1-4-4);
},
colormap={Set1-5}{
  color=(Set1-5-1);
  color=(Set1-5-2);
  color=(Set1-5-3);
  color=(Set1-5-4);
  color=(Set1-5-5);
},
colormap={Set1-6}{
  color=(Set1-6-1);
  color=(Set1-6-2);
  color=(Set1-6-3);
  color=(Set1-6-4);
  color=(Set1-6-5);
  color=(Set1-6-6);
},
colormap={Set1-7}{
  color=(Set1-7-1);
  color=(Set1-7-2);
  color=(Set1-7-3);
  color=(Set1-7-4);
  color=(Set1-7-5);
  color=(Set1-7-6);
  color=(Set1-7-7);
},
colormap={Set1-8}{
  color=(Set1-8-1);
  color=(Set1-8-2);
  color=(Set1-8-3);
  color=(Set1-8-4);
  color=(Set1-8-5);
  color=(Set1-8-6);
  color=(Set1-8-7);
  color=(Set1-8-8);
},
colormap={Set1-9}{
  color=(Set1-9-1);
  color=(Set1-9-2);
  color=(Set1-9-3);
  color=(Set1-9-4);
  color=(Set1-9-5);
  color=(Set1-9-6);
  color=(Set1-9-7);
  color=(Set1-9-8);
  color=(Set1-9-9);
},
colormap={Pastel2-3}{
  color=(Pastel2-3-1);
  color=(Pastel2-3-2);
  color=(Pastel2-3-3);
},
colormap={Pastel2-4}{
  color=(Pastel2-4-1);
  color=(Pastel2-4-2);
  color=(Pastel2-4-3);
  color=(Pastel2-4-4);
},
colormap={Pastel2-5}{
  color=(Pastel2-5-1);
  color=(Pastel2-5-2);
  color=(Pastel2-5-3);
  color=(Pastel2-5-4);
  color=(Pastel2-5-5);
},
colormap={Pastel2-6}{
  color=(Pastel2-6-1);
  color=(Pastel2-6-2);
  color=(Pastel2-6-3);
  color=(Pastel2-6-4);
  color=(Pastel2-6-5);
  color=(Pastel2-6-6);
},
colormap={Pastel2-7}{
  color=(Pastel2-7-1);
  color=(Pastel2-7-2);
  color=(Pastel2-7-3);
  color=(Pastel2-7-4);
  color=(Pastel2-7-5);
  color=(Pastel2-7-6);
  color=(Pastel2-7-7);
},
colormap={Pastel2-8}{
  color=(Pastel2-8-1);
  color=(Pastel2-8-2);
  color=(Pastel2-8-3);
  color=(Pastel2-8-4);
  color=(Pastel2-8-5);
  color=(Pastel2-8-6);
  color=(Pastel2-8-7);
  color=(Pastel2-8-8);
},
colormap={Set2-3}{
  color=(Set2-3-1);
  color=(Set2-3-2);
  color=(Set2-3-3);
},
colormap={Set2-4}{
  color=(Set2-4-1);
  color=(Set2-4-2);
  color=(Set2-4-3);
  color=(Set2-4-4);
},
colormap={Set2-5}{
  color=(Set2-5-1);
  color=(Set2-5-2);
  color=(Set2-5-3);
  color=(Set2-5-4);
  color=(Set2-5-5);
},
colormap={Set2-6}{
  color=(Set2-6-1);
  color=(Set2-6-2);
  color=(Set2-6-3);
  color=(Set2-6-4);
  color=(Set2-6-5);
  color=(Set2-6-6);
},
colormap={Set2-7}{
  color=(Set2-7-1);
  color=(Set2-7-2);
  color=(Set2-7-3);
  color=(Set2-7-4);
  color=(Set2-7-5);
  color=(Set2-7-6);
  color=(Set2-7-7);
},
colormap={Set2-8}{
  color=(Set2-8-1);
  color=(Set2-8-2);
  color=(Set2-8-3);
  color=(Set2-8-4);
  color=(Set2-8-5);
  color=(Set2-8-6);
  color=(Set2-8-7);
  color=(Set2-8-8);
},
colormap={Dark2-3}{
  color=(Dark2-3-1);
  color=(Dark2-3-2);
  color=(Dark2-3-3);
},
colormap={Dark2-4}{
  color=(Dark2-4-1);
  color=(Dark2-4-2);
  color=(Dark2-4-3);
  color=(Dark2-4-4);
},
colormap={Dark2-5}{
  color=(Dark2-5-1);
  color=(Dark2-5-2);
  color=(Dark2-5-3);
  color=(Dark2-5-4);
  color=(Dark2-5-5);
},
colormap={Dark2-6}{
  color=(Dark2-6-1);
  color=(Dark2-6-2);
  color=(Dark2-6-3);
  color=(Dark2-6-4);
  color=(Dark2-6-5);
  color=(Dark2-6-6);
},
colormap={Dark2-7}{
  color=(Dark2-7-1);
  color=(Dark2-7-2);
  color=(Dark2-7-3);
  color=(Dark2-7-4);
  color=(Dark2-7-5);
  color=(Dark2-7-6);
  color=(Dark2-7-7);
},
colormap={Dark2-8}{
  color=(Dark2-8-1);
  color=(Dark2-8-2);
  color=(Dark2-8-3);
  color=(Dark2-8-4);
  color=(Dark2-8-5);
  color=(Dark2-8-6);
  color=(Dark2-8-7);
  color=(Dark2-8-8);
},
colormap={Paired-3}{
  color=(Paired-3-1);
  color=(Paired-3-2);
  color=(Paired-3-3);
},
colormap={Paired-4}{
  color=(Paired-4-1);
  color=(Paired-4-2);
  color=(Paired-4-3);
  color=(Paired-4-4);
},
colormap={Paired-5}{
  color=(Paired-5-1);
  color=(Paired-5-2);
  color=(Paired-5-3);
  color=(Paired-5-4);
  color=(Paired-5-5);
},
colormap={Paired-6}{
  color=(Paired-6-1);
  color=(Paired-6-2);
  color=(Paired-6-3);
  color=(Paired-6-4);
  color=(Paired-6-5);
  color=(Paired-6-6);
},
colormap={Paired-7}{
  color=(Paired-7-1);
  color=(Paired-7-2);
  color=(Paired-7-3);
  color=(Paired-7-4);
  color=(Paired-7-5);
  color=(Paired-7-6);
  color=(Paired-7-7);
},
colormap={Paired-8}{
  color=(Paired-8-1);
  color=(Paired-8-2);
  color=(Paired-8-3);
  color=(Paired-8-4);
  color=(Paired-8-5);
  color=(Paired-8-6);
  color=(Paired-8-7);
  color=(Paired-8-8);
},
colormap={Paired-9}{
  color=(Paired-9-1);
  color=(Paired-9-2);
  color=(Paired-9-3);
  color=(Paired-9-4);
  color=(Paired-9-5);
  color=(Paired-9-6);
  color=(Paired-9-7);
  color=(Paired-9-8);
  color=(Paired-9-9);
},
colormap={Paired-10}{
  color=(Paired-10-1);
  color=(Paired-10-2);
  color=(Paired-10-3);
  color=(Paired-10-4);
  color=(Paired-10-5);
  color=(Paired-10-6);
  color=(Paired-10-7);
  color=(Paired-10-8);
  color=(Paired-10-9);
  color=(Paired-10-10);
},
colormap={Paired-11}{
  color=(Paired-11-1);
  color=(Paired-11-2);
  color=(Paired-11-3);
  color=(Paired-11-4);
  color=(Paired-11-5);
  color=(Paired-11-6);
  color=(Paired-11-7);
  color=(Paired-11-8);
  color=(Paired-11-9);
  color=(Paired-11-10);
  color=(Paired-11-11);
},
colormap={Paired-12}{
  color=(Paired-12-1);
  color=(Paired-12-2);
  color=(Paired-12-3);
  color=(Paired-12-4);
  color=(Paired-12-5);
  color=(Paired-12-6);
  color=(Paired-12-7);
  color=(Paired-12-8);
  color=(Paired-12-9);
  color=(Paired-12-10);
  color=(Paired-12-11);
  color=(Paired-12-12);
},
colormap={Accent-3}{
  color=(Accent-3-1);
  color=(Accent-3-2);
  color=(Accent-3-3);
},
colormap={Accent-4}{
  color=(Accent-4-1);
  color=(Accent-4-2);
  color=(Accent-4-3);
  color=(Accent-4-4);
},
colormap={Accent-5}{
  color=(Accent-5-1);
  color=(Accent-5-2);
  color=(Accent-5-3);
  color=(Accent-5-4);
  color=(Accent-5-5);
},
colormap={Accent-6}{
  color=(Accent-6-1);
  color=(Accent-6-2);
  color=(Accent-6-3);
  color=(Accent-6-4);
  color=(Accent-6-5);
  color=(Accent-6-6);
},
colormap={Accent-7}{
  color=(Accent-7-1);
  color=(Accent-7-2);
  color=(Accent-7-3);
  color=(Accent-7-4);
  color=(Accent-7-5);
  color=(Accent-7-6);
  color=(Accent-7-7);
},
colormap={Accent-8}{
  color=(Accent-8-1);
  color=(Accent-8-2);
  color=(Accent-8-3);
  color=(Accent-8-4);
  color=(Accent-8-5);
  color=(Accent-8-6);
  color=(Accent-8-7);
  color=(Accent-8-8);
}
}

\pgfplotsset{compat=newest }
\pgfplotsset{
    plotstyle1/.style={mark=diamond*,mark size=3,mark options={draw=black,fill=Blues-6-1},semithick},
    plotstyle2/.style={mark=triangle*,mark size=3,mark options={draw=black,fill=Blues-6-2},semithick},
    plotstyle3/.style={mark=*,mark size=3,mark options={draw=black,fill=Blues-6-3},semithick},
    plotstyle4/.style={mark=square*,mark size=3,mark options={draw=black,fill=Blues-6-4},semithick},
    plotstyle5/.style={mark=pentagon*,mark size=3,mark options={draw=black,fill=Blues-6-5},semithick},
    plotstyle6/.style={mark=star,mark size=3,mark options={draw=black,fill=Blues-6-6},semithick},}

\newcommand{\mean}[1]{\overline{#1}}

\usetikzlibrary{patterns}

\pgfplotsset{
  grid style = {
    dash pattern = on 0.15mm off 1mm,
    line cap = round,
    gray,
    line width = 0.5pt
  }
}

\pgfplotsset{compat=1.15,
    /pgfplots/ybar legend/.style={
    /pgfplots/legend image code/.code={%
       \draw[##1,/tikz/.cd,yshift=-0.25em]
        (0cm,0cm) rectangle (3pt,0.8em);},
   },
}

\newcommand{\BfsPerfFigures}[1]{
\begin{tikzpicture}[scale=0.75]
    \begin{axis}[ylabel={MTEPS},
                 legend style={at={(0.5,1.5)},anchor=north,draw=none,column sep=5pt,legend columns=-1,legend to name=throughputBfsReal},
                 legend cell align=left,
                 width=1.3*\columnwidth,
                 height=4.24cm,
                 xlabel={Sessions},
                 ytick pos=left,
                 xtick pos=left,
                 scaled y ticks=false,
                 ymajorgrids,
                 xmajorgrids                 ]
    \addplot[plotstyle1,each nth point=1] table[x=Sessions,y=Sequential] {experiments/data/#1};
    \addplot[plotstyle5,each nth point=1] table[x=Sessions,y=ParallelBuffer] {experiments/data/#1};%
    \addplot[plotstyle6,each nth point=1] table[x=Sessions,y=ParallelOpt] {experiments/data/#1};%
    \legend{sequential, simple, scheduler}
    \end{axis}
\end{tikzpicture}
}

\newcommand{\PrPerfFigures}[1]{
\begin{tikzpicture}[scale=0.75]
    \begin{axis}[ylabel={MPEPS},
                 legend style={at={(0.5,1.5)},anchor=north,draw=none,column sep=5pt,legend columns=-1,legend to name=throughputPrReal},
                 legend cell align=left,
                 width=1.3*\columnwidth,
                 height=4.3cm,
                 xlabel={Sessions},
                 ytick pos=left,
                 xtick pos=left,
                 scaled y ticks=false,
                 ymajorgrids,
                 xmajorgrids                 ]
    \addplot[plotstyle1,each nth point=1] table[x=Sessions,y=SequentialPush] {experiments/data/#1};%
    \addplot[plotstyle3,each nth point=1] table[x=Sessions,y=ParallelBufferPush] {experiments/data/#1};
    \addplot[plotstyle5,each nth point=1] table[x=Sessions,y=SequentialPull] {experiments/data/#1};%
    \addplot[plotstyle6,each nth point=1] table[x=Sessions,y=ParallelPull] {experiments/data/#1};%
    \legend{sequential push, scheduler push, sequential pull, scheduler pull}
    \end{axis}
\end{tikzpicture}
}

\begin{document}


\title{Scheduling of Graph Queries: Controlling Intra- and Inter-query Parallelism for a High System Throughput}

\author{Matthias Hauck}
\affiliation{Heidelberg University/SAP SE}
\email{matthias.hauck@sap.com}
\author{Ismail Oukid}
\authornote{Ismail Oukid contributed to this work while working at SAP SE.}
\affiliation{Snowflake Inc}
\email{ismail.oukid@snowflake.com}
\author{Holger Fr{\"o}ning}
\affiliation{Heidelberg University}
\email{holger.froening@ziti.uni-heidelberg.de}


\begin{abstract}
The vast amounts of data used in social, business or traffic networks, biology and other natural sciences are often managed in graph-based data sets, consisting of a few thousand up to billions and trillions of vertices and edges, respectively.
Typical applications utilizing such data either execute one or a few complex queries or many small queries at the same time interactively or as batch jobs.
Furthermore, graph processing is inherently complex, as data sets can substantially differ (scale free vs. constant degree), and algorithms exhibit diverse behavior (computational intensity, local or global, push- or pull-based).

This work is concerned with multi-query execution by automatically controlling the degree of parallelization, with overall objectives including high system utilization, low synchronization cost, and highly efficient concurrent execution. 
The underlying concept is three-fold: (1) sampling is used to determine graph statistics, (2) parallelization constraints are derived from algorithm and system properties, and (3) suitable work packages are generated based on the previous two aspects.
We evaluate the proposed concept using different algorithms on synthetic and real world data sets, with up to 16 concurrent sessions (queries). 
The results demonstrate a robust performance in spite of these various configurations, and in particular that the performance is always close to or even slightly ahead of the performance of manually optimized implementations.
Furthermore, the similar performance to manually optimized implementations under extreme configurations, which require either a full parallelization (few large queries) or complete sequential execution (many small queries), shows that the proposed concept exhibits a particularly low overhead.
\end{abstract}

\maketitle

%

\section{Introduction}

Graph-based data sets (graphs) are continuously increasing in size, with sizes of currently up to multiple billions of vertices and trillions of edges, commonly found in applications including machine learning and data analytics for social and business networks, biology and many other domains.
Contrary, other applications like route prediction services for road networks with contention operate on rather small graphs, but have to process many queries concurrently. 
Additionally, different types of graphs are commonly found: scale-free graphs show a power-law-based degree distribution with a long tail, like social networks, while other graphs can exhibit a rather constant degree among most or all vertices, such as road networks or electrical power grids.
Furthermore, the behavior of different graph algorithms can be fundamentally different: while some algorithms, like breadth-first search (BFS), are rather operating on a small subset of the overall graph at a given point in time, algorithms like page rank (PR) operate most of the time on the complete graph.
Last, the main objective of a compute system performing graph computations might be unclear at design time: 
if only single queries are processed, usually latency is the key metric.
However, given recent interest in consolidation and hyperscaling, probably the more likely case are large amounts of concurrent queries, which would rather result in throughput being the key metric.

As a result of these orthogonal properties of data and algorithm, it is notoriously difficult to implement high-performance graph computations. 
While it may be desirable to support multi-query processing, actually even accelerating a single query is difficult.
Typical single query optimizations, including algorithm selection (push or pull~\cite{beamer2012direction} based with different requirements on atomic operations) and parallelization~\cite{shun2013ligra, kulkarni07} (multi-threading for faster processing but more memory contention), depend on the properties mentioned before, which are often hard to predict before execution. 
Contrary, many optimizations come with additional overhead, for instance synchronization for parallel execution, which might diminish the gains or even result in lower performance.
Optimizations like NUMA awareness~\cite{zhang2015numa} and code generation~\cite{tetzel2020efficient} can reduce these newly introduced overheads, but are unable to avoid them completely.

For multi-query execution, in the following referred to as number of \emph{concurrent queries}, the problem worsens as an additional dimension is added, which is to select a schedule with the right amount and set of queries to be processed concurrently.
In one of our previous works we showed that multi-query execution using simply multiple parallel engines at the same time can degrade performance~\cite{hauck2017can}.
To overcome this, given enough concurrent queries ready for execution, a promising option might be to process each query internally sequential (intra-query), what minimizes synchronization overhead, and thereby sourcing the required degree of parallelism from many concurrent queries (inter-query).
Another form of multi-query support are solutions that try to share work between different queries as they exist in the literature for relational DBMS~\cite{rehrmann2020sharing, psaroudakis2013sharing}.
While such approaches can reduce the overall work, overhead can be substantial if there is nothing to share.

This work gears to optimize multi-query execution by automatically controlling the degree of parallelization at different levels, with
overall target metrics including high system utilization, low synchronization costs and highly efficient concurrent execution.
We address this problem with resource control system based on knowledge it obtains from the queries that are executed: first, it generates graph statistics by sampling vertices. Second, it takes algorithmic and system properties into account to select certain bounds of parallelization. Last, it generates work packets of certain sizes for execution.

In summary, we hypothesize that the overhead of resource control by efficient statistics generation and simple cost models is outweighed by the improvement of execution. 
While fundamentally being a trade-off, as perfect statistics and cost predictions would certainly result in optimal execution, efficient methods are required that result in minimal overhead during execution but are still accurate enough for a good scheduling decision.
Then, we should observe a highly efficient execution, with i.e. a high number of accumulated operations per time unit, for a variety of scenarios, such as different data (size, type), algorithms (local, non-local), and concurrency (single- or multi-query execution).

In particular, this work makes the following contributions:
\begin{itemize}
\item   An adaptive cost estimation that trades accuracy with overhead based on graph statistics, algorithmic behavior and system properties. In particular, the cost estimator model incorporates traversal behavior estimators (new vertex set, touched memory). 
\item 	The design and implementation of a runtime resource control system that schedules graph queries during execution based on latency-aware parallelization, cost-based work packaging, and selective sequential execution.
\item 	An evaluation based on different algorithms (in total 9 different BFS and PR algorithms), different synthetic and real-world data sets (8 total), and different degrees of concurrent query execution (one to 16 concurrent sessions).
\end{itemize}

\begin{figure}[t]
	\includegraphics[width =0.47\textwidth, 
							]{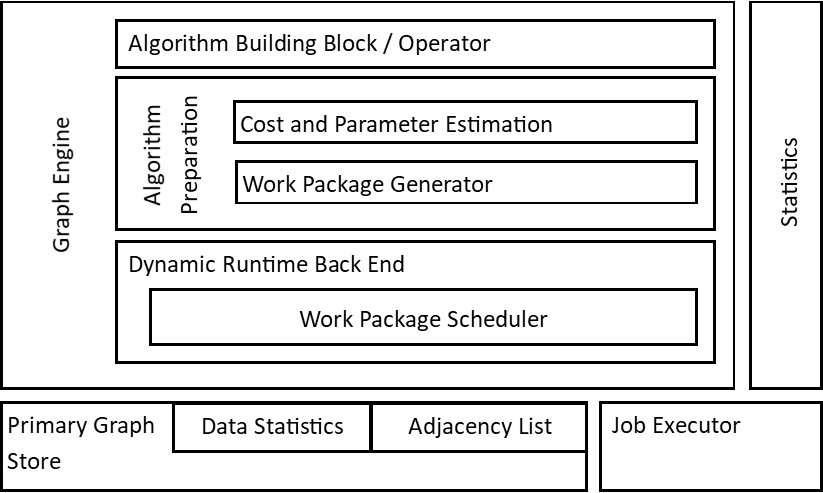}
	\caption{Block diagram of the graph processing engine and its scheduling components}
	\label{fig:system}
\end{figure}

\Cref{fig:system} visualizes the overall organization of the proposed system. 
In the remainer of this work, we will follow up with a section on related work. \Cref{sec:costModel} will provide details on the models and estimators that predict graph- and algorithm-dependent behavior ahead of execution.
In \Cref{sec:system}, details about the runtime responsible for scheduling decisions during execution will be presented.
While up to now, methods and models were not hardware-specific, in \Cref{sec:contention} hardware-specific parameters will be determined. 
The subsequent \Cref{sec:evaluation} will provide a detailed evaluation of the proposed method, based on multiple algorithms and their variants, as well as multiple data sets.
The last section concludes and provides an outlook on future avenues.

\section{Related Work}
\label{sec:related}

There is a plethora of related work on graph processing engines and languages, such as Ligra and derived variants~\cite{shun2013ligra, dhulipala2017julienne, zhang2015numa}, Galois~\cite{kulkarni07}, GraphX~\cite{gonzalez2014graphx}, Emptyheaded~\cite{aberger2015emptyheaded}, Green Marl~\cite{hong2012green}, and many more for accelerators such as GPUs~\cite{wang2016gunrock} or FPGAs. 
Commercial representatives include GDBMS like Neo4j\footnote{\url{http://neo4j.com}} and Sparksee\footnote{\url{http://www.sparsity-technologies.com}}.
However, such work besides GDBMS is usually focused on providing peak performance for the processing of single, possibly large and time-consuming, graph computations.
Instead, our work gears towards concurrent processing of various, possibly small, queries.
In the following, we will therefore rather discuss work related to the intrinsic behavior of graph queries and their properties.

In the related work we found several instances that found graph processing inefficiencies. 
Beamer et al.~\cite{beamer2015locality} analyse different graph algorithm and datasets using performance counters.
They observe a low compute and bandwidth utilization by the algorithm, but also a high last-level cache (LLC) hit rate that indicates the presence of locality.
They also observed that there are diminishing returns for high degrees of multi-threading, which limits its potential.

Several techniques have been proposed that deal with the optimization of structural graph properties on performance.
A well-known technique is graph reordering. 
Balaji and Lucia~\cite{balaji2018when} investigate why reordering often does not amortize.
They analyze therefore the efficiency of reordering techniques on different data sets and algorithms.
They observe among other things that reordering can degrade the performance for push-style operations as it increases false sharing. 
Finally, they propose a metric called packing factor to decide for hub sorting, if it is worth to reorder a graph.

\begin{figure*}[t] 
	\includegraphics[	width =0.97\textwidth, 
							]{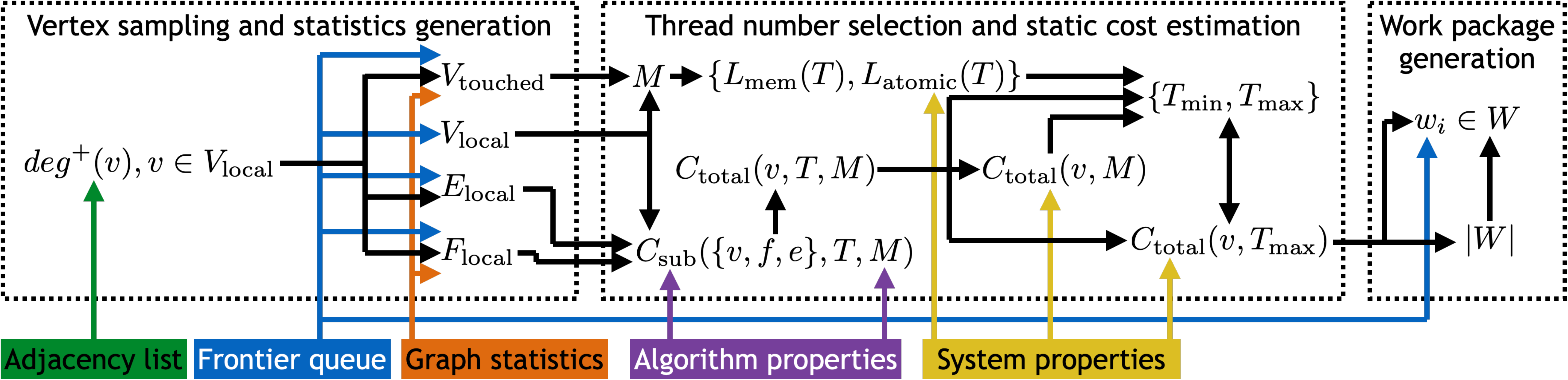}
	\caption{Interaction and dependencies between statistical values, estimations and work package generation}
	\label{fig:estimatorInteraction}
\end{figure*}

There are also some works that try to apply a cost-based optimization in the context of graph queries.
Bonifati et al.~\cite{bonifati2018querying} discuss how classic techniques can be adapted for regular path queries.
Based on the assumptions of uniformity of source/target vertex set size, independence of predicates and inclusion of the vertex domains, they demonstrate the adaptation of a synopsis for graphs.

Verstraaten, Varbanescu and de Laat~\cite{verstraaten2018mix} analyze BFS variants on different types of graphs in the context of GPU processing.
They observe that depending on the graphs different algorithmic variants have an advantage.
This effect has also been observed in other works and resulted in approaches like direction-optimized BFS~\cite{beamer2012direction}.
In order to address this issue they propose to use a decision tree based on graph properties, such as size and degree distribution, and runtime properties like the frontier size to switch between different implementations.
The decision tree is generated upfront via runs on exemplary data.

Zhang, Yunming, et al.~\cite{zhang2018graphit} propose a system comprising of a graph DSL and a scheduling language to formulate graph algorithms.
While the purpose of the graph DSL is to formulate the algorithm, the scheduling language allows to annotate the DSL with different types of optimizations like push/pull or NUMA mappings. 
The system is able to apply combinations of these optimizations to the graph algorithm, while guaranteeing that the algorithm is still valid. 
They also provide an auto tuner, where the system automatically identifies an effective set of optimizations using the elapsed time of the algorithm on the target data.   

There also have been proposed approaches, which specifically optimize multi query support. 
Zhao et. al.~\cite{zhao2019graphm} address multi query execution by sharing the processed graph data structure between several queries. 
They therefore process the graph chunk wise and schedule the work of the different queries so that they work on the same chunk that has already been loaded to the LLC.
Pan and Li~\cite{ pan2017congra} try to increase the execution efficiency by scheduling optimization.
Via up-front profiling runs of each graph and algorithm pair they estimate the bandwidth requirement and which of two thread counts offers the highest performance.
At runtime their system makes sure that only new queries are started with the most effective thread count, if there is available bandwidth not used by other queries.

\section{Cost Model and Parameter Estimation}
\label{sec:costModel}

Graph computations respectively their computational iterations notoriously show small elapsed times.
Thus, any cost model employed to predict behavior during execution has to exhibit a low overhead, effectively resulting in time required for prediction that is only a fraction of the elapsed time of an iteration of a graph computation.
Still, it has to be accurate enough in its predictions to result in reasonable scheduling decisions.

\Cref{fig:estimatorInteraction} shows an overview of the cost model, consisting of vertex sampling and statistic generation, followed by the selection of thread count and the estimation of static costs per vertex, and the final step of work package generation.
In the first part, we will introduce estimators to approximate the size of certain vertex sets, while the second part consists of a cost model for processing vertices including criteria to determine the number of threads for such a processing. 
In the third part, we show how we derive work packages and thread boundaries from the cost model. 

\subsection{Traversal Behavior Estimators}

For graph traversal algorithms typically the number of starting vertices and processed edges are important cost factors.
Besides these often trivially obtainable factors, there are upfront unknown or expensive to obtain factors that have likewise a significant impact on the cost.
For our cost modeling two important factors are the number of first time newly visited vertices ($|F_\text{j}|$) in iteration j and the number of vertices $|U_\text{j}|$ that will be touched during iteration j.
The to-be-visited vertices are directly related to the found vertices, while touched vertices relate to the amount of memory that is shared, such as duplicate filters during a graph traversal.

There exists no general method to obtain these values exactly without executing the algorithm.
In order to estimate them, we model if a vertex is newly visited or touched again as a (conditional) probability process.
Our model (cf. \Cref{tab:traversalEstimator} for the parameters) is based on the general assumption that the probability for all vertices to be visited by an edge is identical.
This means there is neither a correlation between different vertices (pure randomness), nor structural effects like the rich club effect (level-dependent), nor is the graph a multigraph (increased probability for some vertices).

\begin{table}
\caption{Parameters for traversal behavior estimation}
\label{tab:traversalEstimator}
\begin{tabular}{p{0.15\textwidth}p{0.85\textwidth}}
\toprule
Parameter & Description\\
\midrule
$V$  & All vertices in the graph \\
$V_\text{reach}$ & Vertices that are reachable via a graph traversal (i.e., neither isolated nor without an incoming edge)\\
$U_\text{j}$ & Vertices that are touched via edge traversal during each iteration j (cf. \Cref{eq:touchIdea})\\
$S_\text{j}$ & Set of vertices in the current queue at iteration j\\
$F_\text{j}$ & Vertices that are newly found after iteration j (cf. \Cref{eq:newIdea})\\
$E_\text{j}$ & Set of edges traversed by the vertices of the queue $S_\text{j}$ \\
$p_{v\text{visits}}$ & Probability that a vertex will be visited by vertex $v$\\
$V_\text{no visit, j}$ & Reachable vertices that have not been visited before iteration j\\
$p_\text{no visit}$ & Probability that a vertex has not been visited\\
\bottomrule
\end{tabular}
\end{table}

The main idea of our estimation for amount of touched vertices $|U_\text{j}|$ is based on the probability for a reachable vertex to be reached at least once from a vertex from the frontier queue $v \in S_\text{j}$, aggregated for the set of all reachable vertices $V_\text{reach}$ (cf. \Cref{eq:touchIdea}).
As we assume that the graph is not a multigraph and that each outgoing edge of a given vertex $v$ is connected to another vertex, so that the probability for a specific vertex to be reached is $p_{v \text{visits}} = \frac{deg^{+}(v)}{|V_\text{reach}|}$ (cf. \Cref{eq:touchAssumptions}), with $deg^{+}(v)$ being the out degree of a vertex $v$.
It is typically too costly to compute the probability for each vertex of the frontier queue, as the queue can be of substantial size.
To reduce these costs, we consider the difference in between the maximum and mean value of outgoing edges per vertex.
If this difference is small, then the mean value is used as an approximation (cf. \Cref{eq:touchSmallStdev}), otherwise 
we extrapolate the product of the probabilities from a sample of vertices in the queue at the beginning of each cost calculation.
In this work, we use a simple sampling approach that uses essentially up to the first 8192 vertices as this provides a low latency and sufficient quality. 

We approximate also $V_\text{reach}$ with the number of vertices that are neither isolated nor without an incoming edge. This assumption is based on assumption that the graph forms a single connected component and is purely random. This value is also inexpensive to obtainable during the construction of the adjacency list.

\begin{align} 
|U_\text{j}| & = \left( 1 - \prod \limits_{v \in S_\text{j}} \left( 1 - p_{v \text{visits} }\right) \right) \cdot |V_\text{reach}| \label{eq:touchIdea} \\
                                                  & \approx   \left( 1 - \prod \limits_{v \in S_\text{j}} \left( 1 - \frac{deg^{+}(v)}{|V_\text{reach}|}\right) \right) \cdot |V_\text{reach}| \label{eq:touchAssumptions}\\
                                                  & \approx   {\left( 1 - \left( 1 - \frac{\mean{deg^{+}(V)})}{|V_\text{reach}|}\right)^{|S_\text{j}|} \right) \cdot |V_\text{reach}|\label{eq:touchSmallStdev}}
\end{align}

For the estimation of the found vertices $|F_\text{i}|$, we use a similar approach as for $|U_\text{i}|$, but besides being reachable the specific vertex needs also to be unvisited (cf. \Cref{eq:newIdea}). Subsequently, \Cref{eq:newAssumptions} and \Cref{eq:newSmallStdev} are derived analogous to before.
\begin{align} 
|F_\text{j}| & = \left( 1 - p_\text{no visit} \cdot \prod \limits_{v \in S_\text{j}} \left( 1 - p_{v\text{ visits}}\right) \right)\cdot |V_\text{reach}| \label{eq:newIdea} \\
                                                  & \approx  \left( 1 - \frac{|V_\text{no visit, j}|}{|V_\text{reach}|} \cdot \prod \limits_{v \in S_\text{j}} \left( 1 - \frac{deg^{+}(v)}{|V_\text{reach}|} \right) \right) \cdot |V_\text{reach}| \label{eq:newAssumptions} \\
                                                  & \approx  \left( 1 - \frac{|V_\text{no visit, j}|}{|V_\text{reach}|} \cdot \left( 1 - \frac{\mean{deg^{+}(V)}}{|V_\text{reach}|} \right)^{|S_\text{j}|} \right) \cdot |V_\text{reach}| \label{eq:newSmallStdev}
\end{align}

\subsection{Cost Model}

\begin{table}
\caption{Parameters of the cost model}
\label{tab:costModelParameter}
\begin{tabular}{p{0.15\textwidth}p{0.85\textwidth}}
\toprule
Parameter & Description\\
\midrule
$i$ & Item of generic type, including vertex($v$), (new) found vertex($f$) or edge($e$) \\
$I$ & Set of items of generic type, including set of vertices in the current iteration j queue ($S_\text{j}$), the set of found vertices ($F_\text{j}$), or the set of edges related to the current iteration j queue ($E_\text{j}$) \\
$|I|$ & Number of elements in a given set \\
$T$ & Number of threads used \\
$M$ & Amount of accessed data \\
$C_\text{sub}(i,T,M)$ & Sub-cost of execution for a given item, i.e., processing only vertex $v$ or $f$, or edge $e$ \\
$C_\text{total}(v,T,M)$ & Total cost of parallel execution for vertex $v$ \\
$L_\text{op}$ & Latency of an arithmetic operation \\
$L_\text{mem}(M)$ & Latency of a non-atomic memory access, depending on the size $M$ of the accessed data\\
$L_\text{atomic}(T,M)$ & Latency of an atomic operation, depending on the size $M$ of the accessed data and the amount of threads $T$ \\
$N_\text{ops}(i)$ & Number of arithmetic operations used to process item $i$ \\
$N_\text{mem}(i)$ & Number of memory operations (non atomic load \& stores) used to process item $i$ \\
$N_\text{atomics}(i)$ & Number of atomic operations used to process item $i$ \\

\bottomrule
\end{tabular}
\end{table}

In order to model performance (cf. \Cref{tab:costModelParameter} for the parameters), we assume that the cost is proportional to the number of vertices $|S_\text{j}|$ that have to be processed, the related edges $e$ of these vertices that have to be traversed, and the new vertices $f \in F_\text{j}$ that are found as a result of this traversal.
Another fundamental assumption is that the parallel and sequential implementations are identical as the parallel code protects critical sections using atomic operations, while the sequential code can instead simply employ plain memory operations.
We model this assumption by setting the atomic update latency for a single thread $L_\text{atomic}(T=1,M)$ equal to the memory access latency $L_\text{mem}(M)$, with both being dependent on the amount of accessed memory $M$.

In general we have three main cost subcomponents: computations (ops), regular memory access operations (mem) and atomic memory operations (atomics).
Then, we can formulate the sub-cost $C_\text{sub}(i,T,M)$ for the processing of a given item $i$, as shown in \Cref{eq:partsOfCost}.

\begin{align}\begin{split}
\label{eq:partsOfCost}
C_\text{sub}(i,T,M) &= N_\text{ops}(i) \cdot L_\text{op} + N_\text{atomics}(i) \cdot L_\text{atomic}(T,M) + N_\text{mem}(i) \cdot L_\text{mem}(M)
\end{split}\end{align}

In order to compare the estimates for the parallel and the sequential case, we use the total cost per vertex $C_\text{total}(T,M)$. 
This cost is the sum of the costs to process the vertex itself and the cost of its share of edges and the newly found vertices, as shown in \Cref{eq:totalCost}.
\begin{align}\begin{split}
\label{eq:totalCost}
C_\text{j, total}(T,M)         &=  C_\text{sub}(v,T,M)  + \frac{|E_\text{j}|}{|S_\text{j}|} \cdot  C_\text{sub}(e,T,M) \\
                                               &+ \frac{|F_\text{i}|}{|S_\text{j}|} \cdot C_\text{sub}(f,T,M)
\end{split}\end{align}

\subsection{Work Package and Thread Boundaries Estimation }

\begin{table}
\caption{Parameters of the work package and thread boundaries estimation}
\label{tab:WPTBParameter}
\begin{tabular}{p{0.15\textwidth}p{0.85\textwidth}}
\toprule
Parameter & Description\\
\midrule
 $C_\text{T overhead}$ & Start cost for a single thread (typically a few {\textmu}s)\\
 $C_\text{T min}$  & Minimum work per thread (larger than $C_\text{T overhead}$) \\
 $C_\text{para startup}$ & Start cost for parallel execution (typically a few {\textmu}s) \\
 $P$ & Maximum number of cores \\
 $T_\text{min/max}$ & Minimum and maximum thread bound \\
 $J_\text{min/max}$ & Minimum and maximum thread bound per cache level\\
 $W$ & Work package set that partitions the work\\
 $w_\text{i}$ & A work package of a number of vertices that are assigned together to a thread\\
\bottomrule
\end{tabular}
\end{table}

In \Cref{tab:WPTBParameter} we list all parameters related to this section.
Based on the cost model, we decide if it is profitable to execute the code in parallel ($|V| >= |V_\text{min for parallel}|$, see \Cref{eq:minwork}), and for which thread ranges ($T_\text{min}<=T<=T_\text{max}$) this is true. 
In detail, the minimum work per work package $C_\text{w min}$, the start cost for parallel execution $C_\text{para startup}$ and the start cost per thread $C_\text{T overhead}$ are empirically determined, ensuring 
that the overhead remains reasonable in case of a high load.

The computation of these thread boundaries is expensive, as the thread-count-dependent memory latency is non-linear.
In order to do this, we solve the optimization problem for the smallest number of threads ($T_\text{min}$) as shown in \Cref{eq:threadOptProblem} under some side conditions.
For a time-efficient solution, we propose \Cref{alg:packageAndThreadEstimation},
in which 
we continuously double the number of threads and check if we have a valid upper and lower thread bound.

\begin{align}\begin{split}
\label{eq:minwork}
    |V_\text{min for parallel}| = & \frac{C_\text{T min}+ C_\text{para startup}}{C_\text{v total}(T=1, M)}\\
\end{split}\end{align}

\begin{align}\begin{split}
\label{eq:threadOptProblem}
C_\text{v total,seq}(T, M) & > \frac{C_\text{v total,para}(T, M)}{T} + \frac{C_\text{T overhead}\cdot T}{|V|}
\end{split}\end{align}

\begin{algorithm}
\caption{Work Package and Thread Boundaries Estimation Algorithm}\label{alg:packageAndThreadEstimation}
\begin{algorithmic}
\State $\text{minNotSet} \leftarrow  true$
\State $T_\text{min} \leftarrow  0 $
\State $T_\text{max} \leftarrow  0 $

\For{\{$T | 1 \leq T \leq P, T = 2^{n}, n \in \mathbb{N}_0 \}$ }
    \State $J_\text{max} = \text{MAX}(T, \frac{|V| \cdot C_\text{v total}(T, M)} {C_\text{T min}+ C_\text{para startup}} )$
    \State $J_\text{min} = \frac{|V| \cdot C_\text{v total}(T, M) + J_\text{max} \cdot C_\text{T overhead}}{|V| \cdot C_\text{v total,seq}(M)}$

    \Comment{conditions that must be fulfilled}
    \If {$T_\text{max} \leq J_\text{min}$}
        break
    \EndIf

    \If{$J_\text{max} > J_\text{min}$}
        \State $T_\text{max}  \Leftarrow  J_\text{max}$
        \If{minNotSet} 
        	 \State $T_\text{min} \Leftarrow  J_\text{min}$
            \State $\text{minNotSet} \Leftarrow \text{false}$
        \EndIf
    \ElsIf {$\lnot \text{minNotSet}$}
        	break
    \EndIf
\EndFor
\end{algorithmic}
\end{algorithm}

\section{System Design}
\label{sec:system}


\begin{table}
\caption{Qualitative impact of used techniques on design goals.}
\small
\begin{tabular}{p{3.8cm}p{2.3cm}p{2.5cm}p{2.9cm}}
\toprule
 & System Utilization & Synchronization Cost & Concurrent Execution Efficiency \\
\midrule
Latency-aware parallelization	& \multicolumn{1}{c}{-} 	    &  \multicolumn{1}{c}{++} & \multicolumn{1}{c}{+} 	\\
Cost-based work packaging 		& \multicolumn{1}{c}{++}  &  \multicolumn{1}{c}{-}	    & \multicolumn{1}{c}{0}	\\
Selective sequential execution 	& \multicolumn{1}{c}{-}		&  \multicolumn{1}{c}{++} & \multicolumn{1}{c}{++}	\\
\bottomrule
\end{tabular}
\label{tab:effect}
\end{table}

While, as previously stated, there are three main goals for this work (system utilization, synchronization cost reduction, and concurrent execution efficiency), in detail there are additional requirements:
\begin{itemize}
\item 	First, the behavior of the engine has to be predictable. 
Small changes, like a few extra edges and vertices, should not result in a significantly higher or lower performance, and when code is executed in parallel it should always be faster than a sequential alternative. 
\item 	Second, the engine needs to be friendly in its resource consumption towards potential other engines. 
We assume that it is integrated inside of a DBMS along multiple other engines, so the engine cannot assume total control over all system resources.
\end{itemize}

We use three different techniques to achieve these goals: (1) latency-aware parallelization, (2) cost-based work packaging and (3) selective sequential execution.
All of them address the mentioned goals in different ways (cf. \Cref{tab:effect}) and are realized by different engine components (cf. \Cref{fig:system}).
In general, the engine is divided into two major components:
a \textbf{preparation component} and a \textbf{runtime component} that controls the execution of the core graph algorithm.
The idea is to perform a preselection of useful parameters in the preparation step based on data and reasonable assumptions, while the runtime takes care of behavior that is unknown prior to the execution of the algorithm.

\subsection{Latency-aware parallelization}
In the previous sections we introduced a performance model for graph traversal algorithms.
For latency-aware parallelization, we apply this model to our decisions that we take to optimize system parameters. 
It is realized by the subcomponent \textbf{Cost and Parameter Estimation} in the algorithm preparation step (cf. \Cref{fig:system}).

In practice it can be observed that different numbers of threads are optimal in terms of performance depending on the primitive operations performed by an algorithm and the size of the intermediate data.
Also, in many cases parallel execution can even harm the performance, as too many threads might result in work packages that are too small to compensate overhead costs. Thus, using the cost model we also determine an upper and lower bound for parallel execution of a specific query.

\subsubsection{Parameter sets}
As shown in \Cref{fig:systemInteraction}, we have three different data sources respectively parameter sets to compute the cost: 
\begin{enumerate}
\item 
The first type of data are system properties like cache sizes and memory access times, which will be described in \Cref{sec:contention}. 
While many system properties are static, dynamic properties in this work are determined by a single benchmarking run with memoization for future re-use in all queries.
\item The second type are the algorithmic properties like the amount of touched data and the number of operations per edge and vertex, including computations, memory accesses and atomic updates (c.f. $C_{sub}(\{v,f,e\},T,M)$ in \Cref{fig:estimatorInteraction}). 
These numbers can be counted and statically set in a descriptor.
\item The third type are statistics about the processed data, such as the frontier queue or the graph, which will be described in the next subsection. 
\end{enumerate}

These data sets are combined in a linear model to estimate the resulting memory footprint $M$ that is used to determine the cache level the problem can fit into. 
The determined cache level respectively its latency is in turn used as a parameter for the memory cost model (c.f. $L_{mem}$ in \Cref{fig:estimatorInteraction}).
In addition, the level is also used to compute $L_\text{atomic}(T)$ using \Cref{alg:packageAndThreadEstimation}  and the related thread boundaries.
As the estimated thread boundaries are worst cases, our system works well with contention minimization techniques like small local buffers~\cite{hauck2019buffer} even without adaptations.

\subsubsection{Statistics generation}
As stated previously, we need good statistical data related to the vertices we want to process, as the vertex degree and the related work can vary between iterations. 
In order to not outweigh all benefits, in particular statistics generation has to be of low overhead.
An inexpensive way to gather statistical data is to collect them at creation time from the index data structure for the graph topology (adjacency list $V_{local}$ in \Cref{fig:estimatorInteraction}).
Here we can gather statistics for edges and vertices, including the mean and maximum vertex out degrees (cf. $V_{touched}$, $E_{local}$ and $F_{local}$ in \Cref{fig:estimatorInteraction}).

Using these global statistics, we decide at run time if we use global statistics or rather compute local statistics for the current iteration. 
Here, the indicator is the ratio of maximum vertex degree versus mean vertex degree, 
as a simple metric to reflect variance.
In this work, a threshold of $1.1$ was found to be effective.
If we expect a low variance, we compute the cost model parameter for the cost model using global statistics.
If we expect a high variance, we compute $|V_\text{touched}|$ and $|V_\text{new}|$ on a subset (up to the first $4,000$ vertices) using real vertex degrees and extrapolate global values from this subset.
The computation of such statistics is parallelized to minimize overhead.


\subsection{Cost-based work packaging}
While latency-aware parallelization tries to provide optimal parameters, such as thread count bounds for parallel execution and number of work packages, cost-based work packaging tries to optimize the work packages themselves.
On a large scale, the average degree and the related work of sets of vertices can be described statistically.
But if the variance of the edge degrees is high, or the number of vertices in a partition is small, the potential work per partition might be non-uniformly distributed between different partitions, resulting in inefficiencies due to the bulk-synchronous execution.
Then, some work packages take much longer than others and are executed effectively alone for most of their run time.

We address this problem depending on input data statistics. For cases with a high vertex degree variance and a low numbers of vertices, we generate 
work packages that are based on the vertex and edge performance model.
Therefore, we iterate over the vertices in the frontier and obtain the out degree until we exceed the work share for a particular work package. 
The number of work packages is limited to a multiple ($8$ times) of the maximum usable level of parallelism 
to avoid over-parallelization and resulting effects like contention. 
Finally, we reorder the work packages so that work packages with a high cost due to a single dominating vertex are executed first.

If the number of vertices is high or the variance is low, we still use for efficiency reasons a static partitioning. 
In addition, the number of work packages is much larger than the used number of cores, allowing the runtime react on dynamic execution behavior. 

\subsection{Selective Sequential Execution}
While the previously mentioned techniques try to generate optimally-sized work packages, they cannot address runtime effects like the dynamic component of contention.
To deal with these effects and the fact that in some cases the sequential execution is more efficient, we use as a runtime component a work package scheduler.
The work package scheduler has two functions: first, it assigns work to the worker threads; second, it controls if the work is executed sequentially or in parallel.

When the execution of a task starts, the runtime requests worker threads from the system according to the upper thread boundary.
When a worker gets assigned, it registers itself with the scheduler and requests a work package.
The scheduler checks if the number of registered workers is higher than the minimum boundary for parallel execution. If this is true, it assigns a work package to these workers for parallel execution.
If the condition is false, it assigns one worker to execute a package sequentially, while the other threads wait until the package is completed. Then, the scheduler reevaluates the worker situation.
This is repeated for a limited number of sequential packages after which the scheduler releases all but one thread and completes the execution sequentially. 

This approach avoids a central scheduler that needs to deal with many different tasks, which might run a very short time and might be of different type, such as relational and graph combined.
It executes also the tasks using an optimal parallelism strategy, taking into account concurrently executed tasks.
Especially for scenarios with concurrent queries, it can be observed that sequential execution of a single query is preferable over parallel execution of a single query, which follows intuition as such would easily result in over-parallelization.

\subsection{Interaction of the subcomponents}

\begin{figure}[t]
	\includegraphics[	width =0.47\textwidth, 
							]{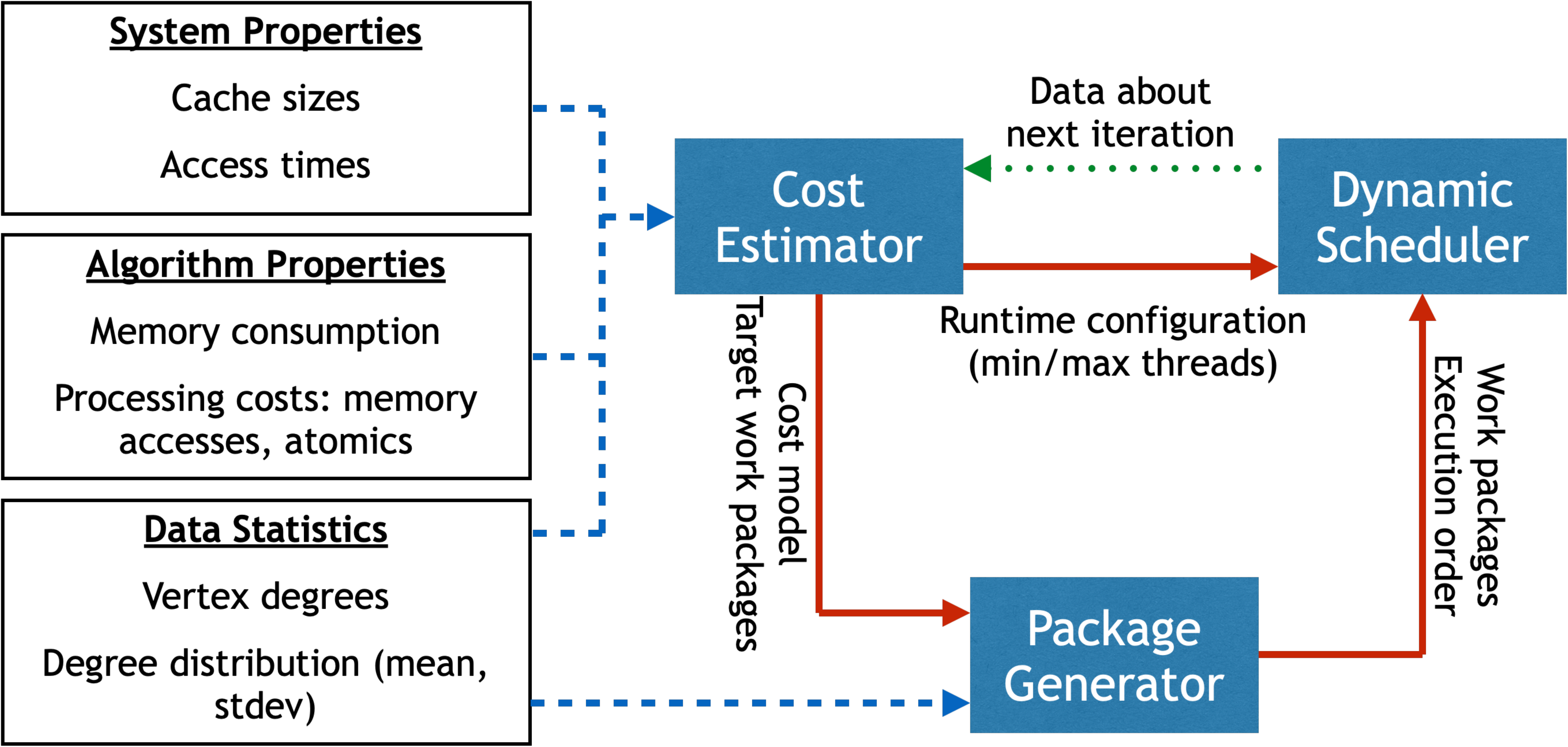}
	\caption{Interaction and dependencies between graph processing engine and scheduler subcomponents. Solid lines represent interaction, dashed lines access to generic data about the problem, and the dotted lines feedback and data about the next iteration.}
	\label{fig:systemInteraction}
\end{figure}

The different interactions in between the subcomponents are summarized in \Cref{fig:systemInteraction}, in particular how the result of one subcomponent is used in the succeeding one and how certain boundaries are set.
The preparation step of the \textit{Cost Estimator} provides information for latency-aware parallelization and therefore about the most-suited execution mode.
Cost-based work packaging (\textit{Package Generator}) provides the work packages that fulfill the assumption of equal work distribution, but also maintains a sufficient amount of work per package.
The runtime in the end (\textit{Dynamic Scheduler}) uses these information for the selective sequential execution, while it also takes into account dynamic system behavior, which is not considered by the previous two subcomponents.

The figure furthermore summarizes which properties of system, algorithm, and data are taken into consideration by which subcomponent. Also, different lines highlight the flow of information, with dashed lines highlighting access to upfront prepared date, solid lines highlight data flow between component, and dotted line highlighting data for the next iterations.
Principally, it is also possible that some stage provides feedback like the measured cost of a work package as shown by the dotted line from dynamic scheduler to cost estimator.
This might allow to optimize later iterations of the graph algorithm, however, for this work we considered this out of scope.

\subsection{Implementation}

The complete system was implemented using C++, a HANA graph engine derived system~\cite{hauck2015highspeed} and HANA's job execution framework~\cite{psaroudakis2013task}.
The algorithms have been implemented using two lambda functions, one sequential and one parallel.
We run the preprocessing for the execution depending on the type of algorithm: 
for topology-centric algorithm like Page Rank, where the vertices stay the same for all iterations, we run it once; 
for data-driven algorithm like BFS, we run it for every iteration as for every iteration a different set of vertices is processed.

For the used data that is not obtained during the preprocessing steps, we have different sources: 
We obtained the \textit{algorithmic properties} by counting the respective operations used and memory touched by the lambdas and stored them for each algorithm as metadata. 
In a productive system a query compiler could do the counting automatically.
The graph \textit{data statistics} are gathered during the construction of the adjacency list.
Dynamic \textit{system properties} like memory access latency are determined prior to experiments using the degree count benchmark.
Similarly, static \textit{system properties} like cache sizes are determined also prior to experiments using appropriate tools such as CPUID.

\section{Modeling Update Contention}
\label{sec:contention}
While in general the compute cores of a multi-core system are independent, 
in reality resources like the CPU's internal interconnection network (system interconnect), some cache hierarchies, or the memory controller are shared among these cores.
Furthermore, some modules are limited in parallelism (e.g., banked caches), or even require serialization (e.g., memory controller).
Typically a shared use of resources increases contention and thus the delay of a particular operation. 
As a result, the CPU has to wait longer until an operation like a memory access or a data transfer completes.

A special case of this resource contention is the use of the same memory address by multiple cores.
Read-only access to a range of addresses causes virtually no contention as multiple shared copies of a given memory address can be installed in different caches.
However, shared copies are not allowed in the case of a write access, thus write operations typically result in a large amount of invalidations, subsequent cache misses, and ultimately contention.
The same applies to atomic operations like mutually exclusive read-modify-write operations, which even intensify the problem as corresponding cache lines will be locked during the operation and no other requests on the same cache line will be served.

Especially for memory sensitive algorithms, like many graph algorithms, 
memory contention has a significant impact on overall performance. 
In general there are many sources on different levels that influence contention and its impact on the overall performance: 
on hardware level, there are (non-conclusive) the CPU cores, system interconnects, memory accesses, cache organization and the specification and implementation of atomic operations.
On software level, there are low-level sources like thread-to-physical-core mapping and high-level sources like the algorithmic design. 
Push-based graph algorithms are typically prone to contention as they update the same address, while pull-based algorithm are not.
Furthermore, the data that is being processed can be a source of contention, and 
between all of these sources there are complex interactions that are highly dynamic in nature.

\subsection{Update Latency Estimation Under Contention}

In practice, the aforementioned sources make it hard to model contention and how it affects latency and throughput of updates.
Essentially, to predict and optimize the performance of parallel graph algorithms, an accurate prediction of update performance is necessary. 
Our solution is to omit analytical modeling of contention entirely, and instead create a simple model based on experiments and measurements. 
Thus, the general idea is to train a parametric model on a given system using a reference algorithm and differently sized data sets. 
Such a training is required any time a new hardware configuration is used, however, the training process is to a large extend automated.
We assume that Little's law
applies and throughput and latency are interchangeable. 
To later predict the latency of a given update, we map it to our reference problem.

As a reference algorithm we select degree count, because its parameters can be varied almost arbitrarily to model different scenarios. Furthermore, it is comparable to many push-based graph algorithms.
This algorithm counts the occurrence of vertex IDs of the vertex set $V$ in an edge list, either as source or target vertex, 
using fetch-and-add atomic operations on a single counter array.
When executed in parallel, the input edge list is partitioned in non-overlapping parts of 16k edges each.
If this results in fewer partitions than cores, we exclude this setting from the experiments.
These partitions are dynamically dispatched as work packages to a set of worker threads.

The reference data set should be of a type that causes high contention, and furthermore should be representative and not of pathological nature.
Thus, RMAT is chosen as being representative for many graph problems. 
The scale-free degree distribution of RMAT graphs causes high contention on vertices with high degree count. 
Furthermore, contention varies with the number of vertices and the related counter array size, as memory accesses will be accordingly distributed.

In our measurements, we focus on two different parameters: 
first, the total number of threads, 
and second, the amount of touched memory, which is the unique set of shared addresses accessed by any memory operation.
To ensure that thread counts are representative with regard to later use in inference, the thread count used here for experimentation is the total number of threads successively divided by two, so that modeled thread counts are exponentially spaced.

In this regard, with $counter$ being a single counter of the counter array, the amount of touched memory, e.g., the counter array size $M_\text{Counters}$ is:

\begin{equation}
M_\text{Counters} = sizeof(counter) \cdot |V|
\end{equation}

\subsection{Measurements and Derived Heuristics}

\begin{figure}
\centering
\begin{minipage}{.49\textwidth}
		\centering
		{\tiny{\ref{updateTimeInvariance}}\\}
		\begin{tikzpicture}[scale=0.75]
    \begin{axis}[ylabel={Mean update time[ns]},
                 legend style={at={(0.5,1.5)},anchor=north,draw=none,column sep=5pt,legend columns=-1,legend to name=updateTimeInvariance},
                 legend cell align=left,
                 width=1.33*\columnwidth,
                 height=4.4cm,
                 xlabel={Counter array [KB]},
                 xmode = log,
                 ytick pos=left,
                 xtick pos=left,
                 scaled y ticks=false,
                 ymajorgrids,
                 xmajorgrids                 ]
    \addplot[plotstyle1,each nth point=1] table[x=Counter1B,y=1B] {experiments/data/updateTimeInvariance.data};%
    \addplot[plotstyle2,each nth point=1] table[x=Counter2B,y=2B] {experiments/data/updateTimeInvariance.data};
    \addplot[plotstyle3,each nth point=1] table[x=Counter4B,y=4B] {experiments/data/updateTimeInvariance.data};%
    \addplot[plotstyle4,each nth point=1] table[x=Counter8B,y=8B] {experiments/data/updateTimeInvariance.data};%
    \legend{1B,2B,4B,8B}
    \end{axis}
\end{tikzpicture}
		\caption{Mean update time as a function of counter array size ($size$), with different data types for a counter, 
		for a constant thread count $T$ of 28. 
 }
	\label{fig:updateTimeInvariance}
\end{minipage}\hfill
\begin{minipage}{.49\textwidth}
	\centering
	{\tiny{\ref{updateContentionThreadScaling}}\\}
	\begin{tikzpicture}[scale=0.75]
    \begin{axis}[ylabel={Relative Cost},
                 legend style={at={(0.5,1.5)},anchor=north,draw=none,column sep=5pt,legend columns=3,legend to name=updateContentionThreadScaling},
                 legend cell align=left,
                 width=1.33*\columnwidth,
                 height=4.4cm,
                 xlabel={Threads},
                 ymode = log,
                 xmode = log,
                 ytick pos=left,
                 xtick pos=left,
                 scaled y ticks=false,
                 ymajorgrids,
                 xmajorgrids                 ]
    \addplot[plotstyle1,each nth point=1] table[x=Threads,y=SF13] {experiments/data/updateContentionThreadScaling.data};%
\addplot[plotstyle2,each nth point=1] table[x=Threads,y=SF15] {experiments/data/updateContentionThreadScaling.data};%
\addplot[plotstyle3,each nth point=1] table[x=Threads,y=SF16] {experiments/data/updateContentionThreadScaling.data};%
\addplot[plotstyle4,each nth point=1] table[x=Threads,y=SF22] {experiments/data/updateContentionThreadScaling.data};%
\addplot[plotstyle5,each nth point=1] table[x=Threads,y=SF23] {experiments/data/updateContentionThreadScaling.data};%
\addplot[plotstyle6,each nth point=1, mark=none] table[x=Threads,y=Threads] {experiments/data/updateContentionThreadScaling.data};%
    \legend{50KB(L2),188KB(L2),364KB(LLC),19MB(LLC),36MB(MM),sequential cost}
    \end{axis}
\end{tikzpicture}					
	\caption{
	Relative Cost for atomics as a function of thread count and counter array size/number of vertices (28 Cores/56 Threads on 2 Sockets).}
	\label{fig:contention}
\end{minipage}
\end{figure}

Experimental results in \Cref{fig:updateTimeInvariance} show that update time is mainly a function of $M_\text{Counters}$.
This particularly holds true as we vary the data type of a single counter ($sizeof(counter)$).
Therefore, we see no dependency to graph parameters like the number of vertices or edges, which allows us to limit experimentation to counter array size and 
to generalize from this measurement. 
As a result, we obtain a set of measurements describing the mean update time, respectively latency $L(M,T)$, which is a function of memory set size $M$ and number of threads $T$.

An additional observation we make in \Cref{fig:updateTimeInvariance} is that with an increasing $M_\text{Counters}$ the update time decreases as the contention is distributed across more memory locations.
This effect seams to be rather a function of the logarithm of the counter array size than depending linearly on the counter array size.

To derive a suitable heuristic that predicts $L(M,T)$ for a given data set size $M$ and thread count $T$, we first identify bounds for the memory access costs.
As cache levels are discrete, we use the highest memory hierarchy level $l$ that can fit the data set of size $M$, i.e. $l=\min\big\{ x : M_x > M\big\}$, with $M_x$ being the capacity of memory hierarchy level $x$.
To approximate the effective access latency, we use a polynomial interpolation between the cache level $l$ and $u=l-1$, with the rationale that higher cache levels will also observe some cache hits.
While this holds true for $l$ referring to main memory (note that we exclude $M>M_m$, $m$ referring to main memory), for $l=1$ a special case is required.
In this case, the problem fits into L1 cache, thus we set $u=l$, effectively an identical lower and upper bound.

Then, $S(M)$ describes the polynomial interpolation, depending on data set size $M$.
Note that we use the logarithm of the data set size, according to the observation made in \Cref{fig:updateTimeInvariance}.

\begin{equation}
S(M) = \frac { \log(M_{l})  - \log(M) }{
                     \log(M_{l})  - \log(M_{u})  }
\end{equation}

The difference $\delta L(T)$ in between the latency of the two memory hierarchy levels $l$ and $u$, with $u$ derived from $l$ as shown above, is:

\begin{equation}
\delta L(T,l) = L(M_{u}, T) - L(M_{l}, T) 
\end{equation}

Last, the update time prediction $L_\text{predict}(M,T)$ is a function of the access cost of the lower bound $L(M_{l}, T)$, but effectively reduced by $\delta L(T)$ multiplied with the previous interpolation $S(M,T)$ cubed:

\begin{equation}
L_\text{predict} = L(M_{l}, T) - \delta L(T) \cdot S(M)^3 
\label{eq:updateTimeSizeEst}
\end{equation}

While cubed seems non-intuitive, it is empirically derived from experiments on multiple systems and showed the best fit in a couple of different regressions.
We leave a detailed understanding of this behavior for future work, but  observe that the found best fit is inline with our observation that modeling update contention is complicated. 
In this sense, the present fit is no surprise.

Another important factor for the update time prediction are dynamic effects that depend on the number of threads.
For the estimators in \Cref{sec:system}, we need thread-dependent estimations, such as atomic update latency, only for the threads that we measure.
In \Cref{fig:contention} we see that there is a clear dependency between the number of threads and the atomic update time.
Furthermore, we also see is that when limiting the problem to higher cache levels (by adapting problem size), thread count has a much higher impact on atomic update time.
To address this dependency, we estimate the atomic update time using the measured access times with the anticipated number of threads.

\section{Evaluation}
\label{sec:evaluation}

In the following, we evaluate the proposed runtime system based on PageRank (PR) and Breadth-First Search (BFS) algorithms, respectively various variants thereof. 
As a test system, we use a two socket system equipped with Intel Xeon E5-2660 v4 processors, each with 14 cores and Hyper-Threading enabled, 35 MB last-level cache for each socket, and 128 GB DDR4 RAM.

PR has been evaluated as a topology-centric algorithm variant using two different types (push, pull).
In the \textit{push} variant the updates to compute the ranks are pushed to the target vertices, which requires in our parallel implementations for each update an atomic operation.
The \textit{pull} variant gathers data from all source vertices to compute a rank, which does not require atomic operations.
For an analysis of these different types, see \cite{whang2015scalable}.
BFS represents the data-driven algorithm and is implemented in a top-down variant.


Both algorithms, PR and BFS and their variants are in addition realized using different schedulers (sequential, simple parallel, scheduler):
\textit{Sequential} executes the code completely sequential. \textit{Simple parallel} executes the code in parallel using a simple work partitioning that partitions the frontier queue in equal sized packages. The package size is determined by the maximum number of threads and a lower limit.
The \textit{scheduler} variant uses our proposed scheduler.
In particular, sequential variants will serve in concurrent query settings as a baseline, as under high concurrency a per-query sequential processing is usually preferable.

We measure the whole execution of a full PR run after the creation of the adjacency list and its dictionary, from the setup of all supporting data structures until the algorithm converges and the result vector can be returned.
Twenty-four times the number of concurrent sessions repeated full PR runs are executed and measured, and we report mean throughput in Processed Edges per Second (PEPS).  The BFS algorithm has been executed from different start points and measured 50 times the number of concurrent sessions, so we report the throughput derived from the mean of these measurements in Traversed Edges per Second (TEPS).


Furthermore, as the behavior of graph computations is highly data-dependent, we cover synthetic data sets based on RMAT with different scale factors (SF), as well as real-world data sets, such as \textit{soc-Live-Journal1}, \textit{as-skitter}, \textit{roadNet-CA}, \textit{cit-Patents}, \textit{roadNet-PA}, \textit{web-BerkStan}, and \textit{soc-pokec-relationships} from the Stanford Large Network Dataset Collection~\cite{snapnets}.

\subsection{Evaluation of single query performance to identify scheduling overhead}

\begin{figure}
\centering
\begin{minipage}{.49\textwidth}
	{\tiny{\ref{throughputPrRmatScaling}}\\}
	\begin{tikzpicture}[scale=0.75]
    \begin{axis}[ylabel={MPEPS},
                 legend style={at={(0.5,0.0)},anchor=north,draw=none,column sep=5pt,legend columns=2,legend to name=throughputPrRmatScaling},
                 legend cell align=left,
                 width=1.33*\columnwidth,
                 height=4.3cm,
                 xlabel={Scale Factor},
                 ytick pos=left,
                 xtick pos=left,
                 scaled y ticks=false,
                 ymode = log,
                 ymajorgrids,
                 xmajorgrids                 ]
    \addplot[plotstyle1,each nth point=1] table[x=Scale,y=SequentialPush] {experiments/data/pr-rmat-scaling.data};%
    \addplot[plotstyle3,each nth point=1] table[x=Scale,y=ParallelBufferPush] {experiments/data/pr-rmat-scaling.data};
    \addplot[plotstyle5,each nth point=1] table[x=Scale,y=SequentialPull] {experiments/data/pr-rmat-scaling.data};%
    \addplot[plotstyle6,each nth point=1] table[x=Scale,y=ParallelPull] {experiments/data/pr-rmat-scaling.data};%
    \legend{sequential push, scheduler push, sequential pull, scheduler pull }
    \end{axis}
\end{tikzpicture}
	\caption{Performance scaling in MPEPS (millions of processed edges per second) for different PR implementations across RMAT graphs of different sizes}
	\label{fig:prRmat}
\end{minipage}\hfill
\begin{minipage}{.49\textwidth}
	{\tiny{\ref{throughputPrRealData}}\\}
	\begin{tikzpicture}[scale=0.75]
    \begin{axis}[ybar,
    		      bar width=4pt,
                 ylabel={MPEPS},
                 legend style={at={(0.5,0.0)},anchor=north,draw=none,column sep=5pt,legend columns=3,legend to name=throughputPrRealData},
                 legend cell align=left,
                 width=1.33*\columnwidth,
                 height=4.3cm,
                 ytick pos=left,
                 xtick pos=left,
                 xtick=data,
                 xticklabels from table={experiments/data/pr-real.data}{Scale},
                 x tick label style={rotate=45,align=center},
                 scaled y ticks=false,
                 ymode = log,
                 ymajorgrids,
                 xmajorgrids                 ]
    \addplot[draw=none, fill=blue!30] table[x expr=\coordindex,y=SequentialPush] {experiments/data/pr-real.data};%
 \addplot[draw=none, fill=blue!65] table[x expr=\coordindex,y=ParallelBufferPushNoOpt] {experiments/data/pr-real.data};
    \addplot[draw=none, fill=blue!100] table[x expr=\coordindex,y=ParallelBufferPushOpt] {experiments/data/pr-real.data};
    \addplot[draw=none, fill=gray!30] table[x expr=\coordindex,y=SequentialPull] {experiments/data/pr-real.data};%
    \addplot[draw=none, fill=gray!65] table[x expr=\coordindex,y=ParallelPullNoOpt] {experiments/data/pr-real.data};%
    \addplot[draw=none, fill=gray!100] table[x expr=\coordindex,y=ParallelPullOpt] {experiments/data/pr-real.data};%
    \legend{sequential push, simple push, scheduler push, sequential pull, simple pull, scheduler pull}
    \end{axis}
\end{tikzpicture}
    \vspace*{-1cm}
	\caption{Performance of different PR implementations on different types of graphs}
	\label{fig:prFiles}
\end{minipage}
\end{figure}

While the intent of the proposed scheduling is to optimize concurrent query execution, we focus in this subsection on single query performance mainly for overhead analysis.
In \Cref{fig:prRmat} we report performance scaling for different PR implementations across RMAT graphs of different scale factors.
We observe that single-query PR performance depends on problem size and algorithm, although pull versions seem to be preferred.
Principally, push and pull algorithms have completely different properties, therefore also differ in their cost models.
Furthermore, we observe that sequential processing is faster for small problem sizes ($SF<=14$ for push and $SF<=18$ for pull), otherwise parallel processing is preferable.

We can assess the overhead of the proposed scheduling method by comparing it to \textit{sequential} to avoid an unrealistic perfect selection of parameter for \textit{simple} parallel.
In this regard, we observe that for the pull-based algorithms, the scheduler makes always the right choice to execute the code in parallel, which prohibits an analysis of the overhead.
Comparing \textit{scheduler push} with sequential push, we see that for a long time the scheduler variant behaves like the sequential variant with only a small (3\% at SF17 - 30\% at SF12) reduction in performance.
Thus, in spite of fundamental differences between push and pull variants, in particular with regard to the use of atomics, the scheduler chooses the right execution strategy, so that the overhead never dominates the elapsed time.


Furthermore, we evaluate the performance of different PR implementations on different types of graphs in Figure~\ref{fig:prFiles}.
We observe that for different data sets and single query PR, the difference between \textit{simple push} and its scheduling-optimized counterpart \textit{scheduler (push)} is negligible, with either one having only small advantages compared to the other, depending on the data set.
The difference for pull variants behaves similarly, even though here for one data set there is a substantial advantage for the scheduler variant.
Thus, for PR it is apparent that our scheduler is stable across different data sets, and that the scheduling overhead is small.

\begin{figure}
\centering
\begin{minipage}{.49\textwidth}
\centering
    {\tiny{\ref{throughputBfsRmatScaling}}\\}
	\begin{tikzpicture}[scale=0.75]
    \begin{axis}[ylabel={MTEPS},
                 legend style={at={(0.5,1.5)},anchor=north,draw=none,column sep=5pt,legend columns=-1,legend to name=throughputBfsRmatScaling},
                 legend cell align=left,
                 width=1.33*\columnwidth,
                 height=4.3cm,
                 xlabel={Scale Factor},
                 ytick pos=left,
                 xtick pos=left,
                 scaled y ticks=false,
                 ymode = log,
                 ymajorgrids,
                 xmajorgrids                 ]
    \addplot[plotstyle1,each nth point=1] table[x=Scale,y=Sequential] {experiments/data/bfs-rmat-scaling.data};
    \addplot[plotstyle5,each nth point=1] table[x=Scale,y=ParallelBuffer] {experiments/data/bfs-rmat-scaling.data};%
    \addplot[plotstyle6,each nth point=1] table[x=Scale,y=ParallelOpt] {experiments/data/bfs-rmat-scaling.data};%
    \legend{sequential, simple, scheduler}
    \end{axis}
\end{tikzpicture}
	\caption{Performance scaling for different BFS implementations across RMAT graphs of different sizes}
	\label{fig:bfsRmat}
\end{minipage}\hfill
\begin{minipage}{.49\textwidth}
\centering
	{\tiny{\ref{throughputBfsRealData}}\\}
	\begin{tikzpicture}[scale=0.75]
    \begin{axis}[ybar,
    			  bar width=7pt,
                 ylabel={MTEPS},
                 legend style={at={(0.5,1.5)},anchor=north,draw=none,column sep=5pt,legend columns=-1,legend to name=throughputBfsRealData},
                 legend cell align=left,
                 width=1.33*\columnwidth,
                 height=4.3cm,
                 ytick pos=left,
                 xtick pos=left,
                 xtick=data,
                 xticklabels from table={experiments/data/bfs-real.data}{Scale},
                 x tick label style={rotate=45,align=center},
                 scaled y ticks=false,
                 ymode = log,
                 ymajorgrids,
                 xmajorgrids                 ]
    \addplot[draw=none, fill=blue!30] table[x expr=\coordindex,y=Sequential] {experiments/data/bfs-real.data};%
    \addplot[draw=none, fill=blue!50] table[x expr=\coordindex,y=ParallelBuffer] {experiments/data/bfs-real.data};%
    \addplot[draw=none, fill=blue!70] table[x expr=\coordindex,y=ParallelOpt] {experiments/data/bfs-real.data};%
    \legend{sequential, simple, scheduler}
    \end{axis}
\end{tikzpicture}
    \vspace*{-1cm}
	\caption{Performance of different BFS implementations on different types of graphs}
	\label{fig:bfsFiles}
\end{minipage}
\end{figure}

Regarding Breadth-First Search (BFS), results for RMAT graphs of different scale factors are reported in Figure~\ref{fig:bfsRmat}.
Similar to PR, data set size determines algorithm choice: for $SF<=16$, \textit{sequential} processing is fastest, otherwise \textit{simple} (parallel) and \textit{scheduler} are almost equally faster.
As a reminder, similar to PR, \textit{simple} is a straight-forward range partitioning of the frontier queue, while \textit{scheduler} is based on our proposed scheduler.
This experiment particularly highlights that these scheduling optimizations come with virtually negligible overhead in comparison to simple parallelization.
Only for small data sets, \textit{sequential} query processing is faster, still our scheduler is performing better than \textit{simple} in this regime.
One reason is that our scheduler reduce false invalidation as they reduce concurrent writes to the same memory locations.

We also assess BFS performance on real-world data sets, as reported in \Cref{fig:bfsFiles}.
For single query execution, BFS performance is highly data-dependent.
In particular, \textit{scheduler} can be slower than \textit{simple}, which is based on naive parallelization, albeit with only small differences and almost only if the difference to \textit{sequential} is substantial.
Presumably, scheduling overhead is virtually constant while BFS processing time is highly dependent on the data set.
Thus, for a small BFS execution time, scheduling overhead can become more dominant.
Furthermore, results suggest that data sets resulting in low performance (roadNet-CA,-PA) prefer \textit{scheduler} over \textit{simple}, probably due to complex dependencies (edges) and little parallelization opportunities.
As a result, \textit{scheduler} is always close to the best algorithm, independent of particular data set characteristics.

In summary, we observe that scheduling overhead seems to be small to negligible, and that sometimes the scheduling optimizations are actually even beneficial for overall performance.
As expected, graph computations are highly dependent on algorithm and data, thus performance results, both in absolute and relative terms, are also highly dependent on this tuple.
In this regard, it is noteworthy to point out that performance of our scheduler is stable across the various evaluated tuples of algorithm variant and data set.

\subsection{Concurrent query performance on synthetic RMAT data}

\begin{figure*}
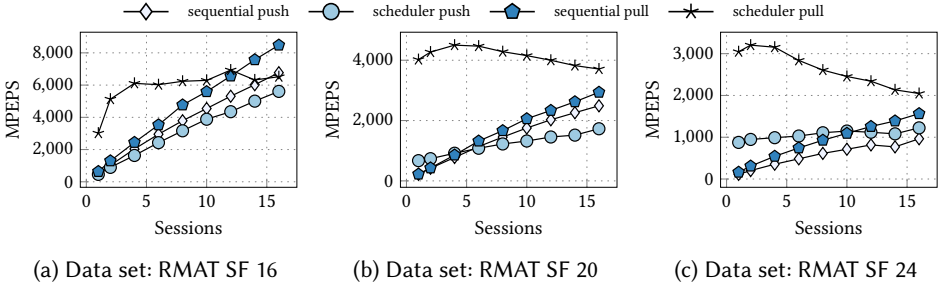
	
\centering
{\tiny{\ref{throughputPrReal}}\\}
\begin{subfigure}{0.3\textwidth}
	\PrPerfFigures{pr-rmat-sf16.data}
	\caption{Data set: RMAT SF 16}
	\label{fig:concurrentPrRmatSf16}
\end{subfigure}
\begin{subfigure}{0.3\textwidth}
	\PrPerfFigures{pr-rmat-sf20.data}
	\caption{Data set: RMAT SF 20}
	\label{fig:concurrentPrRmatSf20}
\end{subfigure}
\begin{subfigure}{0.3\textwidth}
	\PrPerfFigures{pr-rmat-sf24.data}
	\caption{Data set: RMAT SF 24}
	\label{fig:concurrentPrRmatSf24}
\end{subfigure}
\caption{Performance scaling across multiple sessions for different PR algorithms on RMAT}
\label{fig:concurrentPrRmat}
\end{figure*}	

We continue by evaluating the performance scaling across multiple sessions for different PR algorithms, as shown in \Cref{fig:concurrentPrRmat}.
Different scale factors are covered by different subfigures.
We observe that \textit{scheduler pull} is by far fastest, although the advantage depends on data set size and concurrency (number of sessions). 
Only for a small data set size (SF 16) and a high number of sessions, \textit{sequential pull} is faster as these data sets provide very little opportunities for parallelism. 
Furthermore, results show that the break-even point in between these two moves to larger amounts of concurrency with increasing data set size. 
With increasing data set size the still existing overhead of parallel execution becomes more negligible. 

\begin{figure*}
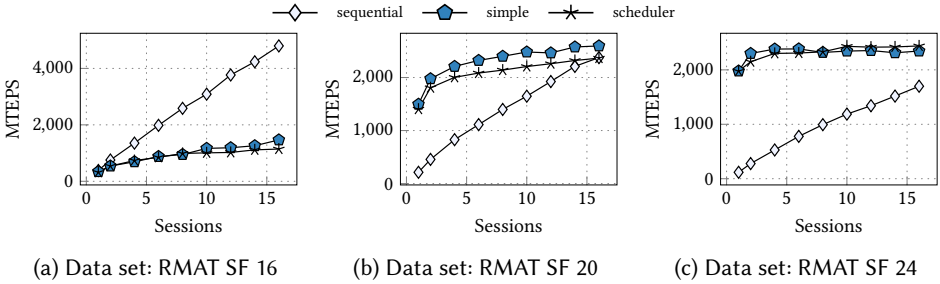
	
\centering
{\tiny{\ref{throughputBfsReal}}\\}
\begin{subfigure}{0.3\textwidth}
    \BfsPerfFigures{bfs-rmat-sf16.data}
	\caption{Data set: RMAT SF 16}
	\label{fig:concurrentBfsRmatSf16}
\end{subfigure}
\begin{subfigure}{0.3\textwidth}
	\BfsPerfFigures{bfs-rmat-sf20.data}
	\caption{Data set: RMAT SF 20}
	\label{fig:concurrentBfsRmatSf20}
\end{subfigure}
\begin{subfigure}{0.3\textwidth}
	\BfsPerfFigures{bfs-rmat-sf24.data}
	\caption{Data set: RMAT SF 24}
	\label{fig:concurrentBfsRmatSf24}
\end{subfigure}
\caption{Performance scaling across multiple sessions for different BFS algorithms on RMAT}
\label{fig:concurrentBfsRmat}
\end{figure*}	

\Cref{fig:concurrentBfsRmat} reports results from similar experiments for BFS on RMAT data.
For small data set sizes, \textit{sequential} processing is by far the fastest, in particular with growing concurrency. As it can be seen in \cref{fig:bfsRmat} our scheduler chooses for the small data set parallel execution as it underestimates the contention.
While for medium sized data sets \textit{simple} parallel is slightly fastest, with growing data set size our \textit{scheduler} becomes advantageous.
This is mainly in line with expectations, as too small data set sizes result in parallel processing being rather inefficient.

\subsection{Concurrent query performance on real-world data}

\begin{figure*}
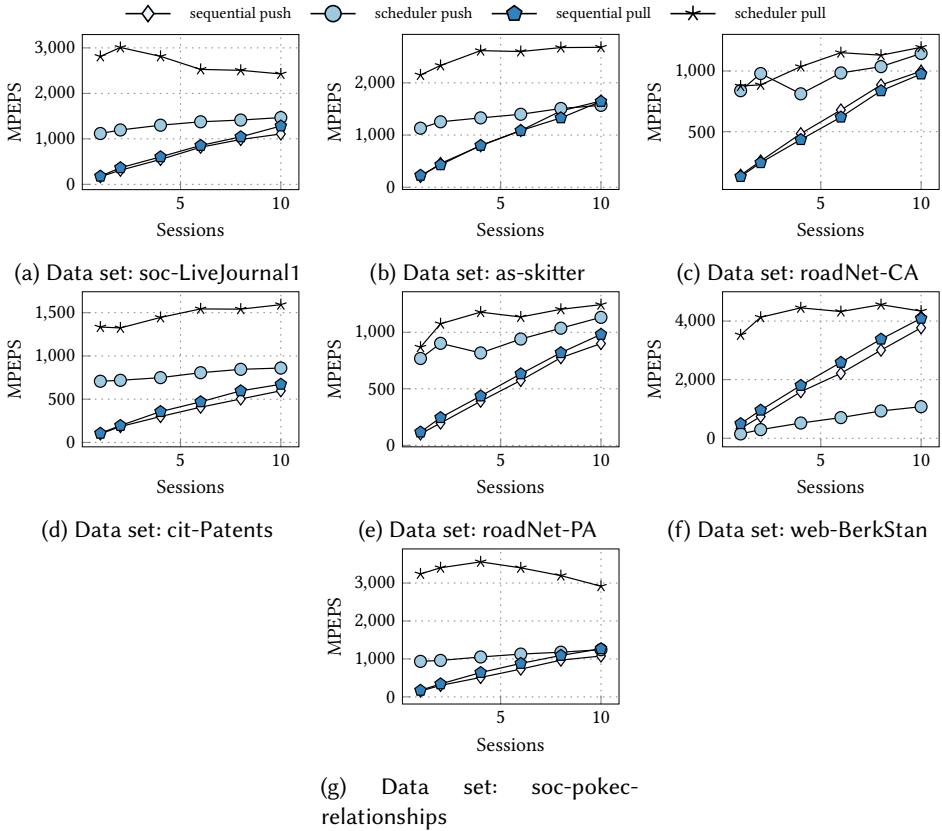

\centering
{\tiny{\ref{throughputPrReal}}\\}
\begin{subfigure}{0.3\textwidth}
	\PrPerfFigures{pr-soc-LiveJournal1.data}
	\caption{Data set: soc-LiveJournal1}
	\label{fig:concurrentPr-soc-LiveJournal1}
\end{subfigure}
\begin{subfigure}{0.3\textwidth}
	\PrPerfFigures{pr-as-skitter.data}
	\caption{Data set: as-skitter}
	\label{fig:concurrentPr-as-skitter}
\end{subfigure}
\begin{subfigure}{0.3\textwidth}
	\PrPerfFigures{pr-roadNet-CA.data}
	\caption{Data set: roadNet-CA}
	\label{fig:concurrentPr-roadNet-CA}
\end{subfigure}
\begin{subfigure}{0.3\textwidth}
	\PrPerfFigures{pr-cit-Patents.data}
	\caption{Data set: cit-Patents}
	\label{fig:concurrentPr-cit-Patents}
\end{subfigure}
\begin{subfigure}{0.3\textwidth}
	\PrPerfFigures{pr-roadNet-PA.data}
	\caption{Data set: roadNet-PA}
	\label{fig:concurrentPr-roadNet-PA}
\end{subfigure}
\begin{subfigure}{0.3\textwidth}
	\PrPerfFigures{pr-web-BerkStan.data}
	\caption{Data set: web-BerkStan}
	\label{fig:concurrentPr-web-BerkStan}
\end{subfigure}
\begin{subfigure}{0.3\textwidth}
	\PrPerfFigures{pr-soc-pokec-relationships.data}
	\caption{Data set: soc-pokec-relationships}
	\label{fig:concurrentPr-soc-pokec-relationships}
\end{subfigure}
\caption{Performance scaling across multiple sessions for different PR implementations}
\label{fig:concurrentPr}
\end{figure*}

We complete our experiments by evaluating concurrent queries on data, which is probably most representative for real-world applications.
We start with PR as reported in \Cref{fig:concurrentPr}.
Overall, it is apparent that \textit{scheduler pull} is dominantly fastest, independent of data set and concurrency. 
While its performance is almost constant with regard to concurrency, performance of alternatives (in particular \textit{sequential push} and \textit{sequential pull}) is usually scaling linearly with concurrency. 
This suggests that for some data sets, in particular roadNet-CA/-PA and web-BerkStan, sequential alternatives can reach the performance of the scheduler implementation given sufficient concurrency. 
However, for other data sets, such as cit-Patents or soc-pokec-relationships, the performance increase by concurrency for sequential seems to be too low to reach the scheduler implementation.
A reason for this behavior is the different internal level of parallelism that these data sets allow.

\begin{figure*}[t]
\centering
{\tiny{\ref{throughputBfsReal}}\\}
\begin{subfigure}{0.3\textwidth}
	\BfsPerfFigures{bfs-soc-LiveJournal1.data}
	\caption{Data set: soc-LiveJournal1}
	\label{fig:concurrentBfs-soc-LiveJournal1}
\end{subfigure}
\begin{subfigure}{0.3\textwidth}
	\BfsPerfFigures{bfs-as-skitter.data}
	\caption{Data set: as-skitter}
	\label{fig:concurrentBfs-as-skitter}
\end{subfigure}
\begin{subfigure}{0.3\textwidth}
	\BfsPerfFigures{bfs-roadNet-CA.data}
	\caption{Data set: roadNet-CA}
	\label{fig:concurrentBfs-roadNet-CA}
\end{subfigure}
\begin{subfigure}{0.3\textwidth}
	\BfsPerfFigures{bfs-cit-Patents.data}
	\caption{Data set: cit-Patents}
	\label{fig:concurrentBfs-cit-Patents}
\end{subfigure}
\begin{subfigure}{0.3\textwidth}
	\BfsPerfFigures{bfs-roadNet-PA.data}
	\caption{Data set: roadNet-PA}
	\label{fig:concurrentBfs-roadNet-PA}
\end{subfigure}
\begin{subfigure}{0.3\textwidth}
	\BfsPerfFigures{bfs-web-BerkStan.data}
	\caption{Data set: web-BerkStan}
	\label{fig:concurrentBfs-web-BerkStan}
\end{subfigure}
\begin{subfigure}{0.3\textwidth}
	\BfsPerfFigures{bfs-soc-pokec-relationships.data}
	\caption{Data set: soc-pokec-relationships}
	\label{fig:concurrentBfs-soc-pokec-relationships}
\end{subfigure}
\caption{Performance scaling across multiple sessions for different BFS implementations}
\label{fig:concurrentBfs}
\end{figure*}

Performance of BFS on real-world data under concurrent execution is reported in Figure~\ref{fig:concurrentBfs}.
Algorithm performance is highly dependent on data set and concurrency, basically forming four patterns:

\begin{enumerate}
\item 	
For \textit{soc-LiveJournal1} and \textit{soc-pokec-relationships},
\textit{simple} parallel and our \textit{scheduler} are almost equally fast, with \textit{sequential} substantially slower. 
However, the difference decreases with increasing concurrency, suggesting that concurrency higher than evaluated might result in a break-even point.

\item 	
For \textit{as-skitter}, \textit{cit-Patents} and \textit{web-BerkStan}, 
performance of the three alternatives is almost equal but with high variance depending on concurrency. 
If \textit{sequential} is much slower, the difference will decrease with growing concurrency, sometimes even outperforming the other algorithms.

\item 	
With regard to the previous pattern, we have to report that \textit{as-skitter} is a rather special case as the overall speed is very high. 
As a result, the overhead is of particular importance.

\item 	
For \textit{roadNet-CA} and \textit{roadNet-PA},
\textit{sequential} and our \textit{scheduler} are almost equally fast, with small advantages for \textit{sequential}, and scaling is almost linear with concurrency. 
\textit{Simple} parallel has a constant performance, resulting in a growing gap with increasing concurrency. 

\end{enumerate}

As a result, we observe that for both PR and BFS the use of our scheduler result in executions that are most efficient, being close to the best performing alternative or even ahead of it.
In detail, for PR, \textit{scheduler pull} or \textit{scheduler push} are substantially faster than alternatives,
while for BFS, our \textit{scheduler} is close or ahead of \textit{sequential} or \textit{simple} parallel performance.


\section{Conclusion}
\label{sec:conclusion}

Considering scalable computing systems, both in the cloud and on-premise, there is a strong trend towards consolidation of resources.
As a result, solutions that are too specialized for a particular task or situation are no longer appropriate, as the overall workload characteristics change to being highly variant in terms of concurrency, workload type, and data set.

In this work, we address some of these needs by presenting a system that focuses on the overall throughput especially when there are multiple active queries.
It increases the overall throughput by controlling the degree of parallelism in each query to reduce the overall synchronization overhead.


The system is split in two parts: a proactive preparation step and a reactive runtime component.
On one side the preparation step estimates the upcoming work for the next iteration and prepares the work in package form for the runtime.
The estimates are based on multiple sources that include information about the system and the algorithm, but also information that is gathered from the processed data.
On the other side the runtime uses the prepared information and combines them with runtime information like the available threads.
Together it uses them to execute the graph algorithm sequential or parallel in an efficient way.

In our experiments, we showed that our approach provides a robust performance for different types of graphs (social network, web graphs road network) and algorithm (BFS, push and pull PR) without the need for manual adjustments of runtime parameters.
The system chooses depending on the circumstances the most effective way to execute a graph algorithm, so that achieved performance is near the optimum.
For multi-query execution we showed that the system is able to efficiently handle multiple queries at the same time. 
For queries with high internal parallelism it keeps the performance constant, while it achieves performance gains with an increasing number of queries when the internal parallelism is low.


\begin{acks}
We would like to thank our colleagues Marcus Paradies and Romans Kasperovics for their support.
\end{acks}


\bibliographystyle{ACM-Reference-Format}
\bibliography{bib/bibliography} 





\end{document}